\documentclass[a4paper,12pt]{report}

\usepackage{lscape}		
\usepackage{geometry}	

\usepackage{cmap}						
\usepackage{fontenc}				
\usepackage[utf8]{inputenc}				
\usepackage[english, russian]{babel}	
\usepackage{mathtools}

\usepackage{amsthm,amsfonts,amsmath,amssymb,amscd} 
\usepackage{amsmath}

\usepackage{indentfirst} 

\usepackage[usenames]{color}
\usepackage{color}
\usepackage{colortbl}

\usepackage{longtable}					
\usepackage{multirow,makecell,array}	

\usepackage[singlelinecheck=on]{caption}	
\usepackage{soul}									

\usepackage{cite} 

\usepackage[linktocpage=true,plainpages=false,pdfpagelabels=false]{hyperref}

\usepackage{graphicx} 
\usepackage{caption}
\usepackage{floatrow}

\usepackage{tocloft}
\makeatletter
\renewcommand{\@biblabel}[1]{#1.}	
\makeatother

\graphicspath{{images/}} 

\definecolor{linkcolor}{rgb}{0.9,0,0}
\definecolor{citecolor}{rgb}{0,0.6,0}
\definecolor{urlcolor}{rgb}{0,0,1}
\hypersetup{
    colorlinks, linkcolor={linkcolor},
    citecolor={citecolor}, urlcolor={urlcolor}
}

\begin{document}

\renewcommand{\abstractname}{Аннотация}
\renewcommand{\alsoname}{см. также}
\renewcommand{\appendixname}{Приложение}
\renewcommand{\bibname}{Литература}
\renewcommand{\ccname}{исх.}
\renewcommand{\chaptername}{Глава}
\renewcommand{\contentsname}{Содержание}
\renewcommand{\enclname}{вкл.}
\renewcommand{\figurename}{Рисунок}
\renewcommand{\headtoname}{вх.}
\renewcommand{\indexname}{Предметный указатель}
\renewcommand{\listfigurename}{Список рисунков}
\renewcommand{\listtablename}{Список таблиц}
\renewcommand{\pagename}{Стр.}
\renewcommand{\partname}{Часть}
\renewcommand{\refname}{Список литературы}
\renewcommand{\seename}{см.}
\renewcommand{\tablename}{Таблица}
\renewcommand{\floatpagefraction}{.8}%

\newcommand{\bra}[1]{\left\langle #1\right|}
\newcommand{\ket}[1]{\left| #1\right\rangle}
\newcommand{\braket}[2]{\left\langle
#1\vphantom{#2}\right|\left.#2\vphantom{#1}\right\rangle}
\newcommand{\ketbra}[2]{\left| #1\right\rangle\!\left\langle#2\right|}
\newcommand{\avg}[1]{\left\langle #1\right\rangle}
\newcommand{\be}[0]{\begin{equation}}
\newcommand{\ee}[0]{\end{equation}}
\newcommand{\intinf}[0]{\int_{-\infty}^{+\infty}}
\newcommand{\ada}[0]{\hat a^\dagger\hat a}
\newcommand{\ad}[0]{\hat a^\dagger}
\newcommand{\ah}[0]{\hat a}
\newcommand{\hv}[1]{\hat{\vec{#1}}}
\newcommand{\tr}[0]{{\rm Tr}}
\newcommand{\re}[0]{{\rm Re}\,}
\newcommand{\im}[0]{{\rm Im}\,}

\newcommand{\eeqref}[1]{ур.~(\ref{#1})}
\newcommand{\figref}[1]{рис.~(\ref{#1})}			

\thispagestyle{empty}

\begin{center}
ФЕДЕРАЛЬНОЕ ГОСУДАРСТВЕННОЕ БЮДЖЕТНОЕ УЧРЕЖДЕНИЕ НАУКИ
ФИЗИЧЕСКИЙ ИНСТИТУТ ИМ. П.Н.ЛЕБЕДЕВА\\
РОССИЙСКОЙ АКАДЕМИИ НАУК\par 
\par
\end{center}

\vspace{20mm}
\begin{flushright}
На правах рукописи

{\sl УДК 535.1}
\end{flushright}

\vspace{30mm}
\begin{center}
{\large ФЁДОРОВ ИЛЬЯ АЛЕКСЕЕВИЧ}
\end{center}

\vspace{5mm}
\begin{center}
{\bf \large НЕЛОКАЛЬНОЕ УПРАВЛЕНИЕ КВАНТОВЫМ СОСТОЯНИЕМ СВЕТА
\par}

\vspace{10mm}
{
Специальность 01.04.05~---

ОПТИКА
}

\vspace{10mm}
Диссертация на соискание учёной степени \\
кандидата физико-математических наук
\end{center}

\vspace{30mm}
\begin{flushright}

Научный руководитель: \\
Кандидат физ.-мат. наук \\
Львовский А.И. \\

\vspace{5mm}

Научный консультант: \\
Доктор физ.-мат. наук, профессор \\
Масалов А.В. \\

\end{flushright}

\vspace{20mm}
\begin{center}
{Москва, 2017}
\end{center}

\newpage			
\tableofcontents
\clearpage
\chapter*{Введение}							
\label{introduction}
\addcontentsline{toc}{chapter}{Введение}	

Диссертация посвящена решению задачи управления состоянием объекта с помощью воздействия на другой -- управляющий -- объект, находящийся в квантово-запутанном состоянии с первым. Воздействие на управляющий объект осуществляется с помощью измерений, имеющих частичный характер и не приводящих к разрушению квантового состояния системы. Разобщенность объектов в пространстве обеспечивает \textit{нелокальный} характер управления. В качестве объектов управления служат световые пучки.

\section{Общая картина}

Квантовая физика -- раздел естествознания, имеющий дело с физическими явлениями, которые не могут быть описаны в рамках механики Ньютона \cite{Newton}, электродинамики Максвелла \cite{Maxwell} и Больцмановской термодинамики \cite{Boltzman} -- вместе с теорией относительности Эйнштейна \cite{OTO} составляющих так называемую ``классическую'' картину физического мира.
Первыми среди таких явлений привлекли к себе внимание научного сообщества фотоэффект \cite{Photoeffect}, поведение спектра теплового излучения \cite{Ultraviolet}, а также феномен атомной стабильности \cite{Exclusion}.

Формирование квантовой физики завершилось в конце 1920-х годов \cite{Schrodinger1935, Bohr}. Последовавшие результаты -- появление ядерного оружия (1945 г.), транзистора (1947 г.) и лазера (1954 г.) -- во многом определили дальнейший ход развития техносферы и человеческой цивилизации в целом. При том, что во всех трёх случаях рабочий процесс -- деление атомного ядра, поведение электронов в полупроводнике и вынужденное излучение света -- является существенно неклассическим, используемый эффект определяется коллективным поведением огромного числа частиц. 
Современник этих событий, Э. Шрёдингер предложил следующую точку зрения на возможность управления индивидуальными физическими системами в микромире: ``\textit{Мы никогда не имеем дела с единственным электроном или молекулой. В мысленных экспериментах, иногда, мы делаем так; это неизбежно приводит к нелепым следствиям... мы не можем экспериментировать с отдельными частицами, как не можем вырастить динозавра в зоопарке}'' \cite{Schrodinger1952}.
Это мнение, хотя и в отсутствие доказательств, утвердилось в научных кругах; в результате, применимость концепций квантовой механики в течение десятилетий была ограничена областью мысленных экспериментов. 

Конец 20 века ознаменовался преодолением этого барьера. Необходимые условия были подготовлены развитием лазерной техники и средств детектирования, методов изготовления, охлаждения и контроля микроскопических объектов, успехами в материаловедении. 
В результате, для физического исследования открылся такой уровень природы, на котором проявлена его фундаментальная многовариантность. Явления природы на этом уровне не укладываются в классическую, механистически-предопределённую картину мироздания.  

Переход с ``классического'' на ``квантовый'' уровень явлений можно проиллюстрировать на примере деления пополам светового луча на полупропускающем зеркале. Получившиеся лучи направлены на детекторы, которые показывают одинаковую интенсивность. Однако, при уменьшении яркости исходного луча, детекторы начинают вести себя иначе: в каждый момент может сработать только один из детекторов, тогда как второй показывает отсутствие сигнала. Таким образом, поглощение света чувствительным материалом детекторов возможно лишь отдельными, неделимыми порциями - ``фотонами'', или ``квантами света'' (лат. \textit{quantum} - сколько); предсказать, какой из детекторов должен сработать, математически невозможно. Подобные дискретные свойства были обнаружены во многих физических системах \cite{Dowling2003, Haroche2012}; термин ``квантовый'' закрепился за всеми физическими явлениями, в которых такая неделимость -- индивидуальность -- играет существенную роль.

Если в описанной системе убрать детекторы и совместить разделённые световые пучки на другом светоделителе, то в квантовом режиме интенсивности обнаружится двойственная картина: одиночный фотон исходного пучка, учитывая доступность двух различных путей, ``интерферирует сам с собой'', проявляя свойства волны. Однако, как только пучок попадает на детектор, опять проявляются дискретные (корпускулярные) свойства. Как справедливо отметил Р. Фейнман, физику этой системы ``невозможно, совершенно, абсолютно невозможно объяснить классическим образом. В этом явлении таится самая суть квантовой механики'' \cite{Feynman1965}.

Остаётся признать, что законы микромира разительно отличаются от законов механики Ньютона. Но каким образом квантовый микромир порождает на привычные нам законы классической физики? И действительно ли на разных масштабах пространства и времени действуют разные законы природы? Точки зрения на эти фундаментальные вопросы в научном сообществе крайне разобщены. 
Не имея возможности осветить их в настоящей работе сколько-нибудь полно и детально, привожу ряд источников, которые -- на мой взгляд -- могут послужить подспорьем для конструктивного исследования названных вопросов в будущем \cite{WheelerZurek1983, Bohm1980, Wheeler1988, Wheeler1989, GellMannHartle, BohmHiley1993, Zurek2002, Kofler2007}.

\section{Актуальность работы}

Сегодня, многие из концепций квантовой физики стали доступны для экспериментального исследования: квантовая запутанность и нелокальные состояния \cite{Freedman1972, Aspect1982}, телепортация \cite{Bouwmeester1997}, квантовые вычисления \cite{Gottesman99}, и даже состояние ``одновременно живой и мёртвой'' Шрёдингеровкой кошки \cite{Arndt2014}. 
Мы наблюдаем переход человечества на очередную ступень технологического развития. Создаются инструменты для инженерии физической реальности на немыслимом до сих пор уровне: хорошо известные физические системы приготавливаются в новых состояниях и демонстрируют невиданные ранее свойства.
Из многообразия этих новых -- квантовых -- состояний, классическая физика охватывает лишь безконечно малую часть. ``Там, в глубине, достаточно места'' \cite{Feynman1959} -- Фейнман указал тренд будущего науки верно.

Основополагающая категория квантовой физики -- физическая система с двумя возможными состояниями -- в информационном смысле является альтернативой бита, имеющей преимущества в области вычислений и передачи данных. Эти преимущества настолько значительны, что привели к рождению новой области знаний -- квантовой информатики \cite{Nielsen2000}, в которой и лежит прикладное значение б\'{о}льшей части квантовых исследований.

Имея ввиду последствия создания транзистора (на момент написания этого текста, человечество ещё не выработало культуры безопасного обращения с ними), значимость новой информационной революции оценить сложно. В обозримой перспективе находится создание квантового компьютера -- машины, которая решает вычислительные задачи, обрабатывая информацию в форме состояний квантовых систем \cite{Deutsch1985, Gottesman99, Lloyd2013}. Создание такого устройства означало бы обесценивание действующей системы защиты информации, основанной на вычислительной сложности задачи о факторизации больших чисел: квантовый компьютер позволит эффективно решать эту задачу.
Одновременно с этим идёт развитие технологий квантовой криптографии, которые призваны информационную безопасность восстановить, обеспечив её фундаментальными законами природы \cite{Bennett84, Gisin2016} \ldots

Свет -- видимое электромагнитное излучение -- является одной из немногих физических систем, имеющих шансы стать основой квантовых технологий второго поколения -- технологий, основанных на квантовых свойствах индивидуальных физических систем \cite{Dowling2003, Pritchard}. Квант света -- фотон -- является единственным носителем квантовой информации, способным без разрушения перемещаться на большие расстояния; это свойство света гарантирует ему ключевое место в технологиях квантовой информатики будущего \cite{OBrien2009}.

Для того, чтобы квантовый свет мог быть использован в полном объёме своих возможностей, должны быть найдены решения ряда задач по приготовлению, управлению, хранению и характеризации квантовых состояний света.
Большая часть препятствий, возникающих при их решении, являются оборотной стороной преимуществ света. Например, световая скорость позволяет с лёгкостью перемещать квантовую информацию в пространстве, и в то же время делает сложной задачу её хранения; слабость взаимодействия с окружением снижает как шумовые искажения сигнала, так и возможности управления квантовыми состояниями света.

Задача управления квантовым состоянием света является особенно сложной. Так, прямое преобразование когерентного состояния света в однофотонное требует создания среды, обладающей значительной нелинейностью на однофотонном уровне интенсивности сигнала \cite{Peyronel2012}. Непосредственная реализация однофотонных логических элементов также невозможно без вовлечения квантовой нелинейной оптики \cite{Chang2014}.
Низкая эффективность этих методов есть следствие переноса квантовой информации в атомную систему и обратно; в результате, наследуется чувствительность последней к шумам окружения. Помимо этого, приготовление экзотических систем, обладающих требуемыми нелинейными свойствами, само по себе является сложнейшей экспериментальной задачей \cite{Peyronel2012, Chang2014, Turchette1995, Firstenberg2013}.

Неклассическим подходом к управлению квантовым состоянием света является использование нелокального взаимодействия запутанных физических систем. Оказывается, что для воздействия на систему А, вовсе не обязательно осуществлять манипуляции непосредственно с ней; если приготовить А в запутанном состоянии со вспомогательной системой Б, то действие над Б приведёт к изменению состояния А. В оптике, наиболее распространённым инструментом такого воздействия является сброс вспомогательной моды в фоковское или квадратурное состояние, осуществляемый с помощью однофотонного или гомодинного измерения -- сравнительно простых экспериментальных процедур.
В основе этого ``действия на расстоянии'' лежит феномен квантовой нелокальности -- возможности мгновенного взаимодействия удалённых физических систем \cite{EPR1935, Bohr}. Практическую мощь описанного метода управления квантовыми состояниями света иллюстрирует тот факт, что используя лишь линейные оптические элементы, а также нелокальное взаимодействие распределённых состояний света, можно построить полноценную квантовую вычислительную машину \cite{Knill2001, Kok07, Ralph10}.

В настоящее время, метод нелокальной проекции лежит в основе многих протоколов инженерии квантовых состояний света; именно этот первый этап является ключевым ноу-хау многих квантово-оптических технологий. Из элементарной базы легко доступных сжатых состояний света, методы нелокальной инженерии сегодня позволяют получать состояния с определённым числом фотонов \cite{Lvovsky2001, Ourjoumtsev2006a, Bimbard2010}, неклассические суперпозиции когерентных состояний \cite{Dakna1997, Ourjoumtsev2006, Nielsen2007, Ourjoumtsev2007, Takahashi2008, Ourjoumtsev2009, Nielsen2010, Etesse2015}, а также их производные: делокализованные $N$-фотонные пакеты \cite{Walther2004, Mitchell2004, Nagata2007}, гибридные состояния \cite{Morin14, Jeong14}, и многие другие.
\vspace{1em}
\begin{center}
	* \quad * \quad *
\end{center}
\vspace{1em}

Квантовые состояния света чрезвычайно чувствительны к потерям; этим объясняется малочисленность реально используемых квантовых технологий. Концентрированию квантовых свойств -- или дистилляции запутанности -- оптических состояний, служит особый класс нелокальных методов манипуляции.
Отлаженной техникой дистилляции является операция вычитания одиночного фотона; применённая к двумодовому состоянию света, она позволяет получить состояние с более сильной запутанностью \cite{Ourjoumtsev2007b, Kurochkin2013}. Недостатком этого и других существующих методов является малость коэффициента усиления. Разработка более эффективного протокола дистилляции имеет важнейшее значение для таких разделов квантовых технологий, как криптография \cite{Pan1998, Waks2002, Gisin2002, Jin2015} и метрология \cite{Giovannetti2001, Edamatsu2002, Mitchell2004, Nagata2007, Kacprowicz2010, Chen2015, Lloyd2008, Ulanov2016}.

Накопление методов приготовления и элементарной обработки квантовых состояний света открывает возможности создания более сложных машин для передачи и обработки квантовой информации. Их сборка требует точной юстировки участвующих строительных блоков; с этим связана постановка задачи о характеризации неизвестного квантового процесса как чёрного ящика \cite{Lobino2008}. Многие важнейшие квантово-оптические процессы --  интерференция, рассеяние, ослабление, все нелокальные операции -- являются многомодовыми, и не могут быть полностью или достаточно эффективно характеризованы с помощью существующих  методов \cite{Poyatos1997, D'Ariano2001, Bongioanni2010}. Разработка удобного метода решения задачи о чёрном ящике является необходимым условием развития и масштабирования квантово-оптических технологий.


\section{Новизна полученных результатов}

\noindent
Настоящая работа включает в себя три результата, каждой из которых отведена отдельная глава:

\begin{enumerate}
	\item[Глава \ref{chapt2}.] \textbf{Эффект Квантового Вампира.}\\
	Продемонстрирована нелокальная природа операции уничтожения фотона в части световой моды. Эффект представляет собой не исследовавшийся ранее тип квантового действия на расстоянии, не приводящий к коллапсу волновой функции оптического состояния. Это свойство оператора уничтожения фотонов является общим для всех бозонных систем. Например, в случае конденсированного состояния Бозе-Эйнштейна, операции уничтожения кванта соответствует удаление группы атомов из конденсата. В соответствие с предсказанием, локальное выполнение такого действия не должно приводить к изменению формы коллективного квантового состояния.

	\item[Глава \ref{chapt4}.] \textbf{Дистилляция запутанности Эйнштейна--Подольского--Розена.}\\
	Впервые осуществлено восстановление уровня запутанности квантового состояния после того, как оно испытало двадцатикратные оптические потери. В отличие от существующих методов \cite{Ourjoumtsev2007b, Kurochkin2013}, разработанная техника использует безшумовое усиления света \cite{Lvovsky2002}; благодаря этому, предложенный протокол является первым, который не имеет принципиальных ограничений на степень усиления квантовой запутанности.
	
	\item[Глава \ref{chapt1}.] \textbf{Томография многомодового квантового процесса.}\\
	Разработана методика, впервые позволившая характеризовать двумодовый квантовый процесс на произвольной области оптического гильбертова пространства; техника представляет собой развитие одномодового метода \cite{Anis2012}. Главной особенностью многомодового случая является необходимость учёта фазовых соотношений между входными и выходными модами процесса. С учётом этого, разработана и продемонстрирована экспериментальная методика проведения многомодовых гомодинных измерений, позволяющая получить томографически полный объём информации, достаточный для достоверной реконструкции тензора квантового процесса.
	
\end{enumerate}

\noindent
Завершает текст глава \ref{methods}, в которой в справочной форме сведены экспериментальные и теоретические методы, являющиеся общими для многих квантово-оптических экспериментов. 
Заключение содержит основные результаты работы. Сведения о публикации и апробации результатов приведены в Выводах соответствующих разделов (\ref{vampire_outlook}, \ref{distillation_outlook}, \ref{sect1_7}).

\section{Практическая значимость полученных результатов}

\begin{enumerate}
	\item[\ref{chapt2}.] \textbf{Эффект Квантового Вампира.}\\
	Эффект является уникальным по области своей действительности: свойство отсутствия тени при вычитании кванта сохраняется независимо от степени классичности / квантововости состояния бозонного поля. При этой универсальности, эффект может быть реализован лишь посредством квантово-механических инструментов, так как в классической физике отсутствует аналог операции уничтожения фотона.	
	Эффект Квантового Вампира имеет практическую значимость в области инженерии квантовых состояний света. Например, с его помощью, широко применяемая операция вычитания фотона \cite{Neergard2011} может выполняться удалённо, либо без обладания полной информацией о структуре моды целевого состояния.
	
	\item[\ref{chapt4}.] \textbf{Дистилляция запутанности Эйнштейна--Подольского--Розена.}\\
	Этап концентрации квантовых свойств оптического состояния необходим для практической реализации технологий квантовой криптографии и вычислений \cite{Halder2007, Takeoka2014}.
	Разработанный метод позволил получить после двадцатикратного оптического ослабления ансамбль состояний более запутанных, чем появились бы в результате аналогичного ослабления бесконечно-запутанного состояния света -- то есть, преодолеть детерминистический предел. Это достижение является ключевым для создания  повторителя квантово-оптического сигнала \cite{Guha2015}.
	
	\item[\ref{chapt1}.] \textbf{Томография многомодового квантового процесса.}\\	
	Разработка метода характеризации многомодового квантового чёрного ящика представляет собой необходимое условие для дальнейшего развития и масштабирования квантовых технологий. Разработанный метод выгодно отличается от существующих: отсутствием ограничений на размер исследуемой области квантового гильбертова пространства -- в сравнению с \cite{Poyatos1997}, неограниченностью числа мод процесса -- в сравнении с \cite{Bongioanni2010, Anis2012}, а также простотой экспериментальной техники, требующей лишь набор пробных когерентных состояний и гомодинных измерений в выходных модах процесса -- в сравнении с \cite{D'Ariano2001}. Предложенная нами техника сегодня представляет собой наиболее мощный и удобный метод характеризации оптических квантовых процессов.
\end{enumerate}

\section{Защищаемые положения}

\begin{enumerate}
	\item Уничтожение фотона в части светового пучка имеет нелокальное действие на квантовое состояние света. При осуществлении такой операции в одной из поляризационных компонент диагонально-поляризованного фоковского состояния света, энергия ортогонально-поляризованной компоненты состояния уменьшается на половину энергии фотона.
	
	\item Разработан метод повышения уровня квантовых корреляций слабо-запутанного ЭПР-состояния света. Метод позволяет восстановить параметр запутанности ЭПР-состояния, испытавшего сколь угодно большие оптические потери, до уровня $\approx0.6$. Эффективность метода продемонстрирована в экспериментальной ситуации, когда одна из мод ЭПР-состояния с исходным уровнем двумодового сжатия 0.65dB, была ослаблена в 5 и 20 раз. В обоих случаях, процедура дистилляции позволила получить ансамбль состояний, имеющий первоначальный уровень квантовых корреляций. 
	
	\item Разработан метод характеризации многомодового квантового процессора; метод продемонстрирован в оптической системе на процессе светоделения. Успешно реконструированы квантовые аспекты процесса, в частности двухфотонная интерференция Хонг-Оу-Манделя. Средняя ошибка определения элементов тензора не превышает 5\%, параметр верности между экспериментально восстановленным тензором процесса и теоретическим ожиданием составил 95\%.
\end{enumerate}

\section{Личный вклад автора}

Описанные в настоящей работе результаты получены в составе научного коллектива, в который кроме автора входили А.И. Львовский, Ю.В. Курочкин, А.К. Фёдоров, А.Е. Уланов, А.А. Пушкина, Т.С. Ральф.
Автор играл ключевую роль в проектировании экспериментов, построении и юстировке оптических схем и электронных схем управления / записи данных, а также в проведении измерений и анализе полученных результатов.

\clearpage	
\chapter{Квантовый вампир: \newline неразрушающее действие на расстоянии оператора уничтожения фотонов} \label{chapt2}

Нелокальность является одним из фундаментальных аспектов квантово-механической картины мира. Это свойство состояния физической системы, в котором часть информации о ней содержится за её пределами -- в другой физической системе. Такая ситуация обратна положению \textit{локального реализма}, имеющего место в классической физике; однако, многие из состояний, допускаемых квантовой механикой -- и наблюдаемых в эксперименте -- демонстрируют нелокальные свойства.

Формализм квантовой механики описывает явление нелокальности на языке запутанного состояния физических систем. В таком состоянии, участвующие системы описываются совместной волновой функцией, однако не все из них имеют свою собственную. 
Примером является пара фотонов, рождённых в разных модах света в процессе спонтанного параметрического рассеяния света (разд. \ref{SPDC}); при измерении, такие фотоны обязаны обнаружить ортогональные поляризации -- при том, что поляризация каждого из фотонов перед измерением не определена. Корреляции в наборе таких измерений не могут быть объяснены в согласии с принципом локального реализма \cite{Bell1964, Zeilinger1999, Gisin2016}; ситуация выглядит так, как будто измерение, проводимое над одним из фотонов, оказывает мгновенное воздействие на второй из пары. За явлениями такого рода закрепилось название: нелокальный эффект действия на расстоянии.

На сегодняшний день, все известные примеры квантово-механического действия на расстоянии основываются на проведении в местной подсистеме распределённого квантового состояния проективного измерения фон-неймановского типа \cite{Neumann1955}. При этом, квантовое состояние коллапсирует, изменяя состояние физической реальности для удалённой подсистемы \cite{EPR1935, Bohr, Fuwa2015}.
Примерами действия этого принципа являются: квантовая телепортация \cite{Bennett1993}, удалённое приготовление квантовых состояний \cite{Bennett2001} и многие другие (см. Введение).
В этом разделе, описывается реализация квантового действия на расстоянии с помощью операции иного типа -- уничтожения фотона.

Бозонный оператор уничтожения играет важнейшую роль как в квантово-механическом формализме \cite{Dirac1927}, так и в квантово-оптических технологиях \cite{Neergard2011}, см. Введение. 
В настоящей работе, мы применяем оператор уничтожения фотонов локально, вычитая фотон в одной из местной части распределённой моды света. В отличие от измерения типа фон-Неймана, акт уничтожения фотона не приводит к редукции квантового состояния, но лишь изменяет его. Далее показано, что это изменение является глобальным; результат соответствует удалению фотона из всего распределённого состояния без возмущения его структуры. 

\section{Концепция} \label{sect2_2}

\begin{figure}[h]
	\includegraphics[width=3.5in]{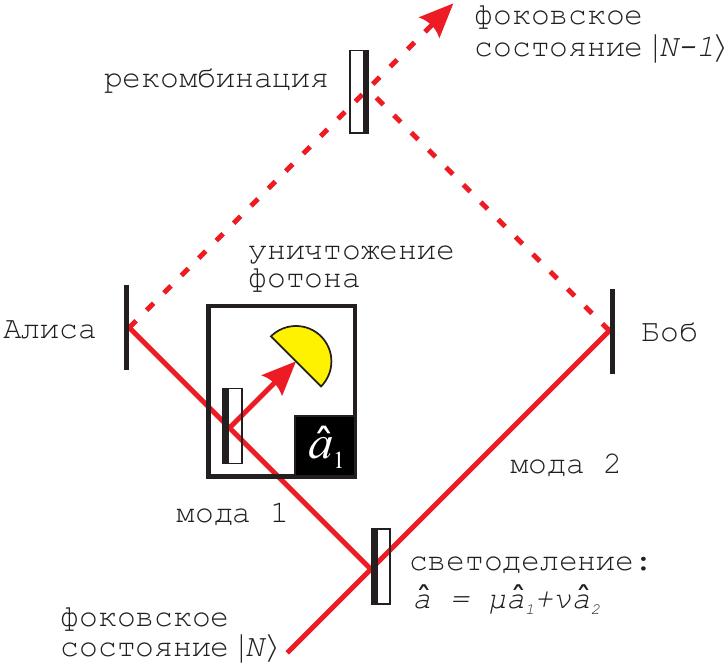}
	\centering
	\caption{Эффект Квантового вампира. Когда состояние $\ket\psi$ в моде света, определяемой оператором уничтожения $\hat a$ распределено между двумя удалёнными сторонами, применение оператора уничтожения фотонов $\hat a_1$ в одной из сторон действует на состояние $\ket\psi$ глобально. Это может быть проверено, например, с помощью анализа результирующего состояния мод $\hat a_1$ и $\hat a_2$, рекомбинировавших на последующем светоделителе.}
	\label{f1}
\end{figure}

Рассмотрим состояние $\ket \psi$, приготовленное в оптической моде, которой соответствует оператор уничтожения фотонов $ \hat a $. Будем полагать, что все остальные моды (ортогональные данной) находятся в вакуумном состоянии. С помошью светоделителя, это состояние распределяется между двумя удалёнными сторонами (Алисой и Бобом), которым соответствуют операторы $\hat a_1 $ и $\hat a_2$, так что 
\begin{equation}
\label{eqv0}
\hat a = \mu\hat a_1 + \lambda\hat a_2,
\end{equation}
где $|\mu|^2$ и $|\lambda|^2$ есть ненулевые коэффициенты отражения и пропускания светоделителя [Рис.~\ref{f1}].
Любое неклассическое состояние $\ket\psi$ -- то есть, не когерентное состояние и не их статистическая смесь -- приведёт в этом случае к перепутанному состоянию мод Алисы и Боба \cite{Caves2013}.

Пусть теперь Алиса вычитает фотон в своей моде. Результирующее двумодовое состояние есть
\begin{equation}
\label{eq1}
\hat a_1\ket\psi_{\hat a}=(\mu^*\hat a+\lambda\hat a_\perp)\ket\psi_{\hat a}=\mu^*\hat a\ket\psi_{\hat a},
\end{equation}
где $\hat a_\perp=\lambda^*\hat a_1-\mu^*\hat a_2$ есть оператор уничтожения моды, ортогональной к $\hat a$. Поскольку эта мода находится в вакуумном состоянии, действие её оператора уничтожения даёт арифметический ноль. Видно, что несмотря на то, что оператор уничтожения применён локально, он действует на всё состояние $\ket\psi$, распределённое между удалёнными сторонами. Если исходное состояние моды $\hat{a}$ содержит определённое число квантов $N$, то соотношение (\ref{eq1}) принимает вид
\begin{equation}
\label{eqVamp}
\hat{a}_1 \left|N\right\rangle_a \propto \left|N-1\right\rangle_a.
\end{equation}

\begin{figure}[h]
	\includegraphics[width=4in]{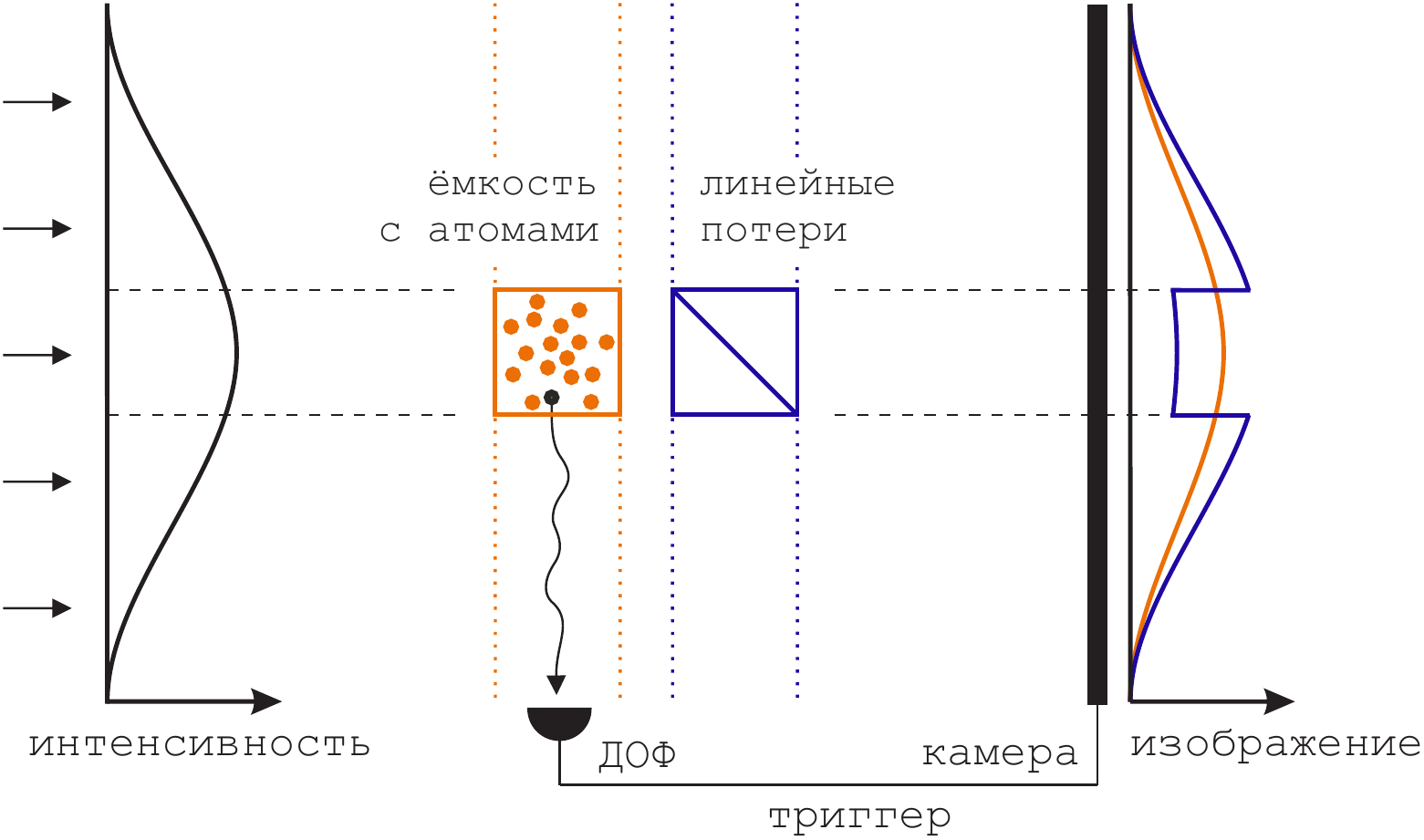}
	\centering
	\caption{Возможная демонстрация эффекта с помощью облака поглощающих атомов \cite{Baragiola2012}. Регистрация единичного переизлучённого фотона сигнализирует об уничтожении фотона в моде, определяемой формой облака, и запускает запись изображения на камере. Вычитание фотона не приводит к появлению тени от облака на фотографии, так что распределение интенсивности на ней (оранжевая линия) не изменяется. Эта ситуация контрастирует с обычным линейным поглощением, которое приводит к появлению тени (синяя линия).}
	\label{f12}
\end{figure}

В качестве примера можно рассмотреть облако слабо-поглощающих атомов, помещённых в широкий оптический луч в моде $\hat a$, как показано на Рис.~\ref{f12}. Вероятность поглощения атомами фотона много меньше $1$. Когда акт поглощение происходит, следующее в случайном направлении переизлучение фотона регистрируется детектором. Его срабатывание сигнализирует об уничтожении фотона в моде $\hat a_1$, соответствующей форме атомного облака. Можно было бы ожидать, что такое вмешательство приведёт к появлению тени -- области пониженной интенсивности внутри лазерного луча. Однако, согласно (\ref{eq1}), это не так; интенсивность уменьшается в профиле луча равномерно, так что мода $\hat a$ сохраняет свою структуру. В результате, невозможно определить положение и форму атомного облака, анализируя выходное состояние света. Отсюда аналогия с мифическим вампиром, давшая имя описываемому эффекту.

Может показаться, что вышеприведённый аргумент противоречит ежедневному опыту наблюдения теней. Объяснение заключается в том, что тени вызываются поглощением света, которое не эквивалентно операции уничтожения фотонов. Поглощение описывается оператором Линдблада 
\begin{equation}
\dfrac{\partial\hat\rho}{\partial z}\propto\hat{a}_1 \hat{\rho} \hat{a}_1^\dag-\dfrac{\hat\rho\hat a_1\hat a_1^\dag+\hat a_1\hat a_1^\dag\hat\rho}{2},
\end{equation}
где $\hat\rho$ есть оператор плотности ослабляемого состояния и $z$ есть направление распространения, который содержит операторы как уничтожения, так и рождения фотонов. Последний, однако, не обладает нелокальным свойством, описанным выше: 
\begin{equation}
\hat{a}_1^\dagger \left|N\right\rangle_a \not{\!\propto} \left|N+1\right\rangle_a.
\end{equation}
Это легко понять с помощью \eeqref{eq1}: если $\hat a$ заменить на $\hat a^\dag$, то член $\lambda\hat a_\perp^\dag\ket\psi_{\hat a}$ не равен более нулю.

Важно, что для корректного вычитания фотона, облако атомов на Рис.~\ref{f12} должно быть слабо поглощающим. Такое облако не должно сильно возмущать входное состояние, или создавать заметную тень на состоянии, когда оно наблюдается вне зависимости от регистрации переизлучённых фотонов (в этом смысле, схема близка к постановке неразрушающих, или слабых измерений \cite{Grangier1998, Aharonov1988, Steinberg2010}).
Однако, когда однофотонный детектор даёт отсчёт, известно, что состояние $\ket\psi$ потеряло фотон. Если энергия состояния сравнима с энергией фотона, то относительная потеря энергии значительна. Интуитивно, можно ожидать, что эта потеря энергии примет форму тени -- и тем не менее, это не так.

\section{Эксперимент}
\label{sect2_4}

\subsection{Идея}

\begin{figure}[h]
	\includegraphics[width=\textwidth]{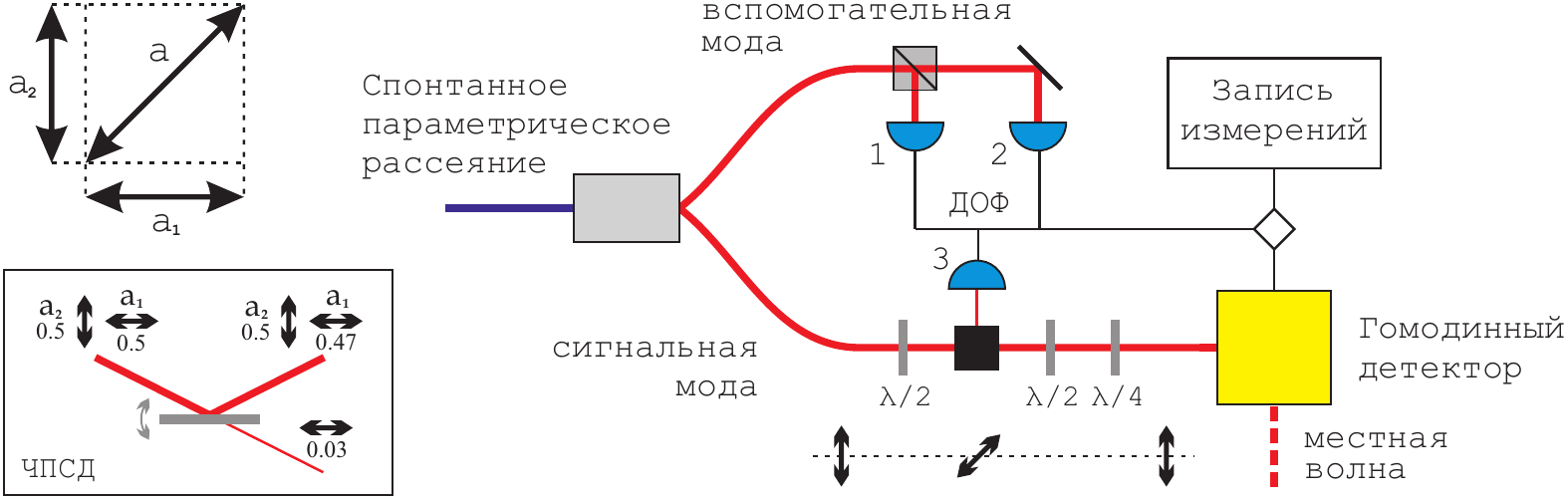}
	\centering
	\caption{Упрощённый вид экспериментальной установки. Мода $\hat a$ приготавливается в сигнальном канале спонтанного параметрического рассеяния  в случае, когда срабатывает один или оба детектора одиночных фотонов (ДОФ 1 и 2) во вспомогательном канале. Волновые пластинки образуют интерферометр Маха-Цейндера в поляризационном базисе, где моды $\hat a_1$ и $\hat a_2$ являются его горизонтально и вертикально поляризованными плечами. Стрелки показывают поляризацию света на ходе сигнала.
	Фотон вычитается из моды $\hat a_1$ с помощью частично-поляризационного светоделителя (ЧПСД). Его импровизированная реализация показана внизу слева. Стрелки и числа показывают поляризации мод и их относительные интенсивности. Пропускание ЧПСД, являющееся компромиссным параметром между качеством вычитания и частотой успешных событий, регулируется с помощью поворота зеркала. 
	Выходное состояние анализируется с помощью гомодинного детектора. Запись результатов измерения производится по условию срабатывания нужной комбинации ДОФ.}
	\label{p2}
\end{figure}

Мода $\hat a$ приготавливается в фоковском одно- или двух-фотонном состоянии, и имеет вертикальную поляризацию. Последняя поворачивается на $45^\circ$ с помощью полуволновой пластинки. Теперь, её вертикальная и горизонтальная компоненты являются модами $\hat a_1$ и $\hat a_2$, что в (\ref{eqv0}) соответствует $\mu = \lambda = 1/\sqrt{2}$. Таким образом, полуволновая пластинка выступает в роль светоделителя, открывающего (поляризационный) интерферометр Маха-Цейндера (Рис.~\ref{f1}). 
Пространственная форма этих мод $\hat a_1$ и $\hat a_2$ совпадает, поэтому разделение между Алисой (горизонтальная поляризация) и Бобом (вертикальная поляризация) в этом случае является поляризационным; не вызывает, однако, сомнения в том что его можно сделать пространственным, разделив моды с помощью светоделителя, как показано на Рис.~\ref{f1}.

Оператор уничтожения фотонов в моде Алисы реализован с помощью асимметричного, частично-поляризационного светоделителя. Для этого, мы использовали диэлектрическое зеркало с покрытием, обеспечивающим высокое отражение для угла падения луча в $45^\circ$. При угле падения $\approx 60^\circ$, вертикальная поляризация всё ещё демонстрирует высокое отражение ($>99.8\%$), тогда как около 6\% горизонтально-поляризованного света пропускается. Это поле регистрируется с помощью детектора ДОФ 3, так что его отсчёт с высокой вероятностью означает уничтожение фотона в моде $\hat a_1$.

Интерферометр Маха-Цейндера завершается волновыми пластинками $\lambda/4$ и $\lambda/2$. Последние устанавливаются так, чтобы в отсутствие вычитания фотона, весь рекомбинировавший сигнал попадал на гомодинный детектор. Мода сигнала для этого должна иметь вертикальную поляризацию, а горизонтально-поляризованная мода должна находиться в состоянии вакуума.
Если Алиса не уничтожила фотон, гомодинный детектор регистрирует результат рекомбинации мод $\hat a_1$ и $\hat a_2$, не испытавших действие оператора $\hat a_1$, то есть исходное состояние моды $\hat{a}$. 
Если же уничтожение фотона произошло, то состояние моды $\hat a$ в этой точке зависит от наличия или отсутствия действия не расстоянии между плечами интерферометра. 

Логика в русле локального реализма говорит о том, что состояние вертикально-поляризованной моды Боба $\hat{\rho}_{\mathrm{Bob}}$ не может быть затронуто происшествиями в моде Алисы; энергия может быть извлечена только из горизонтально-поляризованной моды света.  
Тогда -- как в случае исходного одно-, так и двух-фотонного состояния моды $\hat{a}$ -- вычитание должно оставить горизонтально-поляризованное плечо интерферометра в вакуумном состоянии. Смешиваясь с ним на выходном светоделителе интерферометра, состояние моды Боба испытывает двукратное ослабление и переходит в $\hat{\rho}_{\mathrm{out}}$, являющееся смесью многих фоковских компонент.
В случаях, когда исходное состояние является одно- и двух-фотонным, $\hat{\rho}_{\mathrm{out}}$ находится как
\begin{equation}
\label{eqVnaive}
\begin{aligned}
\hat{\rho}_{\mathrm{Bob}} 
= & \mathrm{Tr}_1 \left[\dfrac{\hat{a}_1^\dagger + \hat{a}_2^\dagger}{\sqrt{2}}\ket{00}\right] 
= \mathrm{Tr}_1 \left[ \dfrac{\ket{01} + \ket{10}}{\sqrt{2}} \right] \\
= & \dfrac{1}{2} \ketbra{0}{0} + \dfrac{1}{2} \ketbra{1}{1} 
\rightarrow \hat{\rho}_{\mathrm{out}} =  \dfrac{3}{4} \ketbra{0}{0} + \dfrac{1}{4} \ketbra{1}{1}
\end{aligned}
\end{equation}
и 
\begin{equation}
\begin{aligned}
\hat{\rho}_{\mathrm{Bob}} 
= & \mathrm{Tr}_1 \left[ \left(\dfrac{\hat{a}_1^\dagger + \hat{a}_2^\dagger}{\sqrt{2}}\right)^2\ket{00}\right] 
= \mathrm{Tr}_1 \left[ \dfrac{\ket{02} + \sqrt{2} \ket{11} + \ket{20}}{2} \right] \\
= & \dfrac{1}{4} \ketbra{0}{0} + \dfrac{1}{2} \ketbra{1}{1} + \dfrac{1}{4} \ketbra{2}{2}
\rightarrow \hat{\rho}_{\mathrm{out}} =  \dfrac{9}{16} \ketbra{0}{0} + \dfrac{3}{8} \ketbra{1}{1} + \dfrac{1}{16} \ketbra{2}{2}.
\end{aligned}
\end{equation}

В контрасте с этим, формула (\ref{eqVamp}) предписывает холостой выходной моде находиться в вакуумном состоянии, а детектируемой -- в чистом $\ket{0}$ или $\ket{1}$.
Таким образом, характеризация состояние вертикально-поляризованной моды на выходе интерферометра даёт возможность подтвердить или опровергнуть предсказание (\ref{eqVamp}).

\subsection{Оптическая схема} \label{alignment_2}
\begin{landscape}
\begin{figure}[]
	\centering
	\includegraphics[width=10in, angle=0]{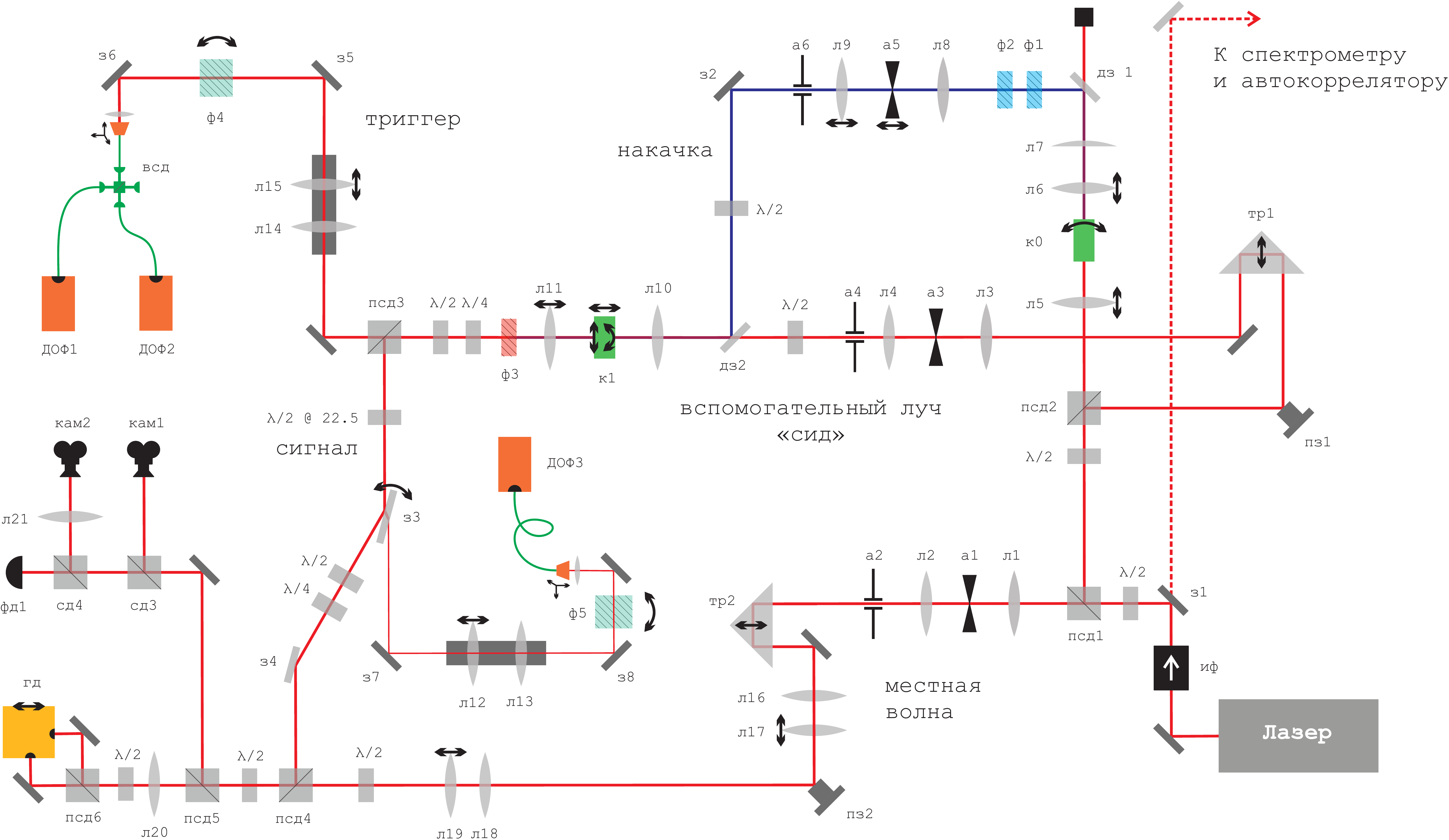}
	\caption{Полная схема экспериментальной установки. ИФ, изолятор Фарадея. Л, линза. А, апертура. (Д/П)З, (дихроическое/пьезо) зеркало. К, нелинейный кристалл. Ф, спектральный фильтр. (П/В)СД, (поляризационный/волоконный) светоделитель. ТР, тромбон. КАМ, камера. ФД, фотодиод. ДОФ, детектор одиночных фотонов. ГД, гомодинный детектор.}
	\label{pVampireFull}
\end{figure}
\end{landscape}

Оптическая схема эксперимента показана на Рис.~\ref{pVampireFull}.
Ключевой частью настройки является приготовление моды $\hat{a}$ (сигнал) в желаемом фоковском состоянии.
Для этого, используется невырожденный процесс спонтанного параметрического рассеяния (СПР) (разд. \ref{SPDC}), который происходит на нелинейном кристалле К1. 

Настройка нелинейного кристалла и приготовление фоковских состояний в моде $\hat{a}$ описаны в разделах \ref{PhaseMatching2} и \ref{SPpreparation}.
Для однофотонного детектирования в триггерной моде использовались кремниевые лавинные фотодиоды Excellitas, имеющие волоконный интерфейс. Последний, с помощью зеркал З5-З6 и телескопической системы линз Л14-Л15 (Рис. \ref{pVampireFull}), настраивается на приём моды DFG, тогда как мода $\hat{a}$ совпадает с модой сида.

По заведении в волокно, сигнал из моды DFG делится между двумя детекторами одиночных фотонов, ДОФ1 и ДОФ2, с помощью симметричного волоконного светоделителя ВСД. Это позволяет регистрировать приход двухфотонного состояния: в процессе симметричного светоделения, фотоны из пары направятся в разные моды с вероятностью 50\%. В остальных случаях, сможет сработать лишь один из детекторов, что будет интерпретировано как однофотонное событие. В случае малой амплитуды параметрического рассеяния -- что соответствует экспериментальной ситуации -- доля ложных срабатываний такого типа среди всех однофотонных событий мала (разд. \ref{SPprojection}).

Частота отсчётов хотя бы одного из детекторов ДОФ1 и ДОФ2 составляет $R_1 = 5 \times 10^4$ Гц. Соответственно, вероятность рождения фотонной пары от одного импульса есть
\begin{equation}
\label{eqProb1}
p_1 = \dfrac{R_1}{\eta \nu} \approx \dfrac{1}{150},
\end{equation}
где эффективность однофотонного детектирования $\eta = 0.12$, а $\nu = 76$ МГц -- частота повторения импульсов. Вероятность срабатывания двух детекторов равна (разд. \ref{SPprojection})
\begin{equation}
\label{eqProb2}
\dfrac{R_2}{\nu} = p_1^2 * \dfrac{\eta^2}{2} = 2.1 \times 10^{-7},
\end{equation}
где $R_2 \approx 16$ Гц есть частота двухфотонных событий. 

\subsection{Уничтожение фотона}

Для уничтожения фотона в моде Алисы, небольшая часть пучка отделяется и направляется на детектор одиночных фотонов ДОФ3. Срабатывание этого детектора свидетельствует об уничтожении фотона в исходной моде \cite{Neergard2011, Kumar2013}.
Формальное описание этой процедуры таково: действие светоделителя $\hat B_2$ на моду Алисы (Рис. \ref{f1}) эквивалентно эволюции пары вовлечённых мод под действием гамильтониана \cite{Jex1995}
\begin{equation}\label{BSHam}
\hat H=i\lambda \hat a_1 (\hat a'_1)^\dag-i\lambda \hat a_1^\dag \hat a'_1,
\end{equation}
где $\hat a'_1$ есть оператор уничтожения фотона во вспомогательной входящей моде, параметр $\lambda$ связан с амплитудным коэффициентом пропускания светоделителя $t$ как $t=\cos(\zeta)$ с $\zeta=\lambda\tau/\hbar$, а $\tau$ есть время эволюции. Если $|\zeta|\ll 1$, этой эволюции соответствует оператор 
\begin{equation}\label{BSevo}
\hat B_2 = \exp\left[-i\hat H\tau/\hbar\right] \approx 1-i\hat H\tau/\hbar.
\end{equation}
Если мода $a'_1$ на входе светоделителя находится в вакуумном состоянии, а состояние моды $a$ есть $\ket \psi$, то результатом эволюции (\ref{BSevo})  будет
\begin{equation}\label{apprBS}
\hat B_2 \left[\ket\psi \otimes \ket 0\right] \approx \ket\psi \otimes \ket 0 + \zeta\left(\hat a_1\ket\psi\right) \otimes \ket 1.
\end{equation}
В случае детектирования одиночного фотона в моде $a'_1$, состояние моды $a_1$ переходит в желаемое $\hat a_1\ket\psi$.

Таким образом, вычитание фотона из моды $\hat a_1$ происходит, когда ДОФ 3 срабатывает одновременно с ДОФ 1 и/или ДОФ 2. С одной стороны, такая постановка подразумевает проективное измерение, и нелокальный эффект вполне ожидаем. Однако, фундаментальное различие между описанной схемой и обычным измерением по фон-Нейману состоит в том, что в рассматриваемой схеме не происходит коллапса исходного состояния.

Как и детекторы в триггерной моде, ДОФ 3 не разрешает число фотонов в импульсе; его срабатывание поэтому эквивалентно детектированию одиночного фотона лишь если члены высших порядков в разложении Тейлора \eqref{BSevo} пренебрежимо малы \ref{SPprojection}. Наибольший из них даёт в конечное состояние поправку с весом порядка 
\begin{equation}
\label{eqVcondition}
(1-t^2) \left\langle \hat{n} \right\rangle_{\psi}\ll 1,
\end{equation}
где $\left\langle \hat{n} \right\rangle_{\psi}$ есть число фотонов в целевом состоянии.

Если условие (\ref{eqVcondition}) не выполняется, то для моделирования действия срабатывания/несрабатывания ДОФ3 на моду Алисы должны использоваться POVM операторы (\ref{SPCM}), модифицированные с учётом наличия шумовых срабатываний \cite{Kok2010}
\begin{equation}
\label{SPCM_noise}
\begin{aligned}
	\hat{\Pi}_{\text{off}}(\eta) &= \sum_{n=0}^\infty \left(1-\dfrac{p_{\rm bg}}{p_{\rm sub}}\right)(1-\eta)^n \ket{n}\bra{n}; \\
	\hat{\Pi}_{\text{on}}(\eta) &= \sum_{n=0}^\infty \left[1-\left(1-\dfrac{p_{\rm bg}}{p_{\rm sub}}\right)(1-\eta)^n\right] \ket{n}\bra{n},
\end{aligned}
\end{equation}
где $\eta$ есть квантовая эффективность детектора, а $p_{\rm bg}$ и $p_{\rm sub}$ есть вероятности фонового и истинного срабатываний ДОФ3. При условии срабатывания ДОФ3, мода $a_1$ переходит из исходного $\hat\rho=\ketbra\psi\psi$ в состояние
\begin{equation}
\label{Sa1}
{\rm Tr}_{a'_1}\left[\hat{\Pi}_{\text{on}}\hat B_2(\hat\rho\otimes\ketbra 00)\hat B_2^\dag\right].
\end{equation}

\section{Обработка данных}

\begin{figure}[h!]
	\includegraphics[width=\textwidth]{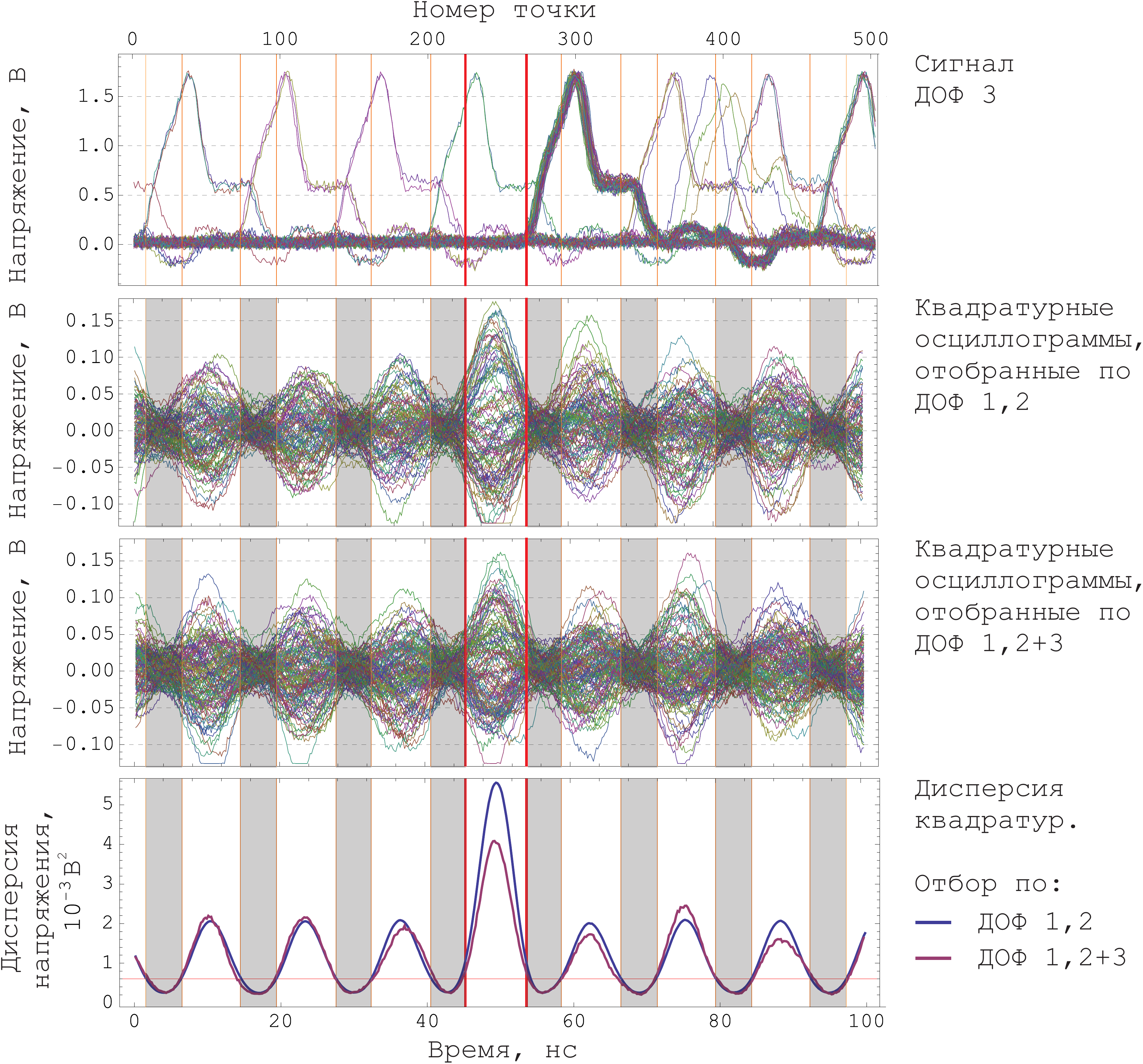}
	\centering
	\caption{Верхняя панель: 8192 осциллограммы сигнала ДОФ 3, записанные по условию срабатывания ДОФ 1 и 2. Пик, соответствующий вычитанию фотона, является пятым слева и имеет место в 145 осциллограммах; срабатывания на других импульсах являются фоновыми. Средние панели: 100 осциллограмм сигнала ГД. Сверху: все записанные. Снизу: отобранные по наличию пика в сигнале ДОФ 3. Нижняя панель: поточечная дисперсия всех квадратурных осциллограмм (синяя линия), и отобранных по срабатыванию ДОФ 3 (красная).}
	\label{pVdata}
\end{figure}

Сохранение квадратурных данных осуществлялось с помощью осциллографа Agilent DSO9254A. Триггерное условие логически связывало сигналы ДОФ 1--3.
В случае, когда исходное состояние моды $\hat{a}$ должно быть двухфотонным, триггерным событием является восходящий фронт в канале 1 при условии, что напряжение в канале 2 превышает 1 В. По этому условию, осциллограф записывает сигнал всех четырёх каналов. Каждая осциллограмма представляет собой 500 квадратурных измерений, проведённых с интервалом $2\times 10^{-10}$ секунды. Примеры осциллограмм ГД и ДОФ 3 показаны на Рис.~\ref{pVdata}. Импульс, содержащий исследуемые квантовые состояния, отмечен красными линиями; извлечение квадратурных данных осуществлялось согласно описанию, данному в разд. \ref{BHD_daq}.

\subsection{Анализ}

\begin{figure}[h!]
	\includegraphics[width=4in]{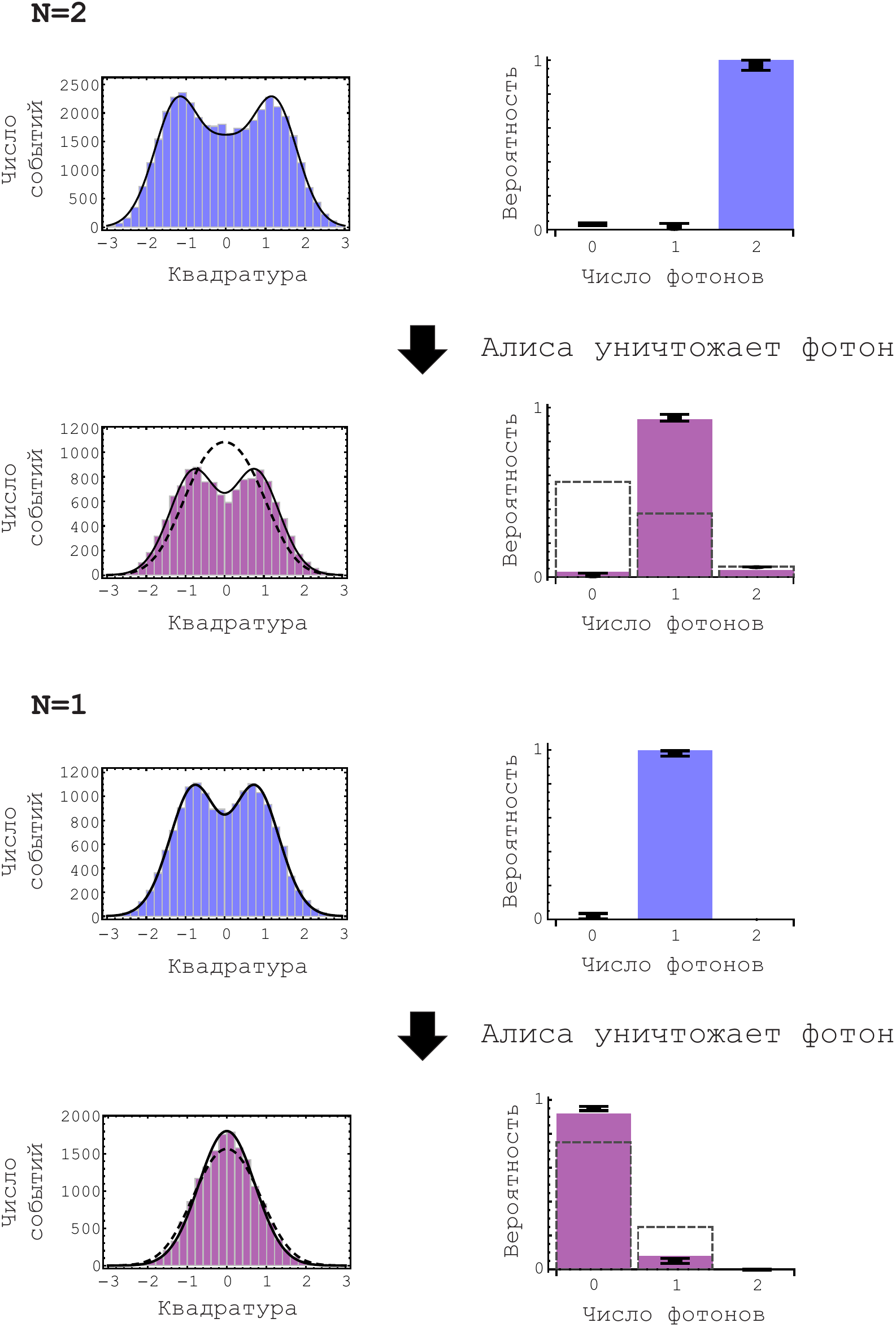}
	\centering
	\caption{Рекомбинировавшее состояние мод Алисы и Боба в случаях, когда мода $\hat a$ первоначально находится в одно- (N=1) и двух-фотонном (N=2) фоковском состоянии. Слева: экспериментальные гистограммы квадратурных данных. Сплошная линия: теоретическое ожидание. Справа: диагональные элементы соответствующих матриц плотности, скорректированные на эффективность детектирования. Результат теоретического моделирования (цветное заполнение) можно сравнить с ожиданием в предположении локального эффекта от вычитания фотона (\ref{eqVnaive}, пунктир). Чёрные точки соответствуют экспериментальным результатам; планки показывают статистические ошибки восстановления.}
	\label{pVresults}
\end{figure}

Гистограмма 40 тысяч квадратурных измерений, записанных по условию срабатывания обоих ДОФ 1,2 и несрабатывания ДОФ 3, показана на верхней левой панели Рис.~\ref{pVresults}. Справа показан результат гомодинной томографии, выполненной по этим данным, с коррекцией на общую квантовую эффективность эксперимента, которая составляла в среднем $\eta = 53\%$. Эта величина определялась с помощью вспомогательного эксперимента по томографии однофотонного состояния (разд. \ref{Fock_BHD}). Помимо обычных источников потерь (разд. \ref{BHD_efficiency}), $\eta$ включает также и эффект конечного пропускания ЧПСД, что эквивалентно примерно $3\%$ дополнительных потерь: шести-процентное пропускание ЧПСД горизонтальной компоненты действует на половину фотонной населённости моды $\hat a$.

Срабатывание ДОФ 3 одновременно с ДОФ 1 и ДОФ 2 сигнализирует о реализации оператора уничтожения фотона в моде Алисы; такое совпадение происходило $\sim 10$ раз в минуту. Гистограмма 12600 соответствующих значений квадратуры показана фиолетовым. 
Правая панель показывает результат гомодинной томографии с коррекцией на исходные $53\%$ потерь -- близкое к чистому однофотонному состоянию. Это согласуется с эффектом Квантового Вампира. Уничтожение фотона произошло во всей моде $\hat a$ без возмущения её структуры, в соответствии с (\ref{eqVamp}). 
Наблюдаемое распределение квадратур и матрица плотности находится в контрасте с наивным ожиданием (\ref{eqVnaive}), показанным штрихованными линиями.

Нижние панели показывают экспериментальные результаты в случае, когда исходное состояние моды $\hat{a}$ является однофотонным. Для его приготовления достаточно срабатывания только одного из ДОФ 1 и 2. Снова, экспериментальный результат свидетельствует о действии эффекта Квантового Вампира.

\subsection{Влияние экспериментальных неидеальностей} \label{sect2_6}

Томография измеряемого состояния требует проведения не менее $10^4$ квадратурных измерений (\ref{Fock_BHD}). Поэтому, в процессе параметрического рассеяния, используемого для приготовления исходных фоковских состояний, вероятность рождения фотонных пар должна быть достаточно высокой; это приводит к фоновым срабатываниям детектора ДОФ 3 даже в отсутствие отсчётов ДОФ 1 и ДОФ 2. Эти фоновые отсчёты являются основной причиной расхождения экспериментальных результатов и идеальной теорией \eeqref{eq1}. 

Так, в двухфотонном эксперименте (Рис.~\ref{pVresults}, N=2) вероятность фонового отсчёта ДОФ 3 есть $p_{\rm bg}=6\times 10^{-4}$ на каждый импульс, тогда как вероятность отсчёта ДОФ 3, обусловленная появлением фотонов на детекторах ДОФ 1 и ДОФ 2, равна $p_{\rm sub}=1.2\times 10^{-2}$. Таким образом, $p_{\rm bg}/p_{\rm sub}=5\%$ отсчётов детектора ДОФ 3 являются фоновыми; вычитания фотона при этом не происходит, что приводит к появлению $\sim 5\%$ двухфотонной компоненты в матрице плотности, показанной на Рис.~\ref{pVresults}, N=2, справа снизу.

\begin{figure}[h]
	\includegraphics[width=4.5in]{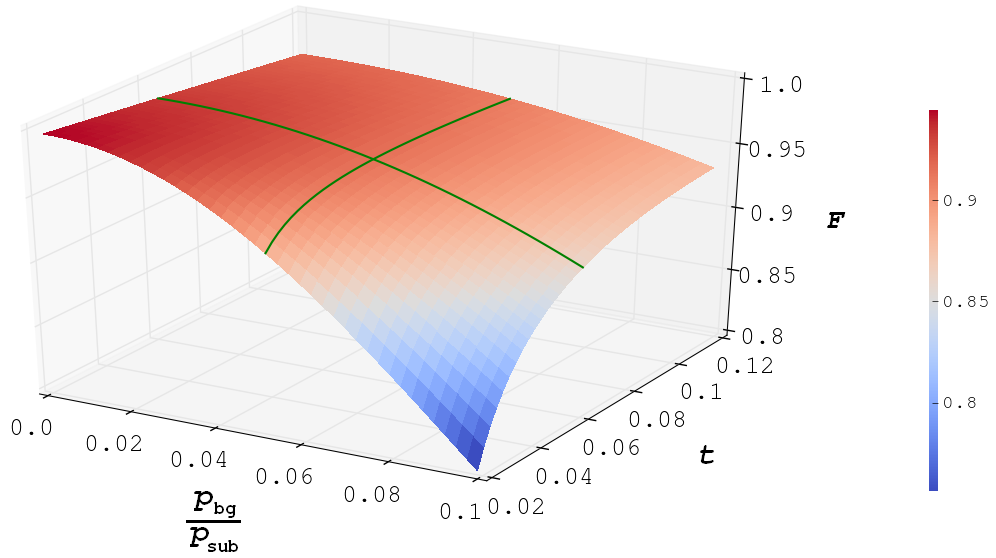}
	\centering
	\caption{Зависимость параметра верности \eqref{eq2} между полной теорией \eqref{Sa1} и предсказанием \eqref{eq1} в зависимости от коэффициента пропускания вычитающего светоделителя $t$ и доли шумовых отсчётов ДОФ 3 для двухфотонного эксперимента. Зелёные линии соответствуют экспериментальным параметрам. Верность в рабочей точке ($t=0.06$, $p_{\rm bg}/p_{\rm sub}=0.05$) равна $0.96$. Расчёт произведён с помощью библиотеки инструментов для квантовой оптики QuTiP \cite{Qutip}.}
	\label{p4}
\end{figure}

Рис.~\ref{p4} показывает параметр верности
\begin{equation}
	\label{eq2}
	\mathcal{F} \left( \hat{\rho}, \hat{\rho}_{\rm ideal} \right) = \left(\mathrm{Tr} \left[ \sqrt{ \sqrt{\hat{\rho}} \hat{\rho}_{\rm ideal} \sqrt{\hat{\rho}} }\right]\right)^2.
\end{equation}
между предсказаниями полной теории \eqref{Sa1} и уравнением \eqref{eq1} для двухфотонного эксперимента (верхние панели Рис. \ref{pVresults}). При отсутствии фоновых срабатываний вычитающего детектора, верность монотонно возрастает с уменьшением пропускания ЧПСД. 
Однако, при наличии фонового шума эта функция начинает иметь экстремум. Это объясняется тем, при малых $t$ фоновый шум подавляет полезные срабатывания ДОФ 3, количество которых пропорционально $t$.

Для значения $p_{\rm bg}/p_{\rm sub} = 0.05$, имевшему место в эксперименте, эффект квантового вампира реализуется наилучшим образом ($\mathcal{F}=0.965$) при $t=0.07$. При малых $t$ верность существенно меньше: при $t \rightarrow 0$ имеем $\mathcal{F} \rightarrow 0.85$.

\section{О передаче информации с помощью действия на расстоянии}

Имея на руках много копий своей части состояния, запутанного с модой Боба, Алиса может установить в своей моде оборудование для вычитания фотона и наблюдать за срабатываниями вычитающего детектора. Несомненно, детектор начнёт время от времени выдавать отсчёты, каждый из которых означает извлечение энергии из моды Боба. Значит ли это, что Алиса таким образом смогла бы с помощью эффекта Квантового Вампира повлиять на моду Боба?

Для ответа рассмотрим ситуацию, когда между Алисой и Бобом распределено состояние
\begin{equation}
\label{eq1001}
\ket{\Psi} = \dfrac{\ket{10}+\ket{01}}{\sqrt{2}}.
\end{equation}
Алиса устанавливает в своей моде неразрешающий однофотонный детектор с квантовой эффективностью $\kappa$. В случае его срабатывания, состояние моды Боба переходит в 
\begin{equation}
\label{eqOutOn}
\hat{\rho}_{\rm on} = \rm Tr_A \left[\hat{\Pi}_{\rm on} \ket \Psi \bra \Psi \right] = \dfrac{\kappa}{2} \ket{0}\bra{0},
\end{equation}
где $\hat{\Pi}_{\rm on}$ есть оператор успешного однофотонного детектирования (\ref{SPCM}), имеющий в фоковском базисе, ограниченном двумя младшими состояниями, вид
\begin{equation}
\label{SPCMon}
\Pi_{\rm on}=
\left[ {\begin{array}{cc}
0 & 0    \\
0 & \kappa \\
\end{array} } \right].
\end{equation}
Вероятность исхода (\ref{eqOutOn}) есть норма этого состояния, $\kappa/2$.

В остальных случаях, т.е. с вероятностью $1-\kappa/2$, детектор Алисы не срабатывает. Это событие также несёт информацию о двумодовом состоянии, в результате чего мода Боба переходит в состояние
\begin{equation}
\label{eqOutOff}
\hat{\rho}_{\rm off} = \rm Tr_A \left[\hat{\Pi}_{\rm off} \ket \Psi \bra \Psi \right] = \dfrac{1}{2} \ket{1}\bra{1} + \dfrac{1-\kappa}{2}\ket{0}\bra{0},
\end{equation}
где оператор $\hat{\Pi}_{\rm off}$ соответствует не-срабатыванию детектора Алисы и равен
\begin{equation}
\hat{\Pi}_{\rm off} = \hat{1} - \hat{\Pi}_{\rm on}.
\end{equation}
Знание состояний (\ref{eqOutOn}) и (\ref{eqOutOff}) позволяет определить среднюю фотонную заселённость моды Боба, не обусловленную результатом измерения Алисы:
\begin{equation}
\label{eqMeanBob}
\left\langle \hat{n}_{\rm Bob} \right\rangle = \mathrm{Tr} \left[\hat{n}_{\rm Bob}  \hat{\rho}_{\rm  on}\right] + \mathrm{Tr} \left[\hat{n}_{\rm Bob}  \hat{\rho}_{\rm  off}\right] = \dfrac{1}{2},
\end{equation}
что совпадает с той же величиной в исходном состоянии (\ref{eq1001}).

Рассмотренная постановка эквивалентна ситуации, когда Алиса пытается реализовать в своей моде эффект Квантового Вампира, организуя вычитание фотона с помощью светоделителя с коэффициентом отражения $r$ и однофотонного детектора с эффективностью $\eta$, при $\kappa = r\eta \ll 1$.
Как видно из (\ref{eqOutOn}), эффект Квантового Вампира, приводящий к извлечению энергии из моды Боба, реализуется с малой вероятностью; в остальном большинстве случаев, несрабатывание детектора Алисы приводит к повышению энергии в моде Боба (\ref{eqOutOff}). Вероятности исходов сбалансированы так, что в среднем энергия моды Боба остаётся на первоначальном уровне (\ref{eqMeanBob}).

Результат (\ref{eqMeanBob}) можно повторить для произвольного исходного состояния $\Psi$:
\begin{equation}
\begin{aligned}
\left\langle \hat{n}_{\rm Bob} \right\rangle_{\rm Alice \, ON} = & \mathrm{Tr} \left[\hat{n}_{\rm Bob} \left(\hat{\Pi}_{\rm on}+\hat{\Pi}_{\rm off}\right)_{\rm Alice} \ket \Psi \bra \Psi \right] = \\
& \mathrm{Tr} \left[\hat{n}_{\rm Bob} \left(\hat{1}\right)_{\rm Alice} \ket \Psi \bra \Psi \right] \quad\quad\quad\quad \,\, = \left\langle \hat{n}_{\rm Bob} \right\rangle_{\rm Alice \, OFF}.
\end{aligned}
\end{equation}
Таким образом, Алиса не может передать сигнал Бобу, влияя с помощью эффекта Квантового Вампира на энергию его моды; ошибочность рассуждения, приведённого в начале этого раздела, состоит в неучитывании изменения квантового состояния вследствие появления информации, содержащейся в не-срабатывании вычитающего фотон детектора.  

Сохранение квантовым оператором $\hat{U}$ следа состояния математически выражается свойством унитарности:
\begin{equation}
\hat{U}^* = \hat{U}^{-1}.
\end{equation}
Оператор уничтожения фотона не является унитарным, а значит, в принципе не может быть реализован детерминистическим образом \cite{Kumar2013}. 
По этой причине, эффект Квантового Вампира -- как и всякое действие на расстоянии -- не означает возможности непосредственной передачи информации \cite{Eberhard1989, Peres2004}.

\clearpage

\section{Выводы}
\label{vampire_outlook}

Природа нелокальности оператора уничтожения фотона, продемонстрированная в этом разделе, является универсальной; эффект Квантового Вампира может быть перенесён на оптические моды в любом базисе: временном, пространственном, спектральном, и т.д., и остаётся в силе при многократном уничтожении фотона.
Эффект является общим для всех бозонных систем. В случае конденсированного состояния Бозе-Эйнштейна, операции уничтожения кванта соответствует удаление группы атомов из конденсата. В соответствие с предсказанием (\ref{eq1}), локальное выполнение такого действия не должно приводить к изменению формы коллективного квантового состояния.

Эффект Квантового Вампира может найти применения в технологиях квантовой информатики; с его помощью, можно манипулировать квантовым состоянием, не обладая полной информацией о структуре его моды. Сюда относятся протоколы по генерации запутанных состояний света и дистилляции запутанности с помощью уничтожения фотонов \cite{Ourjoumtsev2007, Takahashi2008, Kurochkin2013, Morin14}. Возможность извлечь фотон не оставляя тени может оказаться полезной при подслушивании квантовых каналов связи, а также в создании устройств квантовой невидимости \cite{Gevaux2009, Milne2012}.
Эффект интересен также и с фундаментальной точки зрения, так как квантовое действие на расстоянии, не обусловленное коллапсом волновой функции, не исследовалось ранее.

Эффект Квантового Вампира не требует наличия квантовой запутанности между подмодами $\hat a_1$ и $\hat a_2$, которая имеется только в случае, когда их совместное состояние $\ket{\psi}_a$ является неклассическим. Например, в случае когерентного состояния $\ket{\alpha}_a$, состояния мод $\hat a_1$ и $\hat a_2$ также являются когерентными с амплитудами $\mu\alpha$ и $\nu\alpha$, соответственно. Так как последние являются собственными для оператора уничтожения фотонов, то действие $\hat a_1$ оставляет состояние моды $\hat a_1$, как и форму моды $\hat a$, без изменений. 
Аналогичным образом, то же явление будет наблюдаться и для других неклассических, но и не запутанных состояний -- например, состояний нулевого дискорда \cite{Ollivier2001}. Легко видеть, что $\hat a_1  \tilde{\rho}_N \hat a_1^\dag $ по-прежнему даёт $\tilde{\rho}_{N-1}$, где $\tilde{\rho}_N$ есть полностью дефазированная версия состояния $\left| N \right\rangle_a \! \left\langle N \right|$. 
Таким образом, эффект вампира является уникальным по области его действительности: свойства действия на расстоянии и отсутствия тени сохраняются независимо от степени классичности / квантововости бозонного состояния. При этой универсальности, эффект реализуем только с помощью квантово-механического инструмента: в классической физике отсутствует аналог операции уничтожения фотона.

По результатам исследования опубликована журнальная статья: \\
Ilya A. Fedorov, Alexander E. Ulanov, Yury V. Kurochkin, and A.I. Lvovsky, Quantum vampire: collapse-free action at a distance by the photon annihilation operator. Optica 2 (2), 112-115 (2015) \cite{FedorovVAMPIRE2015}. \\

Результаты работы представлены на конференциях:
\begin{enumerate}
	\item [1.] Quantum 2014 -- Workshop ad memoriam of Carlo Novero, 25--31 мая 2014, Турин, Италия
	\item [2.] 23rd annual International Laser Physics Workshop, 14--18 июля 2014, София, Болгария
	\item [3.] Quantum Optics VII, 27--31 октября 2014, Мар--дель--Плата, Аргентина
	\item [4.] Third International Conference on Quantum Technologies, 13--17 июля 2015, Москва, Россия
\end{enumerate}



\clearpage			
\chapter{Дистилляция запутанности Эйнштейна -- Подольского -- Розена} \label{chapt4}

Запутанные состояния физических систем находятся в центре внимания учёных с момента появления квантовой механики. В таких состояниях, физические системы способны демонстрировать явления, не имеющие аналогов в классическом описании природы, и находящиеся за рамками повседневной интуиции (см. введение к гл. \ref{chapt2}). Примером является двумодово-сжатое состояние света, волновая функция которого имеет вид (разд. \ref{TMS})
\begin{equation}
\Psi_\zeta^\mathrm{TMS}(X_a,X_b) = \dfrac{1}{\sqrt{\pi}} \exp \left[{-\left(\dfrac{X_a - X_b}{2\,e^{-\zeta}}\right)^2}\right] * \exp \left[-\left(\dfrac{X_a + X_b}{2\,e^{+\zeta}}\right)^2\right],
\label{eqPsiTMSV}
\end{equation}
где $X_a$, $X_b$ есть канонические координатные переменные двух мод света, а $\zeta$ -- параметр сжатия.
Состояние (\ref{eqPsiTMSV}) получило известность благодаря дискуссии А. Эйнштейна / Б. Подольского / Н. Розена и Н. Бора \cite{EPR1935, Bohr}, в которой авторы обсуждают корректность квантово-механического описания физической реальности в свете возможности мгновенного взаимодействия между удалёнными подсистемами $a$ и $b$, находящимися в состоянии (\ref{eqPsiTMSV}).

Сегодня, когда спор разрешён в пользу квантовой механики \cite{Hensen2015, Wiseman2015}, состояние (\ref{eqPsiTMSV}) (в дальнейшем -- ЭПР) является универсальным ресурсом, используемым в квантовых технологиях. Среди последних -- квантовые телепортация \cite{Furusawa1998}, вычисления \cite{Lloyd1999}, криптография \cite{Ralph1999}, метрология \cite{Grangier1987}, манипулирование атомными состояниями, и другие \cite{Kuzmich2000, BRA05}.

Квантовая теория не накладывает ограничений на параметр сжатия ЭПР-состояния. Однако на практике эта величина, характеризующая уровень запутанности между модами, ограничена -- что является основным препятствием для практического применения означенных технологий. Даже когда сжатие на выходе из источника достаточно велико, оно быстро разрушается за счёт потерь при хранении или передаче. Особенно принципиально эта проблема встаёт в технологиях квантовой связи, когда расстояние между узлами напрямую влияет на безопасность передачи информации \cite{Halder2007, Takeoka2014, Guha2015}.
В этом контексте, принципиальную важность имеет создание квантового дистиллятора \cite{Bennett96, Pan2003} -- устройства, позволяющего восстановить уровень запутанности состояния (\ref{eqPsiTMSV}) после того, как оно испытало потери при хранении или путешествии на большое расстояние.

В этой главе описывается эксперимент, в котором впервые удалось осуществить восстановление запутанности ЭПР-состояния после того, как оно испытало двадцатикратное оптическое ослабление. Разработанная методика применима к сколь угодно сильно ослабленным состояниям и является важным шагом в развитии практических технологий квантовой связи.


\section{Генерация и наблюдение ЭПР-состояния света с помощью двух кристаллов типа 1}
\label{Two_crystal_EPR}

В нашей лаборатории, работа с ЭПР-состоянием в непрерывных переменных началась с разработки нового метода его получения, подразумевающего использование двух процессов вырожденного параметрического рассеяния света, проходящих на разных нелинейных кристаллах. 

Примеры использования пары кристаллов типа 1 для получения неклассических состояний света имеются. Среди них способ фильтрации выходного пучка параметрического усилителя света, позволяющий получить одномодово-сжатое состояния света в единственной пространственной моде \cite{Perez2014}, а также генерация поляризационно-запутанного двумодового состояния света с большой фотонной населённостью \cite{Iskhakov2012}. Последнее, однако, не имеет свойств ЭПР-состояния, так как поляризация каждой из мод света является неопределённой.

Описываемый в этом разделе метод позволяет синтезировать ЭПР-состояния в процессе интерференции двух одномодово-сжатых состояний света. Согласно вычислениям, приведённым в разд. \ref{sect3_TMS_and_SMS}, если исходные одномодовые состояния сжаты по ортогональным направлениям в фазовой плоскости, то моды на выходе процесса их симметричного светоделения окажутся в ЭПР-состоянии (\ref{eqPsiTMSV}).
Такая схема является аналогом метода \cite{Kwiat1999} в непрерывных переменных.

\subsection{Оптическая схема и процедура настройки}

\begin{landscape}
\begin{figure}[h]
	\centering
	\includegraphics[width=10in]{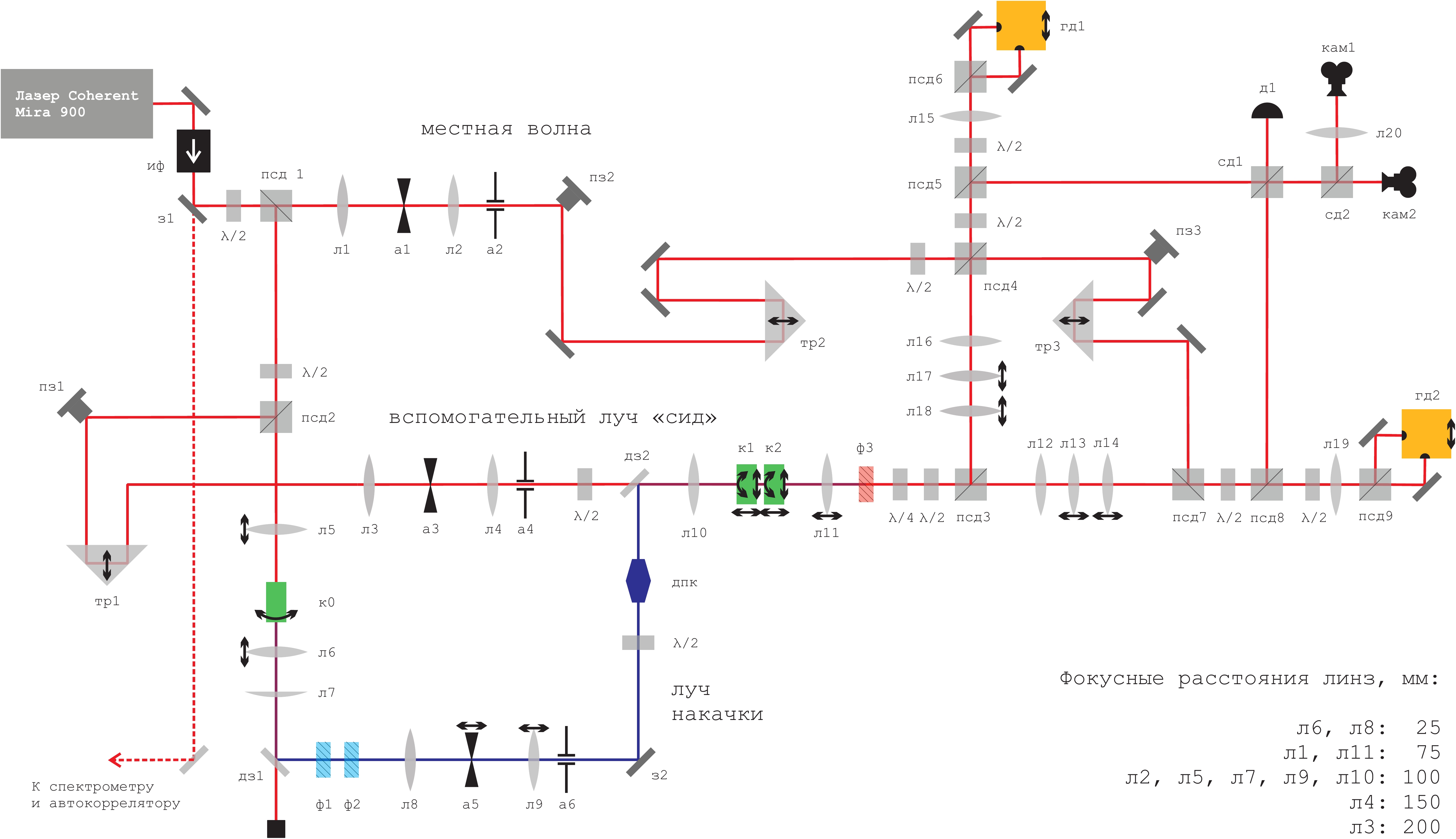}
	\caption{Схема экспериментальной установки для получения ЭПР-состояния с помощью пары кристаллов типа I (К1, К2). ИФ, изолятор Фарадея. Л, линза. А, апертура. (Д/П)З, (дихроическое/пьезо) зеркало. К0, нелинейный кристалл, используемый для удвоения частоты. Ф, спектральный фильтр. (П/В)СД, (поляризационный/волоконный) светоделитель. ТР, тромбон. КАМ, камера. ДПК, двулучепреломляющий кристалл. Д, фотодиод. ГД, гомодинный детектор.}
	\label{pTSMSsetup}
\end{figure}
\end{landscape}

Схема экспериментальной установки для получения сжатых состояний света показана на Рис.~\ref{pTSMSsetup}. Одномодово-сжатые состояния генерируются в процессе вырожденного спонтанного параметрического рассеяния (разд. \ref{SPDC}) луча накачки на двух одинаковых кристаллах титанил-фосфата калия с доменной структурой, имеющей период 2.95мкм (К1 и К2). Имеет место фазовый синхронизм типа 1, при котором пара вторичных фотонов имеет одинаковую поляризации, совпадающую с поляризацией луча накачки (разд. \ref{PhaseMatching1}). 

\begin{figure}[h]
	\centering
	\includegraphics[width=\textwidth]{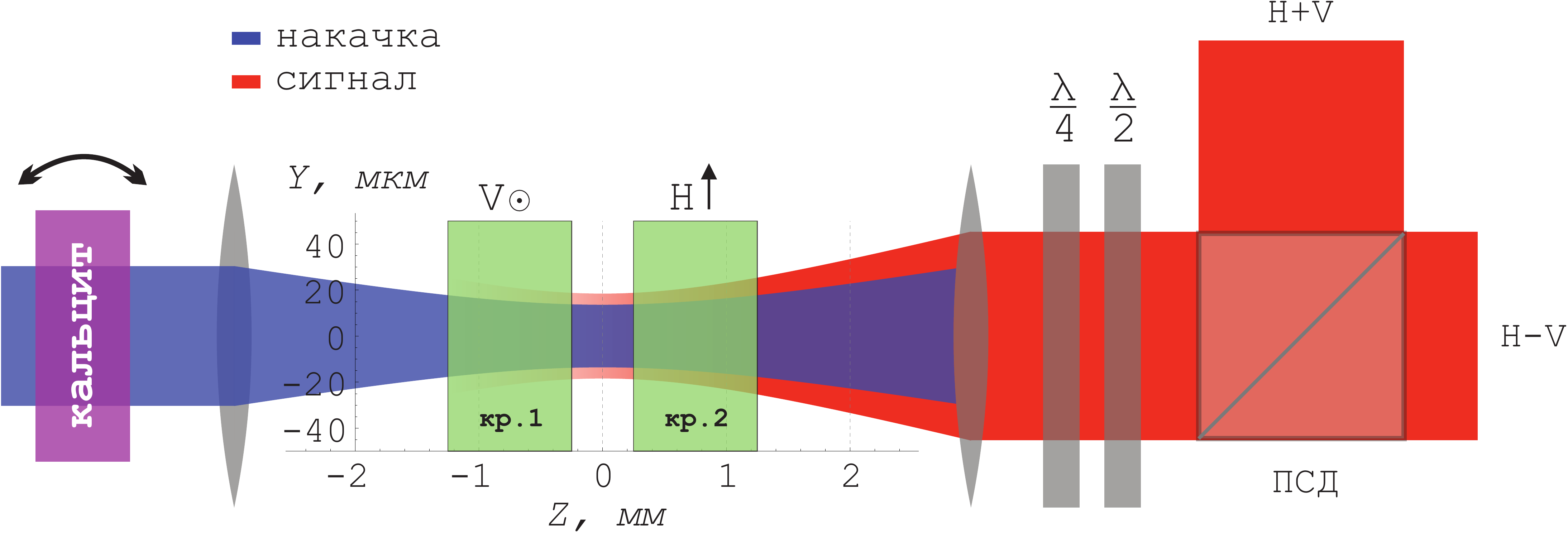}
	\caption{Расположение ортогонально-ориентированных нелинейных кристаллов в пучке. Радиусы перетяжек сида и накачки составляют 13.5 мкм и 9 мкм.}
	\label{pSMSbeams}
\end{figure}

Особенностью установки является то, что кристаллы повёрнуты один относительно другого на $90^\circ$. В результате, сигнальные моды, сжимаемые кристаллами, являются ортогонально поляризованными в вертикальном и горизонтальном направлениях. Кристаллы располагаются в общей перетяжке пучка накачки, как показано на Рис. \ref{pSMSbeams}. Это позволяет минимизировать воздушный дрейф фазы между двумя одномодово-сжатиями состояниями.

Симметричное светоделение этих мод, осуществляемое с помощью последующих четверть- и полу- волновой пластин, позволяет по желанию превратить пару индивидуально-сжатых состояний в ЭПР-состояние света, модами которого являются вертикальная и горизонтальная компоненты пучка. Эти моды пространственно разделяются на поляризационном светоделителе ПСД3, после чего каждая из них подвергается гомодинному детектированию.

Для настройки фазового синхронизма, кристаллы по очереди выводятся в центр перетяжки вспомогательного пучка -- ``сида'' (англ. seed -- сеять), где выполняется техника, описанная в разд. \ref{PhaseMatching1}. Поляризация сида должна быть горизонтальной для К1 и вертикальной для К2, т.е. совпадающей с поляризациями сигнальной моды в каждом случае. Неиспользуемый кристалл в это время может как оставаться в пучке, так и быть выведен из него; в первом случае, следует учитывать $\approx 10\%$ оптических потерь в материале кристалла.

Для независимой юстировки положений кристаллов, каждый был установлен на систему из трёх ортогонально-установленных микрометрических платформы и оправ, позволяющих регулировать 2 угла. Конструкция устроена так, что кристаллы могут быть подведены на расстояние $\sim 1$ мм, сохраняя при этом возможность независимой подстройки положений.

Температуры кристаллов регулировалась независимо, и были обычно существенно ниже лабораторной, см. Рис.~\ref{pTemperatureType1}.
Элемент Пельтье и нагреватель укреплёны вместе с кристаллом на лёгкой поворотной оправе, имеющей слабый контакт с основанием. Теплоотвод обеспечивается дополнительным радиатором.

Одновременное наблюдение одномодово-сжатых состояний, генерируемых на обоих кристаллах, осуществляется, когда они оба выведены в рабочее положение. Поляризация накачки при этом имеет направление около $45^\circ$, так что на каждом из кристаллов работает её горизонтальная или вертикальная компонента. т.е половина полной мощности. 
Кристаллы при этом находятся на минимально допустимом расстоянии, определяемом по касанию оправ. Это расстояние удавалось довести до 0.7 мм, что соответствует отклонению каждого кристалла от оптимального положения на 0.85 мм, или $57\%$ от Рэлеевской длины перетяжки накачки (\ref{eqFWaist}), составляющей 1.5 мм.
Радиус пучка центре кристаллов превышает радиус перетяжки $w_0 = 14$ мкм на $15\%$. Описанная ситуация изображена на Рис.~\ref{pSMSbeams}.

В ходе наблюдение одномодово-сжатых состояний было обнаружено, что для получения ожидаемой величины сжатия на втором кристалле, тромбоны ТР2 и ТР3 требуют дополнительной подстройки, соответствующей задержке местных волн на 0.62 мм. Расчёт показал, что эта задержка вызвана различием групповых скоростей накачки и сигнала СПР в кристаллах (см. разд. \ref{sectDispersion}).
Для компенсации этой задержки в канал накачки был установлен двулучепреломляющий кристалл кальцита $\mathrm{CaCO_3}$ (ДПК) длиной $1.8$ мм, который обеспечивал компенсирующую задержку между вертикальной и горизонтальной компонентами накачки. С помощью малого наклона ДПК, регулируется также и разность фаз между квадратурами, сжимаемыми на кристаллах.

\begin{figure}[h]
	\centering
	\includegraphics[width=4.5in]{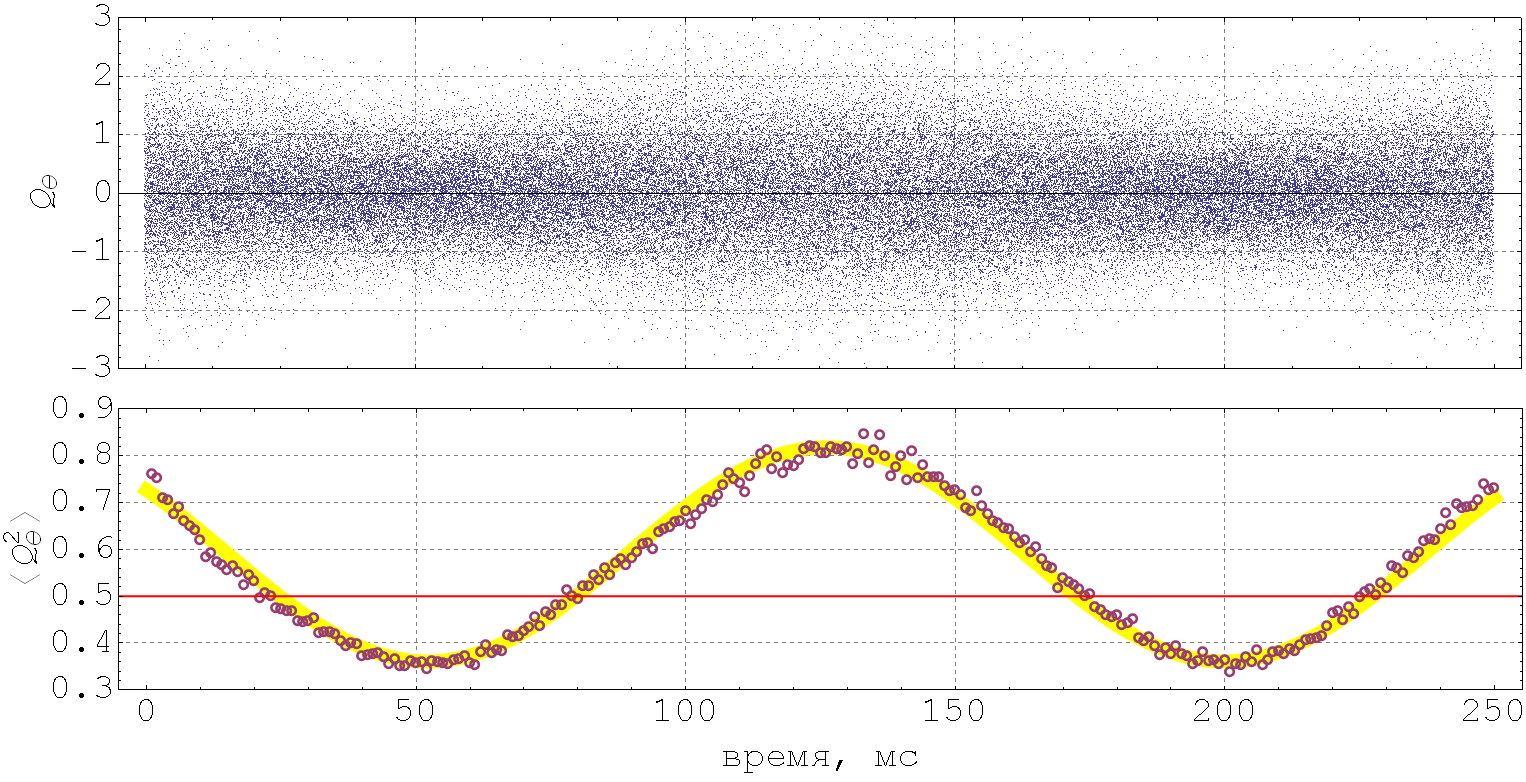}
	\caption{Квадратурные данные гомодинного детектора, принимающего одномодово-сжатое состояние света. Верхняя панель: сырые данные. Величина квадратур нормирована так, что дисперсия вакуумного состояния равна $0.5$. Для удобства показана каждая десятая точка. В течение измерений, относительная фаза между сигналом и местной волной меняется линейно с помощью пьезо-зеркала.	Нижняя панель: дисперсия сырых данных, разбитых на группы по $5000$ точек, в зависимости от времени. Жёлтая линия: теоретическое поведение дисперсии (\ref{eqSMSGeneralNoise}) для линейного движения зеркала. Красная линия: уровень вакуумного шума.}
	\label{pRawSMSdata}
\end{figure}

Когда получены и оптимизированы два одномодово-сжатых состояния, можно переходить к наблюдению двумодово-сжатого состояния. Важно, чтобы степени и эффективности сжатия были одинаковыми, что достигается с помощью подстройки положения полуволновой пластинки в луче накачки, распределяющей мощность между кристаллами, а также продольных положений кристаллов.
Роль светоделителя играет полуволновая пластинка в положении $22.5^\circ$. Признаком правильного положения является отсутствие фазовой зависимости индивидуальных сигналов с каждого из гомодинных детекторов (см. Рис. \ref{pRawSMSdata} и \ref{pRawTMSdata}).

Большое значение имеет также правильный выбор детектируемых мод. В идеальной ситуации пространственные и спектральные наборы сжатых мод, рождающиеся на кристаллах, одинаковы. В ходе оптимизации величины и эффективности одномодовых сжатий, каждый из телескопов Л12--Л14 и Л16--Л18 настраиваются так, чтобы главная мода (правая панель Рис.~\ref{pSpatialSMSmodes}, $i=j=0$), которая является наиболее сжатой, совпадала с модой местной волны (разд. \ref{sect3_realSMS}). В этом случае, настройка телескопов является оптимальной также и для детектирования двумодово-сжатого состояния.
В реальности ситуация другая, и телескопы Л12--Л14 и Л16--Л18 оптимизировались сначала при наблюдении одномодовых, а затем и при наблюдении двумодово-сжатого состояния.

Квадратурные измерения в сигнальных модах осуществляются детекторами ГД1 и ГД2 (разд. \ref{BHD_all}). Для сведения мод местной волны и сигнала, оптимизируется интерференция между местной волной и сидом. При этом временное сведение мод обеспечивается тромбонами ТР2 и ТР3 в путях местных волн, а пространственное сведение настраивается с помощью трёхлинзовых телескопов Л12--Л14 и Л16--Л18 в каналах сигнала.

\subsection{Томография сжатых состояний света}
\label{TSMStomography}

Методика томографии сжатых состояний света \cite{Breitenbach1997, Lvovsky2009} определяется следующими их свойствами (разд. \ref{TMS} и \ref{SMS}): сжатые состояния
\begin{enumerate}
\item Генерируются детерминистическим образом.\\
В противоположность фоковским состояниям (разд. \ref{Fock_BHD}), приготовление сжатых состояний подразумевает лишь действие эрмитового гамильтониана (\ref{eqTMSqueezingHam}). Как и накачка, сжатые состояния света генерируются в каждом лазерном импульсе.

\item Имеют нулевые средние по любой квадратуре (Рис. \ref{pSMSwigner} и \ref{pTMSwavefunction}). \\
Как и в случае фоковской томографии (разд. \ref{Fock_BHD}), снос средней линии сигнала гомодинного детектора не препятствует измерениям; последний может удаляться либо с помощью фильтра низких частот, либо программно при обработке данных.

\item Являются фазово-зависимыми.\\
Для характеризации сжатого состояния, набор квадратурных значений должен быть дополнен знанием фазы каждой измеренной квадратуры. 
\end{enumerate}

\begin{figure}[h]
	\centering
	\includegraphics[width=\textwidth]{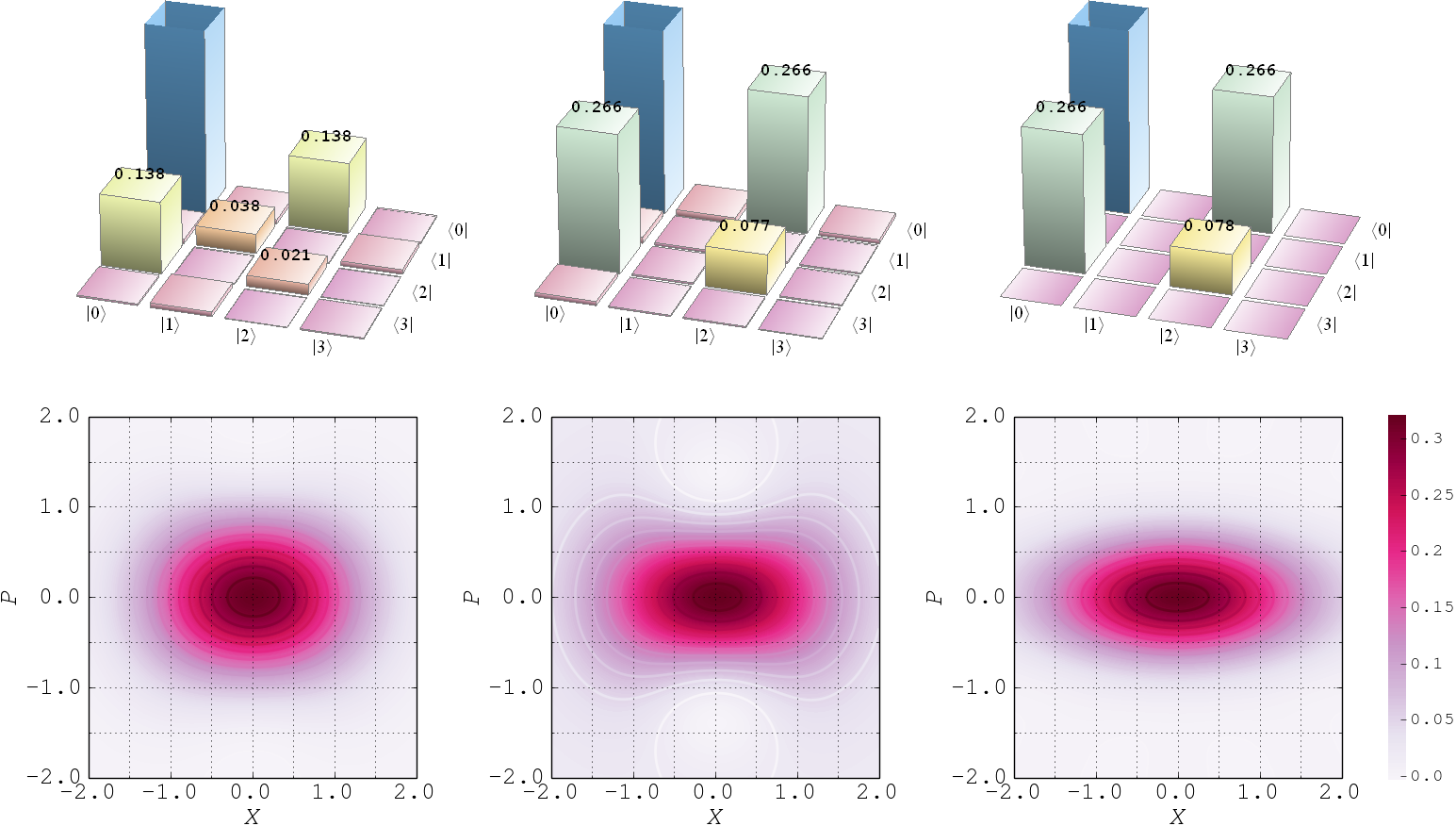}
	\caption{Матрицы плотности и функции Вигнера одномодово-сжатого сосотяния. Слева: экспериментальное состояние. В середине: экспериментальное состояние с коррекцией на $48\%$ потерь. Справа: теоретическое ожидание (Рис. \ref{pSMSwigner}, $\zeta = -0.44$).}
	\label{pSMStomography}
\end{figure}

\subsubsection{Одномодовое сжатие}
\label{sect3_resultSMS}

Рис. \ref{pRawSMSdata} показывает результат $10^6$ последовательно проведённых измерений квадратуры сигнала, находящегося в состоянии одномодового сжатого вакуума. Осцилляции дисперсии соответствуют вращению наблюдаемой квадратуры (\ref{eqGeneralQuadrature}) за счёт фазовой задержки, вносимой пьезо-зеркалом ПЗ3. Минимальная дисперсия квадратуры составляет $\approx0.35$, что соответствует $1.5\mathrm{dB}$ сжатия (\ref{eqDB}) и согласуется с теоретической оценкой ($1.6\mathrm{dB}$, разд. \ref{sect3_1_2}).

Пользуясь соотношением (\ref{eqSMSGeneralNoise}), по наблюдаемой на Рис.~\ref{pRawSMSdata} дисперсии можно вычислить фазу измеряемой квадратуры $\theta$, необходимую для проведения реконструкции состояния (разд. \ref{BHD_all}), в каждый момент времени. Задача становится ещё проще, если пьезо-зеркало движется линейно и настолько быстро, что флуктуациями воздуха можно пренебречь; в этом случае, фаза  пропорциональна времени / номеру измерения. С хорошей точностью, для данных показанных на Рис.~\ref{pRawSMSdata} это условие выполняется.
Реконструированная матрица плотности в фоковском базисе и соответствующее распределение Вигнера показаны слева на Рис.~\ref{pSMStomography}.

В согласии с теоретическим ожиданием, полученное состояние имеет форму (\ref{eqRhoAttSMS}). Соотношение между весами компонент $\ket{1}\bra{1}$ и $\ket{2}\bra{2}$ позволяет оценить коэффициент эффективного ослабления $R$ исходя из теоретического предсказания (\ref{eqRhoAttSMS}) как $0.5$, что согласуется с наблюдаемой дисперсией (см. разд. \ref{sect3_losses}). 
Матрица плотности, которая после прохождения таких потерь дала бы наблюдаемую, показана в средней колонке на Рис. \ref{pSMStomography}. Это состояние близко к чистому состоянию одномодового сжатого вакуума при $\zeta = -0.44$, правая колонка, c параметром верности 
\begin{equation}
\label{eqStateFidelity}
\mathcal{F} = \mathrm{Tr} \left(\left[ \sqrt{ \sqrt{\hat{\rho}_\mathrm{th}} \, \hat{\rho}_\mathrm{exp} \sqrt{\mathcal{\hat{\rho}_\mathrm{th}}} }\right]\right)^2 = 99.9\%.
\end{equation}

\subsubsection{Двумодовое сжатие}
\label{sect3_resultTMS}
\begin{figure}[h]
	\centering
	\includegraphics[width=\textwidth]{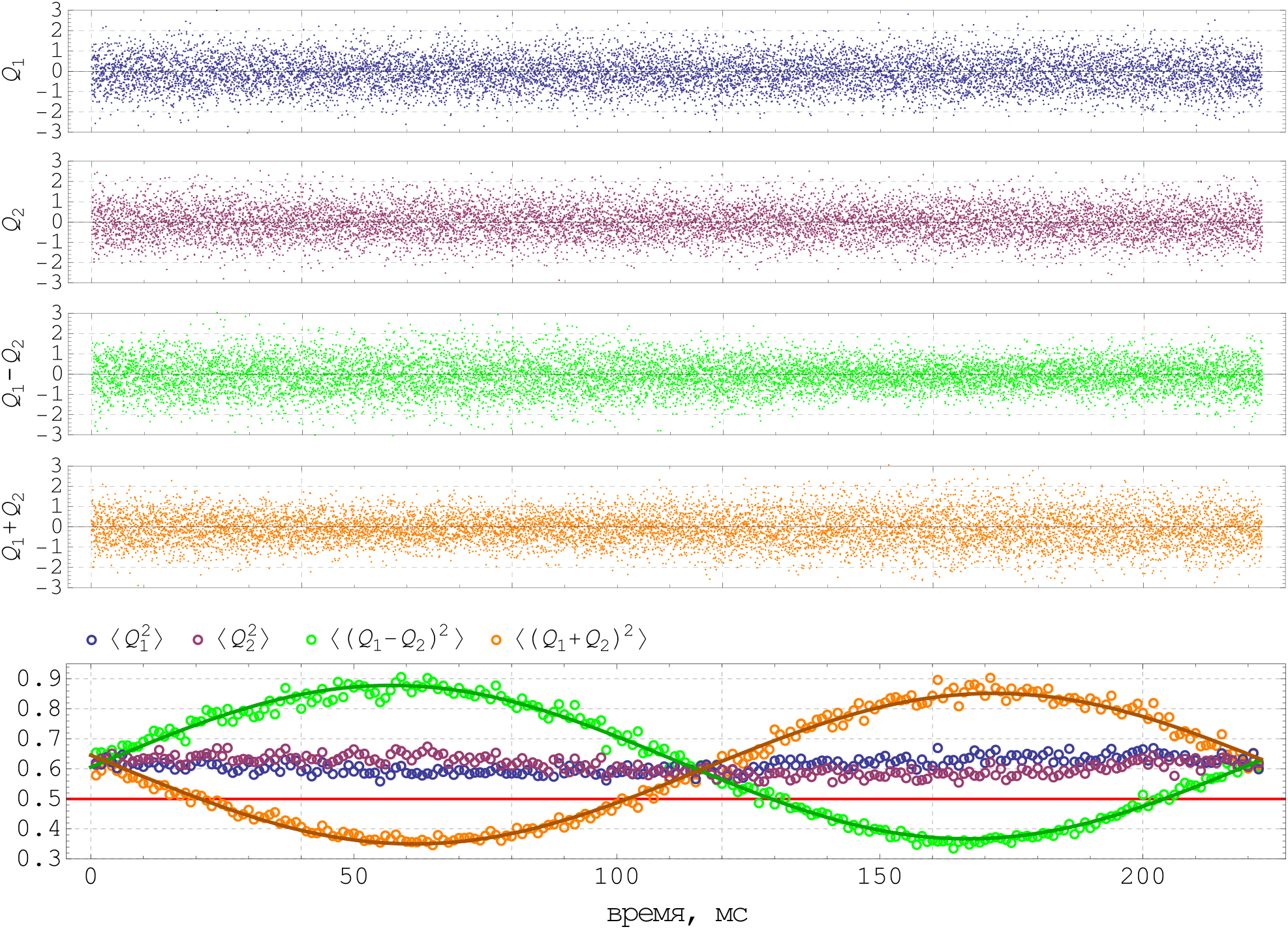}
	\caption{Квадратурные измерения в двух модах ЭПР-состояния света. Верхние панели: сырые данные в зависимости от времени. Для удобства, показана каждая сороковая точка. Фаза обоих местных осцилляторов варьируется с помощью пьезо-зеркал, управляемых пилообразным сигналом. Весь показанный промежуток соответствует линейному движению. Нижняя панель: дисперсия данных, разбитых на группы по 4000 точек. Линии: теоретическая зависимость (\ref{eqTMSRealNoise}).}
	\label{pRawTMSdata}
\end{figure}
	
Рис. \ref{pRawTMSdata} показывает результат $1.33\times10^6$ измерений квадратур сигнала, находящегося в состоянии двумодового-сжатого вакуума. В ходе эксперимента, фаза одной из местных волн варьировалась с помощью пьезо-зеркала ПЗ3.
Две верхние панели показывают значения каждой из квадратур, записанные одновременно. Каждая из мод в отдельности находится в тепловом состоянии и не демонстрирует фазовой зависимости. 
Вторая пара панелей показывает поточечную сумму и разность измерений в двух модах, что соответствует двумодовым наблюдаемым (\ref{eqTMobservables}).

Данные, показанные на нижней панели Рис. \ref{pRawTMSdata} проявляют фазовую зависимость, как и следует ожидать для двумодово-сжатого состояния. Так, при $\theta_1 + \theta_2 = 0$ ($\pi$) происходит наблюдение растянутой (сжатой) двумодовой квадратуры (\ref{eqTMSvar}).
	
Поведение дисперсий всех четырёх наборов данных показано на нижней панели. Минимальное значение для суммы и разности квадратур равно $0.36$, что соответствует 1.4dB сжатия и согласуется с предсказанием (\ref{eqBSvarsConnection}) исходящем из величины исходных одномодовых сжатий.
Дисперсии индивидуальных квадратур, в действительности, также демонстрирует фазовую зависимость, что объясняется неидеальностью реализации симметричного светоделения. Средний их уровень согласуется с теоретическим расчётом (разд. \ref{sect3_losses}) для параметров сжатия $\zeta=-0.44$ и эффективности $1-R=0.52$, найденных в разд. \ref{sect3_resultSMS}:
\begin{equation}
	\label{eqSMSinTMSNoise}
	\overline{\left\langle \Psi^\mathrm{TMS}_\zeta \left| \hat{Q}_\theta^2 \right| \Psi^\mathrm{TMS}_\zeta \right\rangle} = (1-R) \dfrac{e^{-2\zeta}+e^{2\zeta}}{4} + \dfrac{R}{2} \approx 0.6.
\end{equation}

\begin{figure}[t]
	\centering
	\includegraphics[width=5.5in]{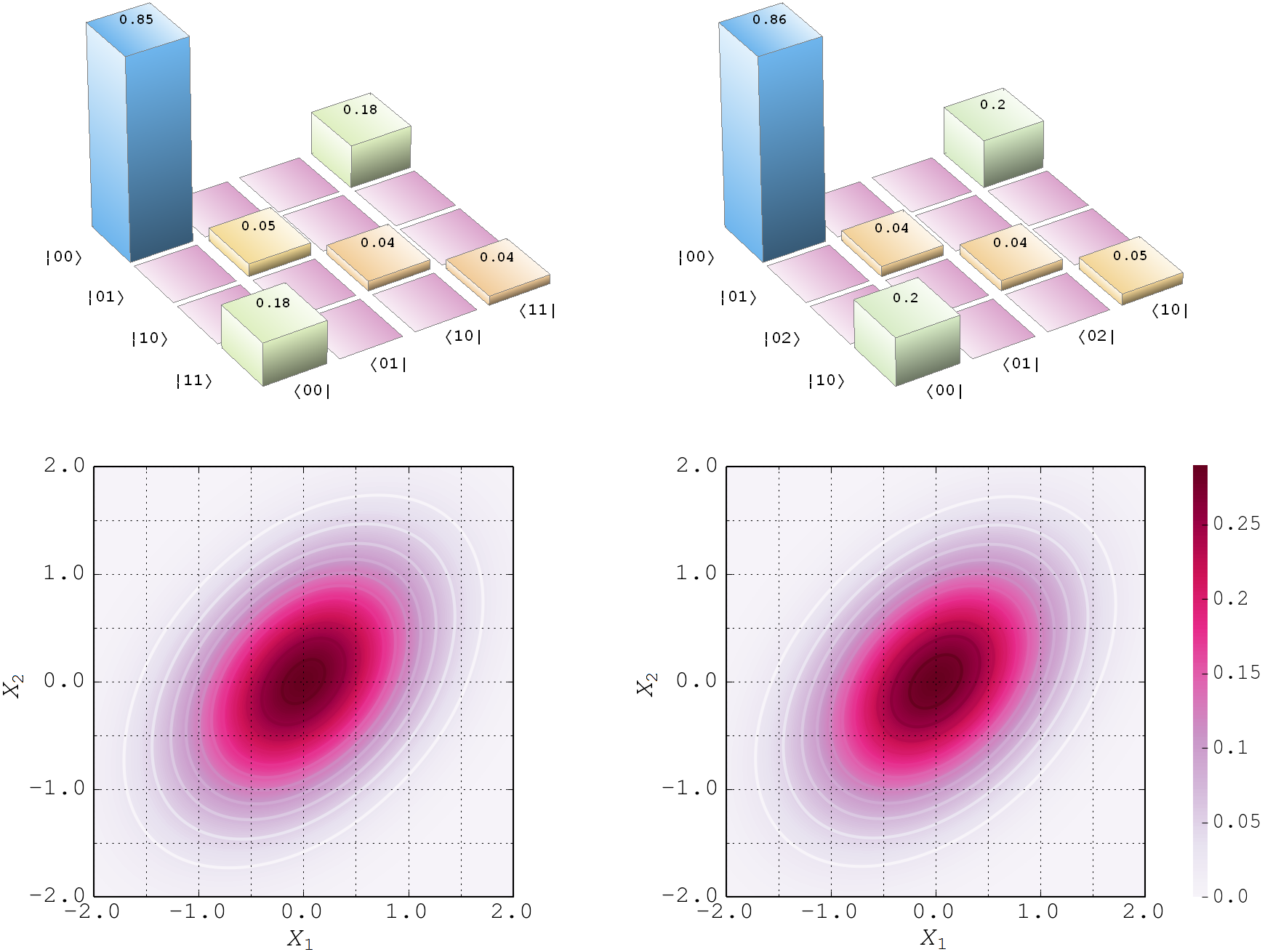}
	\caption{Результат реконструкции экспериментально приготовленного ЭПР-состояния (слева), и теоретическое предсказание (справа).}
	\label{pTMSresults}
\end{figure}

Как и в случае одномодово-сжатого сигнала, необходимая для томографии состояния фазовая информация содержится в самих квадратурных данных. Совмещая теоретическую зависимость (\ref{eqTMSRealNoise}) с экспериментальными данными, как показано на нижней панели Рис.~{\ref{pRawTMSdata}}, позволяет определить $\theta_1 + \theta_2$ в каждый момент времени.
		
Результат томографии двумодово-сжатого состояния показан на Рис.~\ref{pTMSresults}. Верхние панели показывают матрицы плотности, полученные в результате экспериментальной реконструкции (слева) и теоретического расчёта (\ref{eqAttTMSDM}, справа), соответствующего параметрам одномодово-сжатых состояний (разд. \ref{sect3_resultSMS}): $\zeta=-0.44$ и $\tau_1^2=\tau_2=0.52$.
Нижние панели показывают соответствующие этим состояниям квадраты амплитуд волновых функций в координатном базисе. Параметр соответствия (\ref{eqStateFidelity}) между теоретическим предсказанием и экспериментальным результатом составляет $99.5\%$.

\vspace{6em}

\subsection{Выводы}

Описанный в настоящем разделе метод позволяет генерировать двумодово-сжатые состояния света с помощью пары нелинейных процессов в условиях фазового синхронизма типа 1. Преимуществом метода является использование существенно более высокого, по сравнению с процессом поляризационно-невырожденного СПР, коэффициента нелинейного взаимодействия (разд. \ref{sect3_1_2}). Рекордным достижением является 2.36dB двумодового сжатия, что значительно превышает 1.5dB, обычно получаемые с помощью нелинейного процесса типа 2.

Основным недостатком метода является сложная процедура настройки, требующая одновременного поддержания двух нелинейных процессов. Чувствительность метода к флуктуациям фазы между двумя одномодовыми состояниями требует расположения кристаллов в общей перетяжке единственного пучка накачки, мощность которого разделяется между поляризационными компонентами. В результате, через каждый кристалл проходит мощность, вдвое превышающая полезную -- $\approx 80$ мВт; это, в свою очередь, приводит к быстрой деградации нелинейной среды и снижению эффективности. Этим объясняется, например, то, что рекордное значение сжатия не было воспроизведено в дальнейшем.

По этой причине, двух-кристалльный метод был впоследствии заменён более простой техникой.
Как описано в разделе \ref{TMS}, ЭПР-состояние может быть получено с помощью единственного нелинейного кристалла в процессе спонтанного, невырожденного процесса параметрического рассеяния света (СПР). Примером является эксперимент, описанный в главе \ref{chapt2}, в котором генерируемое таким образом ЭПР-состояние применяется для получения фоковских состояний света.

Представленная техника приготовления оптического ЭПР-состояния опубликована в журнальной статье:
Ilya A. Fedorov, Alexander E. Ulanov, Yury V. Kurochkin, and A.I. Lvovsky, Synthesis of the Einstein–Podolsky–Rosen entanglement in a sequence of two single-mode squeezers. Optics Letters 42 (1), 132-134 (2017) \cite{TwoCryst2017}.

\section{Безшумовое усиление света} \label{sect4_nla}
\subsection{Концепция}
Детерминистическое усиление амплитуды бозонного поля неизбежно ведёт к внесению в сигнал дополнительного шума \cite{Haus1962, Caves1981, Caves1982}. Это происходит потому, что для усиления сигнальной моды необходимо осуществить её взаимодействие со сторонней физической системой, которая неизбежно несет с собой дополнительный квантовый шум. Оказывается, что при усилении амплитуды сигнала в $g$ раз, минимальная величина дополнительного шума равна $|g^2-1|$ вакуумных единиц \cite{Clerk2010, Caves2012}.
Обозначая операторы уничтожения входной и выходной мод усилителя $\hat{a}$ и $\hat{b}$, последнее утверждение записывается как
\begin{equation}
\label{eqDetAmpVar}
\left\langle \hat{b}^2 \right\rangle  \geqslant g^2 \left\langle \hat{a}^2 \right\rangle + \dfrac{|g^2-1|}{2}.
\end{equation}
Соотношение операторов $\hat{a}$ и $\hat{b}$ при этом можно записать как
\begin{equation}
\label{eqDetAmpMap}
\hat{b} = g\hat{a} + \hat{\mathcal{F}},
\end{equation}
где последнее слагаемое представляет шум, вносимый усилителем.

Ограничение (\ref{eqDetAmpVar}) не составляет существенной проблемы для усиления классического сигнала, т.к. дополнительный квантовый шум пренебрежимо мал по сравнению как с собственным шумом сигнала, так и с уровнем дополнительных электронных и тепловых шумов.
В квантовых технологиях, напротив, амплитуды сигнальных полей сравнимы с уровнем вакуумного шума, и соотношение (\ref{eqDetAmpVar}) по сути означает фундаментальную невозможность создания сколько-нибудь полезного детерминистического усилителя квантового сигнала.

Таким образом: построение квантового усилителя, лишённого дополнительного шума, требует отказа от условия детерминистичности. Это означает, что безшумовое усиление может завершаться успехом лишь с некоторой вероятностью $P < 1$. То есть, при подаче на вход усилителя когерентного состояния, выходное состояние должно являться статистической смесью успешного и неуспешного исходов:
\begin{equation}
\label{eqNonDeterministicAmp}
\ket{\alpha}\bra{\alpha} \longrightarrow P \ket{g\alpha}\bra{g\alpha} + (1-P) \ket{\emptyset}\bra{\emptyset},
\end{equation}
где $\ket{\emptyset}$ обозначает неудачный исход. Если при этом имеется возможность узнать, какой из исходов реализовался при каждом запуске, то операция (\ref{eqNonDeterministicAmp}) может быть использована для вероятностного (недетерминистического) усиления квантового сигнала. Такой ситуации соответствует запись
\begin{equation}
\label{eqNonDeterministicAmp2}
\ket{\alpha}\bra{\alpha} \longrightarrow P \ket{g\alpha}\bra{g\alpha} \otimes \ket{\rm ON} \bra{\rm ON} + (1-P) \ket{\emptyset}\bra{\emptyset} \otimes \ket{\rm OFF} \bra{\rm OFF},
\end{equation}
где $\ket{\rm ON}$ и $\ket{\rm OFF}$ есть хорошо различимые состояния вспомогательной физической системы -- указателя.

Вероятностное усиление света было впервые реализовано в работах \cite{Xiang2010, Kocsis2012}. В них использовалась схема ``квантовых ножниц'' \cite{Pegg1998}, позволяющая выполнить операцию (\ref{eqNonDeterministicAmp2}) в пределе $\alpha \ll 1$, так что в случае успеха состояние усиливаемой моды преобразуется как
\begin{equation}
\label{eqNLA}
\ket{\alpha} \approx \ket{0} + \alpha\ket{1} \longrightarrow \ket{0} + g\alpha\ket{1} \approx \ket{g\alpha}.
\end{equation}

Теоретический анализ, проведённый А. Улановым, показал, что безшумовое усиление света может быть использовано для усиления запутанности ЭПР-состояния (\ref{eqPsiTMSV}) в пределах малого начального сжатия и существенного сжатия, испытавшего большие потери. Согласно расчётам, применение операции (\ref{eqNLA}) к состоянию (\ref{eqAttTMSDM}) при $\tau_1=1$ позволяет получить состояние 
\begin{equation}
\label{eqDST_result}
\hat{\rho}^\mathrm{TMS \,\, DIST}_{\zeta,\tau_{2},g} = \left[\dfrac{1}{g}\ket{00} + \zeta \tau_2 \ket{11}\right]\left[\dfrac{1}{g}\bra{00} + \zeta \tau_2 \bra{11}\right] + 
\dfrac{\zeta^2 (1-\tau_2^2)}{g^2} \ket{10}\bra{10}.
\end{equation}
При $g = 1/\tau_2$, нормированное состояние (\ref{eqDST_result}) близко к чистому состоянию формы (\ref{eqFockTMS}). Благодаря множителю $1/g$ перед вакуумной компонентой, это состояние обладает большей запутанностью по сравнению с исходным (\ref{eqAttTMSDM}).

\clearpage

\subsection{Квантово -- оптический катализ}
\label{Catalysis}

Для экспериментальной дистилляции ЭПР-запутанности была выбрана схема, известная под названием ``квантовый катализ'' \cite{Lvovsky2002}, которая, как и ``ножницы'', позволяет вероятностным образом реализовать преобразование (\ref{eqNLA}). Преимущество катализа состоит в том, что в случаях несрабатывания усиления, выходное состояние содержит часть исходного состояния света, которая может быть использована для определения фазы дистиллированного состояния.

\begin{figure}[h]
	\centering
	\includegraphics[width=4in]{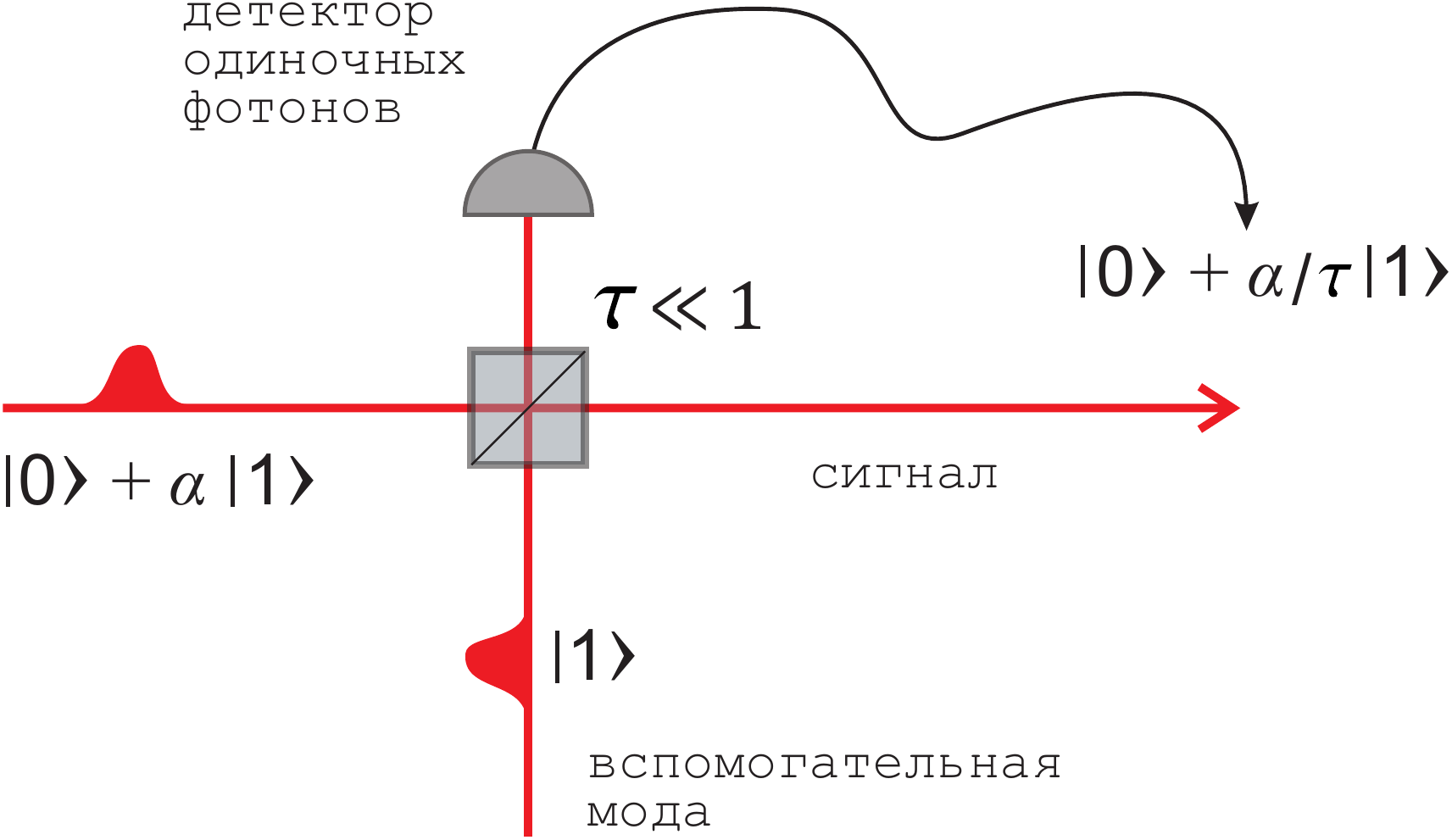}
	\caption{Квантово-оптический катализ \cite{Lvovsky2002}. Сигнальная мода интерферирует со вспомогательной, находящейся в однофотонном состоянии. Амплитуда однофотонной компоненты сигнального состояния изменяется по условию срабатывания детектора одиночных фотонов.}
	\label{pCatalysis}
\end{figure}
Схема квантового катализа представлена на Рис.~\ref{pCatalysis}. В этой схеме, сигнал -- слабое когерентное состояние $\ket{\alpha} \approx \ket{0} + \alpha \ket{1}$ -- интерферирует на светоделителе с пропусканием $\tau \ll 1$ со вспомогательной модой, находящейся в однофотонном состоянии. Одна из выходных мод направляется на детектор одиночных фотонов, срабатывание которого сигнализирует об успехе операции (\ref{eqNLA}).
Качественное понимание принципа квантового катализа можно получить, рассматривая пути, приводящие к срабатыванию детектора:
\begin{enumerate}
	\item[1.] Фотон из когерентного состояния отразился. Амплитуда этого события равна $\alpha\sqrt{1-\tau^2}\approx\alpha$. В выходной сигнальной моде присутствует фотон.
	\item[2.] Фотон из вспомогательной моды прошёл; соответствующая амплитуда равна $\tau$. В выходной моде теперь находится $\ket{\tau\alpha}\approx \ket{0}$.
\end{enumerate}
Неразличимость этих альтернатив означает, что состояние выходной моды будет суперпозицией состояний в каждой из них \cite{Feynman1968}: $\tau\ket{0} + \alpha \ket{1} \propto \ket{0} + \alpha/\tau \ket{1}$.
Таким образом, метод квантового катализа позволяет реализовать не-детерминистическое безшумовое усиление сигнала (\ref{eqNLA}) с регулируемой величиной усиления:
\begin{equation}
\label{eqNLAgain}
g = 1/\tau,
\end{equation}
и осуществить дистилляцию ЭПР-запутанности.

\clearpage
\section{Эксперимент}

\subsection{Обзор}

\begin{figure}[h]
\centering
\caption{Принципиальная схема экспериментальной установки.}
\includegraphics[width=5in]{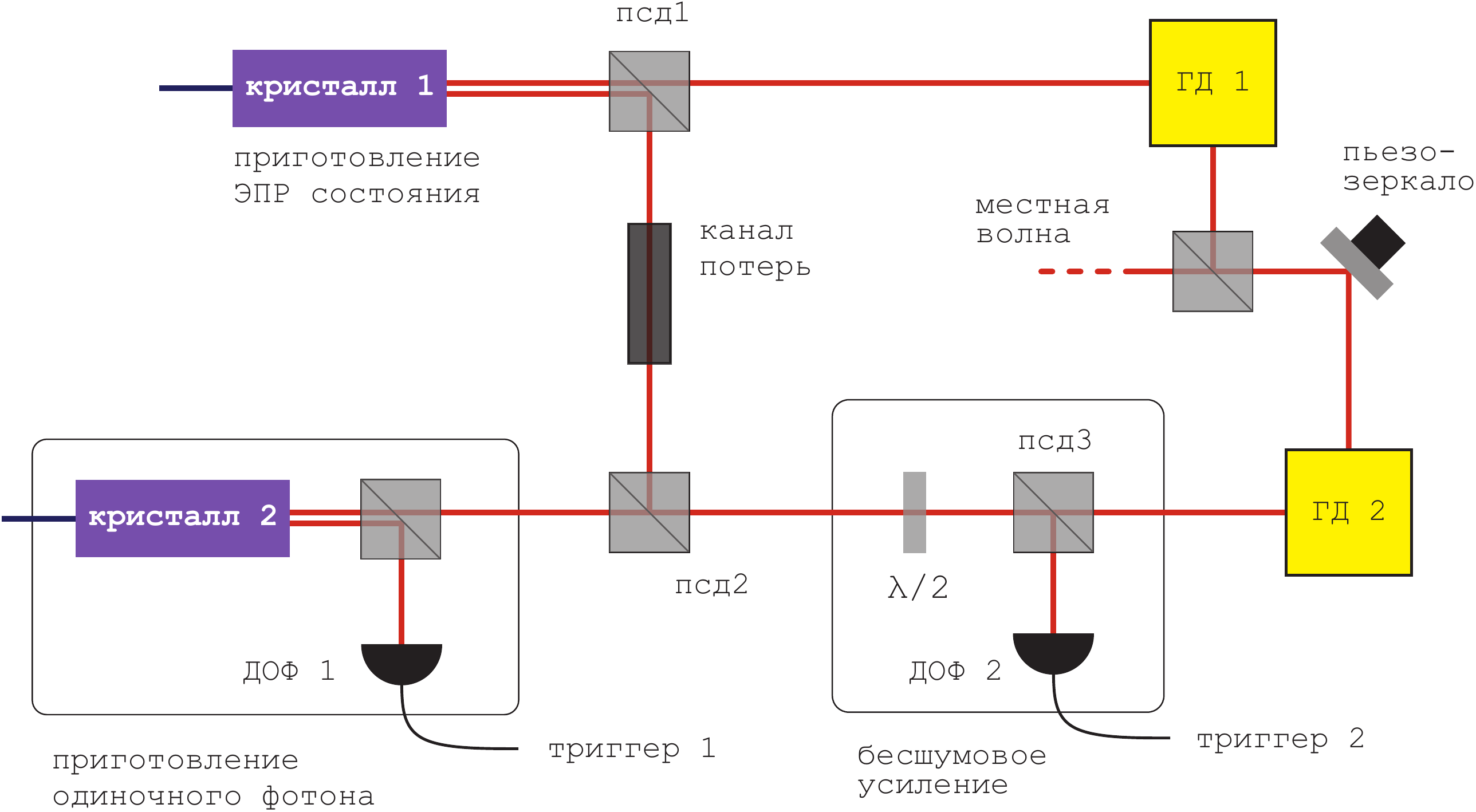}
\label{pDSTschematic}
\end{figure}

Принципиальная схема эксперимента показана на Рис.~\ref{pDSTschematic}. Основой для неё послужила установка для наблюдения сжатых состояний света, Рис. \ref{pTSMSsetup}.
ЭПР-состояние генерируется на кристалле К1, настроенным на фазовый синхронизм типа 2. Моды разделяются пространственно на поляризационном светоделителе ПСД1: мода 1 отражается и направляется на гомодинный детектор ГД1, а проходящая мода 2 следует в сторону ГД2.

В эксперименте моделируется ситуация, когда мода 2 испытывает ослабление при прохождении через канал потерь.  
Как обсуждено в разд.~\ref{sect3_losses}, такое ослабление одной из мод приводит к уменьшению запутанности двумодового состояния. Процедура дистилляции направлена на то, чтобы восстановить уровень запутанности и сделать ЭПР-состояние пригодным для дальнейшего использования.

Ослабленная ЭПР-мода и вспомогательная мода, находящаяся в состоянии одиночного фотона, имея ортогональные поляризации, пространственно перекрываются на светоделителе ПСД2. Операция светоделения для квантового катализа осуществляется на полуволновой пластинке, положение которой определяет пропорции светоделения. Выходные моды разделяются на светоделителе ПСД3, где отражённая проследует на однофотонное детектирование, а прошедшая направляется на ГД2 для характеризации результата дистилляции.
Однофотонное состояние, необходимое выполнения для квантового катализа (разд. \ref{Catalysis}), приготавливается на кристалле К2 (разд. \ref{SPpreparation}).

\subsection{Процедура настройки}
\label{sectDistAlignment}

\begin{landscape}
\begin{figure}
	\centering
	\includegraphics[width=8.7in]{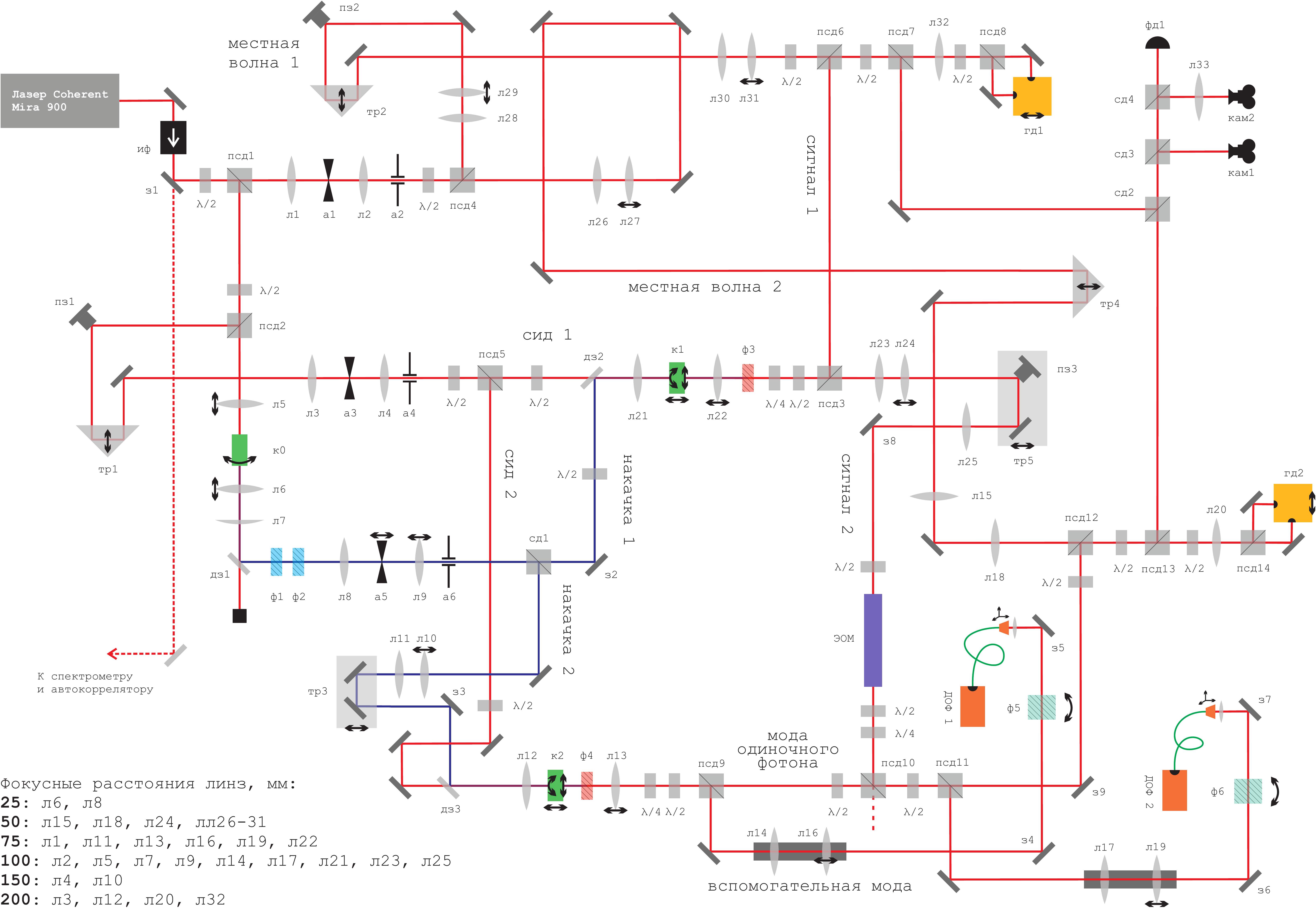}
	\caption{Полная схема экспериментальной установки. ИФ, изолятор Фарадея. Л, линза. А, апертура. (Д/П)З, (дихроическое/пьезо-управляемое) зеркало. ЭОМ, электро-оптический модулятор. К, нелинейный кристалл. Ф, спектральный фильтр. (П)СД, (поляризационный) светоделитель. ТР, тромбон. КАМ, камера. ФД, фотодиод. ГД, гомодинный детектор.}
	\label{pDSTsetup}
\end{figure}
\end{landscape}

Полная оптическая схема эксперимента показана на Рис.~\ref{pDSTsetup}.
Настройка начинается с пространственной фильтрации сида, накачки и местных волн с помощью апертур А1--А6. После фильтрации, пучок сида разделяется на светоделителе ПСД5, и направляется на кристаллы. Пучок накачки, имеющий полную мощность $\approx 80$ мВт, разделяется на светоделителе СД1, имеющем коэффициент отражения по мощности $\sim 2/3$.
Пучок местной волны разделяется на ПСД4 для обеспечения пары гомодинных детекторов ГД1 и ГД2.

Оптимальный порядок дальнейших действий определяется старшинством тромбонов, многие из которых являются связанными. Порядок, описанный ниже, позволяет настроить установку, пройдя каждый из этапов один раз.

\subsubsection{1: Настройка фазового синхронизма для кристалла К1 (разд. \ref{PhaseMatching2}).}

Совмещение мод сида 1 и накачки 1 осуществляется с помощью зеркал З2, ДЗ2 и тромбона ТР1. После этого все пучки, проходящие через кристалл К1, блокируются. 

\subsubsection{2: Настройка кристалла К2 с выполнением эксперимента по наблюдению состояния одиночного фотона (разд. \ref{SPpreparation}).}

Моды сида 2 и накачки 2 совмещаются с помощью зеркал З3, ДЗ3 и тромбона ТР3. 
Сигнальная мода совпадает с модой сида 2, а вспомогательная -- с модой сигнала DFG. После разделения на ПСД9, последний заводится в одномодовое волокно с помощью телескопа Л14--Л16 и зеркал З4 и З5, и направляется на детектор одиночных фотонов ДОФ1. 

Сигнальная мода проследует на детектор ГД2. На этом этапе, обеспечивается максимальное прохождение сигнала через светоделители ПСД10, ПСД11 и ПСД12, с помощью находящихся перед ними полуволновых пластин. 
Перекрытие сида 2 с местной волной 2 происходит на светоделителе ПСД12; совмещение мод обеспечивается зеркалом З9, светоделителем ПСД12 и тромбоном ТР4.

\subsubsection{3: Наблюдение ЭПР-состояния, генерируемого на кристалле К1 (разд. \ref{TSMStomography}). }

Cигнал DFG, определяющий первую моду ЭПР-состояния, используется для настройки гомодинного детектирования ГД1. Перекрытие с местной волной 1 юстируется светоделителями ПСД3,6 и тромбоном ТР2. Юстировочный интерференционный сигнал направляется на фотодиод ФД1 и камеры КАМ1,2 с помощью откидного светоделителя ПСД7.

Сид 1, определяющий вторую моду ЭПР-состояния, направляется на детектор ГД2. 
Для совмещения сида 1 с местной волной 2 достаточно добиться совмещения сидов с помощью зеркала З8, светоделителя ПСД10 и тромбона ТР5. Юстировочная интерференция направляется на фотодиод ФД1 с помощью откидного светоделителя ПСД13.

\subsubsection{\textrm{4}: Финальные приготовления.}

\begin{itemize}
	\item Настройка уровня сжатия исходного ЭПР-состояния. \\ 
	Для финальных экспериментов было выбрано состояние, имеющее минимум дисперсии двумодовой наблюдаемой (\ref{eqTMobservables}) 0.43, что соответствует 0.65dB сжатия (\ref{eqDB}). Максимум дисперсии составлял 0.6, что соответствует средней эффективности в двух модах $1-R=0.5$ при параметре сжатия $\zeta = 0.33$ (\ref{eq329}).
	Настройка производится с помощью полуволновой пластинки, находящейся в пучке накачки 1 перед зеркалом ДЗ2 и регулирующей мощность рабочей компоненты накачки. 
	\item Калибровка канала потерь, величина которых определяется поляризацией сигнальной моды 2 перед попаданием на светоделитель ПСД10. \\
	Амплитудная калибровка производится посредством измерения с помощью фотодиода ФД1 мощностей проходящих пучков в обоих режимах ЭОМ (разд. \ref{sectEOM}).
	\item Настройка однофотонного детектирования триггерного сигнала квантового катализа. Его мода совпадает с модами обоих сидов, значит заведение в волокно перед детектором ДОФ2 может оптимизироваться, используя любой из них.
	\item Настройка коэффициента безшумового усиления. Необходимое положение волновой пластинки перед ПСД11 устанавливается по доле проходящего через ПСД11 сида 1, $\tau^2$ (\ref{eqNLAgain}).
\end{itemize}

\section{Система сбора и обработки данных} \label{chapt4_acquisition}
Как описано в разд. \ref{sect4_nla}, безшумовое усиление света есть вероятностный процесс; кроме того, вероятностным образом создаётся необходимое для него вспомогательное состояние одиночного фотона. Таким образом, ЭПР-состояние дистиллируется при условии одновременного срабатывание двух однофотонных детекторов: ДОФ1 сигнализирует о приготовлении вспомогательного фотона, а ДОФ2 -- об успехе квантового катализа. 

Частота срабатывания ДОФ1 зависит от условий параметрического рассеяния на кристалле К2, и составляла в эксперименте от 50 до 100 кГц, что соответствует вероятности $p_1\approx10^{-3}$. 
При наличии вспомогательного фотона, частота отсчётов ДОФ2 зависит от величин сжатия ЭПР-состояния $\zeta$, пропускания канала потерь $T$, коэффициента усиления $g$, а также эффективности однофотонного детектирования $\eta$. В пределе малого сжатия, вероятность срабатывания ДОФ2 определяется выражением
\begin{equation}
\label{eqDSTp2}
p_2 = \eta \left(\dfrac{\zeta^2}{2} \times T \left(1-\dfrac{1}{g^2}\right) + \dfrac{1}{g^2} \right).
\end{equation}
Для характерных экспериментальных значений $\zeta=0.16$, $T=0.05$, $g=10$, $\eta=0.1$, получается $p_2 = 7.5\times 10^{-4}$. Более $90\%$ от этой величины составляет вклад прошедшего вспомогательного фотона - второе слагаемое в (\ref{eqDSTp2}), т.е. с хорошей точностью $p_2 = \eta/g^2$. 
Итоговая скорость генерации дистиллированных состояний есть
\begin{equation}
\label{eqDSTcount}
R_{\mathrm{dist}} = \nu \times p_1 p_2,
\end{equation}
где $\nu = 76$ МГц есть частота повторения лазерных импульсов. При вероятности успешного эксперимента $p_1 p_2 = 0.75\times10^{-6}$, имеем $R_{\mathrm{dist}} = 50$ Гц.

\subsection{Извлечение фазовой информации}
\label{sectEOM}

Процедура извлечения фаз измеренных квадратур, необходимых для гомодинной томографии ЭПР-состояния, приведена в разд. \ref{TSMStomography}. Согласно этому методу, значение суммы фаз местных волн, необходимое для томографии этого состояния, определяется по текущему уровню дисперсии суммы или разности квадратур. Оказалось, что в интересующем нас режиме больших потерь и последующего безшумового усиления, такой метод связан с дополнительными техническими трудностями. 

Средняя частота генерации дистиллированных событий (\ref{eqDSTcount}) лишь на порядок--два больше скорости воздушного дрейфа фазы ЭПР-состояния. Это означает невозможность восстановления фазовой информации (а значит, и томографии результата дистилляции), используя только дистиллированные состояния.

Фазовая информация, однако, содержится в состоянии, приходящем на гомодинные детекторы, даже если безшумовое усиление света не сработало, так как светоделение квантового катализа пропускает долю $\tau^2 \ll 1$ от исходной населённости второй моды ЭПР-состояния.  
Квадратурные измерения в этой моде проявляют остаточные корреляции с измерениями в моде 1, и могут быть использованы для извлечения фазовой информации. 

Помимо потерь при неудаче усиления, ЭПР-корреляции ослабевают также и на специально вносимой линии потерь; если последние составляют 95\%, а коэффициент усиления $g=12$, то детектора ГД2 достигает лишь $\approx1/3000$ часть первоначального сигнала 2й ЭПР-моды. Остающиеся корреляции квадратур не могли быть измерены с помощью имевшейся техники; расчёт (\ref{eqTMSRealNoise}) предсказывает, что при рабочих значениях параметров $\zeta=0.16$, $\tau_1^2 = \sqrt{1/2}$, $\tau_2 = 1/75$, значения максимума и минимума дисперсии двумодовых наблюдаемых составляют 1.016 и 1.010 от уровня вакуумного шума. 

Решить задачу восстановления фазы удалось с помощью электро-оптического модулятора (ЭОМ) -- двулучепреломляющей среды с регулируемой задержкой. Находясь в канале сигнальной моды 2, ЭОМ был настроен на периодическое переключение между двумя режимами:
\begin{enumerate}
\item[1.] Режим определения фазы. \\ В течение этого периода, ЭОМ обеспечивает максимальное отражение сигнала на ПСД10. С учётом собственных потерь, отражение составляет около $80\%$ по мощности. В этом случае, на ГД2 попадает как минимум $1/180$ от исходного сигнала моды 2;
дисперсия суммы квадратур тогда осциллирует между значениями 0.49 и 0.52, и годится для извлечения фазовой информации (разд. \ref{sect_dataSave}).

Определяемая таким образом фаза, являющаяся фазой исходного ЭПР-состояния, определяет фазу дистиллированного состояния, которая отличается от неё на $\pi/2$. Соответствующий расчёт проведён А. Улановым.

\item[2.] Режим сбора данных. \\ ЭОМ поворачивает поляризацию сигнала 2 в соответствии с необходимой величиной потерь, так что теряемый сигнал проходит через ПСД10 и теряется. В это время, поведение фазы  ЭПР-состояния определяется с помощью интерполяции данных, собранных на соседних интервалах времени. 

Для работы такой схемы необходимо, чтобы  скорость варьирования фазы с помощью пьезо-зеркала ПЗ3 значительно превышала скорость дрейфа воздуха в петле интерферометра, образуемого сигналами и их местными волнами.
\end{enumerate}

\begin{figure}[h]
	\centering
	\includegraphics[width=5in]{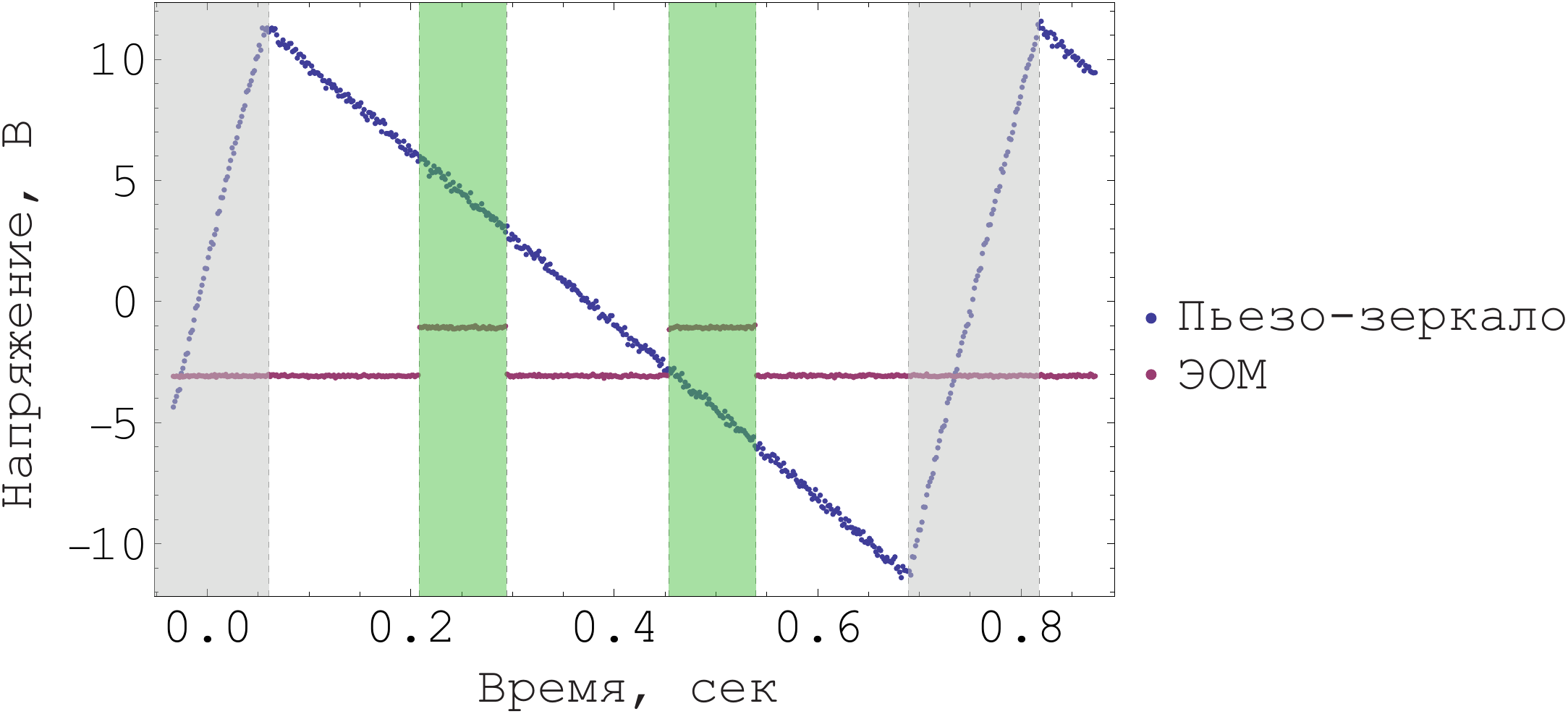}
	\caption{Напряжения, управляющие положением пьезо-зеркала ПЗ3 (синий) и величиной фазовой задержки ЭОМ (бордовый), в течение одного цикла сбора данных. Серые интервалы соответствуют обратному движению пьезо и не используются. Периоды сбора данных отмечены зелёным; остальные могут быть использованы для измерения фазы.}
	\label{pPiezoEOM}
\end{figure}
Зависимость управляющего напряжения ЭОМ от времени представляет собой ступенчатую функцию. Пример её показан на Рис. \ref{pPiezoEOM}.
Для удобства восстановления фазы (разд. \ref{chapt4_acquisition}), управляющее напряжение ЭОМ было синхронизировано с циклами напряжения, управляющего движением пьезо-зеркала ПЗ3. 
Это позволило производить все полезные измерения на участке равномерного движения зеркала. На Рис. \ref{pPiezoEOM}, этот период начинается в момент 0.06 сек и заканчивается в момент 0.69 сек. После этого, в течение $130$ мсек измерения не производятся, а зеркало возвращается в исходное положение.

\clearpage
\subsection{Сохранение данных}
\label{sect_dataSave}

\begin{figure}[h]
	\centering
	\includegraphics[width=4in]{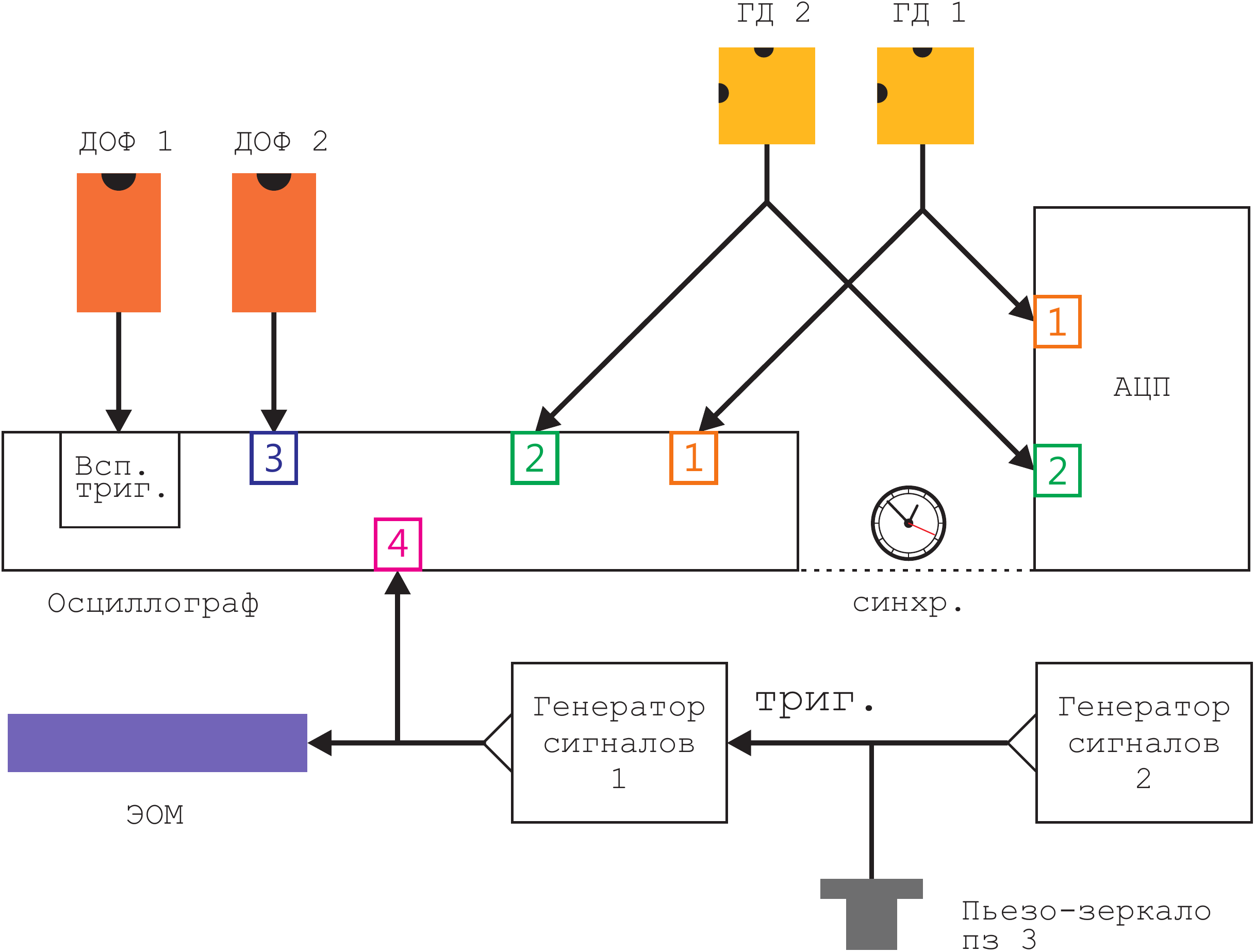}
	\caption{Система записи данных. Генератор сигналов 1: Agilent 33521B, генератор сигналов 2: BK-Precision 4010A--ND, аналого-цифровой преобразователь: Agilent U1084A, осциллограф: Agilent DSO9254A}
	\label{pDSTAcquisition}
\end{figure}

Для характеризации дистиллированного состояния, записывались и анализировались два набора данных: (1) квадратурные данные дистилированных состояний и (2) квадратурные данные всех состояний, дающие информацию о фазе в моменты дистилляции.
Для этого, использовались два записывающих устройства: 
\begin{enumerate}
\item[1.] аналого-цифровой преобразователь (АЦП) Agilent U1084A. Для него была написана программа, сохраняющая величину
\begin{equation}
\label{eqDSTcorr}
C=\left\langle \hat{Q}_1 \hat{Q}_2 \right\rangle - \left\langle \hat{Q}_1 \right\rangle \left\langle \hat{Q}_2 \right\rangle,
\end{equation}
где $\hat{Q}_{1,2}$ есть квадратурные наблюдаемые в двух модах, а $\left\langle ... \right\rangle$ обозначает усреднение по ансамблю из $5\times10^5$ последовательных измерений с периодом $16$ нс; последняя величина не должна быть близко-кратной к лазерному периоду $\nu^{-1} \approx 13.07$ нс.  

Первое слагаемое в (\ref{eqDSTcorr}) соответствует второму слагаемому правой части (\ref{eqTMSRealNoise}), и таким образом несёт в себе фазовую информацию; второе слагаемое в (\ref{eqDSTcorr}) компенсирует дрейф среднего этой величины, возникающий из-за незначительной расстройки гомодинных детекторов со временем. 

Время обработки ансамбля измерений для вычисления (\ref{eqDSTcorr}) составляет $\sim10^{-2}$ сек. Такая производительность позволяет извлекать фазовую информацию из остаточных ЭПР-корреляций, если ЭОМ находится в режиме 1 (разд. \ref{sectEOM}).

\item[2.] осциллограф Agilent DSO9254A. Позволяет сохранять серии последовательно измеренных осциллограмм. Осциллограф осуществляет запись сигналов гомодинных детекторов по тройному триггерному условию: одновременное срабатывание детекторов ДОФ1 и ДОФ2 при условии, что управляющее напряжение ЭОМ превосходит $-2$ В, что соответствует включённой линии потерь (Рис. \ref{pPiezoEOM}).

Максимальное количество квадратурных осциллограмм, сохраняемых осциллографом за один раз, равно 8192. Это гораздо меньше объёма, необходимого для достоверной томографии состояния, который составляет обычно $\sim 10^{5}$ квадратурных значений. Для автоматизированного сохранения нужного количества наборов, использовался специально созданный скрипт. Кроме удобства, последний обеспечивал запись времени сохранения каждого набора осциллограмм, необходимого для синхронизации осциллографа и АЦП.
\end{enumerate}

Полная схема системы записи данных показана на Рис. \ref{pDSTAcquisition}. Генератор сигналов 2 задаёт напряжение, управляющее пьезо-зеркалом ПЗ3. Этот же сигнал подаётся на триггерный вход генератора сигналов 1, управляющего переключением режимов ЭОМ, в результате чего достигается их синхронизм; пример рабочей настройки показан на Рис. \ref{pPiezoEOM}.

Часть сигнала генератора 1 направляется на вход осциллографа №4 и используется в качестве необходимого триггерного условия. Для ДОФ 1 и 2 используется основной вход №3 и вспомогательный триггерный канал осциллографа. 

Два оставшихся канала принимают сигналы гомодинных детекторов. Последние направляются также и на два канала АЦП, который в течение всего времени эксперимента производит измерения без триггерного условия, что обеспечило максимальную скорость сбора данных. В тех же целях, отсутствовала возможность управляемой электронной задержки между потоками данных с разных каналов АЦП; последняя обеспечивалась подбором длин проводов в двух каналах из расчёта 3 нс/метр.

\subsection{Временн\'{а}я синхронизация}

\begin{figure}[h]
	\centering
	\includegraphics[width=4.5in]{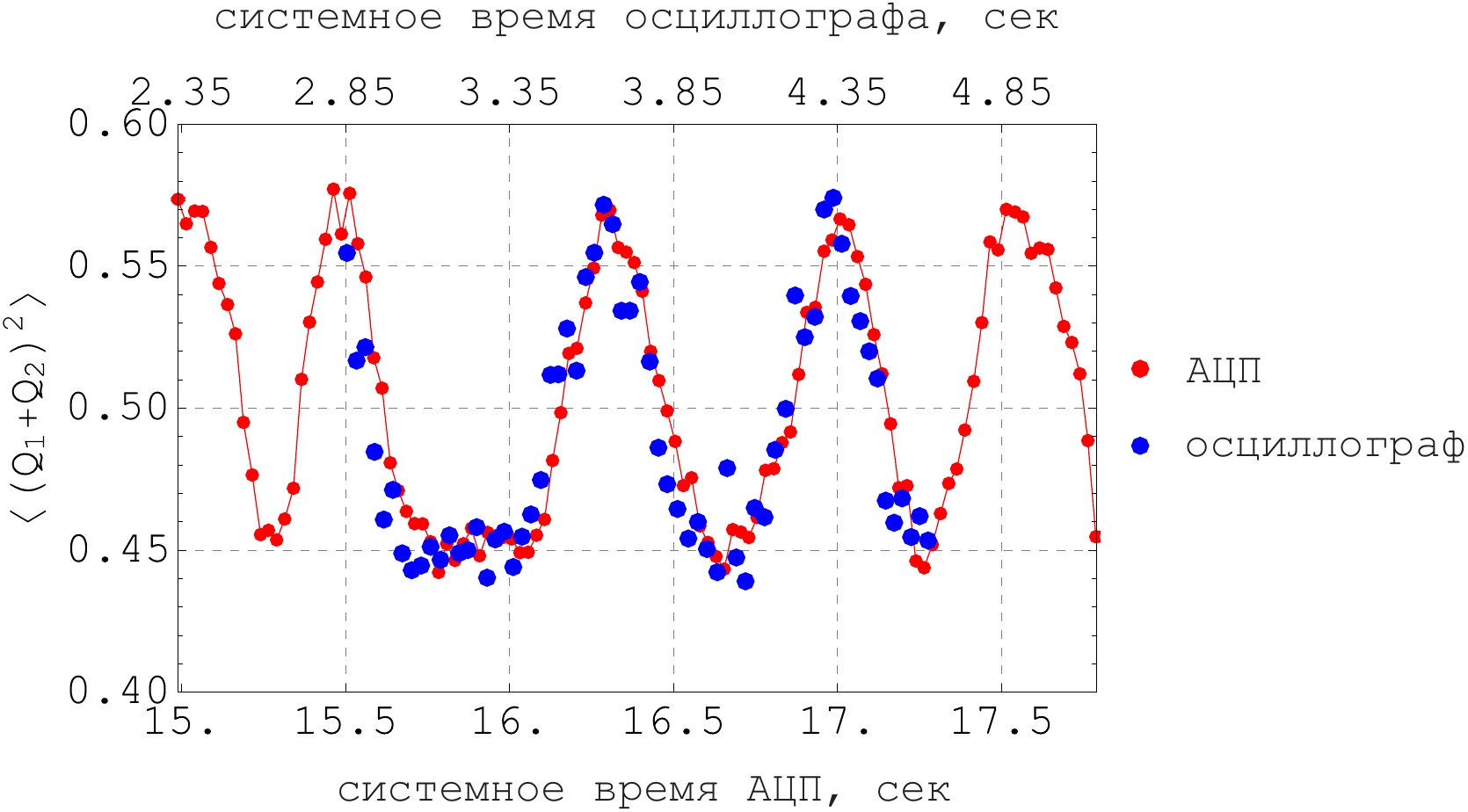}
	\caption{Синхронное детектирование ЭПР-состояния света с помощью осциллографа и АЦП. Задержка между локальными временами, установленная с помощью соединения по локальной сети, равна 12.65 сек. Кроме него, имело место дополнительное запаздывание осциллографа относительно АЦП, равное $50 \pm 5$ мкс, определяемое быстродействием сетевого протокола обмена данными и компенсировавшееся отдельно.}
	\label{pDST_synchro}
\end{figure}
Временная синхронизация измерений осциллографа и АЦП необходима для сопоставления фазовых и квадратурных данных. Для этого, они были соединены по локальной сети, что позволило определять задержку между системными временами обоих машин.
Для проверки качества синхронизации, АЦП и осциллограф запускались одновременно для измерения корреляций не ослабленного ЭПР-состояния. Рис. \ref{pDST_synchro} показывает пример такого измерения. 

\clearpage
\subsection{Обработка данных}

\begin{landscape}
\begin{figure}[]
	\centering
	\includegraphics[width=8in]{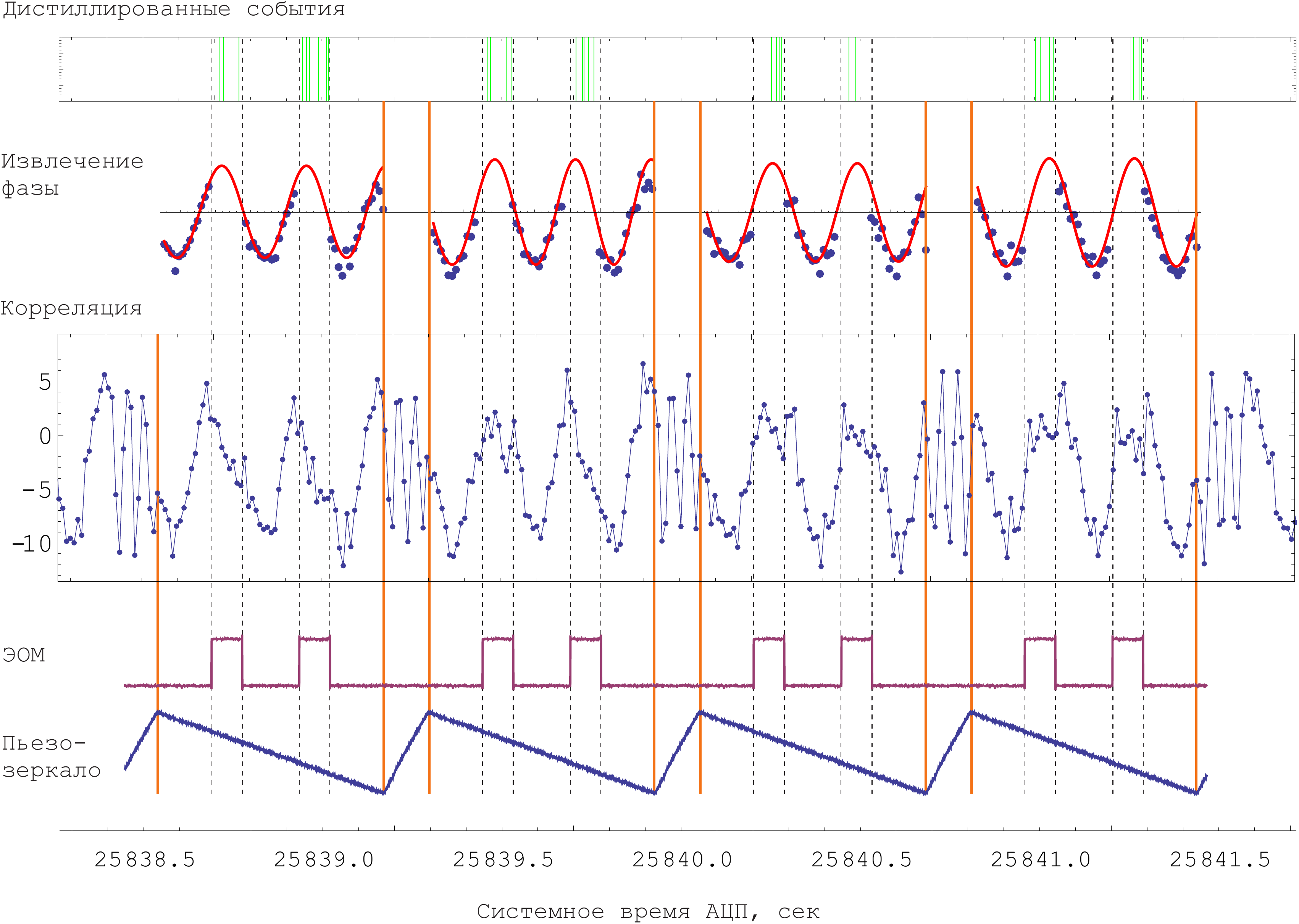}
	\caption{Временн\'{а}я раскладка четырёх циклов измерения. Нижняя панель: циклы пьезо-зеркала ПЗ3 и ЭОМ. Средняя панель: корреляция (\ref{eqDSTcorr}), сохраняемая АЦП. Выше показано извлечение фазы исходного ЭПР-состояния с помощью аппроксимации корреляционных измерений, записанных в периоды выключения потерь, гармонической функцией. Верхняя панель: моменты 34 триггерных событий.}
	\label{pDSTDAQ}
\end{figure}
\end{landscape}

Временн\'{а}я раскладка измерений показана на Рис.~\ref{pDSTDAQ}. Средняя панель показывает результат измерения коррелятора (\ref{eqDSTcorr}) в зависимости от времени АЦП. 
Периоды учащённых осцилляций на этой кривой соответствуют возвращению зеркала ПЗ3 в исходное положение; дополненное знанием порядка синхронизации пьезо-зеркала и ЭОМ (Рис.~\ref{pPiezoEOM}), это соответствие позволяет выбрать из всего набора корреляционных измерений те, которые несут информацию о фазе дистиллируемого состояния. Границы этих интервалов, совпадающие с моментами переключения напряжения ЭОМ, отмечены пунктирными линиями. При правильном наложении этой сетки, дистиллированные события, сохраняемые осциллографом, при синхронизации со временем АЦП попадают в промежутки, отмеченные пунктирными линиями, см. панель Дистиллированные события на Рис.~\ref{pDSTDAQ}.

Описанная процедура хорошо работает на небольшом объёме данных. На практике, однако, набор 8192 измерений может длиться более 10 минут, что соответствует $\sim10^3$ периодов пьезо-зеркала. Дрейф частоты пьезо-зеркала за такое время приводит в смещению сетки на значительную часть периода, что означает необходимость ручной подстройки этого смещения.
Более изощрённые способы нанесения разметки переключения ЭОМ на корреляционные измерения, однако, позволяют извлекать фазовую информацию в автоматическом режиме. К эффективным приёмам относятся:
\begin{itemize}
\item Анализ временн\'{о}й группировки дистиллированных событий. Этот метод эффективен при относительно больших скоростях дистилляции (малых коэффициентах усиления, (\ref{eqDSTcount})), так что на одно окно сбора попадает не менее 10 дистиллированных событий. В этом случае, пунктирные линии можно с хорошей точностью считать совпадающими с первыми/последними точками групп.
\item Комбинация вышеописанного метода с использованием знания о длительности интервалов ЭОМ (которые в действительности являются постоянными с большой точностью). В этом случае, группы дистиллированных событий выступают в роли якорей, на которые одевается местная сетка переключений ЭОМ.
\item Автоматическое нахождение периодов быстрых колебаний корреляции и использование их как якорных событий. Для этого, вычисляется местная дисперсия корреляционных измерений. Максимумы этой функции соответствуют обратному движению пьезо-зеркала. Метод актуален на малых скоростях сбора (больших коэффициентах усиления).
\end{itemize}

Корреляционные данные используются для извлечения информации о фазе дистиллированного состояния. Для этого, результаты корреляционных измерений, выбранные в соответствии с границами интервалов низкого напряжения на ЭОМ и границами разворота пьезо-зеркала (пунктирные и оранжевые линии на Рис.~\ref{pDSTDAQ}), аппроксимируются функцией
\begin{equation}
	\label{eqDSTvarsFit}
	a \cos\left[\phi(t-t_0)\right], \quad \phi(t) = x t^2 + y t + z,
\end{equation}
где $t_0$ есть момент начала цикла данного измерения (ближайшая оранжевая линия слева), а $a$, $x$, $y$, $z$ -- оптимизируемые параметры. В случае быстрого линейного движения пьезо-зеркала, воздушные флуктуации малы, и фаза, меняющаяся линейно со временем, легко аппроксимируется; одновременно, снижается количество корреляционных измерений за период пьезо, что приводит к увеличению случайной ошибки. Оптимальная скорость подбиралась эмпирически.  Выбранный режим является промежуточным: вклад воздушных флуктуаций не-пренебрежимо мал, однако отклонение фазы от линейного закона может быть учтено квадратичным членом в (\ref{eqDSTvarsFit}).

Примеры аппроксимирующей зависимости (\ref{eqDSTvarsFit}) показаны на панели Извлечение фазы Рис.~\ref{pDSTDAQ} красными линиями. Каждому дистиллированному событию, произошедшему в момент времени $t_i$, в соответствии с (\ref{eqTMSRealNoise}), присваивается значение суммы фаз местных волн по правилу
\begin{equation}
\label{eqDSTphase}
\theta_1 + \theta_2 = \mathrm{Mod}\left[2\pi, \phi(t_i-t_0)\right].
\end{equation}

Квадратурные данные дистиллированных событий извлекаются из сохранённых осциллографом осциллограмм, как описано в разд. \ref{BHD_daq}. Дисперсия квадратуры в триггерном импульсе, содержащем дистиллированное состояние (\ref{eqDST_result}), на $\sim15\%$ превышает дисперсию вакуума.
Таким образом, записывается набор квадратурных данных $\{Q_1, Q_2, \theta_1 + \theta_2\}_i$, достаточный для характеризации дистиллированного состояния (разд. \ref{sect3_resultTMS}).

\section{Результаты}

\begin{figure}[h]
	\centering
	\includegraphics[width=\textwidth]{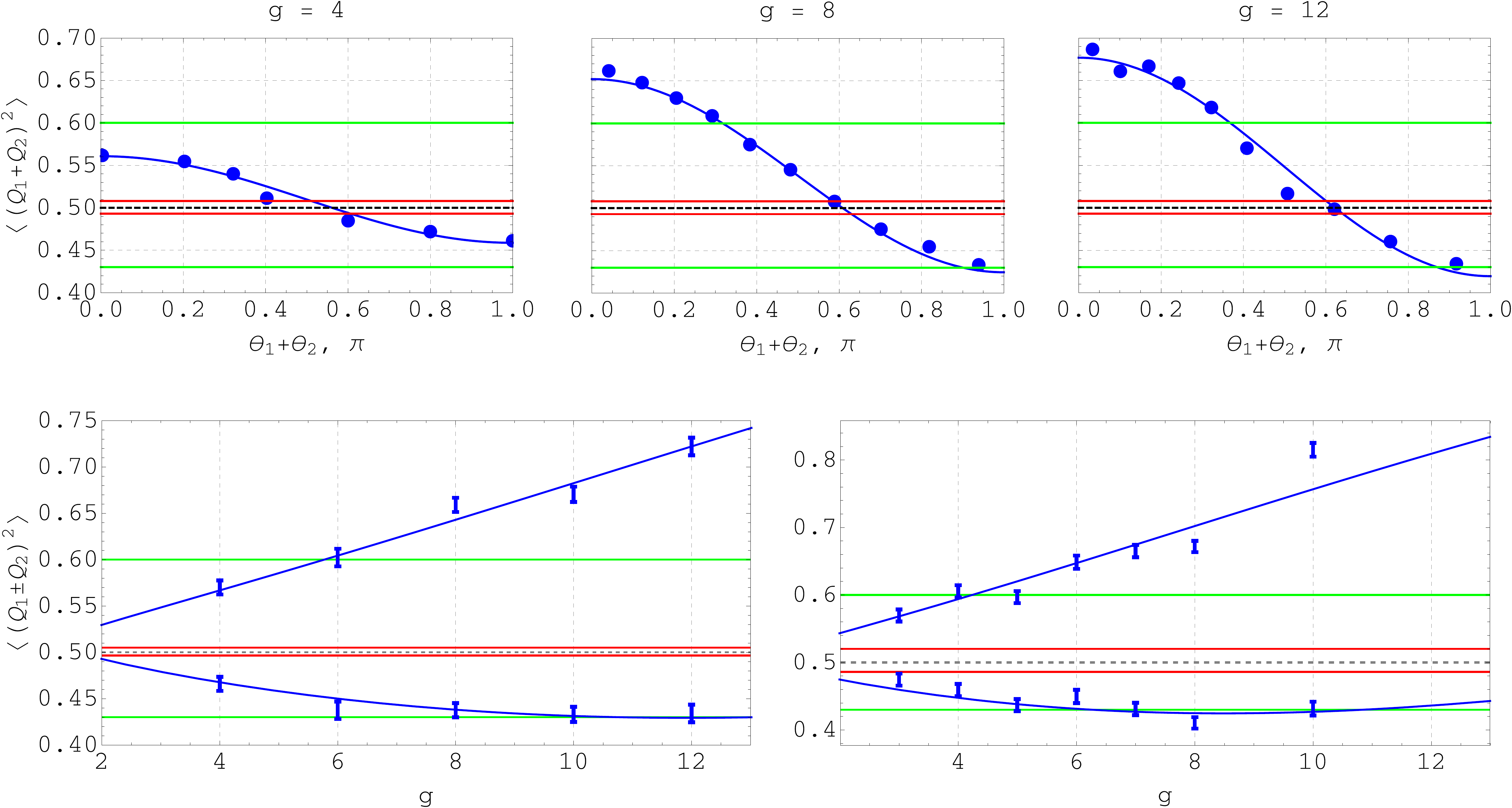}
	\caption{
		Зелёные / красные линии: уровень сжатия и антисжатия исходного состояния / состояния перед дистилляцией. Чёрный пунктир: уровень вакуумного шума.
		Верхние панели: дисперсия суммы квадратур в зависимости от суммы фаз местных волн для состояний, испытавших 20ти--кратное ослабление и дистиллированных при усилениях $g=4$, 8, и $12$. Точки: экспериментальные данные. Кривые: теоретическое ожидание (\ref{eqTMSRealNoise}) для $\zeta=0.19$, 0.35, 0.39 при эффективностях 26, 30, и 30\% соответственно. \\
		Нижние панели: максимальное сжатие и анти-сжатие дистиллированного состояния в зависимости от коэффициента усиления $g$, для ослабления исходного ЭПР-состояния на 95\%, слева и 80\%, справа. Точки: экспериментальные результаты. Кривые: результат расчёта, выполненного А. Улановым.}
	\label{pDSTallvars}
\end{figure}

Экспериментальные результаты показаны на Рис. \ref{pDSTallvars}. Верхние панели показывают дисперсию двумодовой наблюдаемой (\ref{eqTMobservables}) в зависимости от суммы фаз $\theta_1+\theta_2$ для разных коэффициентов усиления в эксперименте, когда одна из мод исходного ЭПР-состояния была ослаблена в 20 раз. Для получения этой зависимости, набор квадратурных данных $\{Q_1, Q_2, \theta_1 + \theta_2\}_i$ разбивался на 10 групп, отсортированных по значению третьей колонки; каждая группа содержит $\sim 10^4$ квадратурных измерений и соответствует одной точке на верхних панелях Рис. \ref{pDSTallvars}.
Уровни сжатия и анти-сжатия состояния перед дистилляцией, а также исходного состояния, показаны красными и зелёными линиями. Видно, что уровень сжатия дистиллированного состояния при высоких коэффициентах усиления близок к исходному состоянию.

Значения максимальной и минимальной дисперсий с верхних панелей Рис. \ref{pDSTallvars}, дополненные результатами дистилляции при $g=6$ и 10, сведены на левой панели нижнего ряда. Правая панель показывает результаты другой серии экспериментов, когда то же самое исходное состояние было ослаблено на 80\%. Как и для 95\%, величину сжатия удалось вернуть на уровень исходного состояния.

\begin{figure}[h]
	\centering
	\includegraphics[width=\textwidth]{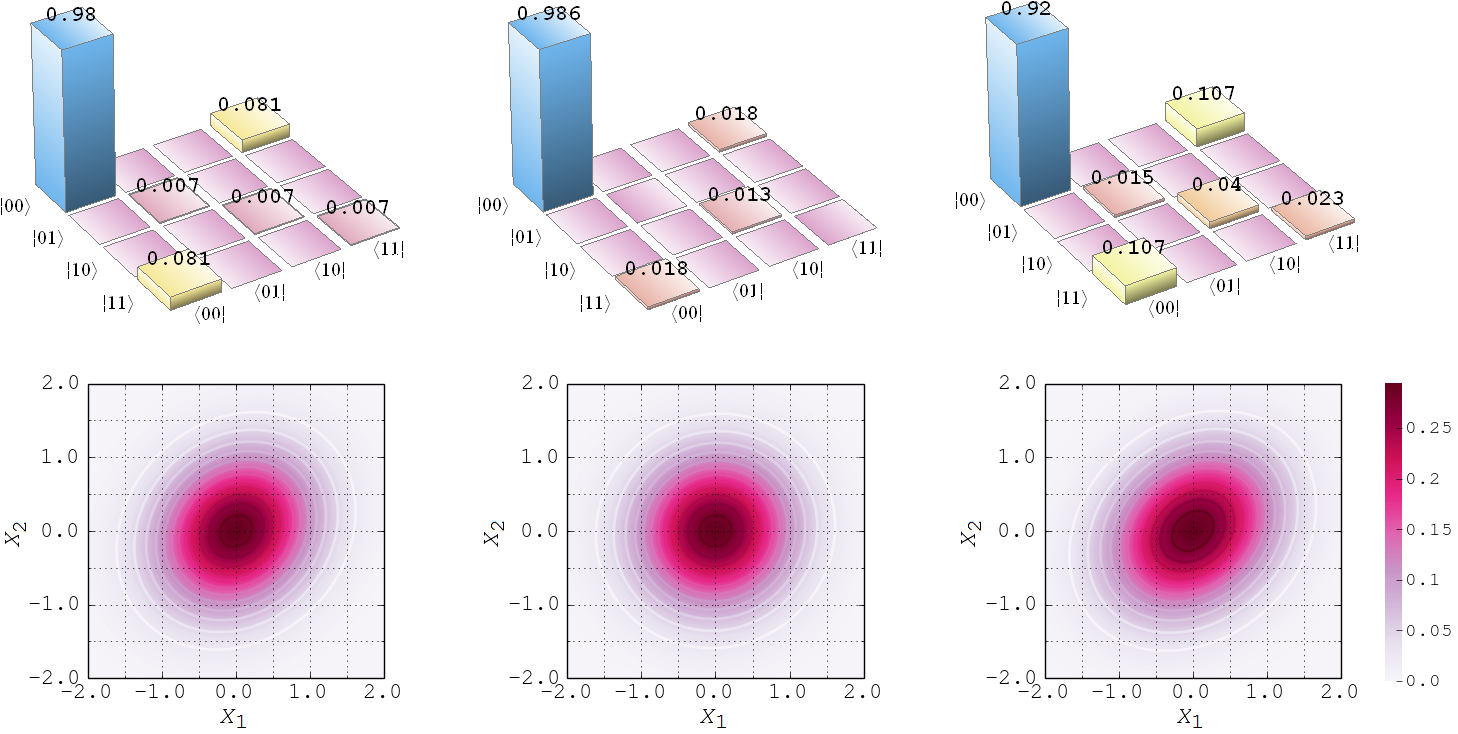}
	\caption{Матрицы плотности и квадраты амплитуд волновых функций для исходного ЭПР-состояния (слева), для того же состояния, прошедшего через 20ти--кратные потери (по центру), и для результата дистилляции при $g=12$ (справа). Параметр верности (\ref{eqStateFidelity}) между исходным и дистиллированным состояниями равен 97\%.}
	\label{pDSTdmswf}
\end{figure}

Состояние (\ref{eqAttTMSDM}), ожидаемое на выходе из ``дистиллятора'', не является в точности гауссовым, так как содержит только вакуумную и однофотонную компоненты; этим объясняется избыточное, по сравнению с исходным, анти-сжатие состояния, дистиллированного при $g=8...12$. Тем не менее, для использовавшихся величин сжатия и потерь, состояние (\ref{eqAttTMSDM}) демонстрируют фазовое поведение, характерное для двумодово-сжатого вакуума, что подтверждается в эксперименте. 

Рис.~\ref{pDSTdmswf} показывает матрицы плотности в фоковском базисе и волновые функции в координатном базисе для состояний: исходного, потерпевшего потери и дистиллированного -- слева направо. Уровень сжатия состояния после потерь настолько мал, что экспериментальная томография не позволяет отличить его от вакуума (см. красную марку на Рис.~\ref{pAttNegativity} и обсуждение в разд. \ref{sectEOM}). По этой причине, в средней колонке представлено теоретическое ожидание.

\vspace{3em}

\section{Выводы}
\label{distillation_outlook}

Процедура дистилляции, по существу, представляет собой метод отбора наиболее запутанных состояний из слабо запутанного ансамбля; чем больше требуемый уровень запутанности, тем меньшее количество состояний проходит отбор. Представленный метод позволил выбрать из ансамбля состояний, показанных красным на Рис.~\ref{pAttNegativity}, состояния, соответствующие зелёной метке, что соответствует эксперименту с уровнем потерь 95\%. Логарифмическая отрицательность  состояния (\ref{eqNeg}) при этом увеличилась с 0.037 до 0.24, т.е. в 6.5 раз. 

Сравнение с другими методами дистилляции запутанности \cite{Bartley2015} показывает, что разработанная нами техника особенно эффективна в случаях слабого исходного сжатия и/или высоких потерь в канале пропускания. В отличие от дистилляции запутанности с помощью пост-селекции в ослабленной моде \cite{Chrzanowski2010}, представленный метод позволяет получить свободно распространяющееся дистиллированное состояние. При этом, метод позволяет восстановить запутанность после прохождения сколь угодно высоких потерь. Это достижение является значительным шагом в сторону практической реализации квантового протокола коррекции ошибок, а в конечном итоге -- повторителя квантовой информации в непрерывных переменных \cite{Ralph2011}.

В силу однофотонной природы операции квантового катализа, дистиллированное состояние (\ref{eqDST_result}) не содержит многофотонных компонент. Это обстоятельство ограничивает достижимые значения запутанности и сжатия. Можно предложить ряд способов преодолеть это препятствие. В частности, не-гауссово состояние (\ref{eqDST_result}) может быть далее подвергнуто гауссифицирующей процедуре дистилляции \cite{Eisert2004, Datta2012}. Теоретически, беcконечно-сжатое ЭПР-состояние может быть получено в результате.

По результатам исследования опубликована журнальная статья: \\
\textit{Alexander E. Ulanov, Ilya A. Fedorov, Anastasia A. Pushkina, Yury V. Kurochkin, Timothy C. Ralph and A. I. Lvovsky. Undoing the effect of loss on quantum entanglement. Nature Photonics 9, 764–768 (2015).} \cite{Ulanov2015}. \\

Результаты работы представлены на конференциях:
\begin{enumerate}
	\item [1.] 14th International Conference on Squeezed States and Uncertainty Relations, Гданьск, Польша, 29 июня -- 3 июля 2015
	\item [2.] 22nd Central European Workshop on Quantum Optics, Варшава, Польша, 6--10 Июля 2015
	\item [3.] Third International Conference on Quantum Technologies, Москва, Россия, 13--17 июля 2015
	\item [4.] XII International Workshop on Quantum Optics, Троицк, Россия,  11--15 августа 2015
	\item [5.] 58 научная конференция МФТИ, Москва, Россия, 23--28 ноября 2015
\end{enumerate}

\clearpage

\chapter{Томография многомодовых квантовых процессов} \label{chapt1}


Детальное понимание функционирования индивидуальных квантовых систем необходимо для создания более сложных, многокомпонентных схем. На основе этой потребности ставится задача об экспериментальной характеризации неизвестного квантового процесса как ``чёрного ящика'': предсказать его отклик на произвольное входное квантовое состояние. Множество техник по решению этой задачи известны под общим названием ``томография квантовых процессов'' (QPT) \cite{Chen2010, Banazhek2013}.

Непосредственный подход к QPT состоит в измерении отклика процесса на набор состояний, образующий базис на рассматриваемом гильбертовом пространстве. Поскольку любой квантовый процесс есть линейное отображение между входным и выходным операторами плотности, этой информации достаточно для полной характеризации процесса \cite{Poyatos1997}. Применимость такого метода, по существу, ограничена процессами, преобразующими простейшие квантовые состояния -- кубиты, так как работа на пространствах большей размерности требовала бы приготовления крайне трудно-достижимых, экзотических пробных состояний.
Другой метод \cite{D'Ariano2001} предполагает использование пробного состояния, являющегося частью запутанного состояния на пространстве, охватывающем входные и выходные моды процесса. В этом случае, благодаря изоморфизму Ямилковского \cite{Jamiokowski1972}, для характеризации чёрного ящика достаточно единственного пробного состояния, однако его приготовление на практике также возможно лишь в простейших случаях.

Для оптических систем, эффективное решение поставленной задачи представляет собой QPT на когерентных состояниях \cite{Lobino2008, Lobino2009, Keshari2011} (csQPT). Основываясь на теореме оптической эквивалентности \cite{Sudarshan1963, Glauber1963b}, этот метод требует лишь когерентных состояний для испытания неизвестного квантового процесса. csQPT отличается простотой в экспериментальной части, однако применение этого метода во многих случаях затруднено обобщённой природой функций Глаубера-Сударшана. Кроме того, результат характеризации процесса методом \cite{Lobino2008, Lobino2009, Keshari2011} может оказаться нефизичным, что связано с индивидуальным восстановлением каждого элемента тензора процесса.
Последний недостаток отсутствует в модификации того же метода, известной как MaxLik csQPT. В ней используется изоморфизм между тензором процесса и оператором плотности на произведении входного и выходного пространств, что позволяет свести QPT к хорошо известной задаче о восстановлении квантового состояния, откуда заимствуется техника оценки тензора процесса с помощью реконструкции Максимального Правдоподобия (MaxLik) \cite{Hradil2004}. Последняя позволяет проводить реконструкцию процесса, не покидая физически осмысленного пространства. MaxLik csQPT был предложен в работе \cite{Anis2012} и успешно применён к недетерминистическому одномодовому процессу \cite{Kumar2013}. 

В этой главе, мы расширяем метод MaxLik csQPT за пределы случая ``один вход -- один выход",  охватывающий лишь малую часть практически важных квантовых процессов. Необходимость этого исследования продиктована растущим объёмом квантовых технологий, многие из которых являются многомодовыми; к последним всецело относятся нелокальные методы манипуляции квантовыми состояниями (разд. \ref{introduction}), которые основаны на использовании корреляций между двумя и более подсистемами общего запутанного квантового состояния света. Важнейшую часть многомодовых процессов составляют логические операции для обработки квантовой информации \cite{Monroe, Pittman2003}.

Описанная в этой главе экспериментальная техника реализована на оптической схеме. Однако, область применимости теории и методологии представленного метода шире. Фактически, он применим для любой физической системы, изоморфной гармоническому осциллятору -- например, сверхпроводящим резонаторам, атомным спиновым ансамблям и наномеханическим системам \cite{Xiang2013, Aspelmeyer2014}. Во всех них, когерентные состояния приготавливаются наиболее просто и, следовательно, наилучшим образом подходят для QPT.

Для демонстрации возможностей метода был выбран процесс светоделения, имеющий первостепенную важность для квантовой оптики: любой линейно-оптические элемент -- интерферометр, волоконное соединение, канал потерь, и т.д. -- может быть представлен в виде комбинации светоделителей \cite{Bouland2014}. Дополненный источниками и детекторами одиночных фотонов, набор светоделителей позволяет реализовать схему квантовых вычислений \cite{Knill2001}; в некоторой форме, светоделитель присутствует едва ли не в любой оптической установке. Более того, гамильтониан процесса светоделения играет ключевую роль при описании квантово-информационных интерфейсов между системами эквивалентными гармоническому осциллятору, например между электромагнитном полем, атомным ансамблем и наноразмерным механическим осциллятором \cite{Hammerer2010, Aspelmeyer2014}. Интерес представляет также процесс светоделения в нелинейных средах \cite{Belinsky2014}.

Действие светоделителя согласуется с классической физикой: когерентные состояния на входе дают когерентные состояния на выходе; вместе с этим, его отклик на неклассические состояния света носит квантовый характер. Ярким примером является эффект Хонг-Оу-Манделя: при подаче на каждый из входов симметричного светоделителя пары фотонов, они всегда появляется вместе в одной из выходных мод, тогда как вторая находится в вакуумном состоянии \cite{Hong1987}. Представляемый метод позволяет реконструировать этот эффект несмотря на то, что в измерениях используются лишь классические состояния света.

В настоящее время, известно о характеризации светоделителя в роли фильтра состояний Белла \cite{Mitchell2003} и канала потерь \cite{Bongioanni2010}. В обоих случаях, томография светоделения как процесса является неполной в силу ограниченности этих методов на специфических подпространствах входных состояний. Представляемый нами метод лишён этого недостатка: он позволяет предсказать результат процесса для любых фоковских состояний или их суперпозиций, вплоть до некоторого фотонного числа отсечки $N$. В настоящем эксперименте использовалось $N=4$, что обусловлено лишь доступной вычислительной мощностью.

В отличие от методов \cite{Rahimi-Keshari2013, Baldwin2014}, представленная в данной работе техника не требует априорных знаний об устройстве чёрного ящика, в частности о его линейно-оптическом либо унитарном характере. Несмотря на то, что для тестирования используется линейный, унитарный процесс, описанный ниже метод может быть применён к квантовому процессу любой природы.
\clearpage

\section{Теоретическое описание} \label{sect1_2}

Наш метод расширяет одномодовый вариант MaxLik csQPT \cite{Anis2012} на произвольное число входных и выходных мод. В этом случае, обобщённый $M$--модовый квантовый процесс $\mathcal{E}$ представляется с помощью тензора $4M$--ранга, отображающего матрицу плотности входного состояния $\hat{\rho}^{\rm in}$ на выходную $\hat{\rho}^{\rm out}$:
\begin{equation}
	\label{eq11}
	\hat{\rho}^{\rm out}_{\underline{j},\underline{k}} = \langle{\underline{j}|\mathcal{E}(\hat{\rho}^{\rm in})|\underline{k}}\rangle = \sum_{\underline{n},\underline{m}}{\mathcal{E}^{\underline{n},\underline{m}}_{\underline{j},\underline{k}}\hat{\rho}^{\rm in}_{\underline{n},\underline{m}}},
\end{equation}
где индексы $|\underline{i}\rangle=|i_1,\dots,i_M\rangle$ нумеруют базисные вектора многомодового состояния. На практике, безконечномерное гильбертово пространство входных и выходных мод процесса ограничивается до конечной размерности. 

Широко распространённым является фоковский базис, т.е. базис собственных состояний оператора числа квантов системы. Именно этот базис выбран для демонстрации метода. Мы ограничиваем рассмотрение процесса до области гильбертова пространства, охватывающего $N+1$ нижних фоковских состояний, так что $i_k \in 0\dots N$. Физически, это соответствует ограничению входных и выходных состояний по энергии величиной $N \times \hbar \omega$, где $\hbar\omega$ есть энергия кванта рассматриваемой системы.

В эксперименте, чёрный ящик испытывается набором когерентных состояний $|{\underline\alpha}\rangle=|\alpha_1,\dots,\alpha_M\rangle$, который должен покрывать выбранную область входного гильбертова пространства. Эти состояния должны также помещаться в этой области, т.е. быть достаточно хорошо представляемы в выбранном ограниченном базисе. Поскольку дисперсия квадратуры фоковского состояния $\ket{N}$ равна $N+1$, объём области покрытия $M$--модового гильбертова пространства соответствует гиперсфере размерности $M$ с радиусом $\sqrt{N+1}$. При этом, каждое пробное когерентное состояние соответствует гиперсфере радиусом $\sqrt{1/2}$. Таким образом, число пробных состояний, необходимое для характеризации рассматриваемого процесса, можно оценить как $(2N+1)^M$.

Представляемый метод реконструкции опирается на изоморфизм Ямилковского \cite{Jamiokowski1972}, который любому супероператору (оператору операторов) $\mathcal{E}$ ставит в соответствие оператор, заданный на произведении входного и выходного пространств $\mathcal{H}$ и $\mathcal{K}$:
\begin{equation}
\label{eq12}
\hat{E} = \sum_{\underline{n},\underline{m},\underline{j},\underline{k}}{
	\mathcal{E}^{\underline{n},\underline{m}}_{\underline{j},\underline{k}} \ket{\underline{n}} \bra{\underline{m}} \otimes \ket{\underline{j}} \bra{\underline{k}}},
\end{equation}
что позволяет свести задачу о характеризации чёрного ящика к задаче о характеризации квантового состояния \cite{Chen2010, Banazhek2013, Lvovsky2009}.
Физичность процесса $\mathcal{E}$ накладывает на оператор $\hat{E}$ ряд ограничений:
\begin{enumerate}
	\item[1.] Процесс является полностью положительным. \newline
	Для всех физически возможных входных состояний (т.е. таких, собственные значения матрицы плотности которых неотрицательны), выходное состояние также является физичным. Это эквивалентно положительной полуопределённости $\hat{E}$.
	\item [2.] Процесс сохраняет след состояния. \newline
	Для всех возможных входных состояний $\hat{\rho}_{\rm in}$ должно выполняться $\mathrm{Tr} \left[\hat{\rho}_{\rm out}\right] = \mathrm{Tr} \left[\hat{\rho}_{\rm in}\right]$. Это эквивалентно $\mathrm{Tr}_\mathcal{K} [\hat{E}] = \hat{I}_\mathcal{H}$, где $\hat{I}$ -- единичный оператор. Для неунитарных процессов, $\mathrm{Tr} \left[\hat{\rho}_{\rm out}\right]/\mathrm{Tr} \left[\hat{\rho}_{\rm in}\right] < 1$ есть вероятность реализации процесса. В этом случае, требование сохранения следа сохраняется посредством расширения выходного пространства $\mathcal{K}_{\rm total} = \mathcal{K} \otimes \mathcal{K}_{\rm fail}$, где $\mathcal{K}_{\rm fail}$ охватывается единственным фиктивным состоянием $\ket{\emptyset}$, которое соответствует не-реализации процесса.
\end{enumerate}

Для каждого входного состояния, выходные каналы подвергаются гомодинному детектированию. Это даёт набор квадратурных данных $\{\underline X_i, \underline\theta_i\}$, где ${\underline\theta}_i=(\theta_{i1},\dots,\theta_{iM})$ есть фазы $M$ локальных осцилляторов в момент $i$--го измерения.

Оптимизация правдоподобия MaxLik состоит в нахождении оператора $\hat{E}$, максимизирующего логарифмическую вероятность реализации собранного набора данных $\{\underline X_i, \underline\theta_i\}$. Математически, это эквивалентно максимизации функционала
\begin{equation}
\label{eqLik}
\mathcal{L}(\hat{E}) = \sum\limits_{j,i} \ln p(\alpha_j, i) - \mathrm{Tr}[\hat{\Lambda}\hat{E}],
\end{equation}
где $\Lambda$ есть эрмитова матрица множителей Лагранжа, отвечающая за выполнение условия сохранения следа, а
\begin{equation}
p(\alpha, i) =
\mathrm{Tr} \left[\mathcal{E} (\ket{\alpha}) \hat{\Pi}_{{\underline\theta}_i}(\underline X_i)\right] =
\mathrm{Tr}
\left[\hat{E} \ket{\alpha}\bra{\alpha} \otimes \hat{\Pi}_{{\underline\theta}_i}(\underline X_i) \right]
\end{equation}
есть вероятность регистрации $i$--го измерения для состояния $\ket{\alpha}$. $\hat{\Pi}_{\underline\theta_i}(\underline X_i)$ есть проектор, соответствующий $i$--му квадратурному состоянию. В случае неунитарного процесса, подпространство $\mathcal{K}_{\rm fail}$ не охватывается квадратурными проекторами; вероятность неудачи $\ket{\emptyset}$ должна быть определена отдельно.

Для унитарных (или детерминистических) процессов оператор $\hat{E}$, максимизирующий правдоподобие (\ref{eqLik}), удовлетворяет экстремальному условию \cite{Anis2012, Hradil2004}
\begin{equation}
\label{eq5}
\hat{E} = \hat{\Lambda}^{-1} \hat{R} \hat{E} \hat{R} \hat{\Lambda}^{-1},
\end{equation}
где
\begin{eqnarray}
\label{eq6}
\hat{R} = \sum\limits_{i,j} \frac{\ket{\alpha_j^*}\bra{\alpha_j^*}\otimes \hat{\Pi}_{\underline\theta_i}(\underline X_i)}{p(\alpha_j, i)}, \\
\hat{\Lambda} = \left(\mathrm{Tr}_\mathcal{K} \left[\hat{R}\hat{E}\hat{R}\right] \right)^{1/2} \otimes \hat{\mathcal{I}}_\mathcal{K}.
\label{eq7}
\end{eqnarray}
Уравнения (\ref{eq5})--(\ref{eq7}) могут быть решены итеративно, начиная с нейтральной начальной точки $\hat{E}^{(0)}{=}\hat{\mathcal{I}}_{\mathcal{H}\otimes\mathcal{K}}/\mathrm{dim} \left[\mathcal{K}\right]$. На каждой итерации для текущего оператора $\hat{E}$ согласно (\ref{eq7}) вычисляется $\hat{\Lambda}$, после чего значение $\hat{E}$ обновляется, следуя (\ref{eq5}).

Эрмитовость операторов $\hat{R}$ и $\hat{\Lambda}$ гарантирует, что $\hat{E}$ остаётся положительно полуопределённым на каждой итерации. Вместе с условием сохранения следа, это обеспечивает физичность процесса реконструкции. Функционал правдоподобия (\ref{eqLik}) является выпуклым на пространстве положительно полуопределённых операторов, что исключает возможность схождения итеративного процесса к локальному минимуму.
Размерность описанной оптимизации растёт экспоненциально с размерностью фоковского подпространства $N+1$ и числом мод $M$; однако, итеративный алгоритм допускает эффективное распараллеливание, что значительно снижает вычислительное время.

Использование когерентных состояний для изучения чёрного ящика обеспечивает масштабируемость описываемой техники в экспериментальной части. Как описано ниже, QPT с использованием ста когерентных состояний так же проста (или сложна) для практической реализации, как и с использованием тысячи.

\clearpage

\section{Эксперимент} \label{sect1_4}

\begin{figure}[h]
\centering
\includegraphics[width=5in]{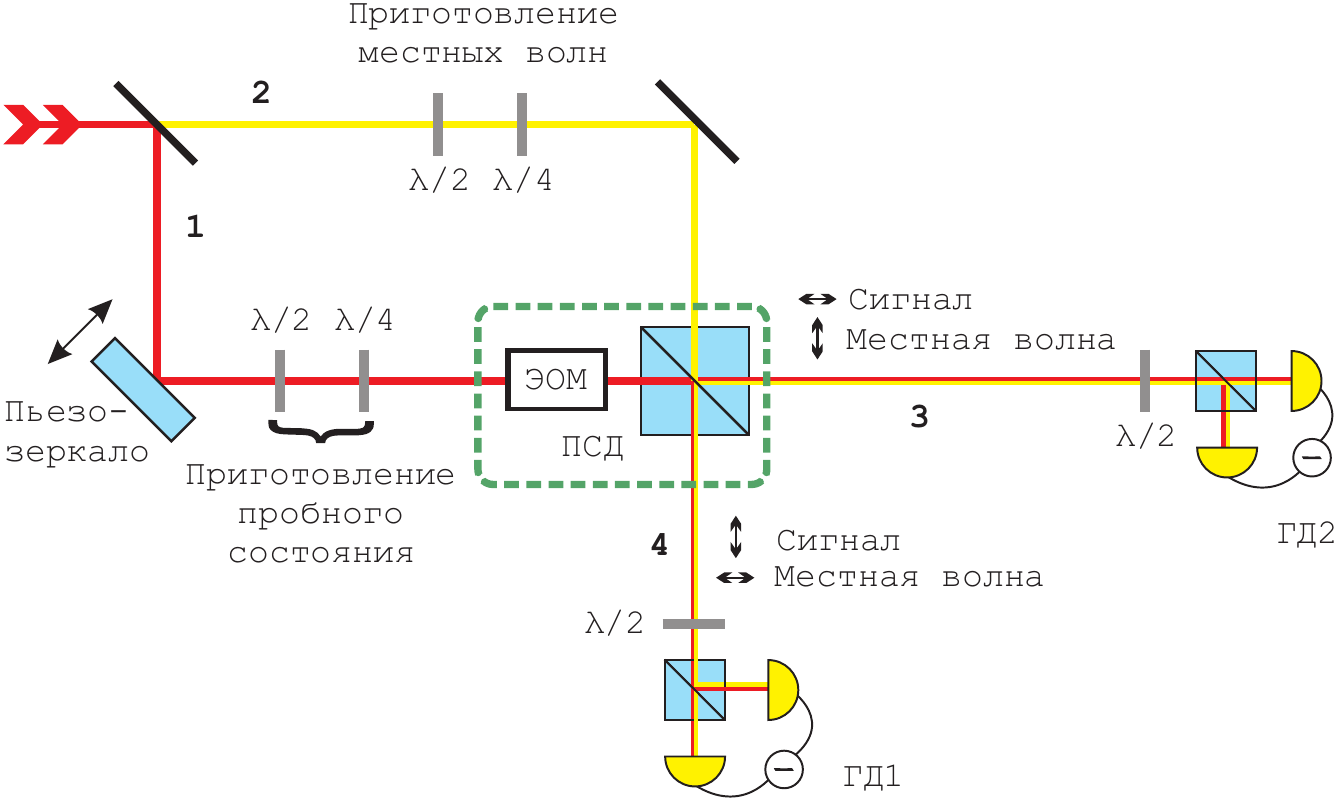}
\caption{Принципиальная схема эксперимента. В роли чёрного ящика выступает процесс светоделения, реализованный в поляризационном базисе с помощью электро-оптического модулятора (ЭОМ) и поляризационного светоделителя (ПСД). Входные каналы процесса являются вертикальной и горизонтальной компонентами пространственной моды 1; выходные каналы разделены на пространственные моды 3 (горизонтальная поляризация) и 4 (вертикальная). Местные волны, показанные жёлтым, необходимы для гомодинного детектирования и подаются на ПСД в поляризационных модах пучка 2. После светоделения, они появляются в пространственных модах 3 и 4 с поляризациями, ортогональными сигналу.}
\label{p11}
\end{figure}

Моды 1 и 2, подвергающиеся светоделению, имеют горизонтальную и вертикальную поляризации, разделяя при этом один пространственный путь, отмеченный 1 на Рис.~\ref{p11}. Такое устройство позволяет избежать фазовых флуктуаций между модами, которые появляются, когда пространственно-разнесённые моды приобретают относительные задержки из-за различных плотностей воздуха. Состояния в модах 1 и 2 при этом определяются положениями волновых пластин в пути 1; для доступа ко всем возможным двумодовым состояниям достаточно комбинации полуволновой и четверть-волновой пластин. 

Процесс светоделения реализован с помощью комбинации электро-оптического модулятора (ЭОМ) и поляризационного светоделителя (ПСД). Оптическая ось ЭОМ направлена под углом $45^\circ$ от горизонтали. ПСД пространственно разделяет выходные поляризационные моды для детектирования, так что горизонтально-поляризованная компонента света проходит, а вертикально поляризованная -- отражается. К ЭОМу приложено четверть-волновое напряжение, так что поляризация на выходе из него становится круговой; мощность при этом поровну разделена между вертикальной и горизонтальной компонентами. В картине Гейзенберга, описанный процесс соответствует следующему преобразованию операторов уничтожения фотонов $\hat{a}_1 $ и $\hat{a}_2$:
\begin{equation}
\label{eq13}
\left[ {\begin{array}{c}
\hat{a}_1^{\mathrm{out}} \\
\hat{a}_2^{\mathrm{out}} \\
\end{array} } \right]
=
\dfrac{1}{2}
\left[ {\begin{array}{cc}
1 + i & 1 - i \\
1 - i & 1 + i \\
\end{array} } \right]
\left[ {\begin{array}{c}
\hat{a}_1^{\mathrm{in}} \\
\hat{a}_2^{\mathrm{in}} \\
\end{array} } \right].
\end{equation}
Относительные амплитуды и фазы входных когерентных состояний регулируются, используя пару волновых пластинок $\lambda/2 + \lambda/4$.

Квадратурные измерения на выходе процесса производятся с помощью балансных гомодинных детекторов (ГД) в обоих выходных модах (разд. \ref{BHD_all}). Пара местных волн приготавливается в ортогональных поляризациях, так что центральный светоделитель направлет их в пространственные моды выходных сигналов (Рис.~\ref{p11}).
После этого, в каждом из выходных каналов процесса присутствуют сигнал и местная волна в ортогональных поляризациях; их интерференция осуществляется с помощью полуволновой пластинки ориентированной под углом $25^\circ$ к горизонтали, и последующего ПСД. 

Набор результатов измерения квадратур $\{ Q_1, ... Q_M \}$, полученный с гомодинных детекторов, должен быть дополнен знанием фазы каждой из них в соответствующий момент времени (разд. \ref{BHD_theory}). Так как фаза измеряемой квадратуры зависит от фазы местной волны и сигнала (\ref{eqQuadSig}), а начало отсчёта фаз можно выбрать произвольно, то знание фаз $\{ \theta_1, ... \theta_M \}$ эквивалентно знанию $2M-1$ фазовых соотношений. Это требование составляет основное различие между томографией одномодового и многомодового процессов. Фазовое поведение многих одномодовых процессов не зависит от интенсивности сигнала, что позволяет не измерять сдвиг фаз между входной и выходной модами. Для многомодовых процессов, напротив, это во многих случаях не так: даже характеризация относительно простого процесс светоделения невозможна без полного контроля над фазами.

Для получения этой информации, напряжение ЭОМ периодически изменяется так, что чёрный ящик становится тождественным процессом; в этом случае гомодинные измерения соответствуют входным состояниям. 
Переключение между этими режимами производится с периодом 0.1с, что много быстрее воздушных флуктуаций в каналах 1 и 2. Благодаря этому, поведение фаз измеряемых квадратур, определённое в период тождественного процесса, может экстраполировано на период измерения чёрного ящика.

\subsection{Настройка установки}
\label{alignment_1}
Главной частью настройки установки является настройка управляющего напряжения ЭОМ. Последнее должно иметь форму ступенчатого сигнала, два уровня которого соответствуют процессам симметричного светоделения и тождественности. 
Для юстировки этих уровней использовались следующие критерии:
\begin{itemize}
\item в режиме тождественности, ЭОМ не должен возмущать поляризацию сигнала, а значит, и соотношение мощностей в выходных модах. 
\item в режиме светоделения, входящий сигнал, имеющий вертикальную/горизонтальную поляризации, должен разделяться между этими поляризационными компонентами поровну. Для этого мощности компонент измеряются фотодиодами ФД1 и ФД2 в каналах 3 и 4.
\item в режиме светоделения входящий сигнал, имеющий диагональную поляризацию, должен направляться в одну из выходных мод.
\end{itemize}
Сигнал создавался с помощью генератора импульсов BK Precision 4010A-ND и последующего усилителя, поднимающего управляющее напряжение на необходимый ЭОМ уровень $\sim 300$ В.

\begin{figure}[h]
\centering
\includegraphics[width=\textwidth]{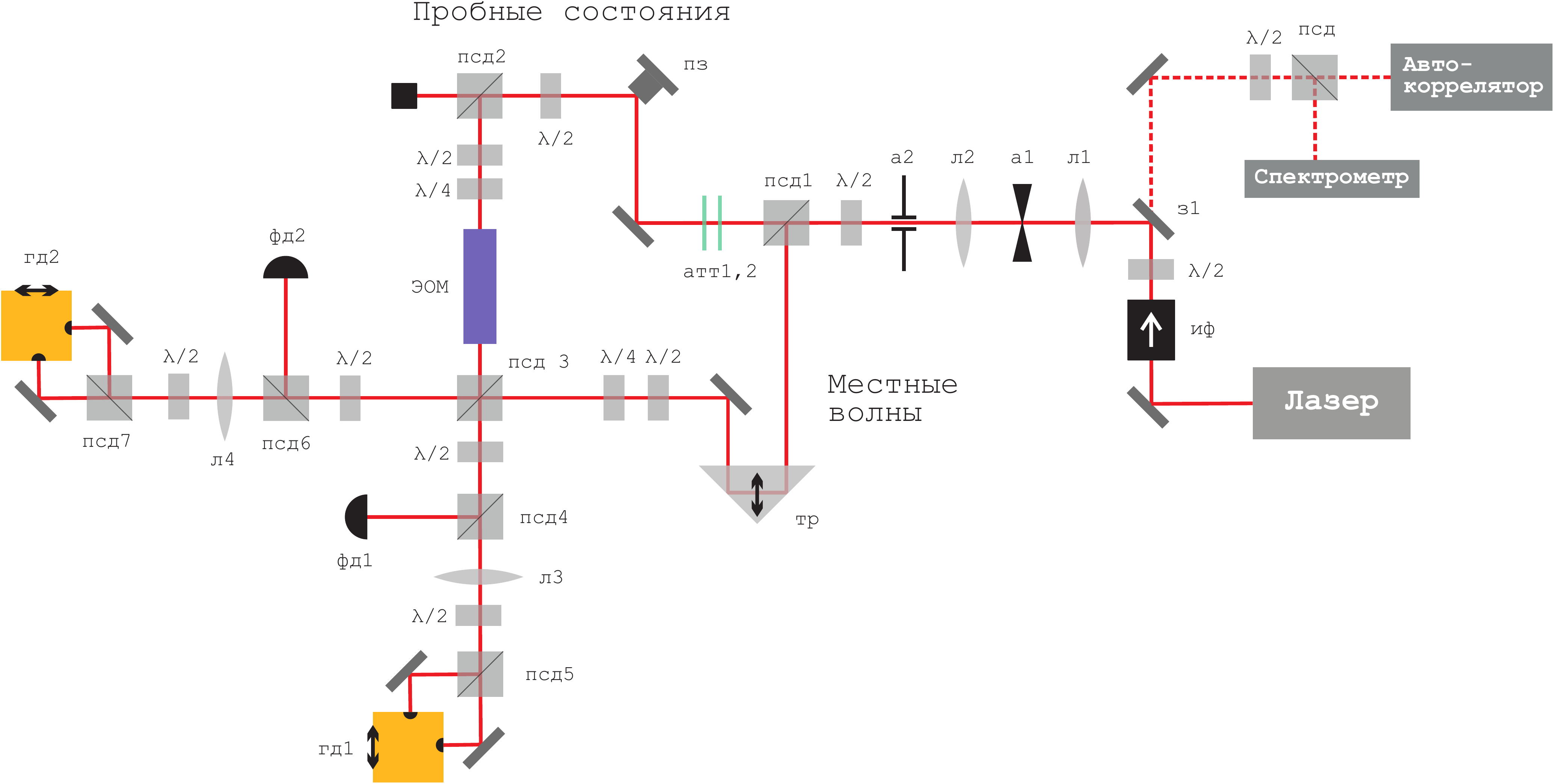}
\caption{Полная оптическая схема. ИФ: изолятор Фарадея. Используется для удаления обратных отражений, препятствующих пульсации лазера. Ширина импульса с длина волны измеряются с помощью автокоррелятора и спектрометра, анализирующих пучок утечки через зеркало З1. ТР: обратно-отражающая призма, закреплённая на микрометрической платформе, тромбон. Л, линза. А, апертура. (П)З, (пьезо) зеркало. Ф, спектральный фильтр. ПСД, поляризационный светоделитель. АТТ, аттенюатор. ФД, фотодиод. ГД, гомодинный детектор.}
\label{p1full}
\end{figure}

Амплитуды когерентных состояний первой (вертикально поляризованной, $\alpha_1$) и второй (горизонтально поляризованной, $\alpha_2$) пробных мод определяется положением полуволновой ($\theta$) и четверь-волновой ($\phi$) пластин. Как описано в разд. (\ref{sect1_5}), использовавшийся алгоритм восстановления не требует предварительной характеризации пробных состояний; поэтому, использовался набор из 16 различных комбинаций
\begin{equation}
\label{eq_angles}
\{\theta, \phi\} \in \{ 0^\circ, \, 15^\circ, \, 30^\circ, \, 45^\circ \}
\end{equation}
отсчитываемых от ``нулевого'' положения, т.е. положения, не возмущающего исходную горизонтальную поляризацию пучка.

\begin{figure}[h]
	\centering
	\includegraphics[width=5.5in]{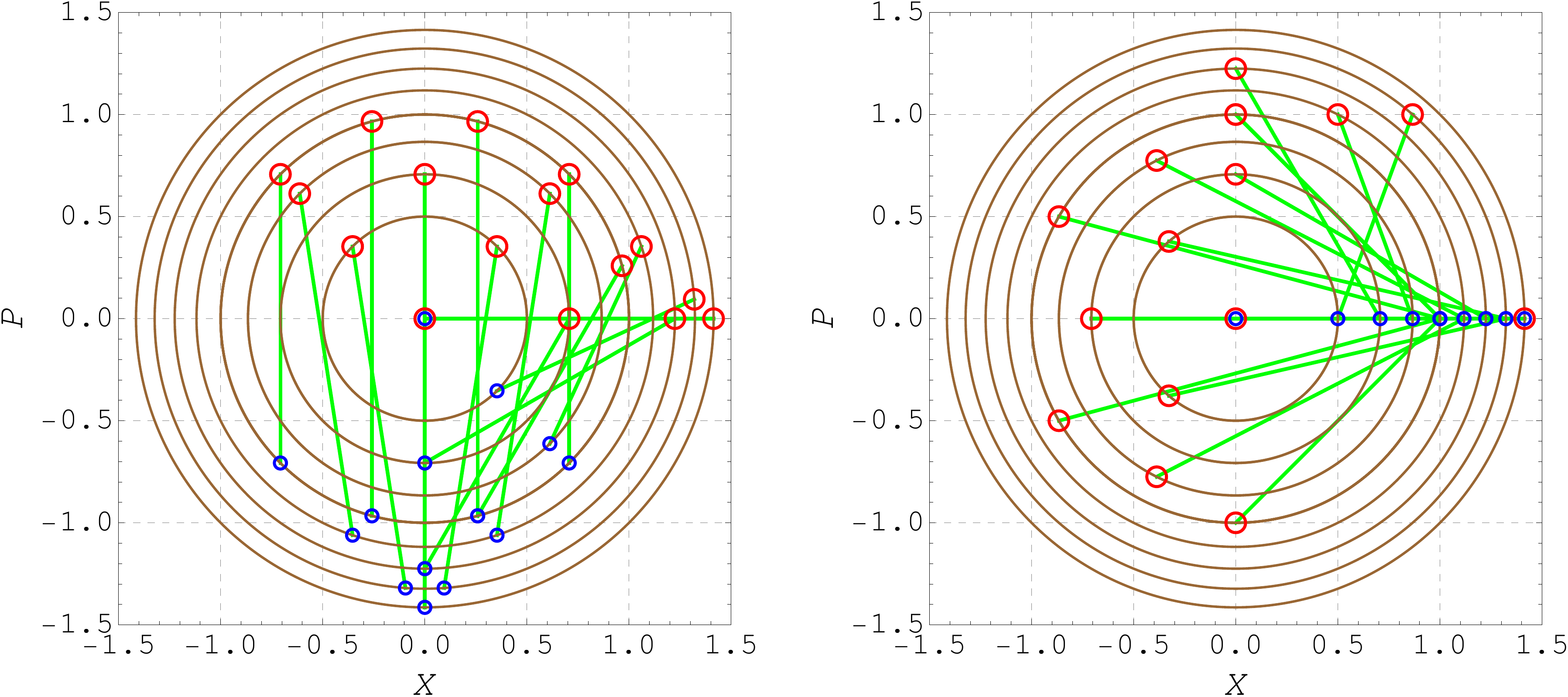}
	\caption{Амплитуды пробных состояний $\alpha_1$ и $\alpha_2$ на фазовой плоскости. Синие точки соответствуют моде 1, красные -- моде 2. Пары соединены линиями. Окружности радиусов $\sqrt{2}\alpha_i$ показывают амплитуды состояний. На левой панели, фаза отсчитывается от фазы входного состояния в первой моде, справа -- от первого выходного. Суммарная энергия одинакова у всех пар и равна $\left| \alpha_1 \right|^2 + \left| \alpha_2 \right|^2 = \left| \alpha \right|^2 = 0.9$.}
	\label{p1probes}
\end{figure}

Амплитуды пробных состояний, соответствующие положениям (\ref{eq_angles}), показаны в фазовом пространстве на Рис.~\ref{p1probes} слева. Как выяснено в разд. \ref{sect1_5}, существенной является разность фаз в паре пробных состояний; поэтому, каждая пара может быть повёрнута на фазовой плоскости на произвольный угол. Правая панель показывает пробные пары, когда угол поворота таков, что амплитуда когерентного состояния в первой моде действительна. Такая операция приводит к совмещению пар, соответствующих $\{\theta, \phi\} = \{30^\circ, 15^\circ\}$, и $\{45^\circ, 45^\circ\}$. Оба этих состояния эквивалентны паре $\{\alpha_1, \alpha_2\} \propto \{1, i\}$. 

До сих пор, настройка требовала использования макроскопически-мощных полей в сигнальных модах. Однако, размер исследуемой части оптического гильбертова пространства ограничен четырьмя младшими фоковскими состояниями. Этой области соответствуют амплитуды пробных состояний порядка единицы. Требуемый уровень ослабления сигнала $\sim 10^{-8}$ (см. разд. \ref{laser}) обеспечивается 40dB аттенюаторами АТ1 и АТ2 в канале 1, тогда как точная подстройка осуществляется с помощью светоделителя ПСД2, на котором лишняя мощность сбрасывается в проходящую моду. Амплитуда пробных состояний определялась экспериментально исходя из качества реконструкции процесса: слишком мощные состояния ведут к погрешностям при описании их на ограниченной области гильбертова пространства, тогда как слишком слабые состояния не несут достаточной информации о работе процесса на старших фоковских компонентах. Оптимальным оказалось ослабление, при котором энергия пробного пучка после ПСД2 составляет $\left| \alpha \right|^2 = 0.9$ энергии фотона.

Вызванное внесением аттенюаторов изменение оптических путей приводит к значительному ухудшению временного перекрытия с импульсами местных волн. Эта задержка пробных состояний должна быть скомпенсирована соответствующей задержкой местных волн с помощью тромбона ТР. Величина смещения последнего может быть подобрана как в ходе максимизации величины квадратурных измерений, так и с помощью отдельного классического интерференционного эксперимента.

\subsection{Гомодинные измерения в выходных каналах процесса}
\label{BHD1}

В двумодовом случае, полнота набора квадратурных данных требует измерений при всех возможных комбинациях фаз $\theta_1$ и $\theta_2$ (разд. \ref{BHD_theory}). Так как моды пробных состояний совпадают пространственно, пьезо-зеркало позволяет обрабатывать лишь одномерное множество точек 
\begin{equation}
\label{eqPhaseRel}
	\theta_1(t) = \theta_2(t) + \Delta
\end{equation}
на плоскости возможных положений. В выражении (\ref{eqPhaseRel}), значение величины $\Delta$ определяется разностью оптических путей сигналов и местных волн. 

Вырожденность пространственных мод сигналов, а также местных волн выше процесса приводит к тому, что воздушные флуктуации в пространственных каналах 1 и 2 (Рис. \ref{p11}) не вносят вклада в $\Delta$. Аналогично, флуктуации ниже процесса варьируют в равной мере фазы перекрывающихся сигналов и местных волн, и также не меняют $\Delta$. Таким образом, величина $\Delta$ не подвержена воздушным флуктуациям и является фиксированной. 

Величина $\Delta$ варьировалась с помощью пары волновых пластинок в канале местных волн. Эксперименты, однако, показали, что восстановление процесса светоделения возможно даже по набору данных, имеющему единственное значение $\Delta$. В ходе финальных измерений, использовались 3 значения этой величины: $\Delta=0.67$, $2.64$ и $5.29$, соответствующие различным комбинациям положений волновых пластин в канале 2, которые были определены во вспомогательном интерференционном эксперименте.

Гомодинное детектирование входных и выходных когерентных состояний производится в соответствии с описанием, представленным в разд. \ref{BHD_all}, однако имеет ряд характерных особенностей, определяемых следующими их свойствами:

\begin{itemize}
	\item Среднее значение квадратуры пропорционально гармонической функции фазы сигнала относительно местной волны:
	\begin{equation}
	\label{eqMeanQuad}
	\bra{\alpha e^{-i\phi}}\hat{Q}_\theta\ket{\alpha e^{i\phi}} = \dfrac{\alpha e^{-i \phi} e^{i\theta} + \alpha e^{i\phi} e^{-i\theta}}{\sqrt{2}} = \sqrt 2 \alpha \cos{\left(\theta-\phi\right)}.
	\end{equation}
	Среднее значение (\ref{eqMeanQuad}) является единственным фазово-зависимым свойством когерентного состояния, и благодаря этому используется для определения фазы наблюдаемой квадратуры. 	
	
	Поскольку движение пьезо-управляемого зеркала ПЗ не параллельно оси луча, то помимо продольного смещения, оно приводит и к поперечному смещению отражённого пучка. Если таким образом смещается местная волна, то это приводит к разбалансировке гомодинного детектора, проявляющейся в результатах измерений как смещение средней линии сигнала. Этот эффект может быть неотличим от поведения среднего значения квадратуры при детектировании когерентного состояния (\ref{eqMeanQuad}), и может таким образом приводить к искажениям результатов измерений.
	По этой причине, пьезо-зеркало устанавливается в сигнальном пучке; смещение последнего приводит к лишь незначительной вариации в перекрытии с модой местной волны, т.е. эффективности измерения.
	
	\item Дисперсия исходов индивидуальных квадратурных измерений когерентного состояния не зависит от его амплитуды и фазы наблюдаемой:
	\begin{equation}
	\label{eqCSnoise}
	\sigma^2_1 = \left\langle \alpha \left| \hat{Q}^2_{\theta} \right| \alpha \right\rangle - \left\langle \alpha \left| \hat{Q}_{\theta} \right| \alpha \right\rangle^2 = \dfrac{\alpha^2 e^{-2i\theta} + 2 \alpha^2 +1 + \alpha^2 e^{2i\theta}}{2} - \left(\sqrt{2}\alpha\cos\theta\right)^2 = \dfrac{1}{2}.
	\end{equation}
	Это свойство, в частности, означает невозможность достоверного определения значения квадратуры слабого когерентного состояния с помощью однократного измерения. Поэтому, применяется усреднение результата по нескольким измерениям. Средне-квадратическая ошибка при этом уменьшается пропорционально квадратному корню от числа измерений $n$:
	\begin{equation}
	\label{eqVarianceOfMean}
	\sigma_n = \dfrac{\sigma_1}{\sqrt{n}}. 
	\end{equation}
\end{itemize}

\begin{figure*}[h]
	\centering
	\includegraphics[width=\textwidth]{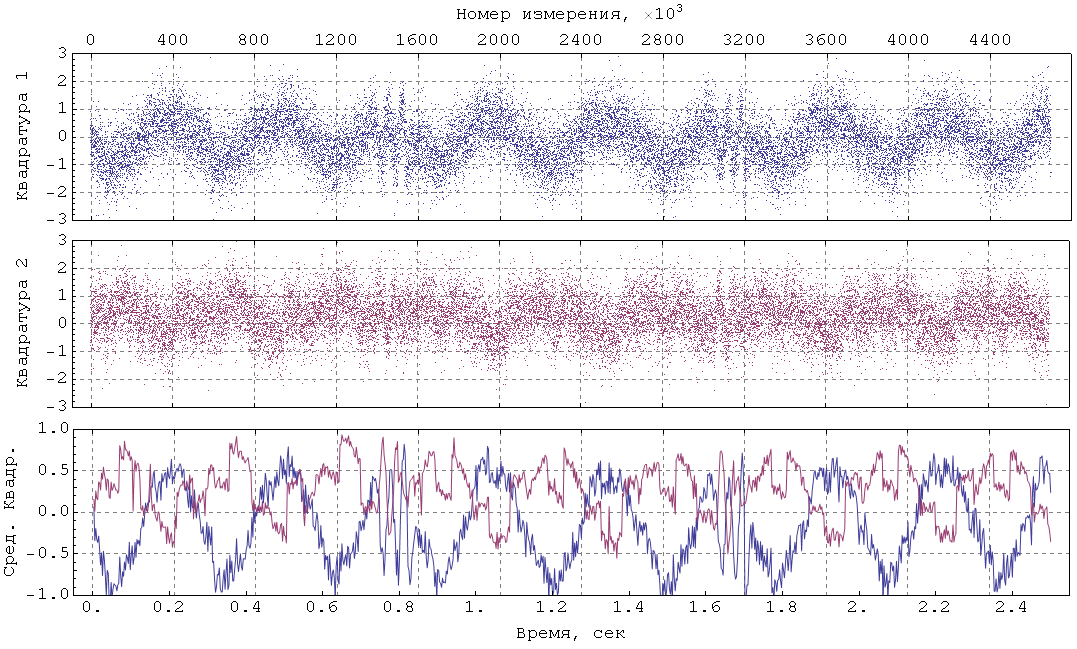}
	\caption{Данные гомодинирования обоих мод процесса.
	Верхние панели: $4.7 \times 10^6$ значений, нормированных на вакуумный шум; каждая точка соответствует одному лазерному импульсу. Для удобства, показано каждое двухсотое измерение. 
	Нижняя панель: усреднённые квадратуры. Каждая точка является средним по ансамблю $4.7\times 10^3$ последовательных измерений. Вертикальные оси нормированы так, что среднеквадратическое отклонение квантового шума вакуума равно $1/\sqrt{2}$.
	Положения полуволновых пластинок в пробном канале: $\lambda/2@0^\circ$, $\lambda/4@15^\circ$. Разность фаз местных волн (\ref{eqPhaseRel}): $\Delta = 3.79$ рад.}
	\label{p13}
\end{figure*}

Пример двумодовых квадратурных данных показан на верхних панелях Рис.~\ref{p13}. Разброс точек обусловлен квантовой неопределённостью исхода каждого измерения (\ref{eqCSnoise}).
Нижний график на Рис.~\ref{p13} показывает поведение среднего значения каждой квадратуры. В результате усреднения по ансамблю $n=4.7\times 10^3$ индивидуальных измерений, квантовый шум становится меньше:
\begin{equation}
\dfrac{\sigma_1}{\sqrt{n}} \approx 0.01. 
\end{equation}

Осцилляции с частотой $\sim 3$ Гц соответствуют движению пьезо-зеркала, за один период которого фаза пробных состояний проходит диапазон $\approx 6\pi$. Управляющее напряжение имеет пилообразную форму, с максимально возможной асимметрией между периодами возрастания и убывания; это позволяет упростить обработку и увеличить скорость сбора данных: на Рис.~\ref{p13}, возвращение зеркала произошло на отметках $0.8$ и $1.67$ сек. Между ними, зеркало движется равномерно в одну сторону.

Модуляция осцилляций с частотой порядка 10 Гц на Рис.~\ref{p13} соответствует переключению напряжения ЭОМ, т.е. замене чёрного ящика на процесс тождественности. 
Управляющее напряжение ЭОМ переключается между двумя значениями напряжения; продолжительности периодов выключенного / включенного процесса были установлены в соотношении 4/5, что позволило определять состояние ЭОМ в каждый момент измерения по его принадлежности к б\'{о}льшей / м\'{е}ньшей группе групп точек на Рис.~\ref{p13}. Эти группы хорошо читаются по красной кривой на нижнем графике Рис.~\ref{p13}; на синей кривой они видны менее явно, так как входное и выходное состояния процесса, для данного набора входных состояний, близки (Рис. \ref{p1quantFitted}).

Такой метод косвенного определения моментов переключения ЭОМ ставит дополнительные требования к экспериментальной технике. Дело в том, что для удовлетворительного определения этих моментов, она требует наличия не менее 10 усреднённых значений, что соответствует $10\times 4700 \sim 10^5$ измерений квадратуры в каждой группе; вместе с этим, процедура обработки данных требует не менее $\sim 5$ групп на один период осцилляций пьезо-фазы (разд. \ref{PracticalPhases}). Максимальная скорость записи квадратурных измерений, определяемая быстродействием программного обеспечения и составляющая около $10^6$ в секунду, таким образом ограничивает сверху допустимую частоту движения пьезо-зеркала на значении $1-3$Гц. С другой стороны, для точного извлечения фазовой информации (разд. \ref{sect1_5}) обратная частота вариации фазы не должна быть больше 1 сек -- характерного времени воздушных флуктуаций.

В течение периодов тождественности, в двух выходных каналах, нумеруемых индексом $k \in \{1,2\}$, измеряются пробные когерентные состояния, имеющие форму 
\begin{equation}
\ket{\alpha_k e^{i \left[\phi_{k} + \phi_p(t)\right]}},
\end{equation}
где $\phi_{k}$ есть фазы пробных состояний в начальный момент времени, а $\phi_p(t)$ есть вариация фаз, вносимая пьезо-зеркалом, а также воздушными и прочими флуктуациями. 
Средние значения квадратур при этом даются выражениями (\ref{eqMeanQuad})
\begin{equation}
\label{eqMeanQuad1}
\left\langle \hat{Q}_{k} \right\rangle (t) = \sqrt{2} \alpha_k \cos\left[\theta_k - \left(\phi_{k} + \phi_p(t)\right)\right],
\end{equation}
где $\theta_k$ есть фазы местных волн. Если пробные состояния известны, то выражения (\ref{eqMeanQuad1}) позволяют определить фазу измеряемых квадратур $\theta_k - \phi_p(t)$ в каждый момент времени.
Экстраполяция полученной зависимости на периоды измерения неизвестного процесса позволит дополнить значения квадратур фазовыми данными, и получить таким образом информацию для составления оператора (\ref{eq6}).

\section{Обработка данных} \label{sect1_5}

\subsection{Практическое определение фаз}
\label{PracticalPhases}

На практике, полезным оказывается подход, отличающийся от описанного выше; его преимуществом является возможность проверки данных на физическую корректность до того, как они загружаются в алгоритм реконструкции тензора процесса. Это значительно ускоряет поиск возможных ошибок и настройку как экспериментальной, так и программной частей. Кроме того, для широкого класса процессов, этот метод не требует предварительной характеризации пробных состояний.

\begin{figure*}[h]
	\centering
	\includegraphics[width=\textwidth]{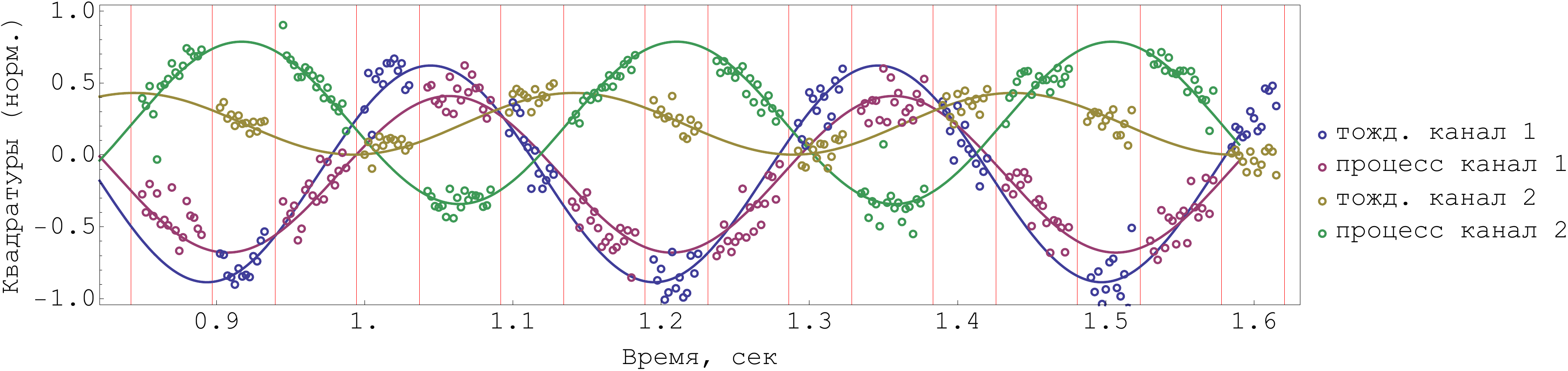}
	\caption{Анализ данных с Рис.~\ref{p13}. Моменты включения и выключения процесса отмечены красными линиями. Малые / большие интервалы между последними соответствуют измерению входных / выходных состояний чёрного ящика.}
	\label{p1quantFitted}
\end{figure*}

На первом этапе, усреднённые квадратурные данные в каждом канале $k$, разделённые на группы, соответствующие режимам работы ЭОМа, аппроксимируются функциями
\begin{equation}
\label{eqFit}
f_k(t) = O_k + A_k \cos\left(\sigma_k -\omega t\right),
\end{equation}
где $O_k$, $A_k$, $\sigma_k$, $\omega$ -- параметры аппроксимации, а $t$ -- время.
Форма (\ref{eqFit}) соответствует теоретическому ожиданию (\ref{eqMeanQuad1}) в случае линейной вариации фазы $\phi_p(t) = \omega t$. Результат такой обработки данных с Рис.~\ref{p13} представлен на Рис.~\ref{p1quantFitted}. 

Если известно, что чёрный ящик обладает свойством фазовой инвариантности (разд. \ref{sect1_6}), то точку отсчёта для измерения фаз входных и выходных когерентных состояний можно выбрать произвольно. Для каждой пары пробных состояний эта точка выбиралась так, чтобы начальная фаза состояния в первой моде была равна нулю, $\phi_1 = 0$. Сопоставляя (\ref{eqMeanQuad1}) с (\ref{eqFit}), фазы местных волн и фаза второго пробного состояния определяются как
\begin{equation}
\begin{aligned}
\phi_1 = 0 \quad \longrightarrow & \quad \theta_1 = \sigma_1 \\
&\quad \theta_2 = \sigma_1 + \Delta \quad \longrightarrow \quad \phi_2 = \sigma_1-\sigma_2+\Delta,
\end{aligned}
\end{equation}
Разность фаз между местными волнами $\Delta$ определяется в отдельном эксперименте по наблюдению интерференций местных волн с двумя синфазными вспомогательными полями. Последние можно приготовить, задав диагональную поляризацию пробного поля с помощью волновых пластинок в пространственном канале 1 на Рис. \ref{p11}, и направив процесс в режим тождественности. Тогда, разность фаз между интерференционными сигналами, наблюдаемыми на фотодиодах ФД1 и ФД2, равна разности фаз между местными волнами.

Описанный метод позволяет аналогичным образом характеризовать выходные когерентные состояния в двух модах. Нумеруя эти состояния индексами $k \in \{3,4\}$, их фазы определяются как
\begin{equation}
\begin{aligned}
& \phi_3 = \sigma_1 - \sigma_3 \\
& \phi_4 = \sigma_1 + \Delta\theta - \sigma_4.
\end{aligned}
\end{equation}
Амплитуды когерентных состояний при этом равны 
\begin{equation}
\alpha_k = \dfrac{A_k}{\sqrt{2}}.
\end{equation}

jХарактеризованный таким образом сигнал сопоставлялся с ``классическими'' экспериментами, в ходе которых наблюдался результат интерференции местных волн $C_k e^{i\theta_k}$ с пробными и выходными когерентными состояниями $B_k e^{i(\phi_k+\omega t)}$. Амплитуды $B_k$ теперь сравнимы с $C_k$, благодаря чему мощность интерференционных сигналов
\begin{equation}
\label{eqQPTinterference}
C_k^2+B_k^2+2C_kB_k\cos{\left[\theta_k - \left(\phi_{k} + \phi_p(t)\right)\right]},
\end{equation}
измеряемая с помощью фотодиодов ФД1 и ФД2 (Рис.~\ref{p1full}), несёт фазовую информацию, идентичную квантовым измерениям (\ref{eqMeanQuad1}).
\begin{figure*}[h]
	\centering
	\includegraphics[width=\textwidth]{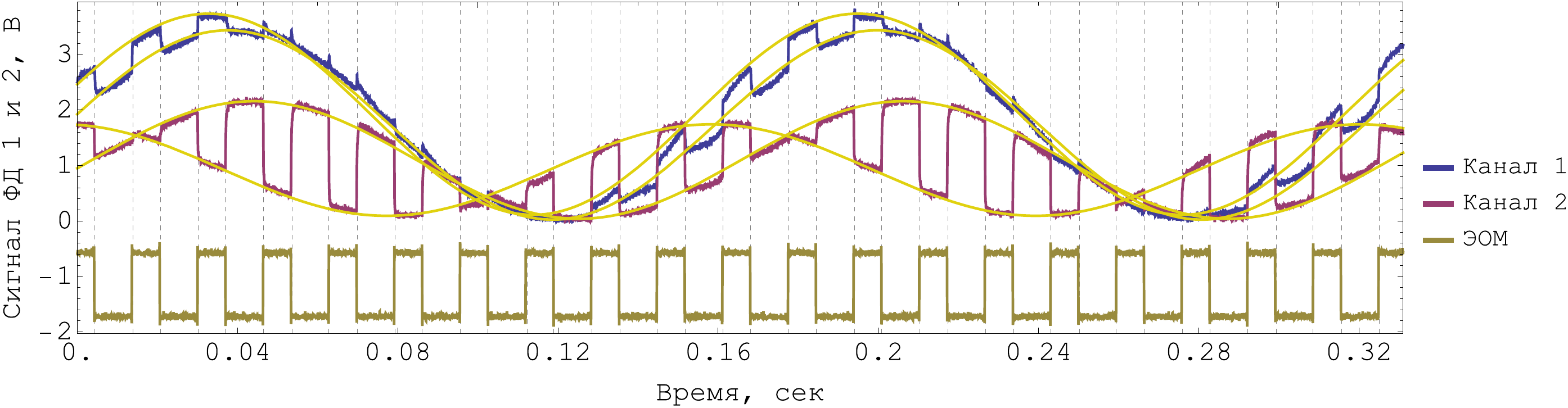}
	\captionsetup{justification=raggedright,singlelinecheck=false}
	\caption{Мощности проинтерферировавших полей местных волн и сигнальных состояний, соответствующих квантовым измерениям, показанным на Рис.~\ref{p1quantFitted}. Жёлтые линии показывают аппроксимацию четырёх сигналов формой (\ref{eqQPTinterference}).}
	\label{p1classic}
\end{figure*}

Пример такого измерения, проведённого с помощью осциллографа, показан на Рис.~\ref{p1classic}. Видно, что соотношение фаз между входными / выходными состояниями в каждом канале соответствует квантовому случаю. При этом, сигнал в канале 2 сдвинут по фазе относительно канала 1 из-за отражения на светоделителях ПСД4 и ПСД6. 

\begin{figure*}[h]
	\centering
	\includegraphics[width=\textwidth]{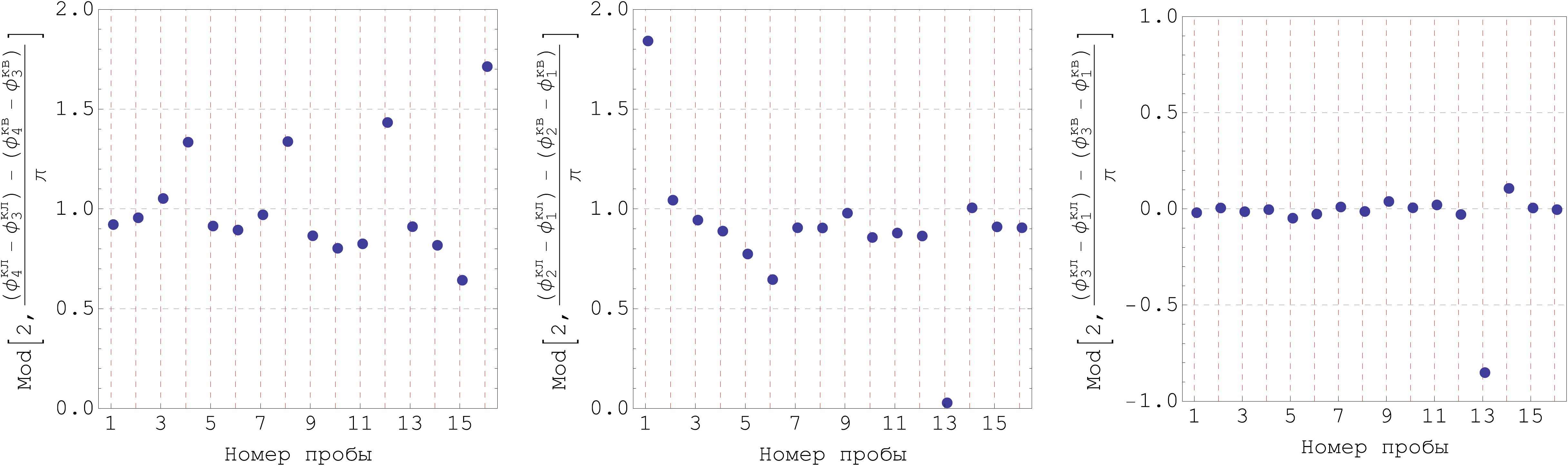}
	\caption{Сопоставление фаз входных и выходных состояний в двух каналах, измеренных в квантовом и классическом режимах.}
	\label{p1compare}
\end{figure*}
Сопоставление фазовых данных между классическими и квантовыми измерениями для всех пробных состояний показано на Рис.~\ref{p1compare}. В случаях, когда амплитуда сигнала мала, фаза определяется с большой погрешностью, что может приводить к значительным расхождениям двух измерений для некоторых пробных состояний.

Только когда соответствие между классическими и квантовыми измерениями, а также предсказаниями (\ref{eq13}) установлено, следует обращаться к алгоритму максимального правдоподобия.
Эта промежуточная между экспериментом и алгоритмом MaxLik проверка позволяет настраивать обе части независимо, что многократно облегчает задачу. Со своей стороны, алгоритм максимального правдоподобия также может отлаживаться на данных, искусственно синтезированных исходя из (\ref{eq13}).

\subsection{Реконструкция} \label{sect1_6}

В отношении томографии процесса, важным фактором является его фазовая инвариантность. Фазово-инвариантный процесс нечувствителен к общему сдвигу фазы: одинаковое смещение фазы пробных состояний приводит к такому же сдвигу выходных состояний. Это свойство является следствием временной симметрии процесса: общий сдвиг фазы на угол $\delta\theta$ эквивалентен сдвигу времени на величину $\delta\theta/\omega$, где $\omega$ есть оптическа частота. Если чёрный ящик не подключён к каким-либо внешним часам то, он отвечает на временной сдвиг входного сигнала таким же сдвигом на выходе.
Эффект фазовой инвариантности на тензор процесса можно определить, исходя из правила преобразования матрицы плотности квантового состояния при внесении фазовой задержки $\theta$:

\begin{eqnarray}
\rho^{\mathrm{in}}_{n_1n_2m_1m_2}&\to&\rho^{\mathrm{in}}_{n_1n_2m_1m_2}e^{i\delta\theta(n_1 + n_2 - m_1 - m_2)}, \nonumber \\
\rho^{\mathrm{out}}_{j_1j_2k_1k_2}&\to&\rho^{\mathrm{out}}_{j_1j_2k_1k_2}e^{i\delta\theta(j_1 + j_2 -k_1 - k_2 )}. 
\label{eqPhaseInv}
\end{eqnarray}
Сравнивая это с (\ref{eq11}), можно получить, что только элементы тензора процесса, удовлетворяющие условию
\begin{equation}
\label{eqZeroEl}
j_1 + j_2 -k_1 - k_2 = n_1 + n_2 - m_1 - m_2,
\end{equation}
могут быть отличны от нуля в тензоре $\mathcal{E}^{\underline n \, \underline m}_{\underline j \, \underline k}$. 

Большинство элементов не удовлетворяют условию (\ref{eqZeroEl}). Так, в случае двумодового ($M=2$) процесса, при выбранном числе отсечки на фоковском пространстве $N=4$ из общего числа $(N+1)^{4M} = 5^8 = 390625$ элементов обнуляется 352460. Помимо методологических преимуществ, описанных в разделе \ref{sect1_5}, использование свойства фазовой инвариантности (\ref{eqPhaseInv}) таким образом снижает размерность задачи нахождения оптимального тензора процесса в $\approx 10$ раз. 
Большинство используемых квантовых процессов обладает свойством фазовой инвариантности; к их числу принадлежит и процес светоделения. 

В соответствии с рекомендациями, изложенными в работе \cite{Anis2012}, мы реализуем двух-ступенчатый процесс реконструкции. На первом шаге, мы выбираем число отсечки на фоковском пространстве $N$ так, чтобы исследуемое явление (например, эффект Хонг-Оу-Манделя \cite{Hong1987}), вместе с предполагаемыми пробными и выходными состояниями, достаточно верно описывалось в её пределах. Это обеспечивает существование тензора процесса, соответствующего экспериментальным данным. В настоящей работе, использовалось $N=4$.

Задача нахождения оптимального тензора процесса решается с помощью численного алгоритма, следующего итеративной процедуре, описанной в разделе \ref{sect1_2}.
Итоговая размерность оптимизационной задачи близка к томографии состояния $8$ ионов описанной в \cite{Haffner2005}, что обуславливает вычислительную сложность реконструкции. Итерационный алгоритм запускался на процессоре Intel Core i7; распараллеленный на 4 из 8 вычислительных ядер, каждая итерация занимает около 2 часов. Алгоритм максимального правдоподобия сходится на $30-80$ итерациях. 

После завершения оптимизации выполняется второй шаг процесса реконструкции, на котором тензор ограничивается более низким максимальным числом фотонов $N'=2$. Это необходимо, так как вес высших фоковских чисел в пробных состояниях низок, и соответствующие им элементы тензора процесса восстанавливаются с ошибками \cite{Anis2012}. Полное число элементов тензора при $N'=2$ составляет $6561$, а при условии фазовой инвариантности -- $1107$.

\begin{figure*}[h!]
	\includegraphics[width=5.5in]{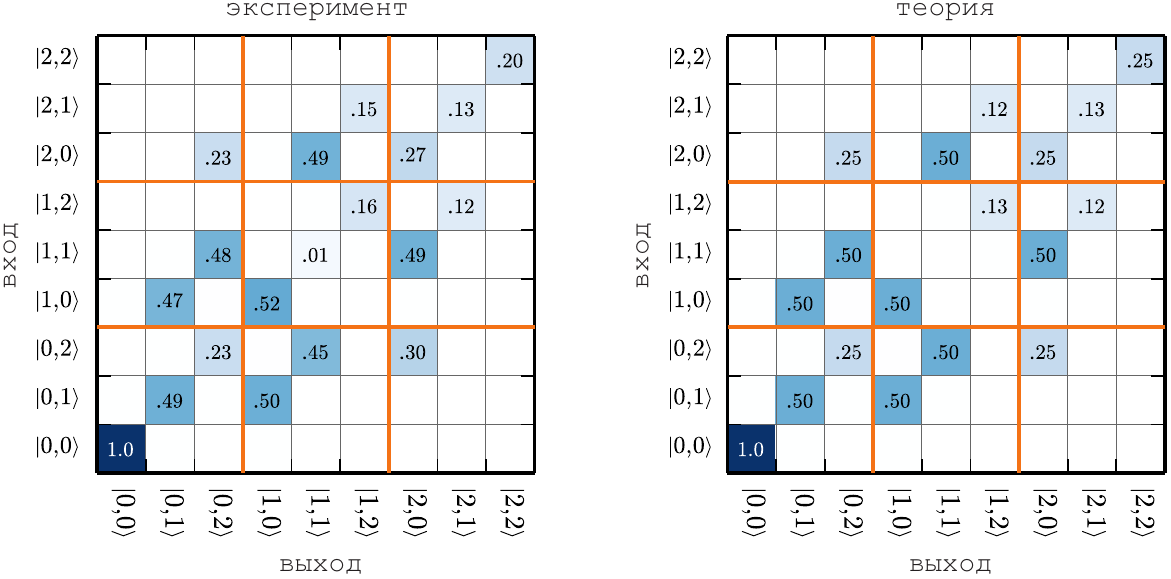}
	\centering
	\caption{Элементы тензора, соответствующие диагональным элементам входных и выходных матриц плотности в фоковском базисе. Ненулевые вероятности переходов даны числами. Элемент $\left| 1,1\right\rangle \rightarrow \left| 1,1\right\rangle$ соответствует вероятности совпадения в эксперименте Хонг-Оу-Манделя \cite{Hong1987}.}
	\label{p1diag}
\end{figure*}
\begin{figure*}[h!]
	\includegraphics[width=\textwidth]{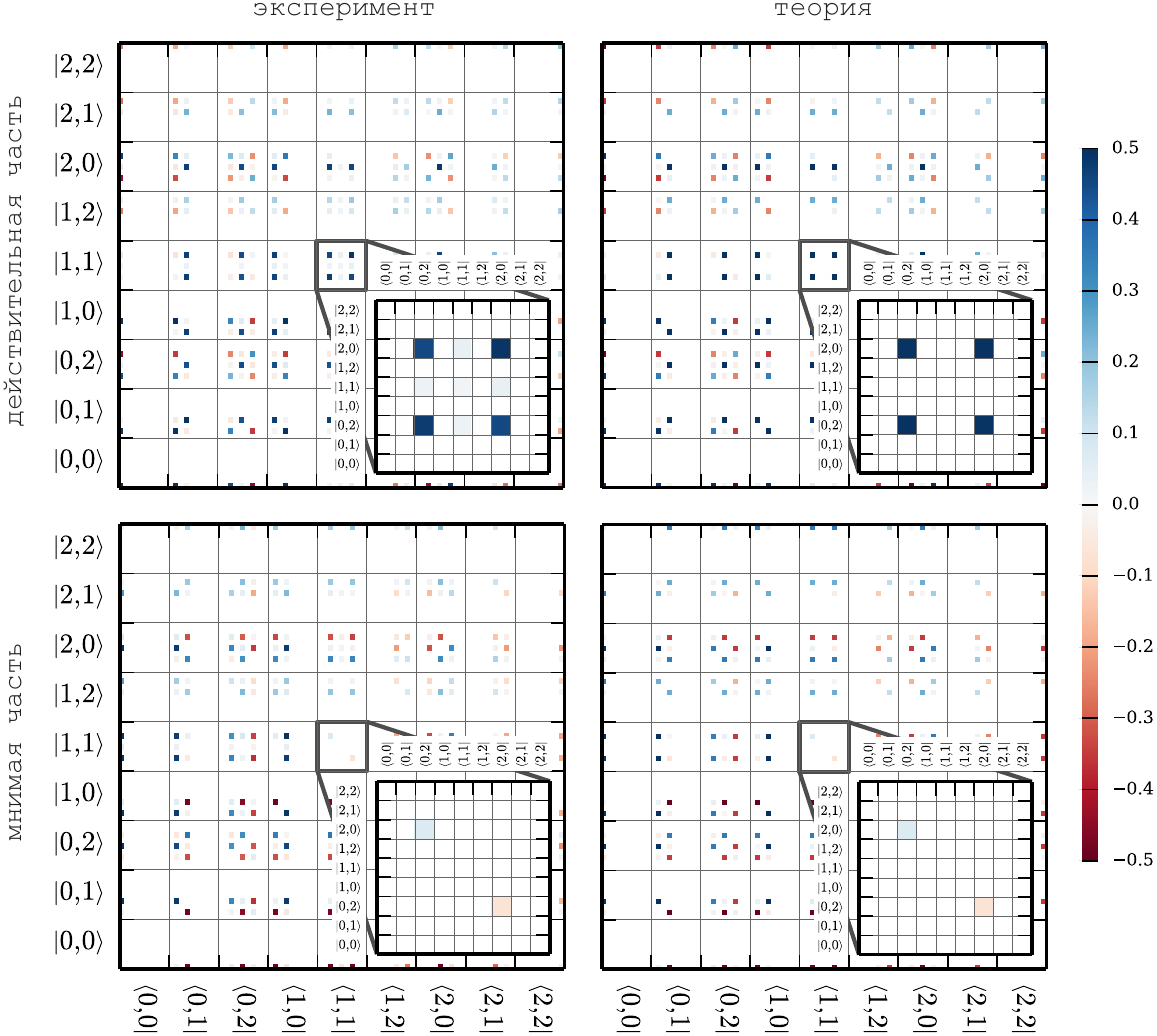}
	\caption{Тензор двумодового квантового процесса в фоковском базисе, ограниченный до $N'=2$. Слева: результат реконструкции. Справа: теоретическое ожидание. Действительной  и мнимой частям соответствуют верхний и нижний ряды. Каждая большая клетка соответствует элементу входной двумодовой матрицы плотности, тогда как её содержимое соответствует выходной матрице плотности. Вставки показывают ответ чёрного ящика на входное состояние $\left| 1,1\right\rangle$.}
	\label{p1all}
\end{figure*}

Частичный результат реконструкции полного тензора процесса, ограниченный максимальным числом фотонов $N'=2$, показан на Рис. \ref{p1diag}. На нём изображены элементы тензора, связанные с диагональными элементами матриц плотности входного и выходного состояний процесса в фоковском базисе. Эти же элементы находятся на главной диагонали в представлении Рис.~\ref{p1all}. Эти элементы имеют смысл вероятностей переходов фоковских компонент входного двумодового состояния в соответствующие компоненты выходного. 
В частности, эффект Хонг-Оу-Манделя представлен вероятностью перехода $\left| 1,1\right\rangle \rightarrow \left| 1,1\right\rangle$, которая равна нулю в идеальном симметричном светоделении (правая панель) и составляет $\approx 0.01$ в реконструированном тензоре (слева).

Данные, представленные на Рис.~\ref{p1diag}, являются лишь малой частью полного тензора, показанного на Рис.~\ref{p1all} сверху. Его элементы являются в общем случае комплексными числами; именно они определяют фазовое поведение чёрного ящика, и являются равно важными для характеризации процесса.
В используемом представлении, внешние индексы соответствуют матрице плотности входного состояния: $\ket{i,j} \bra{k,l} = \ket{i}_1\ket{j}_2\bra{k}_1\bra{l}_2$. Для каждого набора внешних индексов, содержание соответствующей клетки даёт выходную матрицу плотности в том же представлении.

Вставки в каждой панели показывают ответ процесса на входное состояние для эффекта Хонг-Оу-Манделя, $\left| 1,1\right\rangle \left<1,1\right|$. Главная диагональ тензора имеет $(N'+1)^{2M} = 81$ элемент, которые являются полностью действительными, и соответствует диагональным элементам входной и выходной матриц плотности; таким образом, данные показанные на Рис. \ref{p1diag} есть диагональ левой колонки на Рис. \ref{p1all}.

Нижние панели Рис. \ref{p1all} показывают теоретическое ожидание для тензора симметричного светоделения в согласии с (\ref{eq13}), при дополнительной общей фазовой задержке в двух каналах $0.8$ рад. Это расхождение, также как и ошибки в других матричных элементах, вызвана неидеальностью настройки ЭОМа и волновых пластинок, ограниченным набором данных, и неточностью реконструкции вследствие конечного числа итераций.

Для оценки качества реконструкции, мы вычисляем параметр верности между теоретически ожидаемым и восстановленным тензорами процесса в представлении состояний Ямилковского (\ref{eq12}):
\begin{equation}
\label{eq14}
\mathcal{F} \left( \mathcal{E}, \mathcal{E}_{est} \right) = \left(\mathrm{Tr} \left[ \sqrt{ \sqrt{\mathcal{E}} \mathcal{E}_{est} \sqrt{\mathcal{E}} }\right]\right)^2.
\end{equation}

При ограничении восстановленного тензора $(N=4)$ до $N'=3$ и $N'=2$, (\ref{eq14}) даёт $\mathcal{F} \left( \mathcal{E}, \mathcal{E}_{est} \right) =0.88$ и $0.95$, соответственно. Без ограничения ($N'=N=4$), верность равна 0.7.

Для определения причин этих неидеальностей был проведён ряд тестов. 
Во-первых, квадратурные данные (см. раздел \ref{sect1_5}) были напрямую использованы для определения параметров светоделения. Оказалось, что действительный процесс светоделения не являлся в точности симметричным; его пропускание по мощности составляло $0.502$. Верность, соответствующая этому расхождению, равна $0.998$, что свидетельствует о том, что вклад физической неидеальности процесса (по крайней мере, проявляющейся в нарушении симметрии светоделения), является незначительным. 
Далее, были оценены ошибки алгоритма восстановления. Для этого, алгоритм запускался на искуственно смоделированных квадратурных данных, причём объём данных, размерность оптимизационного пространства и количество итераций соответствовали параметрам действительной реконструкции. Для полученного набора восстановленных тензоров $\mathcal{E}'_{{\rm est},i}$, верность $\mathcal{F} \left( \mathcal{E}, \mathcal{E}'_{{\rm est},i} \right)\sim 0.95$ для всех $i$. 
Близкие значения наблюдались и для попарных верностей $\mathcal{F} \left( \mathcal{E}'_{{\rm est},i}, \mathcal{E}'_{{\rm est},j} \right)$, также как и для верности с теоретическим тензором $\mathcal{F} \left(\overline{ \mathcal{E}'_{{\rm est},i}}, \mathcal{E} \right)$. Эти данные свидетельствуют о том, что экспериментально полученная верность $0.95$ обусловлена неточностью работы численного алгоритма, наряду со статистической ошибкой вследствие конечного объёма экспериментальных данных.

\section{Выводы} \label{sect1_7}

В этом разделе, предложен метод экспериментальной характеризации многомодовых квантовых процессов. Его реализация продемонстрирована на примере наиболее распространённого многомодового оптического процесса -- светоделения. 
Метод показал свою эффективность в реконструкции квантовых аспектов процесса, в частности эффекта Хонг-Оу-Манделя. Средняя ошибка определения матричных элементов составляет 5\%, а параметр верности между экспериментально восстановленным тензором процесса и теоретическим ожиданием равен 95\%.
Благодаря простоте требуемых измерений и приготовления пробных состояний, представленный метод может быть применён к другим процессам, обобщён на другие физические системы, а также масштабируем до большего числа каналов и более широких пространств состояний.

По результатам исследования опубликована журнальная статья: \\
\textit{Ilya A. Fedorov, Aleksey K. Fedorov, Yury V. Kurochkin and A.I. Lvovsky, Tomography of a multimode quantum black box. New Journal of Physics \textbf{17}, 043063 (2015)} \cite{FedorovQPT2015}. \\

Работа представлена на международных конференциях:
\begin{enumerate}
	\item [1.] Quantum 2014 -- Workshop ad memoriam of Carlo Novero, 25--31 мая 2014, Турин, Италия
	\item [2.] 23rd annual International Laser Physics Workshop, 14--18 июля 2014, София, Болгария
	\item [3.] Quantum Optics VII, 27--31 октября 2014, Мар-Дель-Плата, Аргентина
	\item [4.] Third International Conference on Quantum Technologies, 13--17 июля 2015, Москва, Россия
\end{enumerate}



\clearpage
\chapter{Методы} \label{methods}

\section{Характеризация лазерного излучения}
\label{laser}

Источником света является титан-сапфировый лазер Coherent Mira 900, выдающий излучение с регулируемой длиной волны $\lambda \approx 780$ нм. Накачка производится с помощью постоянного диодного лазера Coherent Verdi G12, излучение которого имеет длину волны 530 нм и мощность 10-12 Вт.
Лазер Mira работает в импульсном режиме, с длиной импульсов $\tau \approx 2$ пс и интервалом между импульсами $T\approx13.07$ нс, что соответствует частоте $\nu=76$ МГц.   По сравнению со средней мощностью $W\approx 1.5$ Вт, импульсный режим позволяет увеличить плотность энергии в каждом импульсе в $T/\tau \approx 6.5\times 10^3$ раз.

\subsection{Автокорреляционное измерение длительности импульса}
\label{autocorr}

\begin{figure}
	\includegraphics[width=6.2in]{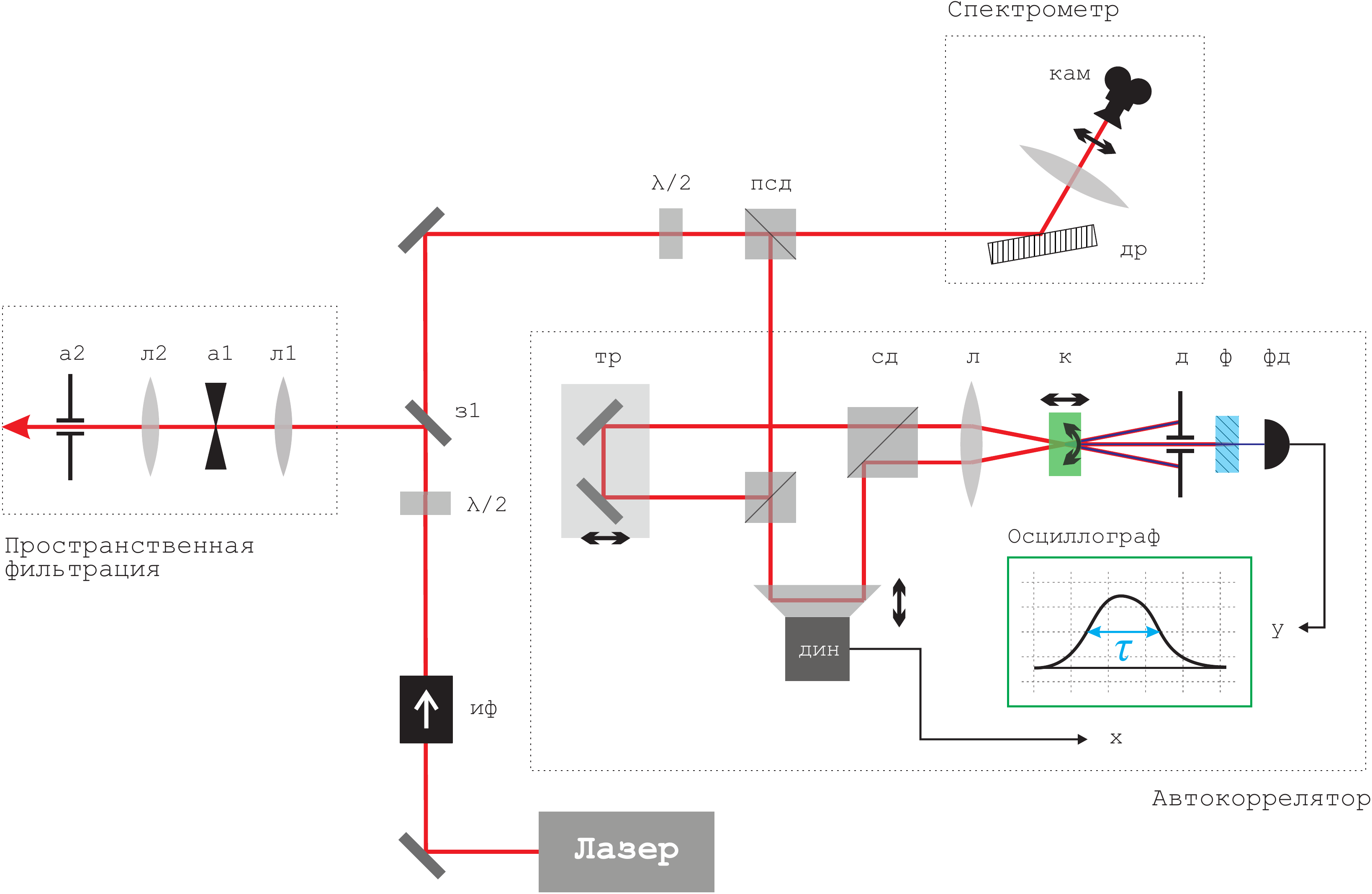}
	\captionsetup{
		justification=raggedright,
		singlelinecheck=false
	}
	\caption{Схема автокорреляционного и спектрального измерений. Используется малое поле утечки, проходящее через зеркало З1. ИФ: изолятор Фарадея. Служит для уничтожения отражений от элементов оптической схемы обратно в резонатор, препятствующих выведению лазера в импульсный режим. Л, линза. А, апертура. З, зеркало. Д, диафрагма. Ф, спектральный фильтр. (П)СД, (поляризационный) светоделитель. ДИН, динамик. К, нелинейный кристалл. ДР, дисперсионная решётка. КАМ, камера.}
	\label{pMethods1}
\end{figure}

Для определения длительности лазерного импульса выполняется автокорреляционное измерение. Его схема, показанная на Рис. \ref{pMethods1}, представляет собой интерферометр Майкельсона. В одном из его плечей установлена микрометрическая платформа ТР с обратно-отражающими зеркалами, в другом находится обратно-отражающая призма, укреплённая на мембране акустического усилителя ДИН. Светоделитель СД делает пучки из обоих плечей параллельными с расстоянием между пучками $\approx 5$ мм, а линза Л с фокусным расстоянием 100 мм пересекает их на нелинейном кристалле типа 1, где происходит генерация сигнала вдвое большей частоты SHG (разд. \ref{PhaseMatching1}). Его амплитуда представляется в виде
\begin{equation}
\label{eqCorrSHG}
E_{\rm shg} \propto \left(E_1+E_2\right)^2 = \left|E_1\right|^2 + \left|E_2\right|^2 + 2 \beta E_1E_2^*,
\end{equation}
где $E_{1,2}$ есть амплитуды полей первой гармоники в плечах интерферометра, а константа $\beta$ учитывает различие эффективностей нелинейного процесса для коллинеарного и неколлинеарного фазового синхронизмов (разд. \ref{PhaseMatching1}). Первые слагаемые в (\ref{eqCorrSHG}) соответствуют сигналу удвоенной частоты от каждого поля индивидуально, а последнее -- генерации суммарной частоты от разных пучков.
Неколлинеарная конфигурация пучков позволяет выделить составляющую, соответствующую последнему слагаемому в (\ref{eqCorrSHG}), с помощью диафрагмы Д. 

Мощность выделяемого сигнала $A$ зависит от задержки между импульсами, приходящими из двух плечей интерферометра. Задержка определяется положением передвижной микрометрической платформы ТР $\Delta$ и смещением мембраны ДИН $\delta$. Эти величины входят в (\ref{eqCorrSHG}) через амплитуды сигналов $E_1$ и $E_2$:
\begin{equation}
\label{eqEshifted}
E_1^2 \propto \exp\left[-\dfrac{(t - 2\Delta)^2}{\sigma^2}\right], \quad E_2^2 \propto \exp\left[-\dfrac{(t - 2\delta)^2}{\sigma^2}\right],
\end{equation}
где $\sigma \approx 2$ пс есть продолжительность импульса. Полная ширина на полувысоте интенсивностей $E_{1,2}^2(t)$ есть
\begin{equation}
\label{eqFWHM1}
\mathrm{FWHM} = 2 \sigma \sqrt{\ln 2}.
\end{equation}

Скорость фотодиода ФД намного меньше чем $\sigma^{-1}$, поэтому его выходной сигнал пропорционален интегральной мощности
\begin{equation}
\label{eqAuto}
A(\Delta,\delta) \propto \int_{t=-\infty}^{\infty} \left[E_1(t,\Delta) E_2^*(t, \delta)\right]^2 dt \propto \exp\left[-\dfrac{(\Delta-\delta)^2}{\sigma^2 / 2}\right].
\end{equation}
Функции $A(\delta)_{\Delta=0}$ и $A(\delta)_{\Delta = \sigma \sqrt{2\ln 2}}$ пересекаются на полувысоте. Соответствующее смещение микрометрической платформы $\sigma \sqrt{2\ln 2}$ в $\sqrt{2}$ раз меньше ширины исходных импульсов (\ref{eqFWHM1}). Их продолжительность таким образом выражается через измеряемое значение $\Delta$ как
\begin{equation}
\label{eqFWHM2}
\tau_\mathrm{FWHM} = \dfrac{\mathrm{FWHM}}{c} = \dfrac{\sqrt{2} \Delta}{c},
\end{equation}
где $c$ есть скорость света в воздухе.

Наблюдение автокорреляционной функции выполнялось с помощью осциллографа GW Instek GOS 620FG, имеющего встроенный генератор сигналов. Созданный с помощью него сигнал треугольной формы с частотой $\approx15$ Гц подавался на акустический усилитель, а также на горизонтальную ось осциллографа. На вертикальную ось направлялся отклик фотодиода ФД. При входящей мощности 30 мВт, автокорреляционный сигнал имеет мощность $\approx 1$ мкВт.
Для характерного смещения микрометрической платформы $\Delta = 0.35$ мм, уравнение (\ref{eqFWHM2}) даёт продолжительность импульса $\tau_{\rm FWHM} = 1.65$ пс.

\subsection{Спектральные измерения}
\label{Spectra}

Длина волны излучения $\lambda$ и ширина линии $\Delta\lambda$ определялись с помощью спектрометра APE PulseCheck. Для характерных значений $\lambda = 780$ нм и $\Delta\lambda = 0.7$ нм, время когерентности лазера есть \cite{Sivukhin2005}
\begin{equation}
\label{eqCohTime}
\dfrac{\Delta\omega}{\omega} = \dfrac{\Delta\lambda}{\lambda} \quad \rightarrow \quad
\tau_{\rm coh} = \dfrac{2 \pi}{\Delta\omega} = \dfrac{\lambda^2}{c\Delta\lambda},
\end{equation}
где $\omega = 2\pi c / \lambda$ есть частота излучения. Используя скорость света в воздухе $c=3\times 10^{11}$ мм/сек, (\ref{eqCohTime}) даёт $\tau_{\rm coh} \approx 2.9$ пс.

Отношение продолжительности импульса к времени когерентности
\begin{equation}
\dfrac{\tau_{\rm FWHM}}{\tau_{\rm coh}} = 0.57
\end{equation}
близко к предельному значению для гауссовых импульсов 0.44 \cite{Akhmatov2004}, что говорит о малости фазовой модуляции внутри импульса. Это значит, что каждый импульс находится в чистом когерентном состоянии \cite{Scully2003}.
Средняя мощность излучения $P$ в этих условиях связана с амплитудой когерентного состояния $\alpha$ как
\begin{equation}
\label{csPower}
P = \nu \left|\alpha\right|^2 \hbar \omega,
\end{equation}
где $\hbar = 10^{-34}$ Дж$*$сек есть постоянная Планка. Для характерного значения $P=6$ мВт, решение (\ref{csPower}) даёт $\alpha \approx 1.8\times 10^4$.

Поломка штатного спектрометра привела к необходимости альтернативного измерения длины волны. Применявшаяся схема, показанная на Рис. \ref{pMethods1}, состоит из дифракционной решётки ДР с периодом $a = 1.5$ мкм и камеры КАМ, настроенной на наблюдение дифракционного максимума порядка $m=1$ в дальнем поле. Разрешающая способность этого прибора равна \cite{Sivukhin2005}
\begin{equation}
R = \dfrac{\lambda}{\delta\lambda} = Nm = \dfrac{d/a}{\sin \alpha}m \approx 3\times 10^3,
\end{equation}
где $N$ есть число задействованных штрихов решётки, $d\approx1$ мм есть диаметр пучка, а $\alpha\approx0.2$ рад -- угол между пучком и плоскостью решётки.

\subsection{Фильтрация пространственной моды}
\label{laser_spatial}
Излучение лазера имеет существенно не-гауссову форму, и перед использованием должно пройти пространственную фильтрацию. Как показано на Рис. \ref{pMethods1}, для этого присутствует телескопическая система (Л1--Л2, 75+100 мм), в фокусе которой установлена круговая апертура А1.

\begin{figure}
	\includegraphics[width=5in]{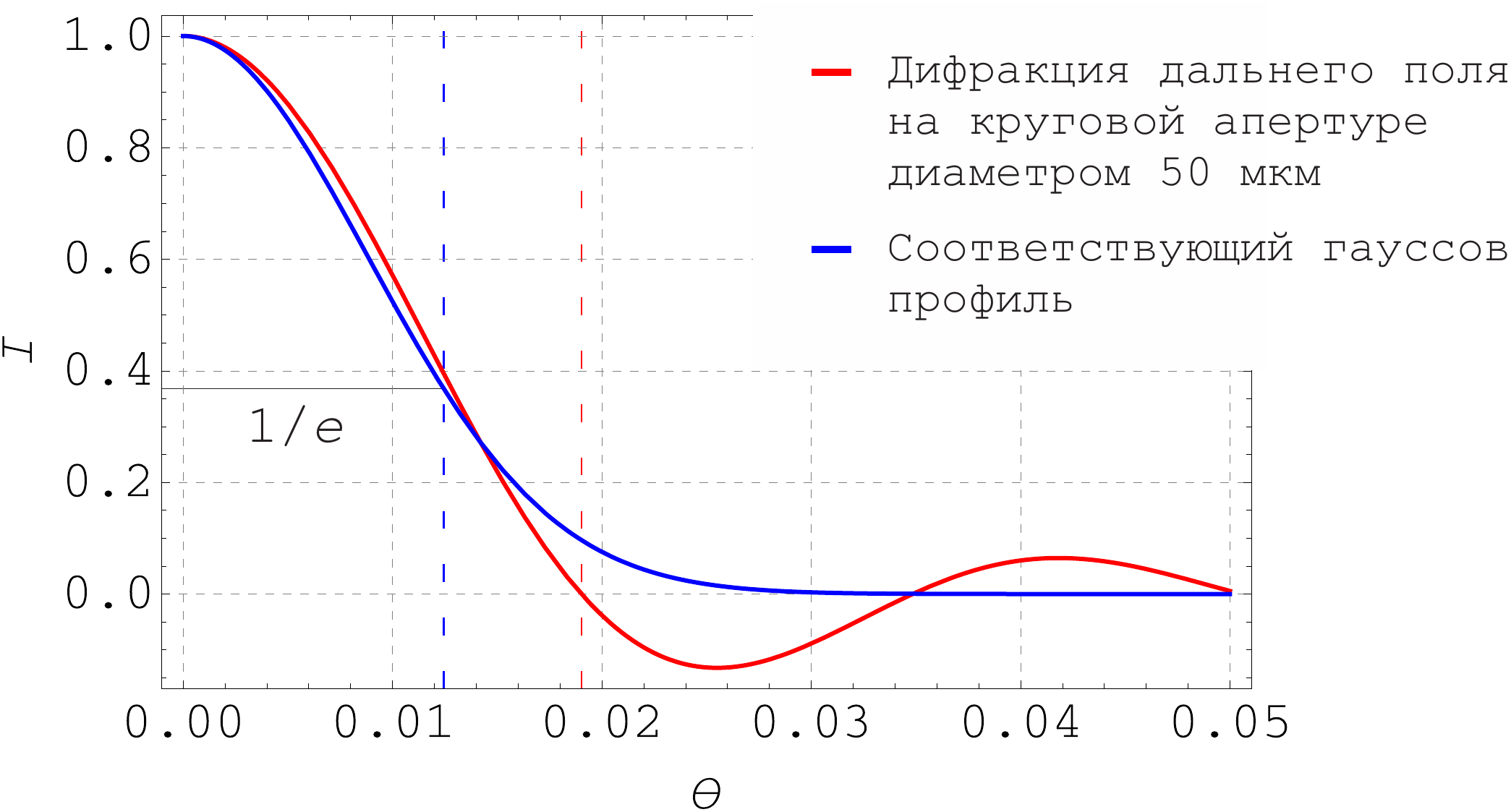}
	\centering
	\captionsetup{justification=raggedright, singlelinecheck=false}
	\caption{Красная линия: амплитуда дальнего поля в результате дифракции на круглом отверстии (\ref{eqAiry}). Красный пунктир показывает положение первого тёмного кольца, где располагается диафрагма А2 (Рис. \ref{pMethods1}). Синяя линия: ближайший гауссов профиль (\ref{eqGauss}) с радиусом, показанным синим пунктиром.}
	\label{pMethods2}
\end{figure}

Обезразмеренная амплитуда дифрагировавшего света описывается в дальнем поле функцией \cite{Sivukhin2005}
\begin{equation}
\label{eqAiry}
A_{\rm airy} = 2 \dfrac{J_1 \left[k a \sin \theta \right]}{k a \sin \theta},
\end{equation}
где $k=2\pi/\lambda$, $a$ -- диаметр апертуры, а $\theta$ -- зенитный угол по отношению к оси пучка. Как показано на Рис. \ref{pMethods2}, форма (\ref{eqAiry}) близка к гауссовой
\begin{equation}
\label{eqGauss}
A_{\rm gauss} = \exp \left[-\left(\dfrac{\sin \theta}{\sqrt{2\pi} / ka}\right)^2\right].
\end{equation}

Завершающая линза $F$ телескопа Л2, находящаяся от апертуры А1 на фокусном расстоянии коллимирует пучок. 
Для отсечения колец Эйри, в коллимированном пучке устанавливается диафрагма радиусом $R$, соответствующим первому нулю функции (\ref{eqAiry}), так что
\begin{equation}
\label{eqBesselMin}
R = \theta_1 F, \quad ka \sin \theta_1 \approx 3.83 \iff \theta_1 \approx \dfrac{1.22 \lambda}{2a}.
\end{equation}
Остающийся центральный максимум имеет перекрытие $99.7\%$ с гауссовым профилем (\ref{eqGauss}).
Радиус последнего определяется по условию спадания амплитуды поля в $e$ раз, и равен
\begin{equation}
\label{eqGaussR}
w = \dfrac{\sqrt{2\pi}}{ka}F \approx 0.65 R.
\end{equation}
Для используемых экспериментальных параметров $F=100$ мм, $a=50$ мкм, $\lambda=780$ нм, (\ref{eqBesselMin}) и (\ref{eqGaussR}) дают $R = 1.9$ мм, $w=1.24$ мм.

Если такой пучок далее фокусируется линзой c фокусным расстоянием $f$, то форма пучка следует уравнению \cite{Sivukhin2005}
\begin{equation}
\label{eqGaussForm}
w(z) = w_0\sqrt{1+\left(\dfrac{z}{z_R}\right)^2},
\end{equation}
где $z$ есть продольная координата, отсчитываемая от центра перетяжки, $w_0$ -- её радиус, а $z_R$ -- Рэлеевская длина. Последние определяются через угол сходимости пучка $\phi$ как 
\begin{equation}
\label{eqFWaist}
\phi = \dfrac{\lambda}{\pi \omega_0} = \dfrac{w}{f} \quad \rightarrow \quad \omega_0 = \dfrac{f\lambda}{\pi\omega}, \quad z_R = \dfrac{\pi\omega_0^2}{\lambda}.
\end{equation}

\clearpage

\subsection{Управление поляризационным состоянием света с помощью фазовой задержки}
\label{waveplates}

Рассмотрим действие пары последовательно расположенных четверть- и полу-волновой пластин, на поляризацию светового пучка. Их быстрые оси составляют с горизонталью углы $\phi$ и $\theta$. Если поляризационные компоненты входящего пучка описываются операторами уничтожения $\hat{a}_\mathrm{h}$ и $\hat{a}_\mathrm{v}$, то операторы выходных компонент равны
\begin{equation}
\label{eqJones1}
\left[ {\begin{array}{c}
	\hat{b}_\mathrm{v} \\
	\hat{b}_\mathrm{h} \\
	\end{array} } \right] = W_{\pi/2}(\theta) \times W_{\pi}(\phi) \times 
\left[ {\begin{array}{c}
	\hat{a}_\mathrm{v} \\
	\hat{a}_\mathrm{h} \\
	\end{array} } \right],
\end{equation}
где $W_{\pi/2}$ и $W_{\pi}$ есть матрицы, соответствующие четверть- и полу-волновым пластинам. Они определяются как
\begin{equation}
\label{eqJones2}
W_{\Delta}(x) = \mathrm{\Pi}(x) + e^{i\Delta} \mathrm{\Pi}(x+\pi/2),
\end{equation}
где $\Delta$ есть величина фазовой задержки, а
\begin{equation}
\label{eqJones3}
\mathrm{\Pi}(x) = 
\left[ 
{\begin{array}{cc}
	\sin^2{x} & \sin(x) \cos(x) \\
	\sin(x) \cos(x) & \cos^2{x} \\
	\end{array} } 
\right]
\end{equation}
есть матрица проектирования вектора на направление $x$, с выходными компонентами представленными в исходном базисе. $x$ и $x+\pi/2$ в выражении (\ref{eqJones2}) есть направления быстрой и медленной осей волновой пластины.

В случае, когда входной пучок находится в когерентном состоянии, амплитуды его входных и выходных поляризационных компонент $\alpha_\mathrm{v/h}$ и $\beta_\mathrm{v/h}$ связаны тем же соотношением (\ref{eqJones1}). Оно использовалось, в частности, для приготовления пробных состояний в эксперименте по томографии квантовых процессов, разд. \ref{chapt1}.

\section{Гомодинная томография}
\label{BHD_all}
\subsection{Концепция}
\label{BHD_theory}
Гомодинная томография представляет собой один из наиболее эффективных методов характеризации квантового состояния света. Метод требует приготовления большого ансамбля физических систем -- в нашем случае, световых импульсов -- в исследуемом состоянии, и проведения над каждой квадратурного измерения. 

Квантовое описание света эквивалентно описанию маятника, в котором энергия осциллирует между электрической и магнитной компонентами поля \cite{Scully2003}. Измерение суперпозиции канонической координаты $\hat{X}$ и импульса $\hat{P}$
\begin{equation}
\label{eqQuad1}
\hat{Q}_\theta = \hat{X} \cos \theta + \hat{P} \sin \theta
\end{equation}
и есть квадратурное, или гомодинное, измерение \cite{Lvovsky2009}.

Для измерения величины (\ref{eqQuad1}), неизвестное сигнальное состояние подвергается интерференции со вспомогательным пучком -- местной волной -- на симметричном светоделителе. В этом процессе, операторы сигнальной моды $\hat{a}_\mathrm{s}e^{i\mu}$ и местной волны $\hat{a}_\mathrm{o}e^{i\nu}$ преобразуются как \cite{Leonhardt1997}
\begin{equation}
\label{eqBHDsplit}
\left[ {\begin{array}{c}
	\hat{a}_{\mathrm{out1}} \\
	\hat{a}_{\mathrm{out2}} \\
	\end{array} } \right]
=
\dfrac{1}{\sqrt{2}}
\left[ {\begin{array}{c}
	\hat{a}_{\mathrm{s}}e^{i\mu} - \hat{a}_{\mathrm{o}}e^{i\nu} \\
	\hat{a}_{\mathrm{s}}e^{i\mu} + \hat{a}_{\mathrm{o}}e^{i\nu} \\
	\end{array} } \right].
\end{equation}
Затем, каждая из выходных мод светоделителя попадает на фотодиод, разность фототоков которых и представляет собой выходной сигнал детектора. Этой величине соответствует оператор разности чисел фотонов
\begin{equation}
\label{eqBHDPower1}
\begin{aligned}
\delta\hat{W} = 
& \hat{a}_{\mathrm{out2}}^\dagger \hat{a}_{\mathrm{out2}} -  \hat{a}_{\mathrm{out1}}^\dagger \hat{a}_{\mathrm{out1}} = 
\sqrt{2} \left( \hat{a}_{\mathrm{o}}^\dagger \hat{a}_{\mathrm{s}} e^{i(\mu-\nu)} + \hat{a}_{\mathrm{s}}^\dagger \hat{a}_{\mathrm{o}} e^{i(\nu-\mu)}\right).
\end{aligned}
\end{equation}

Местная волна приготавливается в когерентном состоянии $\ket{\alpha_\mathrm{o}}$ с $\alpha_\mathrm{o} \sim 10^4$. С учётом этого, оператор (\ref{eqBHDPower1}) сводится к одномодовой форме
\begin{equation}
\label{eqBHDPower2}
\begin{aligned}
\delta\hat{W_\mathrm{s}} = 
& \bra{\alpha_\mathrm{o}} \delta \hat{W} \ket{\alpha_\mathrm{o}} = 
2 \alpha_\mathrm{o} \left( \dfrac{\hat{a}_{\mathrm{s}} e^{i(\mu-\nu)} + \hat{a}_{\mathrm{s}}^\dagger e^{i(\nu-\mu)}}{\sqrt{2}} \right).
\end{aligned}
\end{equation}
В силу определений \cite{Leonhardt1997}
\begin{equation}
\label{eqXP}
\hat{X} = \dfrac{\hat{a}+\hat{a}^\dag}{\sqrt{2}}, \quad
\hat{P} = \dfrac{\hat{a}-\hat{a}^\dag}{i\sqrt{2}},
\end{equation}
выражение (\ref{eqBHDPower2}) можно представить в форме (\ref{eqQuad1}):
\begin{equation}
\label{eqQuadSig}
\delta\hat{W_\mathrm{s}} = 2 \alpha_\mathrm{o} \left[ \hat{X}_\mathrm{s} \cos (\nu-\mu) + \hat{P}_\mathrm{s} \sin (\nu-\mu) \right].
\end{equation} 
Таким образом, разность мощностей (\ref{eqQuadSig}) пропорциональна квадратурной наблюдаемой (\ref{eqQuad1}), где $\theta = \nu-\mu$ есть разность фаз между местной волной и сигнальным полем.
Оптико-электронная схема, производящая манипуляции (\ref{eqBHDsplit}) -- (\ref{eqBHDPower2}), называется гомодинным детектором (ГД) \cite{Lvovsky2009, Kumar2012}. 

Для характеризации $M$--модового квантового состояния с помощью квадратурных измерений, необходимо иметь набор данных в форме
\begin{equation}
\label{eqBHDdata}
\left\{Q_1^i...Q_\mathrm{M}^i, \quad \theta_1^i...\theta_\mathrm{M}^i \right\},
\end{equation}
где $Q_\mathrm{k}^i$ есть результат $i$--го измерения квадратуры в моде $k$, а $\theta_\mathrm{k}^i$ есть фаза соответствующей наблюдаемой в момент измерения. В общем случае, набор (\ref{eqBHDdata}) должен содержать статистически значительный набор измерений квадратур во всех точках фазового гиперкуба $\left[0...2\pi\right]^M$. 

На практике используются конечные массивы данных; шаг фаз при этом должен соответствовать желаемой детализации квантового состояния.  
В эксперименте, фазы варьируются с помощью пьезо-элемента, встроенного в держатель одного из зеркал. Последний преобразует управляющее напряжение в смещение зеркала на несколько длин волн, что покрывает все возможные фазы луча.
Согласно (\ref{eqQuadSig}), фаза наблюдаемой $\theta$ зависит в равной мере как от фазы сигнала $\mu$, так и от фазы местной волны $\nu$; это значит, что пьезо-элемент может быть с одинаковым успехом установлен в любом из этих пучков. 

Измерение фаз для составления массива (\ref{eqBHDdata}) является отдельной задачей и часто оказывается ключевой экспериментальной трудностью. Непосредственное решение заключается в независимом измерении этих величин, например с помощью детектирования референсных состояний. Во многих случаях, однако, априорное знание о свойствах детектируемого состояния позволяет извлекать фазовую информацию из самих квадратурных данных. Именно этим объясняется различие методик гомодинного детектирования разных классов квантовых состояний. Особенности гомодинной томографии когерентных, фоковских и сжатых состояний света описаны в разделах \ref{BHD1}, \ref{Fock_BHD} и \ref{TSMStomography}.

\subsection{Обработка сигнала гомодинного детектора}
\label{BHD_daq}

Оптический сигнал, поступающий на ГД, имеет форму импульсов продолжительностью $\sim2$ пикосекунды (разд. \ref{autocorr}). Электроника детектора, имеющая полосу пропускания $\sim100$ МГц \cite{Kumar2012}, формирует на каждый из них отклик продолжительностью $\sim 5$ наносекунд, и позволяет таким образом разрешать отдельные лазерные импульсы, приходящие с периодом в 13 наносекунд.

\begin{figure*}[ht]
	\centering
	\includegraphics[width=6.65in, angle=0]{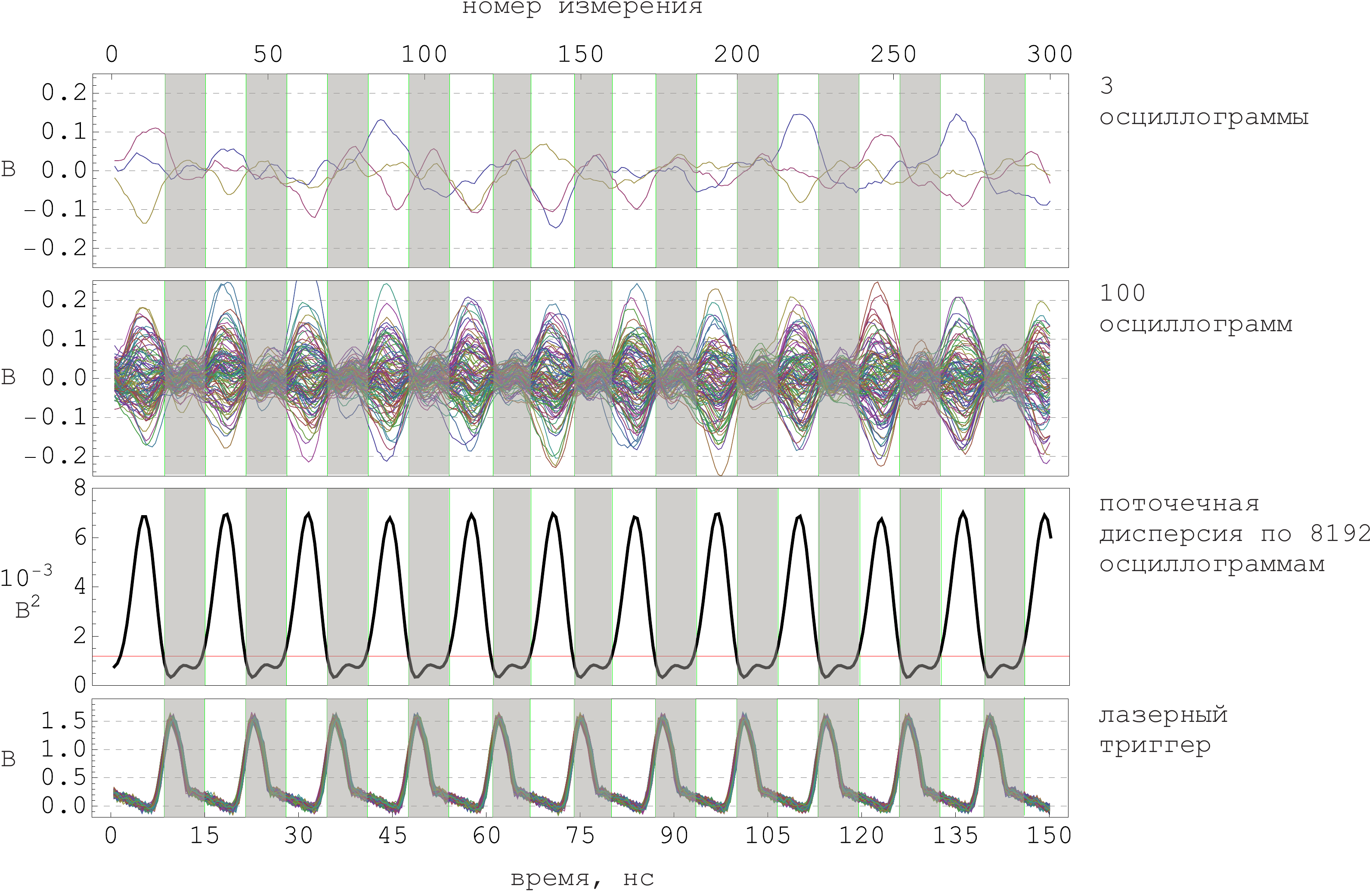}
	\captionsetup{justification=raggedright,singlelinecheck=false}
	\caption{Обработка сигнала ГД, детектирующего вакуумное состояние света. Интервалы, не содержащие квадратурной информации, затемнены. Красная линия на панели 3 соответствует пороговому значению для поточечной дисперсии.}
	\label{pBHDsampling}
\end{figure*}

Измерение и запись сигнала ГД производится с помощью осциллографа Agilent DSO9254A с полосой 2.5 ГГц, или аналогово-цифрового преобразователя Agilent Acqiris U1084A с частотой дискретизации 4 ГГц.
Пример выходного сигнала ГД, измеряющего вакуумное состояние света, показан на верхнем графике Рис. \ref{pBHDsampling}. Сигнал предварительно очищен электронными фильтрами от высокочастотного звона и возбуждений на одинарной и удвоенной частоте повторения лазера (разд. \ref{laser}).

Для извлечения из подобных осциллограмм квадратурных значений необходимо знать временные интервалы, соответствующие отклику детектора на каждый лазерный импульс. Для этого, анализируется набор осциллограмм, записанных так, что первая точка каждой находится в фиксированном положении относительно ближайшего лазерного импульса; этого можно добиться, используя в качестве триггера лазерный сигнал (нижний график на Рис. \ref{pBHDsampling}).

Как показано на третьей панели Рис. \ref{pBHDsampling}, поточечная дисперсия, вычисленная по такому набору осциллограмм, позволяет определить для каждой из них интервалы, содержащие отклик детектора на отдельные лазерные импульсы.
Средние от сигнала ГД, найденные для каждого сигнального интервала, соответствуют результатам измерения квадратуры, проведённого для соответствующего светового импульса.

\begin{figure}[h]
	\centering
	\includegraphics[width=6.65in, angle=0]{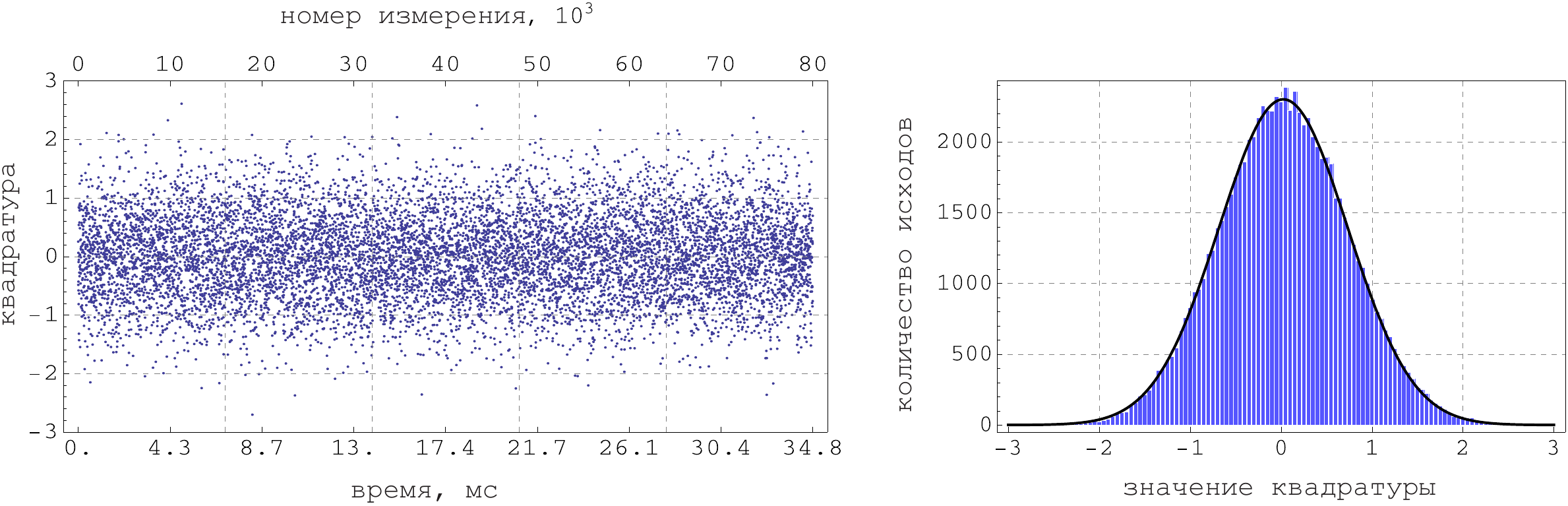}
	\captionsetup{justification=raggedright,			singlelinecheck=false}
	\caption{$8192*10$ нормированных квадратурных значений, извлечённых из осциллограмм, показанных на Рис. \ref{pBHDsampling}. Для удобства, показана каждая пятая точка. Слева: значение квадратуры в зависимости от времени и номера лазерного импульса. Справа: гистограмма квадратурных измерений и теоретическое ожидание \cite{Leonhardt1997}.}
	\label{pBHDdatanorm}
\end{figure}

Дисперсия квадратурных измерений в вакуумном состоянии света, по соглашению равная 1/2 \cite{Leonhardt1997}используется для нормировки квадратурных измерений.
На Рис. \ref{pBHDdatanorm} показаны нормированные таким образом данные, полученные из осциллограмм Рис. \ref{pBHDsampling}.

Восстановление квантового состояния происходит в ходе работы итеративного численного алгоритма \cite{Lvovsky2004, MaxLik07}, использующего принцип максимального правдоподобия (разд.~\ref{chapt2}).

\subsection{Эффективность}
\label{BHD_efficiency}

\subsubsection{Отличие огибающих}

В предыдущем разделе предполагалось, что пространственная и временная форма мод сигнала и местной волны совпадают. Что, если это не так?

Из выражения (\ref{eqQuadSig}) видно, что сигнал гомодинного детектора есть значение наблюдаемой (\ref{eqQuad1}), усиленной пропорционально амплитуде местной волны $\alpha_o$. Если формы сигнала и местной волны различны, то усиливаться будет лишь часть сигнала, перекрывающаяся с местной волной; остальная же часть последней будет усиливать квадратуру иных мод, находящихся в вакуумном состоянии. 
Эта ситуация эквивалентна идеальному гомодинному детектированию сигнала, предварительно претерпевшего потери. Соответствующая эффективность гомодинного измерения есть \cite{Raymer2004}
\begin{equation}
\label{eqBHDefficiency}
\eta_\mathrm{os} = \left| \int_\mathrm{det} \mathrm{\Phi}^*_\mathrm{o} \mathrm{\Phi}_\mathrm{s} d\mathbf{r} dt \right|^2,
\end{equation}
где $\mathrm{\Phi}_\mathrm{s,o}(\mathbf{r},t)$ есть формы мод сигнала и местной волны, а интегрирование ведётся по области детектирования.

Обстоятельство (\ref{eqBHDefficiency}) обуславливает важность тщательного совмещения мод местной волны и детектируемого сигнала. На практике, этого добиваются в два этапа:
\begin{enumerate}
\item[1.] ``Классическое'' совмещение.\\
Используется интерференция местной волны со вспомогательным пучком макроскопической амплитуды, близким к ожидаемой моде квантового сигнала. При равных амплитудах $A$, интерференционный сигнал зависит от разности фаз $\Delta$ между пучками как
\begin{equation}
I = 2A^2(1-\eta \cos \Delta),
\end{equation}
где $\eta$ есть перекрытие (\ref{eqBHDefficiency}) интерферирующих мод. Последнее совпадает с видностью интерференционной картины:
\begin{equation}
\label{eqVisibility}
V = \dfrac{I_\mathrm{max} - I_\mathrm{min}}{I_\mathrm{max} + I_\mathrm{min}} = \eta,
\end{equation}
которая используется как оптимизационный параметр для совмещения мод. Последнее включает в себя:
\begin{itemize}
\item Уравнивание пространственных размеров мод с помощью системы линз
\item Совмещение пространственных мод с помощью пары зеркал
\item Временную синхронизацию мод с помощью передвижной призмы -- ``тромбона''. 
\end{itemize}

Для удобства работы с размером и формой пространственных мод часто применяется пара видеокамер, настроенных на наблюдение пучка в дальнем и ближнем полях; камера дальнего поля располагается в фокусной плоскости собирающей линзы. 
Совпадение пучков на обоих камерах эквивалентно поперечному совмещению пространственных мод. 

\item[2.] ``Квантовое'' совмещение.\\
Мода квантового сигнала обычно отличается от вспомогательного пучка. Поэтому, более точное совмещение мод производится в режиме измерения самого квантового сигнала. В качестве оптимизационного параметра обычно выступает эффективность детектирования квантового состояния, вычисляемая в режиме реального времени в ходе гомодинного измерения.

Этот этап учитывается при постройке экспериментальной установки так, чтобы такая оптимизация не приводила к расстройке гомодинного детектора. Для этого, например, зеркала, используемые для совмещения пространственных мод, располагаются в сигнальном пучке, а не в пучке местной волны. В меньшей мере это касается телескопических систем и тромбонов, прохождение через которые связано со значительными потерями, нежелательными в сигнальном канале. 
\end{enumerate}

В ситуации, когда форма моды сигнального состояния неизвестна, гомодинное детектирование может использоваться для её определения. В работе \cite{Qin2015} показано, как информация о временной моде квантового состояния извлекается из автокорреляционных измерений сигнала ГД.

\subsubsection{Отличие частот}
\label{DiffFreq}

Рассмотрим ситуацию, когда оптическая частота сигнала $\omega_s$ отличается от частоты местной волны $\omega_o$:
\begin{equation}
\omega_s = \left(1 + \delta\right)\omega_o.
\end{equation}
Если амплитуды, а также пространственная и временная формы мод сигнала и местной волны совпадают, то их интерференция на каждом лазерном импульсе порождает сигнал мощностью
\begin{equation}
\label{eqIInter}
I \propto \int_{t=0}^{\tau} \left[\cos (\omega_o t) + \cos(\omega_s t + \Delta)\right]^2 dt,
\end{equation}
где $\Delta$ есть разность фаз в момент времени $t=0$, а
интегрирование ведётся по продолжительности лазерного импульса, составляющего $m$ оптических периодов: $\tau = 2m\pi/\omega_o$.

\begin{figure}
	\centering
	\includegraphics[width=4in, angle=0]{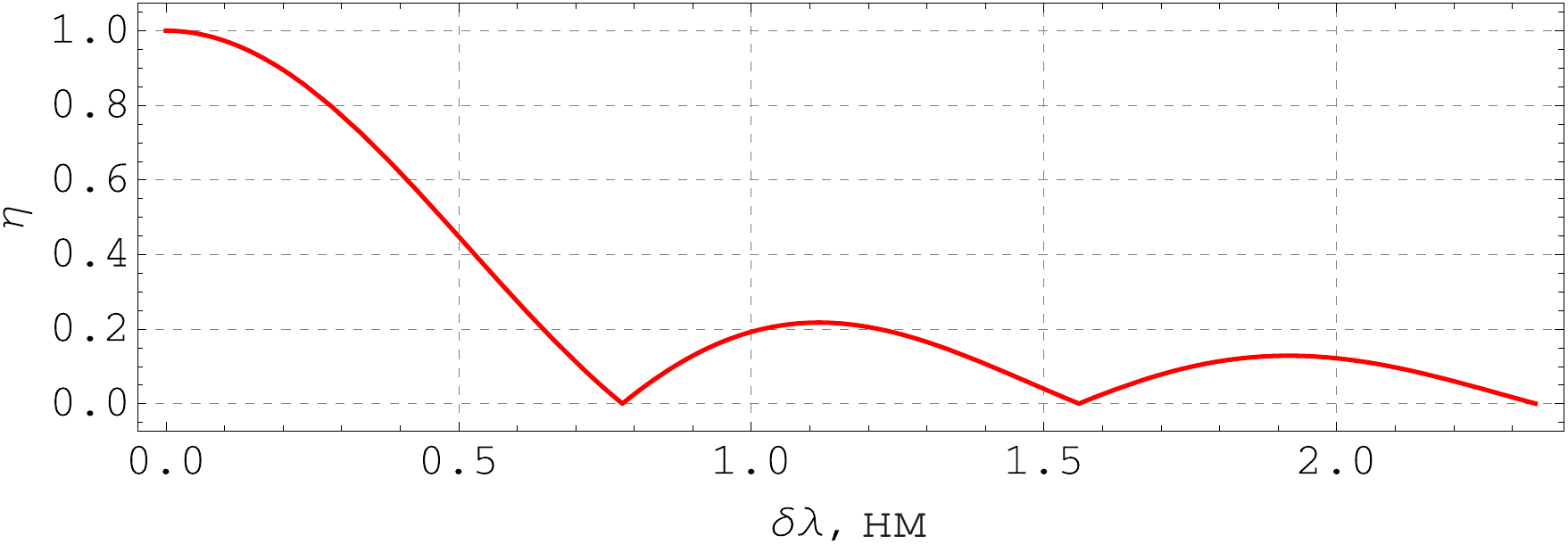}
	\captionsetup{justification=raggedright,singlelinecheck=false}
	\caption{Эффективность гомодинного детектирования, обусловленная различием длин волн местной волны и сигнала $\delta\lambda$ (\ref{eqVisFreq}), $m=10^3$.}
	\label{pSpectralFiltering}
\end{figure}

Видность интерференционной картины $I(\Delta)$, соответствующая эффективности гомодинного измерения (\ref{eqVisibility}), равна
\begin{equation}
\label{eqVisFreq}
V = \eta = \left|\dfrac{\sin\left[\delta m \pi\right]}{\delta m \pi}\right|.
\end{equation} 
Функция $\eta(\delta\lambda)$ показана на Рис. \ref{pSpectralFiltering} для $\tau\approx2$ пс (разд. \ref{autocorr}), что соответствует $m\sim10^3$.

\subsubsection{Электронная эффективность}

Наряду с ``оптической'' эффективностью гомодинного измерения, существенный вклад вносит электронная. По сравнению с качеством сведения мод, она имеет более фундаментальные причины, связанные с наличием собственного шума электронной схемы детектора. Используемый нами детектор обладает отношением сигнал / шум 7-12dB \cite{Kumar2012}. В сумме с квантовой эффективностью фотодиодов, $91\%$, полная эффективность электронной схемы составляет $\sim 80\%$.

\section{Сжатые состояния света: теоретическое описание}
\label{SPDC}
Процесс спонтанного параметрического рассеяния света (СПР) является важнейшим в квантово-оптических технологиях. Он используется как средство приготовления фоковских и сжатых состояний света, которые используются сами по себе или для дальнейшей инженерии более сложных квантовых состояний. Пионерские работы в этом направлении велись в 80х -- 90х годах 20 века \cite{Slusher1985, Wu1986, Ou1988, Shih1988, Brendel1992, Kwiat1993, Kiess1993, Kwiat1995}.

В ходе СПР, фотон накачки распадается на пару дочерних фотонов. Насколько мне известно, этот процесс не имеет места в естественной природе, и наблюдается лишь в лабораторных условиях; в частности, требуется среда, имеющая особые нелинейные свойства. Для экспериментов, описываемых в настоящей работе, использовались кристаллы титанил-фосфата калия \cite{Bierlein1989} с периодической доменной структурой \cite{LvovskyLect}.
В отличие от других нелинейных оптических процессов, СПР не имеет классического описания \cite{LvovskyLect, Boyd2008}; его природа является существенно квантовой. Далее в этом разделе представлено его краткое теоретическое описание.

\subsection{Двумодово--сжатый вакуум}
\label{TMS}
Сигнальные фотоны, родившиеся в результате распада фотона накачки, могут различаться по длине волны, пространственной моде и поляризации. В случае сильного поля накачки, истощением последней можно пренебречь. Этому приближению соответствует полуклассическое описание процесса, рассматривающее лишь эволюцию сигнальных пучков. Эта эволюция описывается Гамильтонианом
\begin{equation}
\label{eqTMSqueezingHam}
\hat{H}_{\mathrm{tms}} = i\hbar \beta \left(\hat{a} \hat{b} - \hat{a}^\dagger \hat{b}^\dagger  \right), 
\end{equation}
где $\hat{a}$ и $\hat{b}$ -- операторы уничтожения фотонов в сигнальных модах \cite{Scully2003}. Второе слагаемое в (\ref{eqTMSqueezingHam}) описывает рождение пары сигнальных фотонов, тогда первое соответствует обратному процессу -- её уничтожению.

В представлении Гейзенберга, взаимодействие (\ref{eqTMSqueezingHam}) приводит к изменению операторов уничтожения, эволюция которых подчиняется законам
\begin{equation}
\label{eqTMSevol}
\begin{aligned}
&\dot{\hat{a}} = \dfrac{i}{\hbar} \left[\hat{H}_{\mathrm{tms}},\hat{a}\right] = -\beta \hat{b}^\dagger, \\
&\dot{\hat{b}} = \dfrac{i}{\hbar} \left[\hat{H}_{\mathrm{tms}},\hat{b}\right] = -\beta \hat{a}^\dagger.
\end{aligned}
\end{equation}
По прошествии времени $t = \zeta / \beta$, соответствующего продолжительности прохождения света сквозь кристалл, результат динамики (\ref{eqTMSevol}) выражается следующим преобразованием операторов:
\begin{equation}
\label{eqTMSHeizenbergAs}
\begin{aligned}
& \hat{a}_{\mathrm{out}} = \hat{a}_{\mathrm{in}} \mathrm{cosh}\,\zeta - \hat{b}_{\mathrm{in}}^\dagger \mathrm{sinh}\,\zeta, \\
& \hat{b}_{\mathrm{out}} = \hat{b}_{\mathrm{in}} \mathrm{cosh}\,\zeta - \hat{a}_{\mathrm{in}}^\dagger \mathrm{sinh}\,\zeta.
\end{aligned}
\end{equation}

\begin{figure}[h]
	\centering
	\includegraphics[width=3.5in]{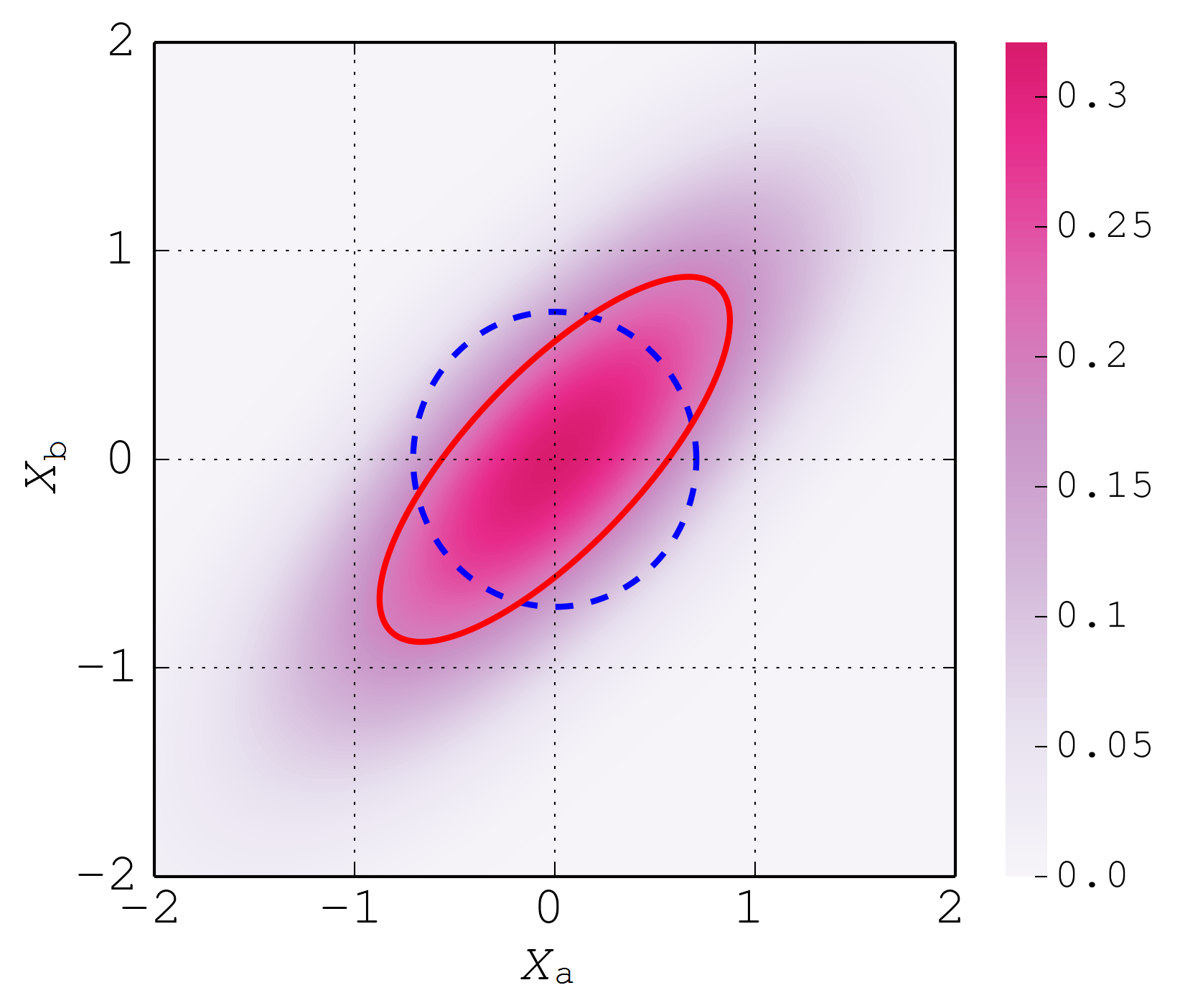}
	\captionsetup{
				justification=raggedright,
				singlelinecheck=false
	}
	\caption{Квадрат модуля волновой функции состояния двумодового сжатого вакуума (\ref{eqTMSWF}) для $\zeta = 0.5$. Красная линия показывает величину среднеквадратического отклонения результатов измерения двумодовых наблюдаемых (\ref{eqTMobservables}) с фазами, соответствующими данному направлению. Синяя линия соответствует вакуумному состоянию обоих мод.}
	\label{pTMSwavefunction}
\end{figure}

Преобразование (\ref{eqTMSHeizenbergAs}) эквивалентно перемасштабированию суммы и разности канонических квадратур (\ref{eqXP}) в двух модах:
\begin{equation}
\label{eqTMSHeizenbergXs}
\begin{aligned} 
& (\hat{X}_a \pm \hat{X}_b)_{\mathrm{out}} \rightarrow (\hat{X}_a \pm \hat{X}_b)_{\mathrm{in}} \times e^{\pm \zeta}, \\
& (\hat{P}_a \pm \hat{P}_b)_{\mathrm{out}} \rightarrow (\hat{P}_a \pm \hat{P}_b)_{\mathrm{in}} \times e^{\mp \zeta}.
\end{aligned}
\end{equation}
Другими словами, некоторые двумодовые наблюдаемые (\ref{eqTMSHeizenbergXs}) оказываются растянутыми, тогда как другие -- сжатыми.
Если моды $a$ и $b$ изначально находились в состоянии вакуума, так что двумодовая волновая функция равна
\begin{equation}
\label{eqTMVacuum}
\begin{aligned}
\Psi_{00}(X_a, X_b) = & \dfrac{1}{\sqrt{\pi}} \exp \left[{-\dfrac{X_a^2 + X_b^2}{2}}\right] = \\
& \dfrac{1}{\sqrt{\pi}} \exp \left[{-\left(\dfrac{X_a - X_b}{2}\right)^2}\right] \exp \left[-\left(\dfrac{X_a + X_b}{2}\right)^2\right],
\end{aligned}
\end{equation}
то в представлении Шрёдингера, преобразование (\ref{eqTMSHeizenbergXs}) приводит их в состояние
\begin{equation}
\label{eqTMSWF}
\Psi^\mathrm{TMS}_{\zeta}(X_a, X_b) =  \dfrac{1}{\sqrt{\pi}} \exp \left[{-\left(\dfrac{X_a - X_b}{2\,e^{-\zeta}}\right)^2}\right] \exp \left[-\left(\dfrac{X_a + X_b}{2\,e^{+\zeta}}\right)^2\right],
\end{equation}
называемое двумодовым сжатым вакуумом. Волновая функция (\ref{eqTMSWF}) показана на Рис.~\ref{pTMSwavefunction}. Её форма есть гауссов колокол, соответствующий двумодовому вакууму, сжатый по диагонали.

Среди всех состояний света, возможных в классической физике, наименьший уровень квадратурного шума достигается вакуумным состоянием света:
\begin{equation}
\label{eqVacuum}
\Psi_0(X) = \dfrac{1}{\pi^{1/4}} \exp \left[{-\dfrac{X^2}{2}}\right],
\end{equation}
в котором дисперсия квадратурных измерений равна
\begin{equation}
\label{eqVacNoise}
\left\langle \hat{X}^2 \right\rangle_{\Psi_0} = \int_{-\infty}^{\infty} dX \, \Psi_0^2 \, X^2 = \dfrac{1}{2}.
\end{equation}
Как можно заключить уже из (\ref{eqTMSHeizenbergXs}), дисперсии разности координат и суммы импульсов (\ref{eqTMSHeizenbergXs}) в состоянии (\ref{eqTMSWF}) меньше классического предела (\ref{eqVacNoise}) в $e^{2\zeta}$ раз:
\begin{equation}
\left\langle \left(\dfrac{\hat{X}_a \pm \hat{X}_b}{\sqrt{2}}\right)^2 \right\rangle_{\Psi^\mathrm{TMS}_{\zeta}} = \left\langle \left(\dfrac{\hat{P}_a \mp \hat{P}_b}{\sqrt{2}}\right)^2 \right\rangle_{\Psi^\mathrm{TMS}_{\zeta}} = \dfrac{1}{2}e^{\pm 2 \zeta}.
\label{eqTMSvar}
\end{equation}

Наблюдаемые (\ref{eqTMSHeizenbergXs}) могут быть обобщены на случай произвольной фазы между квадратурами двух мод:
\begin{equation}
\label{eqTMobservables}
\dfrac{\hat{Q}^{\theta_1}_1 + \hat{Q}^{\theta_2}_2}{\sqrt{2}} = \dfrac{\hat{X}_1 \cos{\theta_1} + \hat{P}_1 \sin{\theta_1} + \hat{X}_2 \cos{\theta_2} + \hat{P}_2 \sin{\theta_2}}{\sqrt{2}}.
\end{equation}
Для наблюдаемой (\ref{eqTMobservables}) в состоянии (\ref{eqTMSWF}), дисперсия зависит от суммы фаз местных волн как
\begin{equation}
\label{eqTMSIdealNoise}
\left\langle \Psi^\mathrm{TMS}_\zeta \left| \left( \dfrac{\hat{Q}^{\theta_1}_1 + \hat{Q}^{\theta_2}_2}{\sqrt{2}} \right)^2\, \right| \Psi^\mathrm{TMS}_\zeta \right\rangle =\dfrac{1}{2}\left(e^{2\zeta}\cos^2{\dfrac{\theta_1+\theta_2}{2}}+e^{-2\zeta}\sin^2{\dfrac{\theta_1+\theta_2}{2}}\right).
\end{equation}

В фоковском базисе, двумодово-сжатый вакуум (\ref{eqTMSWF}) представляется суперпозицией состояний с равным числом фотонов в каждой из мод $\ket{nn} \equiv \ket{n}_a \otimes \ket{n}_b$ \cite{LvovskyLect}:
\begin{equation}
\label{eqTMSfock}
\Psi^\mathrm{TMS}_{\zeta} = \dfrac{1}{\cosh \zeta} \sum_{n=0}^{\infty} \tanh^n \zeta \ket{nn}.
\end{equation}
В случае слабого сжатия, $\zeta \ll 1$, состояние (\ref{eqTMSfock}) приближённо равно
\begin{equation}
\label{eqFockTMS}
\Psi^\mathrm{TMS}_\zeta \propto \ket{00} + \zeta \ket{11}.
\end{equation}

Среднее число фотонов в каждой из мод состояния (\ref{eqTMSfock}) равно
\begin{equation}
\label{eqNPtms}
_{a,b}\!\left\langle \hat{n} \right\rangle_{\Psi^\mathrm{TMS}_{\zeta}} = \sum_{n = 0}^{\infty} n \left|\mathrm{Tr}\left[_{a,b}\!\braket{n}{\Psi^\mathrm{TMS}_{\zeta}}
\right]\right|^2 = \sinh^2 \zeta.
\end{equation}
В случае, если мода $a$ игнорируется, состояние моды $b$ сводится к смеси фоковских компонент с теми же весами (\ref{eqTMSfock}):
\begin{equation}
\label{eqThermal}
\mathrm{Tr}_a\left[\Psi^\mathrm{TMS}_{\zeta}\right] = \dfrac{1}{\cosh^2 \zeta} \sum_{n=0}^{\infty} \tanh^{2n} \zeta \ketbra{n}{n}
\end{equation}
Состояние (\ref{eqThermal}) совпадает с тепловой взвесью фоковских состояний, веса которых подчиняются распределению Больцмана. Эффективная температура последнего $T$ определяется из
\begin{equation}
\exp \left[-\dfrac{\hbar\omega}{kT}\right] = \tanh^2 \zeta.
\end{equation}

\subsection{Одномодово--сжатый вакуум}
\label{SMS}
Если сигнальные моды процесса СПР неразличимы, или вырождены, то результирующее состояние единственной сигнальной моды называется одномодово-сжатым. Несмотря на сходство терминов, свойства такого состояния принципиально отличаются от свойств двумодового сжатия, и стоят отдельного рассмотрения.

По соглашению, гамильтониан вырожденного СПР получается из (\ref{eqTMSqueezingHam}) с помощью замен
\begin{equation}
\label{eqTSsub}
\hat{b} \rightarrow \hat{a}, \quad \beta \rightarrow \dfrac{\beta}{2}: \quad \hat{H}_{\mathrm{sms}} = i\hbar\dfrac{\beta}{2}\left(\hat{a}^2-\hat{a}^{\dagger 2}\right).
\end{equation}
Тогда, эволюция единственной сигнальной моды $a$
\begin{equation}
\dot{\hat{a}} = \dfrac{i}{\hbar} \left[\hat{H}_{\mathrm{sms}},\hat{a}\right] = -\beta \hat{a}^\dagger
\label{eqHeizenbergSQ}
\end{equation}
приводит, по окончании нелинейного взаимодействия через время $t = \zeta / \beta$, к результату
\begin{equation}
\label{eq30}
\hat{a}_{\mathrm{out}} = \hat{a}_{\mathrm{in}} \cosh \zeta - \hat{a}_{\mathrm{in}}^\dagger \sinh \zeta.
\end{equation}
Преобразование (\ref{eq30}), в отличие от (\ref{eqTMSHeizenbergAs}), соответствует сжатию или растяжению собственных квадратур (\ref{eqXP}):
\begin{equation}
\label{eqSqueezedQuads}
\begin{aligned}
& \hat{X}_{\mathrm{out}} = \hat{X}_{\mathrm{in}}e^{-\zeta}, \quad \hat{P}_{\mathrm{out}} = \hat{P}_{\mathrm{in}}e^{+\zeta}.
\end{aligned}
\end{equation}

В координатном представлении Шрёдингера, если сигнальная мода изначально находилась в состоянии вакуума (\ref{eqVacuum}), то в соответствии с (\ref{eqSqueezedQuads}) конечное состояние принимает вид
\begin{equation}
\Psi^\mathrm{SMS}_\zeta(X) = \Psi_0(e^\zeta X) = \dfrac{e^{\zeta/2}}{\pi^{1/4}} \exp \left[{-\dfrac{(e^\zeta X)^2}{2}}\right].
\label{eqSMS}
\end{equation}

\begin{figure}[h]
	\centering
	\includegraphics[width=3.5in]{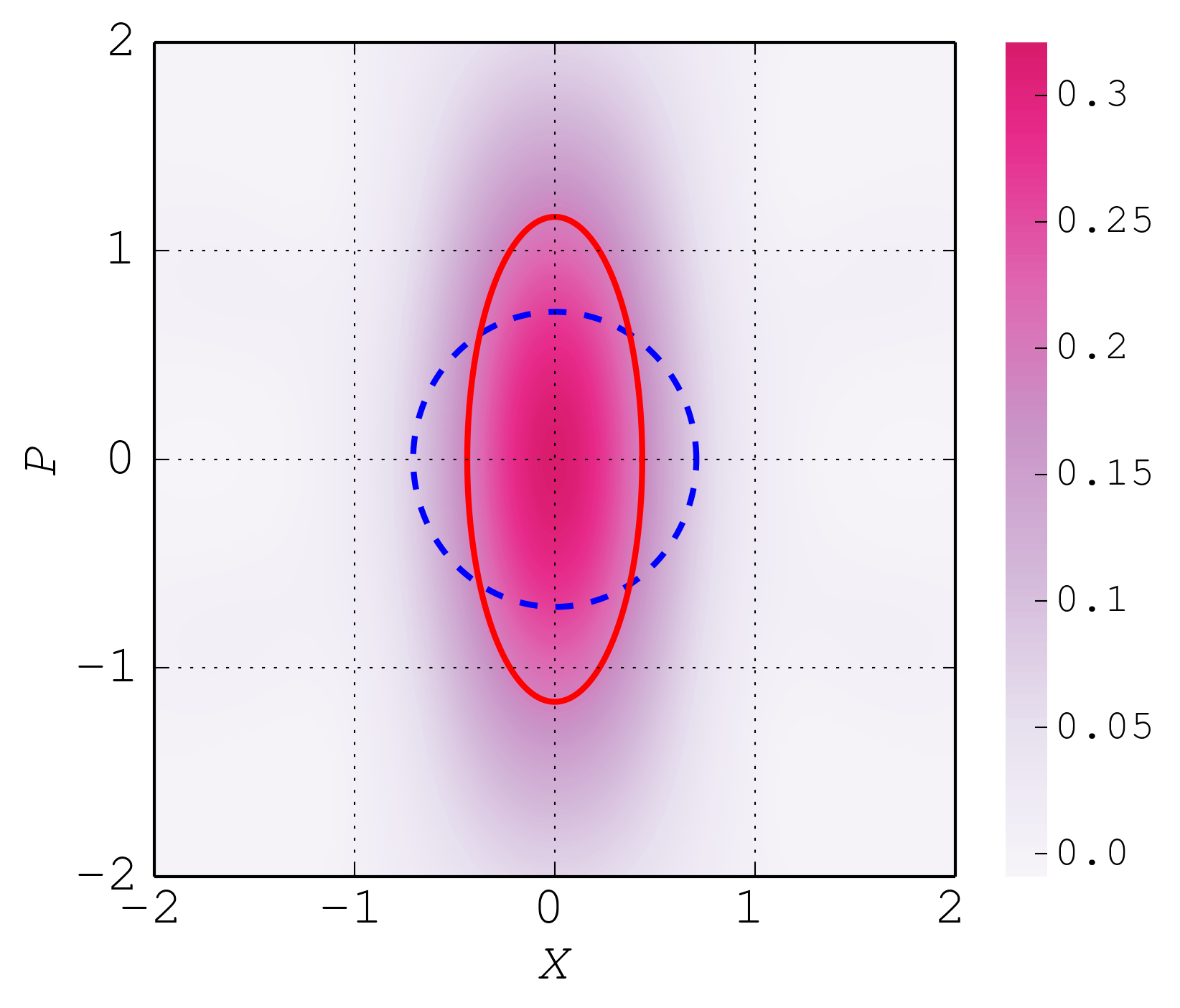}
	\captionsetup{justification=raggedright, singlelinecheck=false}
	\caption{Распределение Вигнера одномодово-сжатого состояния (\ref{eqSMS}) для параметра сжатия $\zeta = 0.5$. Средний шум показан красной линией. Синяя линия соответствует шуму вакуумного состояния.}
	\label{pSMSwigner}
\end{figure}
Функция Вигнера \cite{Schleich2005} состояния (\ref{eqSMS}) показана на Рис.~\ref{pSMSwigner}. Она имеет форму Гауссового колокола, сжатого по направлению $X$. В каждый момент времени, функция распределения измерений обобщённой наблюдаемой
\begin{equation}
\label{eqGeneralQuadrature}
\hat{Q}_\theta = \hat{X} \cos \theta + \hat{P} \sin \theta = \dfrac{\hat{a} e^{-i\theta} + \hat{a}^\dag e^{i\theta}}{\sqrt{2}}.
\end{equation}
есть интегральная проекция этого распределения на направление, составляющее угол $\theta$ с осью $X$ в фазовой плоскости \cite{Lvovsky2009}.
Отсюда видно, что дисперсия результатов квадратурных измерений в состоянии сжатого вакуума пропорционален $\cos{2\theta}$, что согласуется с предсказанием, основанным на выражениях (\ref{eqSqueezedQuads}) и (\ref{eqGeneralQuadrature}):
\begin{equation}
\left\langle \Psi^\mathrm{SMS}_\zeta \left| \hat{Q}_\theta^2 \right| \Psi^\mathrm{SMS}_\zeta \right\rangle = \dfrac{\cos^2\!{\theta} \,\, e^{-2\zeta} + \sin^2\!{\theta} \,\, e^{2\zeta}}{2}.
\label{eqSMSGeneralNoise}
\end{equation}

При $\zeta > 0$, дисперсия сжатой квадратуры $X$ меньше классического предела (\ref{eqVacNoise}):
\begin{equation}
\left\langle \Psi^\mathrm{SMS}_\zeta \left| \hat{X}^2 \right| \Psi^\mathrm{SMS}_\zeta \right\rangle = \dfrac{1}{2e^{2\zeta}}.
\label{eqSMSvar}
\end{equation}

Для характеризации величины сжатия квантового состояния широко используется логарифм отношения дисперсий сжатой квадратуры и вакуумного шума. Эта величина измеряется в децибелах и определяется как
\begin{equation}
\label{eqDB}
\mathrm{dB} = 10 \log_{10} \left[\dfrac{1/2}{\min_\theta \left\langle Q_\theta^2 \right\rangle}\right].
\end{equation}
Для состояния, показанного на Рис.~(\ref{pSMSwigner}), величина (\ref{eqDB}) равна $\mathrm{dB} = \dfrac{20\zeta}{\ln{10}} \approx 8.7\zeta$.
Аналогичным образом, величина (\ref{eqDB}) может быть вычислена и для состояния двумодово-сжатого вакуума (\ref{eqTMSWF}).

Фоковское представление состояния (\ref{eqSMS}) даётся выражением \cite{LvovskySQ}
\begin{equation}
\label{eqSMSfock}
\Psi^\mathrm{SMS}_\zeta = \dfrac{1}{\sqrt{\cosh \zeta}} \sum_{n=0}^{\infty} (-\tanh \zeta)^n \dfrac{\sqrt{(2n)!}}{2^n n!} \ket{2n}.
\end{equation}
В пределе малого сжатия $\zeta \ll 1$, в разложении (\ref{eqSMSfock}) часто можно учитывать лишь первые два слагаемых:
\begin{equation}
\label{eqSMSfock2}
\Psi^\mathrm{SMS}_\zeta \approx \ket{0} + \dfrac{\zeta}{\sqrt{2}}\ket{2}.
\end{equation}

Среднее число фотонов одномодово-сжатом состоянии (\ref{eqSMSfock}) совпадает с населённостью каждой из мод двумодово-сжатого состояния с тем же значением $\zeta$ (\ref{eqNPtms}):
\begin{equation}
\label{eqNPsms}
\left\langle \hat{n} \right\rangle_{\Psi^\mathrm{SMS}_{\zeta}} = \sinh^2 \zeta.
\end{equation}


\subsection{Импульсное сжатие в трёх измерениях} \label{sect3_realSMS}

Идеальная ситуация, описанная в предыдущем разделе, не соответствует типичной экспериментальной практике; в результате вырожденного параметрического рассеяния, сигнальные фотоны населяют большое количество временных мод, каждая из которых характеризуется собственной фазой и коэффициентом сжатия. В ходе гомодинного приёма такого сигнала при помощи местной волны, имеющей временные параметры волны накачки, детектируется некоторая суперпозиция этих мод, что эквивалентно снижению эффективности измерения сжатого состояния \cite{Wasilewski2006}. 

Этот результаты, полученные для одномерного случая, могут быть обобщены и на случай трёх пространственных измерений. Представленный ниже подход применим также и к процессу невырожденного параметрического рассеяния \cite{Christ2013}.

Назовём направление распространения сигналов осью $z$. Электрическое поле сигнальных фотонов параметрической конверсии представляется в виде
\begin{equation}
\label{eq31}
\hat{E}(t,x,y,z) = \int \!\! \,d\omega\,dk_x\,dk_y\, N(\omega) e^{-i\omega t} e^{-i(k_x x + k_y y)} \hat{a}(\omega,k_x,k_y,z),
\end{equation}
где $N(\omega)$ -- нормирующий множитель, а $\hat{a}(\omega,k_x,k_y,z)$ есть оператор уничтожения фотонов, зависящий от координаты. Уравнение его распространения в кристалле имеет вид \cite{Wasilewski2006}
\begin{equation}
\label{eq32}
\begin{aligned}
& \dfrac{\partial \hat{a}(\omega,k_x,k_y,z)}{\partial z} = ik_z(\omega,k_x,k_y) \hat{a}(\omega,k_x,k_y,z) + \\
&  \dfrac{1}{L_{\rm NL}E_0} \int \!\! \,d\omega'\,dk_x'\,dk_y'\, E_p(\omega+\omega', k_x+k_x', k_y+k_y') e^{ik_{pz}(\omega+\omega', k_x+k_x', k_y+k_y')} \hat{a}^\dagger(\omega',k_x',k_y',z),
\end{aligned}
\end{equation}
где $k$ и $k_p$ это волновые вектора сигнала и накачки, а $E_0=\int \!\! \,d\omega \,dk_x \, dk_y \, E_p(\omega,k_x,k_y)$ есть интеграл от классического поля накачки $E_p$. 
$L_{\rm NL}$ есть характерная длина нелинейного взаимодействия:
\begin{equation}
\label{eq33}
\dfrac{1}{L_{\rm NL}} = \dfrac{\omega_p^2 d_{\rm eff} E_0}{8c^2 k(\omega_p/2)},
\end{equation}
где $d_{\rm eff}$ есть эффективный коэффициент нелинейности кристалла для выбранных направлений поляризации, и $\omega_p$ -- центральная частота поля накачки. В случае, когда поле сигнала ограничено единственной модой $a$ (\ref{eq30}), степень её сжатия на выходе из кристалла длиной $L$ равна $\zeta = L/L_{\rm NL}$.

Поскольку уравнение распространения (\ref{eq32}) линейно по сигнальному полю, состояние последнего на выходе из кристалла ($z=L/2$) может быть выражено через состояние на входе ($z=-L/2$) как
\begin{equation}
\label{eq34}	
\begin{aligned}
\hat{a}_{\mathrm{out}}(\omega, k_x, k_y) & = 
\int \!\! \,d\omega' \, dk_x' \, dk_y' \, C(\omega,k_x,k_y,\omega',k_x',k_y')\hat{a}_{\mathrm{in}}(\omega', k_x', k_y') \\
& + \int \!\! \,d\omega' \, dk_x' \, dk_y' \, S(\omega,k_x,k_y,\omega',k_x',k_y')\hat{a}_{\mathrm{in}}^\dagger (\omega', k_x', k_y') .
\end{aligned}
\end{equation}

Преобразование (\ref{eq34}) может быть приведено к канонической форме c помощью разложения Блоха-Мессии \cite{Braunstein2005}. Последнее основывается на разложении функций Грина $S$ и $C$
\begin{equation}
\label{eq35}
\begin{aligned}
& S(\omega,k_x,k_y,\omega',k_x',k_y') = \sum_{n} \psi_n^*(\omega,k_x,k_y) \phi_n(\omega',k_x',k_y') \mathrm{cosh}\zeta_n \\
& C(\omega,k_x,k_y,\omega',k_x',k_y') = \sum_{n} \psi_n^*(\omega,k_x,k_y) \phi_n^*(\omega',k_x',k_y') \mathrm{sinh}\zeta_n,
\end{aligned}
\end{equation}
где функции $\psi_n$ и $\phi_n$ составляют два ортонормированных базиса, а $\zeta_n$ есть неотрицательные действительные числа.
Это разложение позволяет определить новый набор мод
\begin{equation}
\label{eq36}
\begin{aligned}
& \hat{b}_n^{\mathrm{in}} = \int \!\! d\omega \, \phi(\omega) \hat{a}^{\mathrm{in}} \\
& \hat{b}_n^{\mathrm{out}} = \int \!\! d\omega \, \psi(\omega) \hat{a}^{\mathrm{out}},
\end{aligned}
\end{equation}
для которых преобразование (\ref{eq34}) принимает вид
\begin{equation}
\label{eq37}
\hat{b}_n^{\mathrm{out}} = \hat{b}_n^{\mathrm{in}} \mathrm{cosh}\,\zeta_n + \hat{b}_n^{\dagger \mathrm{in}} \mathrm{cosh}\,\zeta_n.
\end{equation}
Таким образом, уравнения (\ref{eq36}) определяют набор собственных пространственно-временных мод параметрической конверсии, каждая из которых на выходе из кристалла выглядит так, как будто независимо от остальных прошла эволюцию, описываемую соотношением (\ref{eq30}), со своим собственным коэффициентом сжатия. В случае, когда поле накачки $E_p$ действительно, входной и выходной базисы связаны соотношением $\psi_n = \phi_n^*$ \cite{Wasilewski2006}.

Разложение (\ref{eq35}) может быть найдено аналитически в пределе слабой нелинейности, т.е. $L\ll L_{\rm LN}$. Функции Грина в этом случае аппроксимируются следующим образом:
\begin{equation}
\label{eq38}
\begin{aligned}
& C(\omega,k_x,k_y,\omega',k_x',k_y') = \delta(\omega-\omega') \delta(k_x - k_x') \delta(k_y - k_y') e^{ik_z(\omega,k_x,k_y)} \\
& S(\omega,k_x,k_y,\omega',k_x',k_y') = e^{i(k_z(\omega, k_x, k_y)-k_z(\omega', k_x', k_y'))L/2} \frac{L E_p(\omega + \omega', k_x+k_x', k_y+k_y')}{L_{NL} E_0} \mathrm{sinc} \dfrac{\Delta k L}{2},
\end{aligned}
\end{equation}
где
\begin{equation}
\Delta k = k_{pz}(\omega+\omega', k_x+k_x', k_y+k_y') - k_z(\omega, k_x, k_y) - k_z(\omega', k_x', k_y').
\end{equation}
Предполагая фазовый синхронизм идеальным, $k_p(\omega_p,0,0) = 2k(\omega_p/2,0,0)$, и работая в первом порядке по малым величинам $\left(k_x, k_y\right) \ll \left(k, k_p\right)$, пространственные и временные переменные разделяются, и можно записать
\begin{equation}
\label{39}
\begin{aligned}
\Delta k & \approx (\beta_{1p} - \beta_1)(\omega+\omega'-\omega_p) - \dfrac{\beta_2}{2} \left[\left(\omega-\dfrac{\omega_p}{2}\right)^2 + \left(\omega' - \dfrac{\omega_p}{2}\right)^2\right] \\
& + \dfrac{\beta_{2p}}{2}\left(\omega + \omega' - \omega_p\right)^2 + \dfrac{\left(k_x-k_x'\right)^2 + \left(k_y-k_y'\right)^2}{2k_p(\omega_p)},
\end{aligned}
\end{equation}
где 
\begin{equation}
\label{310}
\begin{aligned}
\beta_n = \dfrac{d^n k}{d\omega^n}|_{\omega = \omega_p/2}, \,\,\,\,\,\,\, \beta_{n,p} = \dfrac{d^n k_p}{d\omega^n}|_{\omega = \omega_p}. 
\end{aligned}
\end{equation}
Полагая форму импульса накачки гауссовой c пространственной и временной ширинами $\tau_p$ и $d_p$ соответственно,
\begin{equation}
\label{eq311}
E_p(\omega,k_x,k_y) \propto \mathrm{exp}\left[-\dfrac{\tau_p^2}{2}\left(\omega-\omega_p\right)^2\right]
*\mathrm{exp}\left[-\dfrac{d_p^2}{2}\left(k_x^2 + k_y^2\right)\right],
\end{equation}
и приближая функцию $\mathrm{sinc}$ как
\begin{equation}
\label{eq312}
\mathrm{sinc} \, x \approx e^{-x^2/5}, \,\,\,\,\,\,\, \mathrm{sinc} \, x^2 \approx e^{-x^2/3},
\end{equation}
можно получить Гауссово приближение функции $S$ (\ref{eq38}):
\begin{equation}
\label{eq313}
\begin{aligned}
S(\omega,k_x, & k_y,\omega',k_x',k_y') = e^{i\left[k_z(\omega, k_x, k_y)-k_z(\omega', k_x', k_y')\right]L/2} \sqrt{\dfrac{8N}{\pi^3 \delta_\omega \Delta_\omega \delta_k^2 \Delta_k^2}}\\
& *\mathrm{exp}\left[-\dfrac{\left(\omega + \omega' -\omega_p\right)^2}{2\delta_\omega^2}-\dfrac{\left(\omega-\omega'\right)^2}{2\Delta_\omega^2}\right]\\
&*\mathrm{exp}\left[-\dfrac{\left(k_x+k_x'\right)^2 + \left(k_y+k_y'\right)^2}{2\delta_k^2}-\dfrac{\left(k_x-k_x'\right)^2 + \left(k_y-k_y'\right)^2}{2\Delta_k^2}\right],
\end{aligned}
\end{equation}
где
\begin{equation}
\label{eq314}
\begin{aligned}
&\dfrac{1}{\delta_\omega^2} = \tau_p^2 + \dfrac{L^2}{10}\left(\beta_1 - \beta_{1,p}\right)^2, \,\,\,\,\,\,\,
\dfrac{1}{\Delta_\omega^2} = \dfrac{L\beta_2}{12},\\
&\dfrac{1}{\delta_{k}^2} = d_{p}^2, \,\,\,\, \quad \dfrac{1}{\Delta_k^2} = \dfrac{L/6}{k_p(\omega_p)}, \,\,\,\,\,\,\,
N = \dfrac{L^2}{4L_{\rm NL}} \tau_p^2 \delta_\omega \Delta_\omega d_p^4 \delta_k^2 \Delta_k^2. 
\end{aligned}
\end{equation}

\begin{figure}
	\centering
	\includegraphics[width=6.65in]{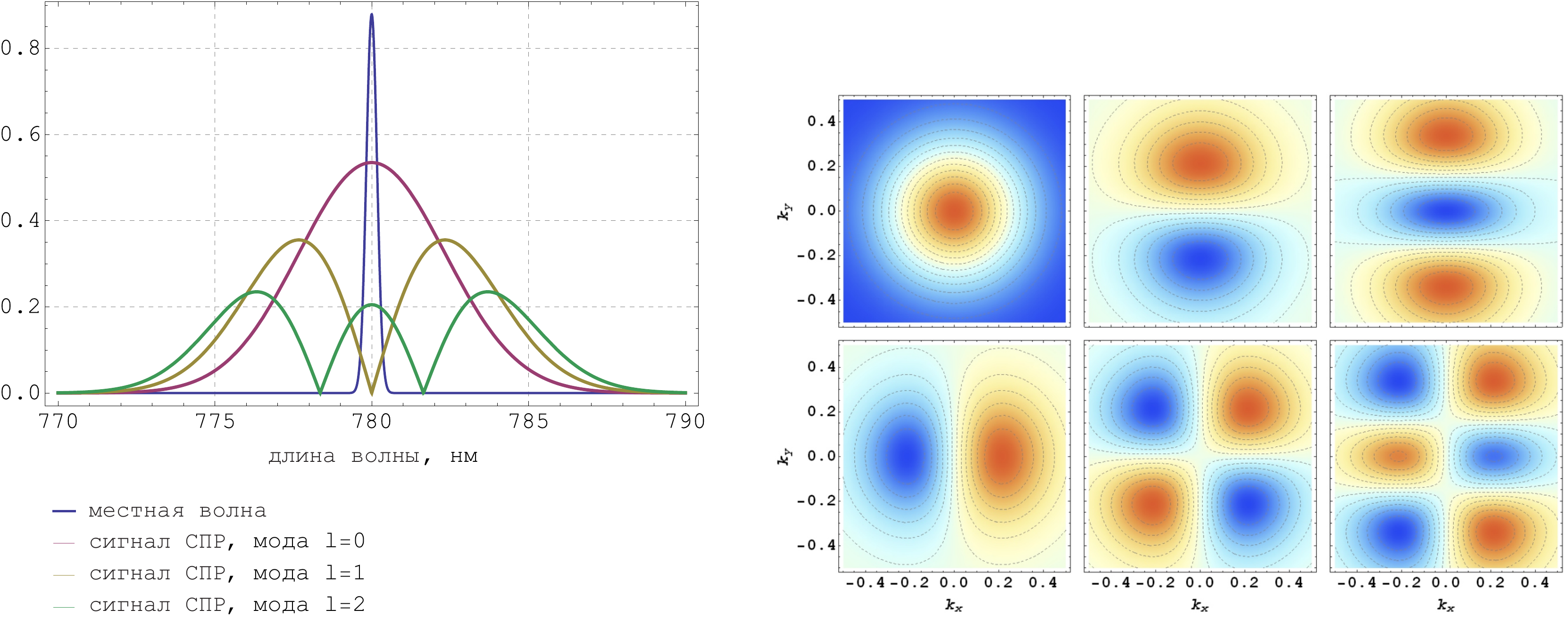}
	\captionsetup{justification=raggedright, singlelinecheck=false}
	\caption{Слева: модуль спектральной части собственных функций (\ref{eq315}) в сравнении со спектром исходного лазерного света. Справа: модуль пространственной части собственных функций (\ref{eq315}), $i=0..1$, $j=0..2$.}
	\label{pSpatialSMSmodes}
\end{figure}
Гауссова форма функций Грина позволяет найти в явном виде разложение (\ref{eq35}) \cite{Grice2001}.
Собственные функции, нумеруемые тройным индексом ${\bf n} = \{l,i,j\}$, даются выражением
\begin{equation}
\label{eq315}
\begin{aligned}
e^{-ik_z(\omega, k_x, k_y)}&\psi_{\bf n}(\omega, k_x, k_y) = e^{ik_z(\omega, k_x, k_y)} \phi_{\bf n}(\omega, k_x, k_y) = \\
& \sqrt{\dfrac{\tau_s}{\pi^{1/2}2^ll!}}H_l\left[\tau_s(\omega-\omega_p/2)\right]
\mathrm{exp}\left[-\dfrac{\tau_s^2}{2}\left(\omega-\omega_p\right)^2\right] * \\
&
\dfrac{d_s}{\sqrt{\pi2^{i+j}i!j!}}
H_i\left[d_s k_x \right]H_j\left[d_s k_y \right]
\mathrm{exp}\left[-\dfrac{d_s^2}{2}\left(k_x^2+k_y^2\right)\right],
\end{aligned}
\end{equation} 
где
\begin{equation}
\label{eq316}
\tau_s^2 = \dfrac{2}{\delta_\omega \Delta_\omega}, \,\,\,\,\,\,\, d_s^2 = \dfrac{2}{\delta_k \Delta_k}.
\end{equation}
Соответствующие им собственные значения есть
\begin{equation}
\label{eq317}
\begin{aligned}
\mathrm{sinh} \, \zeta_{\bf n} = \sqrt{N}\dfrac{\mathrm{tanh}^l \, r_\omega}{\mathrm{cosh}\,r_\omega} \dfrac{\mathrm{tanh}^i \, r_k}{\mathrm{cosh}\,r_k} \dfrac{\mathrm{tanh}^j \, r_k}{\mathrm{cosh}\,r_k}, \quad
r_\omega = \dfrac{\mathrm{ln(\Delta_\omega/\delta_\omega)}}{2}, \quad r_k = \dfrac{\mathrm{ln(\Delta_k/\delta_k)}}{2}.
\end{aligned}
\end{equation}
Функции (\ref{eq315}) младших порядков показаны на Рис.~\ref{pSpatialSMSmodes}.

Для используемого нами кристалла титанил-фосфата калия \cite{Bierlein1989}, имеем $\beta_1 \approx 6.4\times 10^{-9}$ сек/м, $\beta_{1,p} \approx 8.2\times 10^{-9}$ сек/м, $\beta_{2} \approx 2.8\times 10^{-31}$ сек$^2$/м, $L=1$мм. Длина импульса накачки составляет $\tau_p \approx 2$ псек, ширина луча накачки есть $d_p\approx 10$мкм. Это даёт 
\begin{equation}
\label{eq318}
\begin{aligned}
&\,\, \delta_\omega = 5\times 10^{11}\,\mathrm{rad/sec}, \quad \delta_k = 2.4\times 10^{5}\,\mathrm{m}^{-1}, \\
&\Delta_\omega = 2\times 10^{14}\,\mathrm{rad/sec}, \quad \Delta_k = 3.8\times 10^{5}\,\mathrm{m}^{-1}.
\end{aligned}
\end{equation}

Поскольку $\Delta_\omega \gg \delta_\omega$, главная мода сигнала параметрической конверсии имеет значительно меньшую продолжительность импульса $\tau_s$ по сравнению с накачкой $\tau_p$ (\ref{eq314}, \ref{eq316}), и значит гораздо более широкий спектр. Это приводит к наличию множества временных мод в окне наблюдения, задаваемом импульсом местной волны, и неизбежному снижению эффективности наблюдаемого состояния. Спектр трёх младших собственных мод по сравнению со спектром исходного лазера (местной волны) показан на Рис.~\ref{pSpatialSMSmodes} слева. Существенное расхождение временных (спектральных) форм ограничивает максимально возможную эффективность детектирования сжатия величиной $86\%$ \cite{Wasilewski2006}.

\begin{figure}
	\centering
	\includegraphics[width=5.5in]{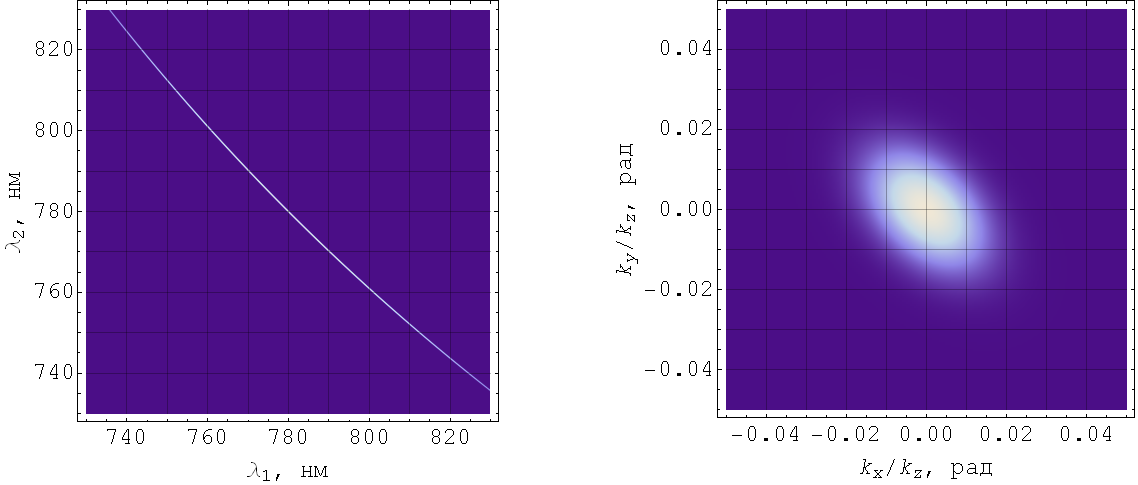}
	\captionsetup{justification=raggedright, singlelinecheck=false}
	\caption{Слева: временная часть функции Грина (\ref{eq313}), показанная в зависимости от длин волн сигнальных фотонов. Справа: пространственная часть, показанная в безразмерных единицах углового отклонения $k_x/k_z$.}
	\label{pGreens}
\end{figure}

Пространственные величины $\delta_k$ и $\Delta_k$, напротив, имеют один и тот порядок; это значит, что поперечные размеры сигнального луча $d_s$ близки к размерам местной волны $1/\delta_k$ (\ref{eq316}). Следовательно, с помощью телескопической системы возможно настроиться на детектирование главной пространственной моды сигнала параметрической конверсии (\ref{eq315}, $i=j=0$).
Таким образом, учёт пространственных координат не изменяет ожидаемую эффективность детектирования сжатого света, полученную в работах \cite{Wasilewski2006, Lvovsky2007}. 

Временная и пространственная части функции Грина (\ref{eq313}), рассчитанные для используемых экспериментальных параметров, показаны на Рис.~\ref{pGreens}. 
В обоих случаях, ширина максимума по диагонали (например, $\lambda_1 = \lambda_2$) определяется соответствующей шириной пучка накачки, $\delta_\omega$ и $\delta_k$. 
В перпендикулярном направлении, ширины $\Delta_\omega$ и $\Delta_k$ определяется условием фазового синхронизма. В отличие от временной компоненты, для которой $\Delta_\omega / \delta_\omega \approx 400$, $\Delta_k / \delta_k \approx1.6$, что означает отсутствие сильной запутанности сигнальных фотонов по направлению.


\subsection{Связь одномодово-- и двумодово-- сжатых состояний}
\label{sect3_TMS_and_SMS}

Рассмотрим ситуацию, когда моды двумодово-сжатого состояния (\ref{eqTMSWF}) интерферируют на симметричном светоделителе. В представлении Гейзенберга, такому процессу соответствует следующее преобразование операторов координат:
\begin{equation}
\label{eqSMSbsTMS}
\left[ {\begin{array}{c}
	\hat{X}_a' \\
	\hat{X}_b' \\
	\end{array} } \right]
=
\dfrac{1}{\sqrt{2}}
\left[ {\begin{array}{cc}
	1 & -1 \\
	1 & 1 \\
	\end{array} } \right]
\left[ {\begin{array}{c}
	\hat{X}_a \\
	\hat{X}_b \\
	\end{array} } \right].
\end{equation}
Согласно ур.(\ref{eqTMSvar}), дисперсии операторов координат на выходе светоделения равны
\begin{equation}
\label{eqBSvarsConnection}
\left\langle \hat{X}_{a/b}'^2 \right\rangle = \left\langle \left(\dfrac{\hat{X}_a \mp \hat{X}_b}{\sqrt{2}}\right)^2 \right\rangle_{\Psi_\zeta} = \dfrac{1}{2}e^{\mp 2 \zeta},
\end{equation}
что означает, что моды $a$ и $b$ оказались в состоянии одномодового сжатия(\ref{eqSMS})! При этом, мода $a$ оказывается сжатой по координатному, а мода $b$ -- по импульсному направлению фазовой плоскости.
В силу обратимости операции \ref{eqSMSbsTMS} \cite{Leonhardt1997}, пара состояний одномодового сжатого вакуума может быть использована для получения двумодового сжатия (разд. \ref{Two_crystal_EPR}, \cite{LvovskySQ}). 

\subsection{Влияние потерь}
\label{sect3_losses}

Сжатые состояния света теряют свои неклассические свойства при воздействии оптических потерь. Основными их источниками являются:
\begin{itemize}
	\item Потери на оптических элементах. Обычно составляют не более $\sim5\%$.
	\item Потери при детектировании. К обычным источникам потерь, приведённых в разд. \ref{BHD_efficiency}, при работе с сигналом нелинейного кристалла добавляется несовпадение временных длин мод сигнала и местной волны, ограниченное величиной $86\%$ (разд. \ref{sect3_realSMS}). Суммарная эффективность детектирования составляет $\sim 70\%$. 
	\item Потери в кристалле вследствие эффекта ``gray tracking'' (Рис. \ref{pDegradedCrystal}), которые зависят от мощности накачки, состояния кристалла, температуры и не поддаются простой оценке.
\end{itemize}
Таким образом, суммарный коэффициент потерь ограничен снизу значением $35 \%$.

\subsubsection{Одномодово--сжатый вакуум}

После прохождения одномодово-сжатым состоянием света линейных потерь $R$, зависимость дисперсии квадратуры от фазы (\ref{eqSMSGeneralNoise}) принимает вид
\begin{equation}
\label{eqSMSLssNoise}
\left\langle \hat{Q}_\theta^2 \right\rangle_{\mathrm{att}} = (1-R) \left\langle \hat{Q}_\theta^2 \right\rangle + \dfrac{R}{2} = \dfrac{1-R}{2} \left[\cos^2\!{\theta} \, e^{-2\zeta} + \sin^2\!{\theta} \, e^{2\zeta}\right] + \dfrac{R}{2}.
\end{equation}
Дисперсии максимально сжатой / растянутой квадратур теперь равны
\begin{equation}
\label{eq328}
V_{\rm max/min} = \left\langle \hat{Q}^2_{0/\pi:2} \right\rangle_{\mathrm{att}} = \dfrac{\exp{\left[\pm 2\zeta\right]}(1-R) + R}{2}.
\end{equation}

Выражение (\ref{eq328}) позволяет по наблюдаемым значениям максимума и минимума дисперсии найти величину исходного сжатия и эффективность его детектирования:
\begin{equation}
\begin{aligned}
& \zeta = \dfrac{1}{2} \ln{\dfrac{1-2V_{\rm max}}{2V_{\rm min} -1}}, \quad
& R = \dfrac{2 V_{\rm max} V_{\rm min} -1/2}{V_{\rm max} + V_{\rm min} -1}.
\end{aligned}
\label{eq329}
\end{equation}
Эти выражения особенно полезны при юстировке установки в ``квантовом'' режиме, когда рассчитываемые в реальном времени, они служат оптимизационными параметрами.

\begin{figure}[h]
	\centering
	\includegraphics[width=6in]{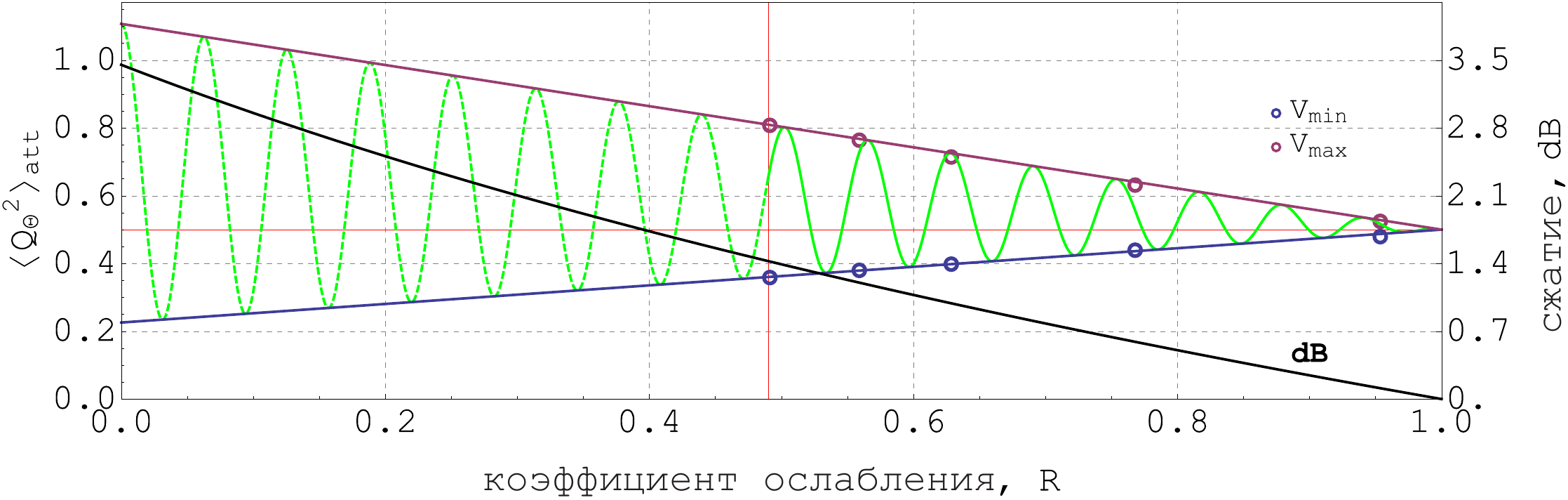}
	\captionsetup{justification=raggedright, singlelinecheck=false}
	\caption{Зелёная линия: осцилляции дисперсии квадратуры (\ref{eqSMSLssNoise}) при варьировании фазы $\theta$ при различных значениях потерь $R$. Красная и синяя линия есть максимальное и минимальное значения (\ref{eq328}). Вертикальная линия соответствует наивысшему значению эффективности $1-R=0.52$, для которого в данном эксперименте наблюдалось сжатие 1.4 dB. Остальные измерения проведены при внесении дополнительных потерь на пути между кристаллом и гомодинным детектором.  Серая линия: соответствующее сжатие в децибелах (\ref{eqDB}). Горизонтальная линия показывает уровень вакуумного шума.}
	\label{pSMSloss}
\end{figure}
Рисунок (\ref{pSMSloss}) иллюстрирует эффект потерь при наблюдении одномодово-сжатого вакуума (разд. \ref{sect3_resultSMS}). Красные и синие точки, соответствующие экспериментально полученным значениям максимума и минимума дисперсии, согласуются и теоретическим ожиданием (\ref{eq328}) показанным линиями.

Оператор плотности одномодового, слабо-сжатого состояния (\ref{eqSMSfock2}), претерпевшего ослабление с коэффициентом $R$ равен
\begin{equation}
\label{eqRhoAttSMS}
\hat{\rho}^\mathrm{SMS}_{\zeta, R} = \left(\ket{0}+\dfrac{\zeta}{\sqrt{2}}(1-R)\ket{2}\right) \left(\bra{0}+\dfrac{\zeta}{\sqrt{2}}(1-R)\bra{2}\right) + \zeta \sqrt{R(1-R)} \ket{1}\bra{1} + \dfrac{\zeta R}{2}\ket{0}\bra{0}.
\end{equation}
Наличие в матрице плотности компоненты $\ket{1}\bra{1}$ вызвано потерями при генерации и детектирования состояния, что может приводить к утрате одного фотона из пары.

\subsubsection{Двумодово--сжатый вакуум}

После прохождения модами двумодово-сжатого состояния света (\ref{eqTMSWF}) каналов потерь с пропусканиями $\tau_1$ и $\tau_2$, дисперсия суммы квадратур сохраняет зависимость лишь от суммы фаз местных волн $\theta_1+\theta_2$ (\ref{eqTMSIdealNoise}) .
Если $\tau_1=\tau_2$, то в силу соотношения (\ref{eqBSvarsConnection}), двумодовое сжатие наследует свойство (\ref{eqSMSLssNoise}) одномодового; следовательно, величина и эффективность могут вычисляться согласно выражениям (\ref{eq328})--(\ref{eq329}).
Если же $\tau_1 \neq \tau_2$, то зависимость дисперсии от фаз принимает вид
\begin{equation}
\label{eqTMSRealNoise}
\left\langle \Psi^\mathrm{TMS}_{\zeta,\tau_{1,2}} \left| \left( \dfrac{\hat{Q}^{\theta_1}_1 + \hat{Q}^{\theta_2}_2}{\sqrt{2}} \right)^2\, \right| \Psi^\mathrm{TMS}_{\zeta,\tau_{1,2}} \right\rangle =
\dfrac{1}{2} + \dfrac{ \left(\cosh{\left[2\zeta\right]}-1\right)\left(\tau_1^2+\tau_2^2\right) + 2 \tau_1\tau_2 \cos\left(\theta_1+\theta_2\right) \sinh{\left[2\zeta\right]}}{4}.
\end{equation}

В случае малого сжатия $\zeta \ll 1$, оператор плотности двумодово-сжатого вакуума (\ref{eqFockTMS}), претерпевшего ослабление, имеет в фоковском базисе вид
\begin{equation}
\label{eqAttTMSDM}
\begin{aligned}
\hat{\rho}^\mathrm{TMS}_{\zeta,\tau_{1,2}} = \left[\ket{00} + \zeta \tau_1 \tau_2 \ket{11}\right]&\left[\bra{00} + \zeta \tau_1 \tau_2 \bra{11}\right] + \\
& \zeta^2 (1-\tau_1^2) \ket{01}\bra{01} + \zeta^2 (1-\tau_2^2) \ket{10}\bra{10}.
\end{aligned}
\end{equation}
Видно, что ослабление ведёт к уменьшению амплитуды компоненты $\ket{11}$ и связанных с ней недиагональных элементов $\ket{00}\bra{11}$ и $\ket{11}\bra{00}$, а также к примеси компонент $\ket{01}\bra{01}$ и $\ket{10}\bra{10}$, соответствующих потере фотона в одной из мод.

\begin{figure}[h]
	\centering
	\includegraphics[width=4in]{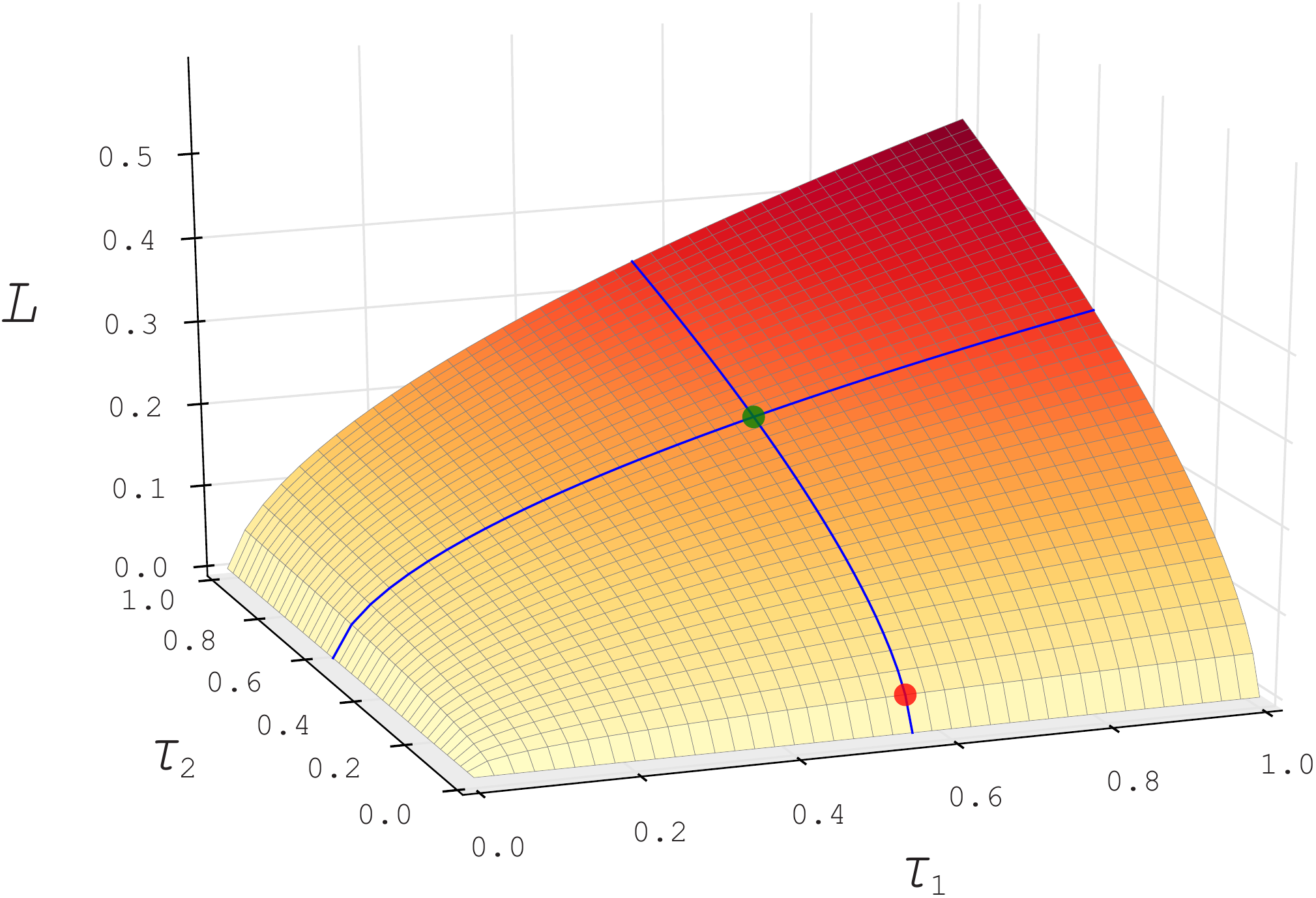}
	\captionsetup{justification=raggedright, singlelinecheck=false}
	\caption{Логарифмическая отрицательность состояния (\ref{eqAttTMSDM}) при $\zeta=0.16$ в зависимости от пропускания оптических каналов. Линии соответствуют наилучшей суммарной эффективности $\tau_{1,2}^2=0.55$. В точке $\tau_{1,2}^2=0$ $L=0.47$, на пересечении линий $L=0.24$. Красная точка соответствует состоянию перед дистилляцией (разд. \ref{chapt4}), $L=0.037$. Расчёт произведён с помощью библиотеки инструментов для квантовой оптики QuTiP \cite{Qutip}.}
	\label{pAttNegativity}
\end{figure}
Рис.~\ref{pAttNegativity} показывает зависимость запутанности двумодового состояния (\ref{eqAttTMSDM}) от $\tau_1$ и $\tau_2$. В качестве меры выбрана логарифмическая отрицательность 
\begin{equation}
\label{eqNeg}
L = \ln{\mathrm{Tr}\left[\sqrt{  \hat{\rho}_\mathrm{pt}^* \hat{\rho}_\mathrm{pt}  }\right]},
\end{equation}
где $\hat{\rho}_\mathrm{pt}$ есть матрица плотности двумодового состояния, транспонированная относительно одной из мод \cite{Vidal2002}.
Если $\tau_{1} = \tau_2$, то величина сжатия пропорциональна логарифмической отрицательности:
\begin{equation}
\label{eqZetaLDB}
\mathrm{dB} = 3L.
\end{equation}

\section{Наблюдение сжатых состояний света}
\label{NLCalignment}
Интенсивность нелинейного взаимодействия трёх световых пучков, участвующих в процессе параметрического рассеяния света (\ref{SPDC}), определяется точностью выполнения для них следующих соотношений:
\begin{enumerate}
\item[1.] Закон сохранения энергии: \\
\begin{equation}
\label{eqEnergy}
\hbar \omega_p = \hbar \omega_a + \hbar \omega_b,
\end{equation}
где $\omega_p$ есть частота пучка накачки, а $\omega_{a,b}$ есть частоты сигнальных пучков.
Соотношение (\ref{eqEnergy}) означает, что параметрическое рассеяние света является упругим процессом.
\item[2.] Условие синхронизации фаз: \\
\begin{equation}
\label{eqMomentum}
\Delta \vec{\mathbf k} = \vec{\mathbf k}_p - \vec{\mathbf k}_a - \vec{\mathbf k}_b = 0,
\end{equation}
где $\vec{\mathbf k}_{p,a,b}$ есть волновые вектора накачки и сигнальных мод в кристалле.
Уравнение (\ref{eqMomentum}) выражает условие того, что то по ходу продвижения в кристалле, накачка и сигнал набирают фазу с равными скоростями. Если это не так, то процесс происходит намного менее эффективно: когда разность фаз между накачкой и сигналом достигает половины длины волны, нелинейный процесс начинает действовать в обратную сторону, ослабляя сигнал \cite{LvovskyLect}.
\end{enumerate}

Во всех экспериментах, описываемых в настоящей работе и использующих СПР, желаемой была ситуация, когда
\begin{equation}
\label{eqNPconditions}
\begin{aligned}
& \omega_a = \omega_b = \dfrac{\omega_p}{2} = \omega, \\
& \vec{ \mathbf k}_p \parallel \vec{\mathbf k}_a \parallel \vec{\mathbf k}_b.
\end{aligned}
\end{equation}
Условие (\ref{eqEnergy}) тогда выполняется автоматически, а (\ref{eqMomentum}) формулируется в скалярной форме как
\begin{equation}
\label{eqPhaseMatching}
\dfrac{\Delta k}{2\pi} = \dfrac{n_{2\omega,p}}{\lambda/2} - \left( \dfrac{n_{\omega,a}}{\lambda} + \dfrac{n_{\omega,b}}{\lambda} \right) = 0,
\end{equation}
где $\lambda=780$ нм есть сигнальная длина волны, $n_{\omega,i}$ есть показатель преломления материала кристалла для света с частотой $\omega$, а индекс $i$ обозначает поляризацию пучков.
Соотношение (\ref{eqPhaseMatching}) называется условием фазового синхронизма \cite{Boyd2008, LvovskyLect}. Две распространённые конфигурации сигнальных мод, позволяющие выполнить условие (\ref{eqPhaseMatching}), таковы:
\begin{enumerate}
\item[1.] Фазовый синхронизм типа 1. \\
Поляризации сигнальных мод совпадают с поляризацией накачки. Сигнальные моды тогда неразличимы, и СПР ведёт к генерации одномодово-сжатого состояния света в сигнальной моде (\ref{SMS}).
\item[2.] Фазовый синхронизм типа 2. \\
Сигнальные моды имеют ортогональные поляризации, одна из которых совпадает с поляризацией накачки. В этом случае, СПР ведёт к двумодовому сжатию сигнальных мод (\ref{TMS}).
\end{enumerate}

Средняя населённость сигнала параметрического рассеяния определяется с помощью однофотонного детектора, и при характерной мощности накачки $P_{\mathrm{pump}}=50-100$ мВт (разд. \ref{pump_alignment}) составляет $\sim 1.5\times 10^5$ фотонов в секунду. 
Юстировать нелинейный процесс, используя сигнал такой мощности, сложно; ещё большие трудности возникли бы при настройке однофотонного и гомодинного детектирований. 
Для решения этих задач применяются методы настройки, использующие вспомогательный пучок с исходной длиной волны лазера -- ``сид'' \cite{Aichele2002}. Детальное экспериментальное и теоретическое описание этих техник дано в диссертационных работах \cite{HansenThesis, KumarThesis}. 
Их краткое описание, достаточное для практической работы, приведено в разделах \ref{PhaseMatching2} и \ref{PhaseMatching1}.

\subsection{Настройка пучка накачки}
\label{pump_alignment}
\begin{figure}[h]
	\centering
	\includegraphics[width=6.6in]{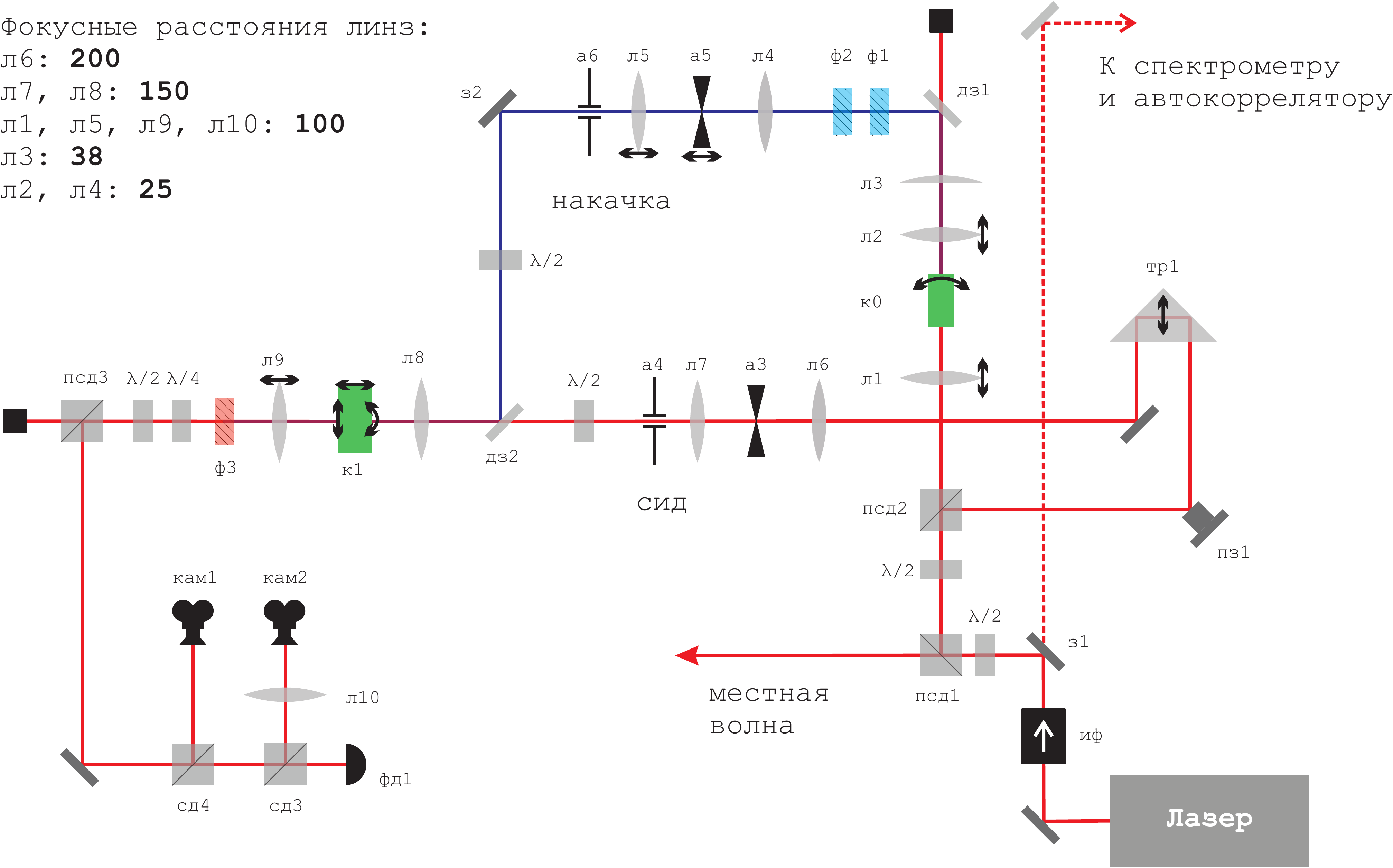}
	\captionsetup{justification=raggedright, singlelinecheck=false}
	\caption{Часть установки, используемая при настройке фазового синхронизма. ИФ, изолятор Фарадея. Л, линза. А, апертура. (Д/П)З, (дихроическое/пьезо) зеркало. К, нелинейный кристалл. Ф, спектральный фильтр. (П)СД, (поляризационный) светоделитель. ТР, тромбон. КАМ, камера. ФД, фотодиод.}
	\label{pPhasematching}
\end{figure}

Первым этапом настройки СПР является получение пучка накачки в определённой пространственной моде; если это не так, то и сигнальные моды унаследуют нерегулярность её пространственной структуры, что осложнит детектирование итоговых состояний.
Генерация луча накачки, имеющего длину волны 390 нм, происходит в процессе удвоения частоты исходного лазерного излучения. Последний осуществляется с помощью кристалла трибората лития (К0, Рис.~\ref{pPhasematching}).

Кристалл К0 имеет размеры $2\times2\times17$ мм и расположен в фокусе сферической линзы Л1 с фокусным расстоянием $F=100$ мм; излучение распространяется вдоль длинной оси кристалла. Для максимизации мощности сигнала второй гармоники, накачка кристалла осуществляется нефильтрованным излучением, следующим напрямую от лазера (разд. \ref{laser}) и составляющем б\'{о}льшую часть его мощности. В результате, $\sim 1$ Вт импульсного излучения, преобразуются в $200-300$ мВт сигнала второй гармоники. 

После кристалла К0, сигнал второй гармоники коллимируется линзами Л2 и Л3 -- сферическая F25 и цилиндрическая F38, соответственно, позволяющими независимо управлять сходимостью пучка в вертикальном и горизонтальном направлениях. Далее, пучок проходит пространственную фильтрацию с помощью методики, описанной в разд. \ref{laser_spatial}. 

Получаемая мощность накачки, во многих случаях, критически важна для успеха эксперимента. Поэтому, фильтрация апертурой диаметра много меньшего, чем ширина перетяжки, в данной ситуации недопустима; для построения телескопа Л4--Л5 используется фокусирующая линза F25, дающая перетяжку с размером порядка диаметра апертуры А5, 25 мкм.
Положение апертуры А5, таким образом, является компромиссным параметром между пропускаемой мощностью и качеством пространственной моды накачки; последняя наблюдается с помощью камер ближнего и дальнего полей КАМ1 и КАМ2. 

Достижимые значения мощности и качества моды зависят от положений и фокусных расстояний коллимирующих линз Л2--Л3. В оптимальном положении последних,  фильтрация обычно позволяет получить из $200-300$ мВт ``грязного'' излучения второй гармоники $P_{\mathrm{pump}}=50-100$ мВт фильтрованной накачки.


\subsection{Фазовый квази-синхронизм и дисперсионные свойства нелинейного кристалла}
\label{sectDispersion}

В настоящей работе в качестве нелинейных сред использовались кристаллы титанил-фосфата калия с периодической доменной структурой. Эта структура состоит в периодической инверсии кристаллической решётки кристалла в направлении распространении накачки и сигнала. Такая организация доменов позволяет компенсировать отклонение от условия (критического) фазового синхронизма (\ref{eqMomentum}). В рабочем режиме, рассинхронизация фаз в каждом из доменов равна $\pi$ \cite{LvovskyLect}, что выражается в изменении условия (\ref{eqMomentum}) на
\begin{equation}
\label{eqQuasiPhase}
\Delta \vec{\mathbf k} D/2 = \pi \iff \Delta \vec{\mathbf k} = \vec{\mathbf k}_D,
\end{equation}
где $D$ есть период доменной структуры.
В коллинеарном режиме, условие (\ref{eqQuasiPhase}) раскрывается как
\begin{equation}
\label{eqQPcollinear}
n_p - \dfrac{n_a + n_b}{2} = \dfrac{\lambda}{D},
\end{equation}
где $\lambda$ есть длина волны накачки, $n_p$ и $n_{a/b}$ есть показатели преломления для накачки и двух сигнальных мод удвоенной длины волны.

Кристалл титанил-фосфата калия является анизотропным: испытываемый светом коэффициент преломления зависит от его поляризации.
Для трёх основных оптических направлений эта зависимость имеет вид
\begin{equation}
\label{eqSellmeyer}
\begin{aligned}
& n^2_x = 3.29100 + \dfrac{0.04140}{\lambda^2 - 0.03978} + \dfrac{9.35522}{\lambda^2 - 31.45571} \\
& n^2_y = 3.45018 + \dfrac{0.04341}{\lambda^2 - 0.04597} + \dfrac{16.98825}{\lambda^2 - 39.43799} \\
& n^2_z = 4.59423 + \dfrac{0.06206}{\lambda^2 - 0.04763} + \dfrac{110.80672}{\lambda^2 - 86.12171},
\end{aligned}
\end{equation}
для длины волны $\lambda$ в микрометрах \cite{Kato2002, Bierlein1989}.

Использовавшиеся нами кристаллы были вырезаны по главным оптическим осям, так что $\vec{x}$ есть направление распространения волн. В случае фазового квази-синхронизма типа II, накачка поляризована по оси $\vec{y}$, а сигнальные моды занимают поляризации $\vec{y}$ и $\vec{z}$. Для такой конфигурации, условие (\ref{eqQPcollinear}) принимает вид
\begin{equation}
\label{eqQPcollinear2}
n\left[\lambda, \, y\right] - \dfrac{n\left[2\lambda, \, y\right] + n\left[2\lambda, \, z\right]}{2} = \dfrac{\lambda}{D}.
\end{equation}
Для $D = 7.825$ мкм, равенство (\ref{eqQPcollinear2}) выполняется при $\lambda = 0.38902$ мкм.

Вследствие дисперсионных свойств кристалла, групповые скорости распространения сигналов первой и второй гармоник отличаются, что приводит к отставанию накачки от сигнала при прохождении кристалла.
Групповой скорость излучения равна \cite{Sivukhin2005}
\begin{equation}
\label{eqVgroup}
v = \dfrac{n}{c} - \dfrac{\lambda}{c}\dfrac{dn}{d\lambda}.
\end{equation}
Для горизонтальной поляризации сигналов первой и второй гармоник, уравнения (\ref{eqSellmeyer}) и (\ref{eqVgroup}) дают
\begin{equation}
\label{eqVGR}
\begin{aligned}
n_{2\lambda, \, y} = 1.76 & \quad \quad v_{2\lambda, \, y} = 0.52 c \\
n_{\lambda, \, y} = 1.85 & \quad \quad v_{\lambda, \, y} = 0.41 c, \\
\end{aligned}
\end{equation}
где $c$ есть скорость света в воздухе. Результирующее отставание равно
\begin{equation}
\label{eqSMSdelay}
\delta = L c \left( \dfrac{1}{v_{\lambda, \, y}} - \dfrac{1}{v_{2\lambda, \, y}} \right),
\end{equation}
где $L$ есть длина кристалла. При $L=1$ мм, уравнение (\ref{eqSMSdelay}) при помощи (\ref{eqVGR}) даёт $\delta = 0.58$ мм.


\subsection{Фазовый синхронизм типа 2}
\label{PhaseMatching2}

Первым этапом настройки фазового синхронизма типа 2 является эксперимент по генерации удвоенной частоты (SHG) от вспомогательного пучка -- сида, имеющего длину волны основного лазера. Этот процесс обратен СПР: в один синий объединется пара красных фотонов; так как последние имеют ортогональные поляризации -- вертикальную и горизонтальную -- то для эффективного удвоения частоты, пучок сида должен быть поляризован диагонально; для этого используется полуволновая пластина, находящаяся диафрагмой А4 и дихроическим зеркалом ДЗ2, Рис.~\ref{pPhasematching}.

\clearpage
Основные условия эффективной генерации сигнала SHG:
\begin{itemize}
	\item Кристалл должен быть помещён вблизи перетяжки сида, сфокусированного линзой Л8.
	\item Луч должен входить в кристалл по нормали к передней грани. Это условие выполняется, когда отражённый от кристалла пучок совмещён с падающим.
	\item Кристалл должен быть правильно ориентирован. Добиться этого можно экспериментально (генерация наблюдаемого количества SHG возможна только при правильной ориентации), или определив направление полинга с помощью микроскопа (при наблюдении перпендикулярно рабочему направлению, полирование с периодом 7.85 мкм выглядит как лёгкая штриховка).
\end{itemize}

\begin{figure}[h]
	\includegraphics[width=6.7in]{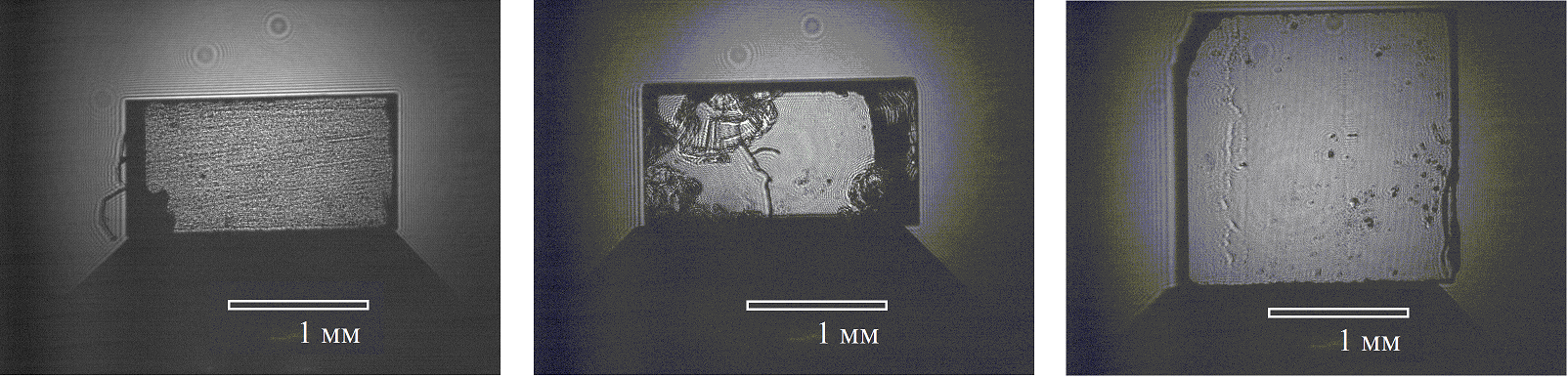}
	\centering
	\captionsetup{justification=raggedright, singlelinecheck=false}
	\caption{Фотографии кристалла титанил-фосфата калия с периодической доменной структурой, деградировавшего в результате интенсивного использования. Изображение получено в проходящем свете на длине волны 780 нм. Слева: вид сбоку, направление лучей слева направо. В середине: вид спереди. Справа: вид сверху.}
	\label{pDegradedCrystal}
\end{figure}

Когда эти условия выполнены, сигнал SHG имеет мощность $\sim1$ мкВт и видим невооружённым глазом. Далее, с помощью вспомогательного зеркала, сигнал направляется на измеритель мощности, показание которого максимизируется с помощью вариации следующих параметров:
\begin{enumerate}
	\item[1.] Положение кристалла относительно перетяжки. Подстраивается с помощью смещения микрометрической платформы, на которой закреплён кристалл. При наличии дефектов (Рис.~\ref{pDegradedCrystal}), актуальна оптимизация поперечного положения кристалла относительно пучка.
	 
	\item[2.] Температура кристалла, от которой зависят коэффициенты преломления в (\ref{eqPhaseMatching}). Регулировка осуществляется нагревателем и элементом Пельтье, управляемыми термоконтроллером. Оптимальное значение зависит от текущей длины волны; при $\lambda=780.1$ нм, типичное значение равно $15^\circ$C.
	
	При настройке температуры, следует учитывать возможность выпадения росы на охлаждённых поверхностях. Преломления на каплях приводит к разрушению формы световых мод, что губительно сказывается на экспериментальных результатах. При выборе длины волны следует учитывать температуру и влажность воздуха в лаборатории, см. Рис.~\ref{pTemperatureType1}; оба параметра устанавливаются в настройках системы кондиционирования. В тёплую и влажную погоду, однако, система может не справляться с просушкой воздуха, и влажность может подниматься с обычных $40-50$ до $85\%$.
	
	\item[3.] В малых пределах, оптимизируются поляризация сида и наклоны кристалла. 
\end{enumerate}

При мощности сида 6 мВт, характерная мощность сигнала SHG составляет 10-20 мкВт. В результате этого эксперимента, достигается фазовый синхронизм между горизонтальной и вертикальной компонентами моды сида и модой SHG.

На втором этапе, мода накачки совмещается с модой SHG. Сделать это можно непосредственно, наблюдая интерференцию между двумя пучками, либо косвенно, выполнив эксперимент по генерации разностной частоты (DFG) между пучками сида и накачки. 
Во втором случае, поляризация накачки должна совпадать с поляризацией второй гармоники от сида, а поляризация сида поворачивается из диагонального в горизонтальное положение. DFG тогда имеет вертикальную поляризацию, благодаря чему может быть отделена от луча сида на поляризационном светоделителе ПСД3 и направлена на фотодиод ФД1.

Необходимым условием взаимодействия пучков сида и накачки являются:
\begin{itemize}
	\label{dfg_optimize}
	\item Одновременное появлению их импульсов на кристалле. Обеспечивается подбором положения ТР1 в канале сида, представляющего собой укреплённую на микрометрической платформе обратно-отражающую призму.
	\item Пространственное совмещение моды накачки с модой сида. Настраивается зеркалами З2 и ДЗ2 в канале накачки.
	\item Сравнимые диаметры пучков сида и накачки, определяемых радиусами $a$ апертур А3 / А5 и фокусными расстояниями $F$ линз Л4 / Л9 как $F\lambda/a$.
\end{itemize}

После того как сигнал DFG получен, с помощью тех же степеней свободы по его мощности оптимизируется фазовый синхронизм между сидом, накачкой и DFG. В результате, становится определённой тройка мод, удовлетворяющих условиям (\ref{eqEnergy}) и (\ref{eqMomentum}), а значит и взаимодействующих в ходе нелинейного процесса:
\begin{enumerate}
	\item[1.] Мода накачки
	\item[2.] Мода сида
	\item[3.] Мода сигнала DFG.
\end{enumerate}

При работе на мощностях пучка накачки $\gtrsim 40$ мВт, мода сигнала DFG имеет тенденцию к уменьшению её размера, а также некоторому смещению в пространстве, со временем. После характерного времени 5-10 мин, такое плавание приводит к новой, квази-стабильной пространственной моде. Поэтому, перед переходом к следующим этапам, (например, настройке интерференции с местной волной, или заведения в волокно) важно выждать это время. В некоторых случаях, в результате удаётся добиться повышения эффективности квантового сигнала на $\sim5\%$.

Высокая мощность накачки приводит также к постепенному разрушению кристалла и ухудшению его нелинейных свойств. Фотографии кристалла К1 после завершения эксперимента по эффекту Квантового Вампира (разд. \ref{chapt2}) показаны на Рис. \ref{pDegradedCrystal}. В одной из проекций явно видны следы так называемого эффекта ``gray tracking'', видимые как затемнённые дорожки, соответствующие использовавшимся положениям луча накачки; на другой грани очевидно наличие скола, образовавшегося, предположительно, из-за чрезмерного механического или теплового напряжения.
При работе вблизи дефектов, наравне с мощностью оптимизируемых сигналов следует уделять внимание форме их мод, наблюдая их с помощью камер КАМ1 и КАМ2. 

\clearpage
\subsection{Фазовый синхронизм типа 1}
\label{PhaseMatching1}

Первым этапом настройки является наблюдение генерации удвоенной частоты (SHG) от сида. Процедура близка к описанной в разделе \ref{PhaseMatching2}, с отличиями в следующем:
\begin{itemize}
	\item так как сигнальные моды имеют теперь одинаковые поляризации, поляризация сида должна совпадать с этим направлением.
	\item нелинейный коэффициент для процесса типа 1 в 5-10 раз выше соответствующей величины для процесса типа 2. Это приводит к пропорциональному увеличению характерных мощностей SHG до 70-200 мкВт (см. разд. \ref{sect3_1_2}).
\end{itemize}
Результатом этого этапа должно явиться выполнение условия фазового синхронизма (\ref{eqPhaseMatching}) между сигналом SHG и сидом \cite{LvovskyLect, Boyd2008}, а также определение оптимального положения кристаллов в поле сида.

\begin{figure}[h]
	\centering
	\includegraphics[width=5in]{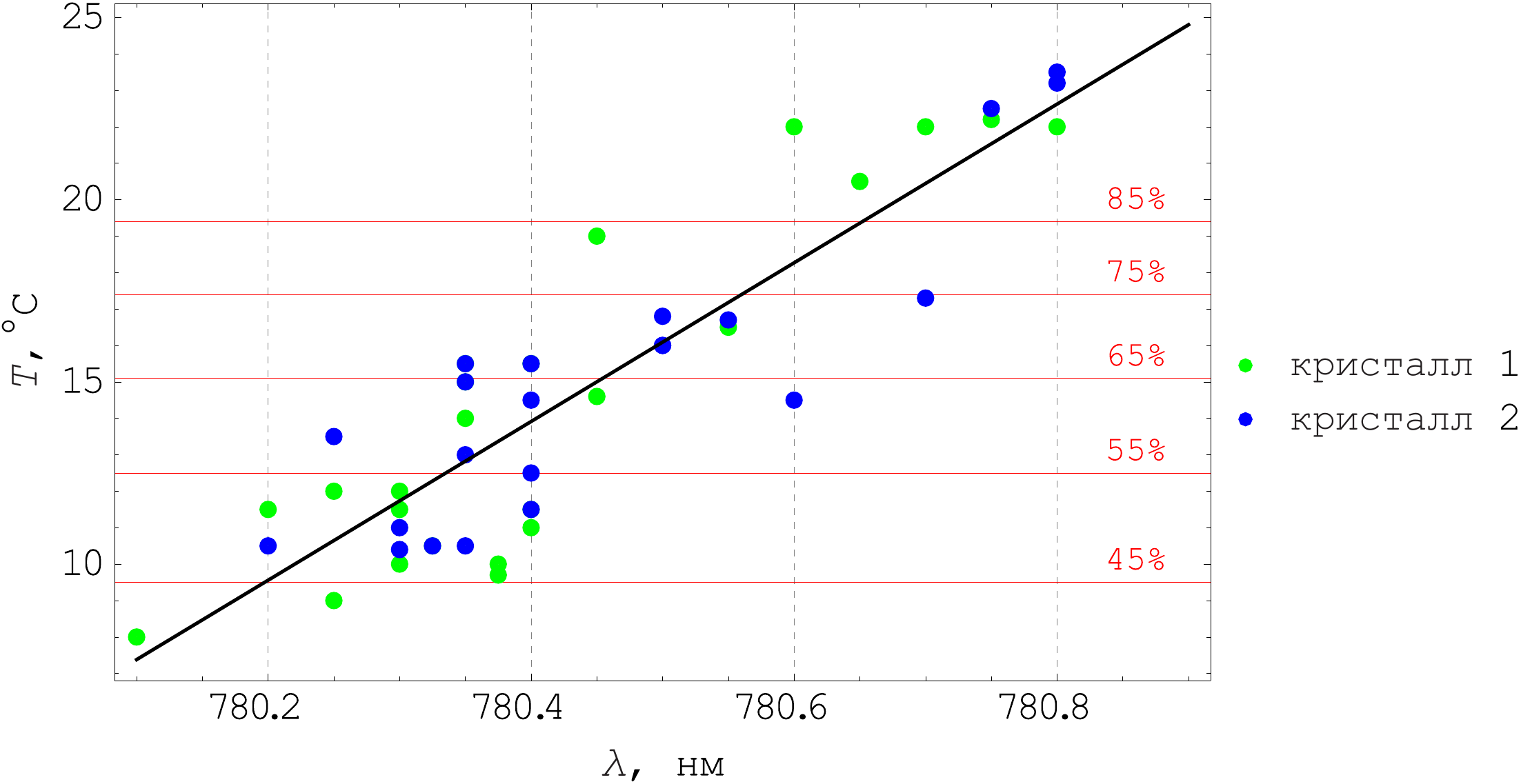}
	\captionsetup{justification=raggedright, singlelinecheck=false}
	\caption{Зависимость оптимальной температуры $T_{\mathrm{opt}}$ для двух кристаллов типа 1 от длины волны. Наклон общей линейной регрессии составляет $22 ^\circ$С/нм. Точки росы при температуре окружения $22^\circ$C для различных значений относительной влажности воздуха показаны красными линиями.}
	\label{pTemperatureType1}
\end{figure}

Рис.~\ref{pTemperatureType1} показывает рабочие температурные точки для фазового синхронизма типа 1, в зависимости от текущей длины волны, использовавшиеся в эксперименте по генерации двумодово-сжатого вакуума с помощью двух кристаллов (разд. \ref{Two_crystal_EPR}).

Мощность сигнала второй гармоники $W_{\rm shg}$ обратно пропорциональна квадрату фокусного расстояния линзы Л10 $F$ и не имеет прямой корреляции с выбранной длинной волны и температурой. При $F=100$ мм и мощности сида 6 мВт, в среднем $W_{\rm shg} = 100$ мкВт.

На втором этапе настройки достигается совмещение мод накачки и SHG. В результате этого мода, которая испытывает сжатие под действием накачки, совмещается с модой сида. Как и в случае нелинейного процесса типа 2, вспомогательным экспериментом для этого является оптимизация сигнала разностной частоты между сидом и накачкой (DFG, см. разд. \ref{PhaseMatching2}); отличие в случае типа 1 заключается в том, что мода DFG неотделима от моды сида, так как их поляризации теперь совпадают. В результате, имеет место интерференция, которая наблюдается как изменение интенсивности сида; интерференция является конструктивной или деструктивной в зависимости от фазы накачки по отношению к сиду, которая варьируется с помощью пьезо-зеркала ПЗ1 в канале сида. Видность этой интерференции, наблюдаемой с помощью фотодиода ФД1, является параметром для максимизации мощности DFG.

\section{Классическое измерение нелинейности. Оценка степени сжатия} 
\label{sect3_1_2}

Помимо генерации сжатых состояний света, нелинейные свойства кристаллов приводят к ряду ``классически'' наблюдаемых явлений, которые используются в процессе юстировки оптической схемы (см. разделы \ref{PhaseMatching2}, \ref{PhaseMatching1}). Среди таких явлений генерация сигналов удвоенной (SHG) и разностной (DFG) частот, по отношению к частотам исходных полей.

\subsection{Генерация второй гармоники}
\label{sectSHG}

Генерация сигнала SHG является процессом слияния двух фотонов исходной моды света в один.
Пренебрегая истощением исходного поля $E_{\mathrm{seed}}$ с частотой $\omega$ и полагая фазовый синхронизм идеальным, нарастание поля второй гармоники $E_{\rm SHG}$ на частоте $2\omega$ происходит по закону \cite{LvovskyLect, Boyd2008}
\begin{equation}
\label{eq319}
\dfrac{dE_{\rm shg}}{dz} = \dfrac{2i \omega d_{\rm eff}}{n_2c}E_{\rm seed}^2,
\end{equation}
где $d_{\rm eff}$ есть эффективный коэффициент нелинейности. Его можно определить с помощью уравнения (\ref{eq319}), если известны мощности сида и SHG:
\begin{equation}
\label{eqDeff}
d_{\rm eff} = \dfrac{E_{\rm shg}}{E_{\rm seed}^2} \dfrac{n_2 c}{2\omega L},
\end{equation}
где $L$ есть длина кристалла. 

Амплитуды электрических полей связаны со средними их мощностями $W$ как
\begin{equation}
\label{eq320}
\dfrac{W}{\nu \tau} = 2n \epsilon_0 c \left|E\right|^2 S,
\end{equation}
где $\nu=76$ МГц есть частота повторения лазера, $\tau\approx2$ пс есть ширина импульса, $n$ -- показатель преломления кристалла для соответствующей волны (\ref{eqVGR}), а площадь перетяжки пучка
\begin{equation}
\label{eqBeamSize}
S = \pi w^2, \quad w = \dfrac{\lambda}{\pi \theta} = \dfrac{\lambda}{\pi r/F},
\end{equation}
где $\theta$ есть половинный угол сходимости, а $r$ -- радиус коллимированного пучка перед фокусирующей линзой $F$.

В силу (\ref{eq319}), пространственный профиль пучка сида определяет в перетяжке профиль сигнала SHG, который имеет в $\sqrt{2}$ раз меньший радиус:
\begin{equation}
\label{eqSeedSpace}
E_\mathrm{seed} \propto \exp \left[-\dfrac{r^2}{w_\mathrm{seed}^2}\right] 
\quad \rightarrow \quad 
E_\mathrm{shg} \propto E_\mathrm{seed}^2 = \exp \left[-\dfrac{r^2}{w_\mathrm{seed}^2/2}\right].
\end{equation}
Так как дифракционная расходимость сигнала SHG вдвое меньше чем у сида, то и для произвольной плоскости наблюдения имеет место
\begin{equation}
\label{eqSHGseed}
w_{\rm shg}(z) = \dfrac{w_{\rm seed}(z)}{\sqrt{2}}.
\end{equation}

С учётом (\ref{eq320}), (\ref{eqBeamSize}) и (\ref{eqSHGseed}), выражение (\ref{eqDeff}) преобразуется в
\begin{equation}
\label{eqDeff2}
d_{\rm eff} = \sqrt{\dfrac{\epsilon_0 c^3 \nu \tau n_1^2 n_2}{\omega^2L^2}} * \dfrac{\sqrt{W_{\rm shg}} S_{\rm seed}}{W_{\rm seed}}.
\end{equation}

Для параметров эксперимента из разд. \ref{Two_crystal_EPR} (фазовый синхронизм типа 1) имеем $L=1$ мм, $F=100$ мм, $W_{\rm seed} = 6$ мВт, $r_{\rm seed}=2.5$ мм, $W_{\rm shg} = 100$ мкВт, выражение (\ref{eqDeff2}) даёт $d_{\rm eff} =6.05$ пм/В, что согласуется с ожиданием
\begin{equation}
\label{eqDeff3}
d_{\rm eff} = \dfrac{2}{\pi} d_{33},
\end{equation}
где $d_{33} = 10.7$ пм/В есть коэффициент нелинейности чистого титанил-фосфата калия \cite{Dmitriev1993}. Множитель $2/\pi$ учитывает эффект периодической доменной структуры \cite{LvovskyLect}.

В случае фазового синхронизма типа 2, коэффициент нелинейности значительно ниже. Для экспериментальной схемы из разд. \ref{sectDistAlignment}, нелинейный кристалл имеет длину $L=2$ мм, а характерная мощность второй гармоники $W_{\rm shg} = 10$ мкВт, при остальных параметрах, данных в предыдущем параграфе. Уравнение (\ref{eqDeff2}) в этом случае даёт $d_{\rm eff} = 0.96$ пм/В.

\subsection{Генерация разностной частоты}
\label{sectDFG}

Амплитуда сигнала разностной частоты $E_{\rm dfg}$ между накачкой $E_{\rm pump}$ и сидом $E_{\rm seed}$ подчиняется закону \cite{LvovskyLect}
\begin{equation}
\label{eq321}
\dfrac{dE_{\rm dfg}}{dz} = \dfrac{2i\omega d_{\rm eff}}{n_1c}E_{\rm seed}^* E_{\rm pump}.
\end{equation}

\begin{figure}[h]
	\includegraphics[width=5in]{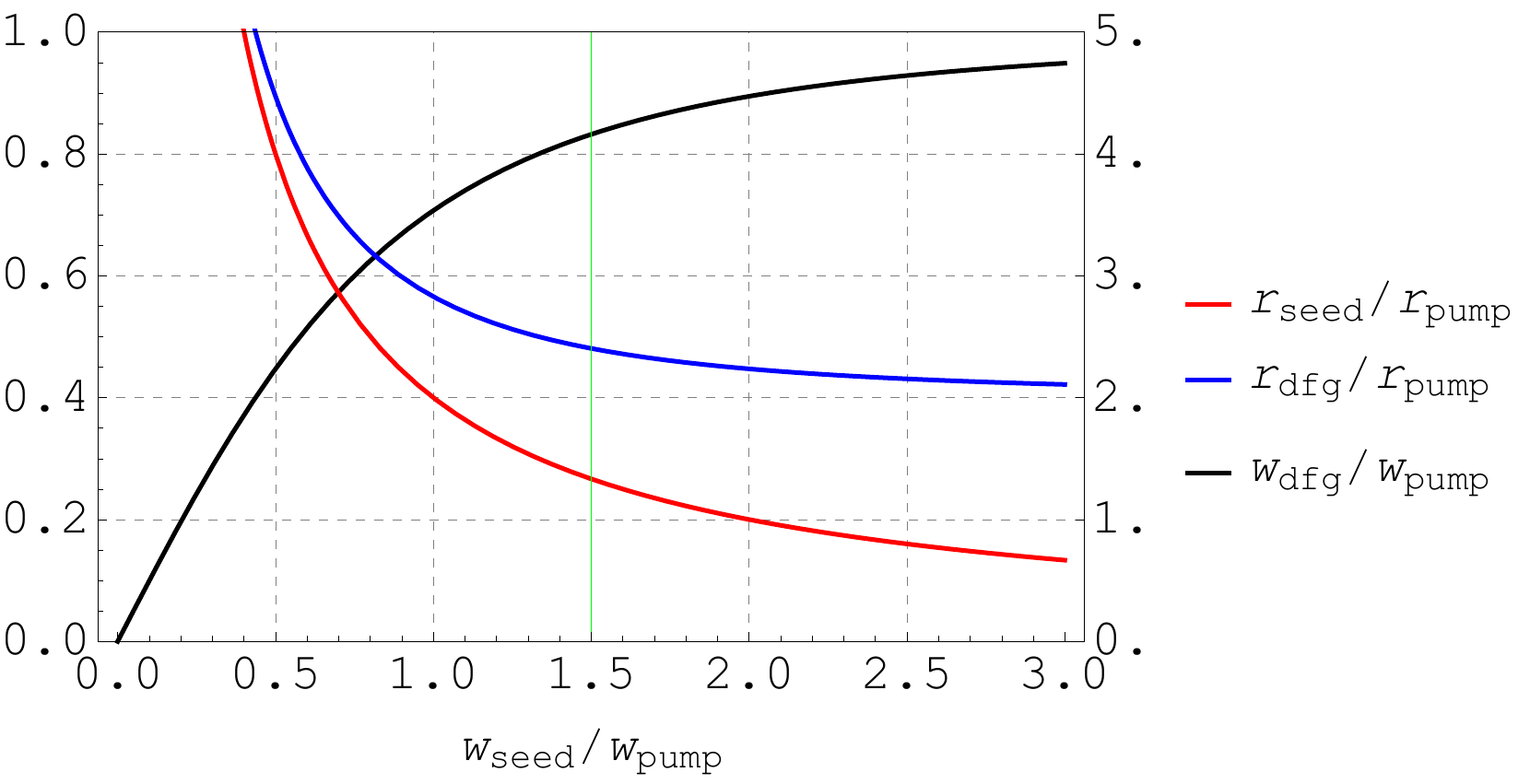}
	\centering
	\captionsetup{justification=raggedright, singlelinecheck=false}
	\caption{Чёрная линия, левая шкала: радиус перетяжки сигнала DFG в зависимости от размера перетяжки сида. Красный и синий, правая шкала: размеры сида и DFG  в дальнем поле. Все величины показаны в отношении с соответствующей величине пуска накачки. Зелёная линия: рабочее положение.}
	\label{pVSizes}
\end{figure}

Пространственный профиль сигнала DFG определяется через (\ref{eq321}) профилями сида и накачки:
\begin{equation}
\label{eqSPsizes}
\begin{aligned}
E_\mathrm{seed} \propto \exp \left[-\dfrac{r^2}{w_\mathrm{seed}^2}\right]&, \quad
E_\mathrm{pump} \propto \exp \left[-\dfrac{r^2}{w_\mathrm{pump}^2}\right] \rightarrow \\
& E_\mathrm{dfg} \propto E_\mathrm{seed}*E_\mathrm{pump} =
\exp\left[-\frac{r^2}{w_{\rm dfg}^2}\right], \quad w_{\rm dfg}^2 = \dfrac{w_{\rm pump}^2w_{\rm seed}^2}{w_{\rm pump}^2 + w_{\rm seed}^2}.
\end{aligned}
\end{equation}
Зависимость $w_{\rm dfg}$ от $w_{\rm seed}$ для фиксированной величины $w_{\rm pump}$ показана на Рис.~\ref{pVSizes} чёрным. 

\subsubsection{Синхронизм типа 2}

В случае фазового синхронизма типа 2, уравнение (\ref{eq321}) есть классический аналог квантового закона (\ref{eqTMSevol}), где операторы $a$ и $b$ соответствуют модам сида и DFG, с эффективной частотой Раби 
\begin{equation}
\label{eqRabi}
\beta = \dfrac{2\omega d_{\rm eff} v_g}{n_1 c}E_{\rm pump},
\end{equation}
где $v_g$ есть групповая скорость сигнала параметрической конверсии в кристалле (разд. \ref{sectDispersion}).

В силу слабой нелинейности при фазовом синхронизме типа 2, классический сигнал DFG мал, и его мощность на выходе из кристалла можно вычислить, полагая
\begin{equation}
\label{eqDFGpower}
E_{\rm dfg} = \dfrac{dE_{\rm dfg}}{dz} L.
\end{equation}
C помощью (\ref{eq320}, \ref{eqBeamSize}), (\ref{eqDeff2}) и (\ref{eqSPsizes}), (\ref{eqDFGpower}) даёт
\begin{equation}
\label{eqDFGpower2}
W_{\rm dfg} = \dfrac{2 S_{\rm seed}}{S_{\rm pump} + S_{\rm seed}}* 
\dfrac{W_{\rm pump} W_{\rm shg}}{W_{\rm seed}}.
\end{equation}
Для параметров, данных в разд. \ref{sectSHG}, а также $W_{\rm pump}=40$ мВт и $r_{\rm pump}=2$ мм, уравнение (\ref{eqDFGpower2}) даёт $W_{\rm dfg} = 96$ мкВт, что согласуется с экспериментально наблюдаемой величиной $W_{\rm dfg} = 74$ мкВт.

Знание $d_{\rm eff}=0.96$ пм/В (разд. \ref{sectSHG}) позволяет спрогнозировать величину двумодового сжатия, ожидаемую в квантовом режиме. Частота Раби (\ref{eqRabi}), умноженная на время прохождения сигнала через кристалл, есть параметр сжатия (\ref{eqTMSHeizenbergAs}):
\begin{equation}
\label{eqZeta1}
\zeta = \beta \dfrac{L}{v_g} = \dfrac{2\omega d_{\rm eff} L}{n_1 c}E_{\rm pump}.
\end{equation}
С помощью (\ref{eqDeff2}), параметр $\zeta$ может быть выражен через измеряемые величины:
\begin{equation}
\label{eqZeta}
\zeta = \dfrac{\sqrt{2}r_{\rm pump}}{r_{\rm seed}} \dfrac{\sqrt{W_{\rm shg}W_{\rm pump}}}{W_{\rm seed}}.
\end{equation}
Для характерных параметров, уравнение (\ref{eqZeta}) даёт $\zeta = 0.24$.

\subsubsection{Синхронизм типа 1}

В случае фазового синхронизма типа 1, моды сида и разностной частоты имеют одинаковые поляризации. Их амплитуды складываются, что приводит усилению или ослаблению мощности сигнала в моде сида, в зависимости от относительной фазы между этими амплитудами. 
Если пространственные моды сида и DFG близки, то такой процесс описывается уравненем (\ref{eq321}) c заменой 
\begin{equation}
E_{\rm dfg} \quad \rightarrow \quad E_{\rm seed},
\end{equation}
что эквивалентно эволюции моды сида, описываемой оператором $a$, под действием гамильтониана одномодового сжатия (\ref{eqHeizenbergSQ}) с той же частотой Раби (\ref{eqRabi}).
Таким образом, явление параметрического усиления / ослабления света можно интерпретировать и как интерференцию с сигналом разностной частоты, и как преобразование фазового пространства моды сида под действием накачки. В квантовом режиме, именно это преобразование проявляет себя как изменение дисперсии квадратурных измерений в зависимости от направления в фазовой плоскости.

По величине параметрического ослабления сида можно оценить степень сжатия, ожидаемую в квантовом режиме:
\begin{equation}
\label{eq325}
\zeta = \dfrac{1}{2}\mathrm{ln}\dfrac{W_{\rm seed}}{W_{\rm min}} =  \dfrac{1}{2}\mathrm{ln}\dfrac{W_{\rm seed}}{\left(\sqrt{W_{\rm seed}}-\sqrt{W_{\rm dfg}}\right)^2},
\end{equation}
где $W_{\rm min} = 3.7$ мВт есть экспериментально измеренная мощность наиболее ослабленной квадратуры сида.  
Это даёт $\zeta = 0.24$ (\ref{eq325}), что соответствует $2.1$ dB сжатия. Учёт $35\%$ неизбежных потерь при детектировании (разд. \ref{sect3_losses}) даёт для сжатия, наблюдаемого в квантовом режиме, оценку в $1.3$ dB.

Как и в случае синхронизма типа 2, зная $d_{\rm eff}=6.05$ пм/В (разд. \ref{sectSHG}), с помощью (\ref{eqZeta}) можно оценить величину одномодового сжатия, ожидаемую в квантовом режиме.
Для параметров, данных в разд. \ref{sectSHG} и $W_{\rm pump}=40$ мВт и $r_{\rm pump}=2$ мм, уравнение (\ref{eqZeta}) даёт $\zeta = 0.75$.


\section{Приготовление фоковских состояний света}

\subsection{Концепция}
\label{SPpreparation}

Наиболее распространённый метод генерации фоковских состояний света основан на использовании невырожденного процесса спонтанного параметрического рассеяния света (СПР). 
В таком процессе, фотоны накачки вероятностным образом распадаются на пары сигнальных фотонов, населяющие разные световые моды, 1 и 2. Как показано в разделе \ref{TMS}, в результате пара сигнальных мод оказывается в состоянии
\begin{equation}
\label{eq_tms}
\ket{\Psi} = \dfrac{1}{\sqrt{1-\gamma^2}} \sum_{n=0}^{\infty} \gamma^n \ket{nn}.
\end{equation}

Из множества сигнальных фотонов, сгенерированных в ходе СПР, половина направляется в моду 1.
Если известно, какой импульс накачки привёл к появлению фотона в моде 1, то с уверенностью можно сказать, что тот же самый импульс в пространственной моде 2 также содержит фотон, парный первому. На этом основывается метод приготовления однофотонного состояния света \cite{Lvovsky2001, HansenThesis}. 

Этот принцип обобщается также и на случай многофотонных состояний; например, одновременное детектирование пары фотонов в триггерной моде соответствует рождению двух фотонных пар от одного импульса накачки. В результате, сигнальная мода окажется в двухфотонном состоянии \cite{Bimbard2010}.

\subsection{Проективное измерение в триггерном канале}
\label{SPprojection}

Для получения информации о рождении фотонов в моде 1 используются детекторы одиночных фотонов (ДОФ), имеющие чувствительность, достаточную для распознавания одиночных фотонов. В большинстве случаев, эти детекторы реагируют одинаково на любое число пришедших фотонов, и называются неразрешающими. 

ДОФы характеризуются величиной квантовой эффективности $\eta$, которая определяется как вероятность срабатывания детектора при подаче на вход однофотонного состояния света. Если на такой детектор поступило $n$--фотонное состояние, то вероятность не-срабатывания есть
\begin{equation}
(1-\eta)^n.
\end{equation} 
Срабатыванию такого детектора соответствует POVM оператор \cite{Kok2010}
\begin{equation}
\label{SPCM}
	\hat{\Pi}(\eta) = \sum_{n=0}^\infty \left[1-(1-\eta)^n\right] \ket{n}\bra{n},
\end{equation}

Если ДОФ расположен в моде 1 состояния (\ref{eq_tms}), то его срабатывание ведёт к переходу моды 2 в состояние
\begin{equation}
\label{eqSPprojection}
\hat{\rho}_{\mathrm{proj}} = \mathrm{Tr}_1\left[ \hat{\Pi} \ket{\Psi}\bra{\Psi}\right] \propto \sum_{n=0}^{\infty} \gamma^n \left[1-(1-\eta)^n\right] \ket{n}\bra{n},
\end{equation}
В пределе малого сжатия исходного состояния (\ref{eq_tms}), компоненты $n \geq 2$ в (\ref{eqSPprojection}) имеют пренебрежимо малый вес, а значит в правой части (\ref{eqSPprojection}) находится чистое однофотонное фоковское состояние:
\begin{equation}
\gamma \ll 1 \rightarrow \hat{\rho}_{\mathrm{proj}} \approx \ket{1}\bra{1}.
\end{equation}

Если эффективность однофотонного детектора также мала, $\eta \ll 1$ (как для используемых в описываемых здесь экспериментах детекторах Perkin-Elmer SPCM-AQR-14-FC), то результат (\ref{eqSPprojection}) сводится к ещё более простому выражению
\begin{equation}
\label{eqSPprojection2}
\hat{\rho}_{\mathrm{proj}} \propto 
\sum_{n=0}^{\infty} n \gamma^n \ket{n}\bra{n}.
\end{equation}

Для синтезирования многофотонных фоковских состояний $\ket{N}$, триггерная мода должна анализироваться с помощью детектора, отличающего, например, $N$ от $N-1$. Распространённой схемой для этого является деление триггерного пучка на $N$ равных частей, каждый из которых направляется на неразрешающий число фотонов ДОФ.

Рассмотрим случай $N=2$. Триггерный сигнал разделяется на симметричном светоделителе, выходные моды которого направлены на неразрешающие число фотонов ДОФы.
Если в сигнале имеется $n=2$ фотона, то вероятность срабатывания только одного из двух детекторов складывается из двух альтернатив:
\begin{enumerate}
\item оба фотона направились на один и тот же ДОФ. \\
В этом случае вероятность срабатывания этого ДОФ есть $1-(1-\eta)^2$.
\item фотоны направились на разные ДОФ. \\
Тогда, вероятность срабатывания ровно одного из них есть $2\eta(1-\eta)$.
\end{enumerate}
Так как вероятность реализации каждой из двух альтернатив равна $1/2$, полная вероятность одиночного срабатывания в этом случае есть
\begin{equation}
\label{eqProbEx}
n=2 \rightarrow p\left[\rm 1 \, click\right] = \dfrac{1}{2} \left(1-(1-\eta)^2\right) + \dfrac{1}{2}2\eta(1-\eta) = 2\eta - \dfrac{3\eta^2}{2}.
\end{equation}
Вычисленные аналогичным образом величины приведены в таблице (\ref{tabPOVM}). 

\begin{table} [htbp]
	\centering
	\parbox{15cm}{\caption{Вероятности срабатывания одного / двух ДОФ при симметричном разделении сигнала.}\label{tabPOVM}}
	\begin{tabular}{| p{1cm} || p{3cm} | p{3.5cm} | p{2.5cm}l |}
		\hline
		\hline
		\centering $n$ & \centering $p\left[\rm 0\,clicks\right]$  & \centering $p\left[\rm 1\,click\right]$ & \centering $p\left[\rm 2\,clicks\right]$ &\\
		\hline
		\hline
		\centering 0 &\centering  $(1-\eta)^0$  &\centering  0   		&\centering  0    &  \\
		\hline
		\centering 1 &\centering  $(1-\eta)^1$  &\centering  $\eta$ 	&\centering  0    &  \\
		\hline
		\centering 2 &\centering  $(1-\eta)^2$  &\centering  $2\eta-\dfrac{3\eta^2}{2}$&\centering  $\dfrac{\eta^2}{2}$ & \\
		\hline
		\centering 3 &\centering  $(1-\eta)^3$  &\centering  $3\eta-\dfrac{9\eta^2}{2}+\dfrac{7\eta^3}{4}$&\centering  $\dfrac{3\eta^2}{2} - \dfrac{3\eta^3}{4}$ & \\
		\hline
		\hline
	\end{tabular}
\end{table}

Аналогично (\ref{SPCM}), POVM срабатывания обоих детекторов записывается как
\begin{equation}
\hat{\Pi}_{\rm 2} = \sum_{n=0}^{\infty} p_n\left[\rm 2\,clicks\right] \ketbra{n}{n},
\end{equation}
где $p_n\left[\rm 2\,clicks\right]$ обозначает вероятность срабатывания обоих детекторов от $n$--фотонного состояния. В случае успеха такого измерения в триггерной моде 1 двумодово-сжатого вакуума (\ref{eq_tms}), состояние сигнальной моды перейдёт в
\begin{equation}
\label{eqSPdouble}
\hat{\rho}_{\rm proj \, 2} = \mathrm{Tr}_1\left[ \hat{\Pi}_\mathrm{\rm 2} \ket{\Psi}\bra{\Psi}\right] 
\propto \ket{2}\bra{2} + \gamma \left(3 - \dfrac{3\eta}{2}\right) \ket{3}\bra{3} + O\left[\gamma^2\right].
\end{equation}
Как и в случае (\ref{eqSPprojection}), в пределе малого сжатия, на выходе оказывается целевое состояние:
\begin{equation}
\gamma \ll 1 \rightarrow \hat{\rho}_{\mathrm{proj \, 2}} \approx \ket{2}\bra{2}.
\end{equation}

В эксперименте, эффективность однофотонного детектирования $\eta = 0.12$ складывается из следующих компонент:
\begin{itemize}
\item Коэффициент пропускания фильтра Andover, $\approx 0.25$ (разд. \ref{SpecFiltering}).
\item Качество заведения в волокно, $\approx 0.8$.
\item Собственная эффективность детектора, $\approx0.60$.
\end{itemize}

Распределение сигнала по нескольким детекторам одиночных фотонов используется также для снижения доли нежелательных старших фотонных компонент в синтезируемом состоянии; в описанном случае с $N=2$ детекторами, вероятность единственного срабатывания при наличии двух фотонов на входе (\ref{eqProbEx}) ниже, чем соответствующая величина для единственного неразрашающего детектора (\ref{SPCM}) $\bra{2}\hat{\Pi}\ket{2} = 2\eta - \eta^2$. Ясно, что в пределе $N\rightarrow\infty$, вероятность срабатывания какого-либо из детекторов по двум из двух фотонов стремится к нулю; таким образом, система ведёт себя как разрешающий число фотонов ДОФ.

Анализ более сложных схем однофотонного детектирования можно найти в работах \cite{Rohde2007, Hogg2014}.

\subsection{Фильтрация и выбор мод}

\subsubsection{Пространственные моды}

Как показано в разд. \ref{sect3_realSMS}, фотоны, родившиеся в процессе СПР, населяют широкий спектр пространственных мод.
В такой ситуации, ловить в триггерной моде фотон наугад бессмысленно, так как неизвестно, в какой моде следует ожидать парный ему.

В этом отношении, большую роль играет тщательная настройка фазового синхронизма (\ref{PhaseMatching2}), так как в результате определяется пара сигнальных мод, эффективно взаимодействующих с накачкой. Именно для этой пары мод записывается гамильтониан (\ref{eqTMSqueezingHam}), заселяющий их коррелированными парами фотонов.
Такой парой коррелирующих мод оказываются мода сида и мода сигнала DFG (разд. \ref{PhaseMatching2}, \ref{sectDFG}); эти моды имеют ортогональные поляризации, что позволяет разделять из пространственно с помощью поляризационного светоделителя.

Теоретически, в качестве триггерной или сигнальной моды может быть использована каждая из них. На практике, мода DFG часто имеет менее регулярную пространственную форму по сравнению с модой сида, а юстировочный сигнал в ней ограничен по мощности значением $\sim 50$ мкВт (разд. \ref{sectDFG}). Последний фактор, например, в эксперименте по эффекту Квантового Вампира (разд. \ref{chapt2}) обусловил использование в качестве сигнальной моду сида, так как сигнал DFG не имеет мощности, достаточной для оптимизации заведения в волокно вычитающего детектора. 

\subsubsection{Спектральная фильтрация}
\label{SpecFiltering}

\begin{figure}[h]
	\includegraphics[width=6in]{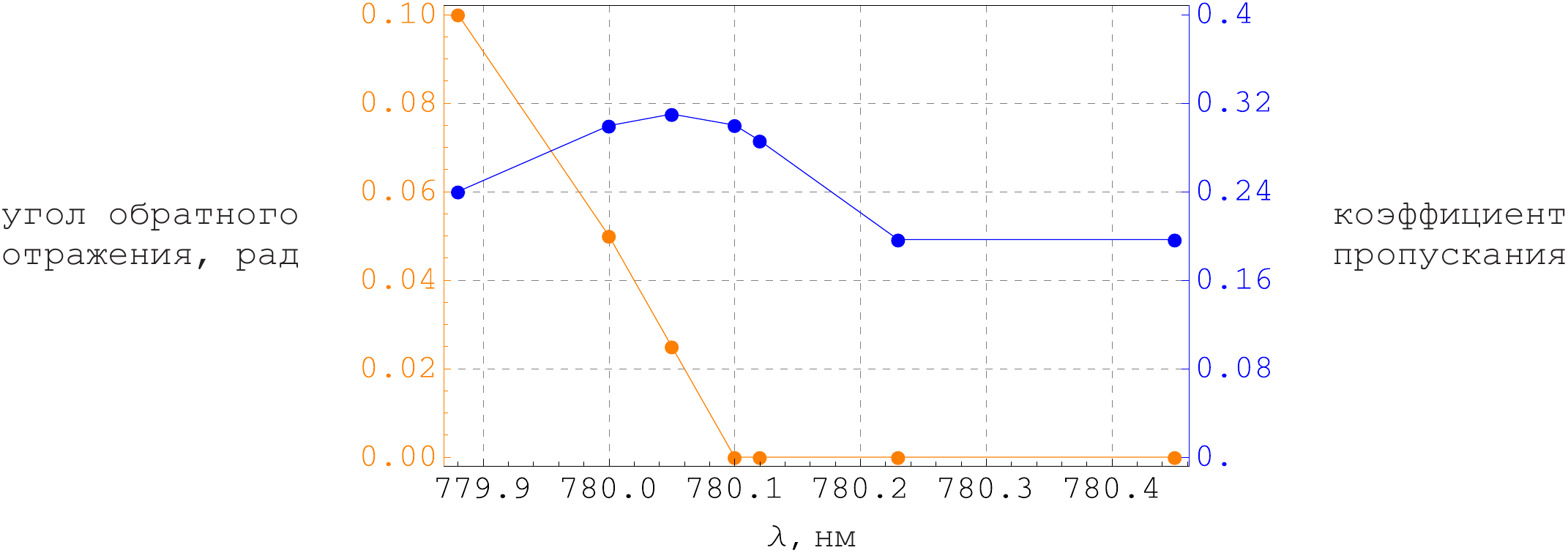}
	\centering
	\captionsetup{justification=raggedright, singlelinecheck=false}
	\caption{Угол обратного отражения и коэффициент пропускания фильтра Andover в зависимости от длины волны.}
	\label{pVAndover}
\end{figure}

Соответственно разд. \ref{sect3_realSMS}, ширина спектра сигнала СПР намного превышает ширину спектра местной волны, используемой для гомодинного детектирования. 
Чтобы избежать связанных с этим потерь (разд. \ref{DiffFreq}), перед подачей на ДОФ триггерная мода подвергается спектральной фильтрации с помощью интерференционного фильтра Andover, имеющего полосу пропускания шириной $\sim 0.3$ нм, близкой к ширине спектра лазерного излучения (разд. \ref{Spectra}). Центровка полосы пропускания этого фильтра критически важна для достижения высокой эффективности \cite{HansenThesis}; если её центр смещён относительно частоты лазера, смещённой окажется также и частота фотонов в сигнальной моде.

Положение полосы пропускания фильтра зависит от направления проходящего луча. Благодаря этому, центровка пропускания на частоту лазера может настраиваться с помощью поворота фильтра; для этого достаточно направить на фильтр пучок сида и максимизировать мощность проходящего сигнала. Зависимость угла отклонения обратного отражения от поверхности фильтра в оптимальном положении, в зависимости от длины волны, показана на Рис.~\ref{pVAndover}. Видно, что подстройка возможна только на длину волны, меньшую 780.1 нм; это обстоятельство должно учитываться при выборе длины волны лазера.

\subsection{Гомодинное детектирование фоковских состояний}
\label{Fock_BHD}

Особенности гомодинного приёма (разд. \ref{BHD_all}) фоковских состояний определяются следующими их свойствами: 
фоковские состояния
\begin{enumerate}

\item являются фазово-инвариантными:
\begin{equation}
\label{eqFockPhase}
\ketbra kk \xrightarrow[e^{i\hat{n}\phi}]{} e^{ik\phi} \ketbra kk e^{-ik\phi} = \ketbra kk,
\end{equation}
благодаря чему нет необходимости изменять и контролировать фазу сигнала относительно местной волны. Это обстоятельство также снижает объём квадратурных данных, достаточный для томографии состояния в $\sim 10$ раз по сравнению с фазово-неинвариантными состояниями. Для томографии фоковского состояния достаточно иметь набор $\sim 10^4$ квадратурных измерений; фаза, соответствующая каждому квадратурному измерению, при необходимости может быть сгенерирована случайным образом.

\item имеют нулевое среднее по любой квадратуре:
\begin{equation}
\label{eqFockMean}
\bra{k} \hat{Q}_\theta \ket{k} = 0 \quad \forall \theta, \quad \hat{Q}_\theta = \dfrac{\hat{a} e^{i \theta} + \hat a^\dagger e^{-i \theta}}{\sqrt{2}}.
\end{equation}
В результате можно игнорировать дрейф средней линии сигнала ГД, который возникает из-за малых разбалансировок последнего. На практике, используется фильтр верхних частот, отсекающий компоненты сигнала с частотами ниже 300 кГц.

\item генерируются вероятностным образом (см. разд. \ref{SPpreparation}). \\ 
Большая часть сигнала ГД соответствует вакууму. Лишь малая доля лазерных импульсов коллапсирует при однофотонном измерении в фоковское состояние; только такие осциллограммы сигнала ГД должны отбираться для дальнейшего анализа. Для этого, триггерное условие осциллографа для записи данных устанавливается как нужная комбинация сигналов детекторов одиночных фотонов. 
\end{enumerate}

Получаемое в результате состояние, обычно, не является чистым фоковским вследствие его частичного разрушения следующими факторами:
\begin{itemize}

\item Оптические потери при генерации и распространени:
\begin{enumerate}
\item в материале кристалла
\item в ходе следования по оптической схеме, $\sim 5$\%.
\end{enumerate}

\item Потери при детектировании:
\begin{enumerate}
\item неоптимальное совмещение мод сигнала и местной волны, $\lesssim 75$\%.
\item эффективность гомодинного детектора. Складывается из соотношения квантового / электронного шумов, $\approx 12\mathrm{dB}$, и квантовой эффективности фотодиодов -- 91\% \cite{Kumar2012}.
\end{enumerate}
\end{itemize}

Влияние всех перечисленных явлений на детектируемое состояние эквивалентно единственному линейному оптическому ослаблению \cite{LvovskyLect, Lvovsky2009}. Определяемая таким образом суммарная эффективность гомодинного измерения $\eta$ (типично $\sim 50\%$) может быть найдена в ходе эксперимента по томографии однофотонного состояния \cite{Lvovsky2001, Huisman2009}.

\subsubsection{Эффективность}

Рассмотрим моду $\hat{a}$, находящуюся в однофотонном состоянии, подвергающуюся ослаблению с коэффициентом по мощности $1-\eta$. 
В представлении Гейзенберга, эта ситуация соответствует следующему преобразованию оператора уничтожения \cite{LvovskyLect}:
\begin{equation}
\label{eqVbs}
\hat{a}'=\sqrt{\eta}\hat{a}+\sqrt{1-\eta}\hat{a}_{\mathrm{aux}},
\end{equation}
где $\hat{a}_{\mathrm{aux}}$ соответствует вспомогательной моде, находящейся в состоянии вакуума. Состояние пары вовлечённых мод при этом есть
\begin{equation}
\label{eqVbsIn}
\hat{a}^\dagger \ket{0}_a\otimes \ket{0}_\mathrm{aux} = \ket{1}_a\otimes \ket{0}_\mathrm{aux} = \ket{10}.
\end{equation}

В картине Шрёдингера, состояние провзаимодействовавших мод находится как
\begin{equation}
\label{eqVbsOut}
\hat{a}'^\dagger \ket{00} = \sqrt{\eta} \ket{10} + \sqrt{1-\eta}\ket{01} = \ket{\psi}.
\end{equation}
В силу того, что информация о состоянии вспомогательной моды теряется, итоговое состояние моды $\hat{a}$ есть
\begin{equation}
\label{eqVbsOutTr}
\hat{\rho}_a = \tr_\mathrm{aux}{\ket{\phi}} = \eta \ketbra 11 + (1-\eta) \ketbra 00.
\end{equation}

Состояние вида (\ref{eqVbsOutTr}) является обычным результатом томографии состояния одиночного фотона. Доля фотонной компоненты определяет таким образом суммарную эффективность приготовления, доставки и детектирования квантового состояния, т.е. эффективность работы установки в целом.

В случае, когда заранее известно, что принимаемое состояние имеет форму (\ref{eqVbsOutTr}), существует способ определить эффективность, не прибегая к затратной процедуре томографии. Он основан на однозначной зависимости между эффективностью $\eta$ и дисперсией квадратурных измерений. Так как для фоковских состояний последняя даётся выражением
\begin{equation}
\bra{n} \hat{Q}_\theta^2 \ket{n} = n + \dfrac{1}{2}, 
\end{equation}
дисперсия состояния (\ref{eqVbsOutTr}) равна
\begin{equation}
\label{eqVvar}
\sigma^2(\eta) = \tr \left[ \hat{Q}_\theta^2 \hat{\rho}_a \right] = \dfrac{3}{2} \eta + (2-\eta)\dfrac{1}{2}.
\end{equation}
При скорости счёта фотонов $R=100$ кГц, правая часть выражения (\ref{eqVvar}) может вычисляться с погрешностью $\sigma \sqrt{\dfrac{2\nu}{R}} = 0.014*\sigma$ и частотой $\nu=10$ Гц, что позволяет в ходе юстировки измерять эффективность $\eta$ в реальном времени \cite{Huisman2009}.

\chapter*{Заключение}	 \label{conclusion}					
\addcontentsline{toc}{chapter}{Заключение}	

В диссертации получены следующие основные результаты:
\begin{enumerate}
\item[1.] 
\begin{enumerate}
\item[(а)] Показано, что уничтожение кванта в бозонном поле, находящемся в перепутанном состоянии с другими, имеет нелокальный эффект. Проявляющееся воздействие на удалённые физические системы представляет интерес как инструмент управления их квантовыми состояниями. Представленный нелокальный эффект не обусловлен коллапсом распределённого квантового состояния света, что отличает его от обычного квантово-механического действия на расстоянии. 

\item[(б)] Эффект подтверждён экспериментально на примере уничтожения фотонов в перепутанных состояниях световых пучков. В частности, продемонстрировано нелокальное изменение энергии удалённого светового пучка.
\end{enumerate}

\item[2.] 
\begin{enumerate}
\item[(а)] Разработан метод повышения запутанности оптического состояния Эйнштейна--Подольского--Розена, позволяющий вероятностным образом поднять сколь угодно низкий уровень запутанности до макроскопической величины. 

\item[(б)] Эффективность метода продемонстрирована в экспериментальной ситуации, когда одна из мод ЭПР-состояния с исходным уровнем двумодового сжатия 0.65dB, была ослаблена в 20 раз -- что соответствует оптическим потерям в $\sim70$--километровой оптоволоконной линии. Процедура дистилляции позволила восстановить квантовую запутанность состояния до первоначального уровня; при этом, логарифмическая отрицательность состояния увеличилась с 0.037 до 0.24.
\end{enumerate}

\item[3.] 
\begin{enumerate}
\item[(а)] Разработан метод характеризации неизвестного квантового процесса. Метод может быть применён к многомодовому квантовому процессу любой природы, не требуя при этом априорных знаний о его устройстве. Метод отличается простотой экспериментальной техники, а также высоким качеством реконструированного тензора процесса.
\item[(б)] С помощью предложенного метода характеризован двумодовой процесс симметричного светоделения. Метод показал свою эффективность в реконструкции квантовых аспектов процесса, в частности эффекта Хонг-Оу-Манделя. Параметр верности между экспериментально восстановленным тензором процесса и теоретическим ожиданием составил 95\%.
\end{enumerate}
\end{enumerate} 

\chapter*{Благодарности}						
\addcontentsline{toc}{chapter}{Благодарности}	

\setlength{\parindent}{0cm}
Здесь перечислены люди, которые сделали эту работу возможной. 
\vspace{3.5em}
\newline
\textbf{Львовский Александр Исаевич.} \newline 
Шеф.
\vspace{3em}
\newline
Масалов Анатолий Викторович. \newline
Его участию я обязан своим обучением в аспирантуре.
\vspace{3em}
\newline
Сарычев Андрей Карлович. \newline
Его энергии я обязан своим присутствием в профессиональной науке.
\vspace{3em}
\newline
Химин Николай Михайлович. \newline
На его уроках началось моё увлечение физикой.
\vspace{4em}
\newline
\centerline{Спасибо!}
\vspace{1em}\newline
\centerline{ИФ}\\
\centerline{Февраль 2016, Москва.}

\clearpage
\addcontentsline{toc}{chapter}{\bibname}	


\end{document}